\documentclass[aps,prd,a4paper]{revtex4}

\usepackage{graphicx}
\usepackage{bm}
\usepackage{amsfonts}
\usepackage{amsmath}

\usepackage[russian,ngerman,english]{babel}

\newcommand{\muas}[0]{\hbox{\rm $\mu$as}}

\newcommand{\ve}[1]{\mbox{\boldmath$#1$}}

\let\oldbibitem\bibitem
\renewcommand\bibitem[2][]{\oldbibitem{#2}}

\begin{document}

\title{Light propagation in 2PN approximation in the monopole and quadrupole field of a body at rest:  
Initial value problem}

\author{Sven Zschocke}

\affiliation{Lohrmann Observatory,
Dresden Technical University, Helmholtzstrasse 10, D-01069 Dresden, Germany}


\begin{abstract}
The light trajectory in the gravitational field of one body at rest with monopole and quadrupole structure is determined in the 
second post-Newtonian (2PN) approximation. The terms in the geodesic equation for light rays are separated into time-independent tensorial coefficients 
and four kind of time-dependent scalar functions. Accordingly, the first and second integration of geodesic equation can be reduced in each case 
to only four kind of scalar master integrals. These integrals can be solved in closed form by recurrence relations. The 2PN terms   
of monopole and quadrupole contribute less than $1$ nano-arcsecond to the total light deflection. There are, however, enhanced terms in the 
2PN light deflection, both in case of monopole and quadrupole. These enhanced 2PN terms are caused by the use of an impact vector which is 
indispensable for modeling of real astrometric measurements. In case of grazing light rays at Jupiter and Saturn, the enhanced 2PN terms,  
caused by the quadrupole structure of the body, amount up to $0.95$ micro-arcseconds and $0.29$ micro-arcseconds, respectively.
Thus, the 2PN quadrupole terms are relevant for high-precision astrometry on the sub-micro-arcsecond scale of accuracy.
\end{abstract}

\pacs{95.10.Jk, 95.10.Ce, 95.30.Sf, 04.25.Nx, 04.80.Cc}

\maketitle




\newpage
\tableofcontents
\newpage

\section{Introduction}\label{Introduction}

Space-based astrometry missions of the European Space Agency (ESA) have initiated a renewed advance of highly precise astrometric measurements.  
The Hipparcos mission (nominal mission: 1989-1993) \cite{Hipparcos,Hipparcos1,Hipparcos2} has reached the milli-arcsecond (mas)
level in astrometric precision, while the Gaia mission (nominal mission: 2013-2018) \cite{GAIA,GAIA_ESA,GAIA1}
has reached the micro-arcsecond (\muas) scale of accuracy in angular observations \cite{GAIA_DR1_1,GAIA_DR2_1,Gaia_Archive}. 

While micro-arcsecond astrometry has been realized both theoretically and technologically within the Gaia mission, the dawning of
sub-micro-arcsecond (sub-\muas) or even nano-arcsecond (nas) astrometry has come in the strategic focus of
astronomers \cite{Proceeding_Sub_Microarcsecond_Astrometry,article_sub_micro_4}. In particular,
ESA has recently called for new science themes to be investigated in terms of feasibility and technological developments, some of which may become candidates
for possible astrometry missions within the ESA Science Program in future \cite{ESA}. For instance, among several others,  
the M-5 mission Gaia-NIR \cite{Gaia_NIR} has been proposed to ESA to operate in the near infrared spectrum and in the \muas$\,$ and sub-\muas$\,$ domain in 
astrometric accuracy. Recently, in the final recommendations of the ESA Senior Science Committee \cite{Voyage_2050} 
the Gaia-NIR mission has been selected as one possible candidate for M-sized astrometry mission. 
Another astrometry mission is the proposed M-5 mission Theia \cite{Theia}  
which is operating in the domain of sub-\muas$\,$ in astrometric precision.  
A further candidate is the space-based mission NEAT (Nearby Earth Astrometric Telescope) \cite{NEAT1},
originally designed for an astrometric precision of about $50$ nas. 
The proposed NEAT mission consists in adopting a long focal length realized by a formation flight configuration 
of two satellites. 
Also feasibility studies of earth-bounded telescopes are under consideration aiming at
accuracies up to about 10 nas \cite{nas_telescopes}.

The impressive progress of astrometric accuracies necessitates the use of general relativity to model the reduction of 
observational data. Such relativistic modeling of observational data consists of several aspects and issues which need to be solved 
\cite{Proceeding_Sub_Microarcsecond_Astrometry}: 
a set of several astronomical reference systems, relativistic description of the process of observations, relativistic equations of motions 
of Solar System bodies, etc. A scheme of general relativistic modeling of observational data is given by Fig.~$1$ in  
\cite{Klioner2003b}. One important aspect thereof concerns the relativistic modeling of propagation of a light signal from the celestial 
light source (e.g. star, quasar, exoplanet) through the gravitational field of the Solar System towards the observer. The present-day 
relativistic models, like the Gaia Relativistic Model (GREM) \cite{Brumberg1991,Klioner2003a,KlionerKopeikin1992}
and the Relativistic Astrometric Model (RAMOD) \cite{RAMOD1,RAMOD2,RAMOD3,RAMOD4,RAMOD5} as used for data reduction  
of ESA astrometry mission Gaia, are designed for relativistic astrometry at \muas-level of accuracy. 
These models need considerably to be refined for modeling observational data on the sub-\muas-level of accuracy.  
In particular, unprecedented accuracies on the sub-\muas-level of accuracy necessitate considerable advancements in the theory of light propagation 
in the Solar System. 

For determining the light trajectory in curved space-time  
one needs to solve the geodesic equation \cite{Einstein2,MTW,Brumberg1991} 
(e.g. Eqs.~(1.2.46) - (1.2.47) in \cite{Brumberg1991}))
\begin{eqnarray}
	&& \frac{d^2 x^{\alpha}\left(\lambda\right)}{d \lambda^2}
+ \Gamma^{\alpha}_{\mu\nu}\,\frac{d x^{\mu}\left(\lambda\right)}{d \lambda}\,
\frac{d x^{\nu}\left(\lambda\right)}{d \lambda} = 0\,,
	\label{Geodetic_Equation}
	\\ 
	&& g_{\alpha\beta}\,\frac{d x^{\alpha}\left(\lambda\right)}{d \lambda}\,\frac{d x^{\beta}\left(\lambda\right)}{d \lambda} = 0\;,  
\label{Null_Condition}
\end{eqnarray}

\noindent
where $x^{\alpha}\left(\lambda\right)$ are the four-coordinates of the light signal which depend on some affine curve-parameter $\lambda$.  
The solutions of geodesic equation (\ref{Geodetic_Equation}) for light rays must satisfy the null-condition (\ref{Null_Condition}) as additional constraint 
along the light trajectory. 
The Christoffel symbols in (\ref{Geodetic_Equation}) are given by 
\begin{eqnarray}
\Gamma^{\alpha}_{\mu\nu} = \frac{1}{2}\,g^{\alpha\beta} 
\left(\frac{\partial g_{\beta\mu}}{\partial x^{\nu}} + \frac{\partial g_{\beta\nu}}{\partial x^{\mu}} 
- \frac{\partial g_{\mu\nu}}{\partial x^{\beta}}\right) ,  
\label{Christoffel_Symbols2}
\end{eqnarray}

\noindent 
so they are functions of the metric tensor, 
where $g^{\alpha\beta}$ and $g_{\alpha\beta}$ are the contravariant and covariant components of the metric tensor of space-time; the metric signature  
is assumed to be $\left(-,+,+,+\right)$.

The metric tensor of some Solar System body, which enters the geodesic equation (\ref{Geodetic_Equation}) via the Christoffel symbols,  
cannot be given in it's exact form and one has to resort rather than to approximation schemes. 
Such an approximation scheme is provided by the post-Newtonian expansion of the metric tensor, which
is an expansion in terms of inverse powers of the speed of light, given by:
\begin{equation}
g_{\alpha \beta} = \eta_{\alpha \beta} + h^{(2)}_{\alpha\beta} + h^{(3)}_{\alpha\beta} + h^{(4)}_{\alpha\beta} + {\cal O} \left(c^{-5}\right),
\label{post_Newtonian_metric_C}
\end{equation}

 \noindent
where $h^{(2)}_{\alpha\beta} = {\cal O} \left(c^{-2}\right)$ and $h^{(3)}_{\alpha\beta} = {\cal O} \left(c^{-3}\right)$ are the 1PN and 1.5PN 
corrections and $h^{(4)}_{\alpha\beta} = {\cal O} \left(c^{-4}\right)$ is the 2PN correction of the light-ray metric;   
see also endnote \footnote{In celestial mechanics another notation is used (in celestial mechanics the term {\it post-Newtonian correction} includes 
$h^{(2)}_{\alpha\beta}$ and $h^{(3)}_{\alpha\beta}$ and $h^{(4)}_{00}$) and, therefore, 
our notation for the terms in the expansion (\ref{post_Newtonian_metric_C}) is sometimes called {\it light-ray metric}. In our investigation 
we will call (\ref{post_Newtonian_metric_C}) just metric and will drop the term {\it light-ray metric}.}.  
The justification of such an expansion is based on the fact that the gravitational 
fields in the Solar System are weak and the velocities of matter are slow, 
\begin{equation}
	\frac{m}{P}  \ll 1 \quad {\rm and} \quad \frac{v}{c} \ll 1\;, 
\label{Weak_Field}
\end{equation}

\noindent
where $m = G M /c^2$ is the Schwarzschild radius of some Solar System body with mass $M$ and radius $P$, and $v$ stands representatively for   
possible velocities of matter (orbital velocities, rotational motions, inner circulations, oscillations, etc.). 

Let the curved space-time assumed to be covered by the four-coordinates $x^{\mu} = \left(ct,\ve{x}\right)$, where $\ve{x}$ are the spatial coordinates 
and $t$ is the coordinate time.  
A unique interpretation of astrometric observations requires the determination of light trajectory $\ve{x}\left(t\right)$ as function of coordinate time. 
Consider a light signal which is emitted by some light source having four-coordinates $\left(ct_0,\ve{x}_0\right)$. In the Newtonian theory that signal would 
travel along some straight line: $\ve{x}_{\rm N}\left(t\right) = \ve{x}_0 + c \left(t-t_0\right) \ve{\sigma}$ in some unit-direction $\ve{\sigma}$.  
In curved space-time the light trajectory is given by a complicated curvilinear trajectory 
\begin{eqnarray}
\ve{x}\left(t\right) = \ve{x}_0 + c \left(t-t_0\right) \ve{\sigma} + \Delta \ve{x}\left(t,t_0\right),  
        \label{Introduction_1} 
\end{eqnarray}

\noindent 
where $\Delta \ve{x}\left(t,t_0\right)$ are corrections to the unperturbed light ray. 
More precisely, these statements are valid in case of a weak gravitational field and ordinary topology of space-time.
Inserting the post-Newtonian expansion  of the metric tensor (\ref{post_Newtonian_metric_C}) into the 
geodesic equation (\ref{Geodetic_Equation}) implies a corresponding post-Newtonian expansion of the light trajectory, 
\begin{eqnarray}
	\ve{x}\left(t\right) &=& \ve{x}_0 + c \left(t-t_0\right) \ve{\sigma} 
	   + \Delta\ve{x}_{\rm 1PN}\left(t,t_0\right) 
	   + \Delta\ve{x}_{\rm 1.5PN}\left(t,t_0\right) 
	   + \Delta\ve{x}_{\rm 2PN}\left(t,t_0\right) 
	   + {\cal O}\left(c^{-5}\right) ,  
	\label{Introduction_5} 
\end{eqnarray}

\noindent
where the post-Newtonian corrections $\Delta\ve{x}_{\rm 1PN} = {\cal O}\left(c^{-2}\right)$ and $\Delta\ve{x}_{\rm 1.5PN} = {\cal O}\left(c^{-3}\right)$ 
are the 1PN and 1.5PN terms, while the post-post-Newtonian correction $\Delta\ve{x}_{\rm 2PN} = {\cal O}\left(c^{-4}\right)$ is the  
2PN term of the light ray. 

Remarkable advancement in determining the 1PN and 1.5PN terms $\Delta\ve{x}_{\rm 1PN}$ and $\Delta\ve{x}_{\rm 1.5PN}$ has been achieved during recent decades. 
These advancements were also triggered by the ESA astrometry missions Hipparcos and Gaia. The problem of light propagation in  
1PN and 1.5PN approximation has been solved for the following systems: monopoles at rest \cite{Brumberg1991}, 
monopoles in motion \cite{Klioner1991,Klioner2003b,KS1999}, spin-dipoles at rest \cite{Klioner1991}, spin-dipoles in motion \cite{KopeikinMashhoon2002},  
quadrupoles at rest \cite{Klioner1991}, quadrupoles in motion \cite{KopeikinMakarov2007}, full set of mass-multipoles and spin multipoles at rest  
\cite{Kopeikin1997,KopeikinKorobkov2005,KopeikinKorobkovPolnarev2006}, full set of mass-multipoles and spin multipoles in slow motion \cite{Zschocke1,Zschocke2} 
(the calculations in \cite{KS1999,KopeikinMashhoon2002,KopeikinMakarov2007} are performed in the first order of the post-Minkowskian scheme, that means 
they are valid for arbitrary motions, especially also for ultra-relativistic motions of the bodies). 

However, much less is known about light propagation in 2PN approximation, that means for the terms $\Delta\ve{x}_{\rm 2PN}$ in (\ref{Introduction_5}). 
In fact, the light trajectory in 2PN approximation has only been determined in the gravitational field
of spherically symmetric bodies, namely either monopoles at rest or monopoles in slow motion:
\begin{enumerate}
\item[$\bullet$] 2PN light trajectory in the field of one monopole at rest \cite{Brumberg1991,Brumberg1987,KlionerKopeikin1992,Teyssandier,AshbyBertotti2010,Deng_2015,Xie_Huang,Minazzoli2}, 
\item[$\bullet$] 2PN light trajectory in the field of two point-like bodies of a binary system \cite{Bruegmann2005},
\item[$\bullet$] 2PN light trajectory in field of one arbitrarily moving monopole \cite{Zschocke3,Zschocke4,Zschocke5}.
\end{enumerate}

\noindent 
These results are by far not sufficient for astrometry on the sub-\muas-level of accuracy \cite{Proceeding_Sub_Microarcsecond_Astrometry}. 
Especially, it is a well-known fact that astrometric accuracies on sub-micro-arcsecond or nano-arcsecond level cannot be achieved
without accounting further second post-Newtonian terms in the theory of light propagation
\cite{Xu_Wu,Xu_Gong_Wu_Soffel_Klioner,MC2009,Deng_Xie,2PN_Light_PropagationA,Deng_2015,Xie_Huang,Minazzoli2,XWKS,KS,Conference_Cambridge,Talk_Klioner}.
In particular, real Solar System bodies are not simply monopoles but can be of arbitrary shape. To account for the arbitrary shape of the bodies 
one utilizes a decomposition of the gravitational field in terms of multipoles, which are certain integrals over the stress-energy tensor of these 
bodies \cite{MTW,Poisson_Will,Kopeikin_Efroimsky_Kaplan,Poisson_Lecture_Notes}. 
The first term in such an expansion is the monopole term. After the dipole term, which vanishes if the origin of the spatial coordinates is 
located at the center of mass of the body, the next term in the multipole-decomposition  
is the quadrupole term. Accordingly, in this investigation the geodesic equation for light rays  
in the quadrupole field of one massive body at rest is considered in 2PN approximation and the  
impact of the quadrupole structure on light deflection is determined. 

The manuscript is organized as follows:    
The metric of one Solar System body being of arbitrary shape and inner structure is presented in Section \ref{Section_Metric} and the exact geodesic equation 
is given in Section \ref{Section0}.  The unperturbed light ray is described in Section \ref{Section1}. In Section \ref{Section2} and \ref{Section3} the light 
propagation in the gravitational field of one massive body at rest with monopole and quadrupole structure is determined in 1PN and 2PN approximation. The total 
light deflection is calculated in Section \ref{Total_Light_Deflection}. The problem of {\it enhanced terms} is considered in Section \ref{Enhanced_Terms}. Finally, 
a summary and outlook can be found in Section \ref{Summary}. The notation and some details of the calculations are relegated to a set of several appendices.

\section{Metric of a Solar System body}\label{Section_Metric}

In this investigation we will consider the propagation of a light signal in the gravitational field generated by one Solar System body at rest.  
The metric tensor enters the geodesic equation (\ref{Geodetic_Equation}) via the Christoffel symbols (\ref{Christoffel_Symbols2}). 
Hence, the first step is to consider the metric of a Solar System body.  
If the gravitational fields are weak, then the so-called post-Minkowskian (PM) expansion (weak-field expansion)  
of the metric tensor can be utilized, which is an expansion of the metric tensor in powers of the gravitational constant   
\begin{eqnarray}
g_{\alpha\beta}\left(t,\ve{x}\right) &=& \eta_{\alpha\beta} + \sum\limits_{n=1}^{\infty} G^n h_{\alpha\beta}^{\left({\rm nPM}\right)}\left(t,\ve{x}\right),  
\label{PM_Expansion_1}
\end{eqnarray}

\noindent
where $\eta_{\alpha\beta} = {\rm diag} \left(-1,+1,+1,+1\right)$ is the flat Minkowskian space-time and $h_{\alpha\beta}^{\left({\rm nPM}\right)}$ 
are small metric perturbations, which vanish at Minkowskian past infinity ${\cal J}_M^{-}$ (cf. Fig.~$34.2$ of Section $34$ in \cite{MTW})  
\begin{eqnarray} 
\lim_{x \rightarrow \infty \atop t  + \frac{x}{c} = {\rm const}} g_{\alpha\beta}\left(t,\ve{x}\right) = \eta_{\alpha\beta}\;. 
\label{PM_Expansion_1A}
\end{eqnarray}

\noindent 
In reality the massive Solar System bodies are of arbitrary shape and inner structure and the bodies can be in arbitrarily rotational and oscillating motions.
The Multipolar Post-Minkowskian (MPM) approach provides a robust framework to solve Einstein's field equations for such complex gravitational systems.  
The MPM formalism has originally been introduced in \cite{Thorne} and later considerably been extended in a series
of articles \cite{Blanchet_Damour1,Blanchet_Damour2,Blanchet_Damour3,Blanchet_Damour4,2PN_Metric1,Multipole_Damour_2} 
to determine the metric density, $\overline{g}^{\alpha\beta} = \sqrt{-g}\,g^{\alpha\beta}$,
in the exterior of an isolated source of matter, where $g = {\rm det} \left(g_{\mu\nu}\right)$ is the determinant of the metric tensor. In the MPM approach 
the space-time is covered by harmonic coordinates, in line with the IAU (International Astronomical Union) resolutions \cite{IAU_Resolution1}; 
harmonic coordinates are Cartesian-like coordinates which are Cartesian at spatial infinity and they are defined by the 
gauge-condition $\overline{g}^{\alpha\beta}_{\;\;\;\,,\, \beta} = 0$ \cite{MTW,Poisson_Will,Kopeikin_Efroimsky_Kaplan,Poisson_Lecture_Notes}.
An isolated gravitational system is defined by the Fock-Sommerfeld boundary conditions, which imply flatness of the metric at spatial infinity 
(\ref{PM_Expansion_1A}) and the constraint of no-incoming gravitational radiation is imposed at Minkowskian past null infinity  
\cite{Fock,KlionerKopeikin1992,IAU_Resolution1,Kopeikin_Efroimsky_Kaplan,Radiation_Condition} (e.g. Eq.~(10) in \cite{Radiation_Condition}).  

The MPM approach was mainly developed for theoretical understanding of the generation of gravitational waves by some isolated source of matter, 
like inspiralling binary stars which consist of compact objects like black holes or neutron stars. The compact source of matter can of course also 
be interpreted as some massive Solar System body. Thus, the MPM approach allows us to obtain the metric perturbations in (\ref{PM_Expansion_1}).  

The MPM formalism utilizes the post-Minkowskian expansion of metric density. The most general solution for the metric density is given  
in terms of six so-called source-multipoles \cite{Blanchet4,Blanchet6}. Harmonic coordinates are not unique but allow for a residual gauge transformation. 
Using this freedom in the choice of harmonic coordinates, it has been demonstrated in \cite{Thorne,Blanchet_Damour1} that the metric density can 
finally be expressed in terms of only two symmetric trace-free (STF) multipoles: mass-multipoles $\hat{M}_L$ (describing shape, inner structure and 
oscillations of the body) and spin-multipoles $\hat{S}_L$ (describing rotational motions and inner currents of the body). For an explicit form of 
these multipoles in the first post-Minkowskian approximation it is referred to Eqs.~(5.33) and (5.35) in \cite{Multipole_Damour_2}. 

The metric density $\overline{g}^{\alpha\beta}$ and the metric tensor $g_{\alpha\beta}$ contain the same physical information, because  
they are related bijective to each other \cite{MTW,Poisson_Will,Kopeikin_Efroimsky_Kaplan,Poisson_Lecture_Notes}; some relations 
are also given in Appendix D in \cite{Zschocke_2PM_Metric}. 
From the metric density $\overline{g}^{\alpha\beta}$ one may deduce the metric tensor $g_{\alpha\beta}$. A detailed description  
about how to obtain the metric tensor $g_{\alpha\beta}$ in (\ref{PM_Expansion_1}) from the metric density 
$\overline{g}^{\alpha\beta}$ of the MPM formalism can be found in our recent article \cite{Zschocke_2PM_Metric}.   
In this way one arrives at the following post-Minkowskian expansion of the metric tensor for a Solar System body in the so-called canonical harmonic gauge 
\cite{Thorne,Blanchet_Damour1,Book_PN,Poujade_Blanchet,Blanchet_Damour3,Zschocke_2PM_Metric,Proceeding_Zschocke} 
(cf. Eqs.~(78) and (79) in \cite{Zschocke_2PM_Metric})   
\begin{eqnarray}
	g_{\alpha\beta}\left(t,\ve{x}\right) &=& \eta_{\alpha\beta} + \sum\limits_{n=1}^{\infty} G^n  
	h_{\alpha\beta}^{\left({\rm nPM}\right)}\!\left(\hat{M}_L\left(s\right),\hat{S}_L\left(s\right)\right).  
\label{PM_Expansion_2}
\end{eqnarray}

\noindent
In (\ref{PM_Expansion_2}) the origin of the spatial axes of the coordinate system can be assumed to be located at the center 
of mass of the body and $s = t - x/c$ is the retarded time (travel time of a light signal from the field point $\left(t,\ve{x}\right)$ to the center of the 
coordinate system). For the concrete form of the 1PM term $h_{\alpha\beta}^{\left({\rm 1PM}\right)}$ and 2PM term 
$h_{\alpha\beta}^{\left({\rm 2PM}\right)}$ we refer to Eqs.~(109) - (111)  and Eqs.~(115) - (117), respectively, in \cite{Zschocke_2PM_Metric}. 

If the speed of the matter (e.g. oscillations, rotational motions and inner currents of the body) is small compared to the speed of light, 
then one may utilize the post-Newtonian (PN) expansion (weak-field slow-motion expansion), which is a expansion of the metric tensor in inverse powers of the 
speed of light, and reads up to terms of the order ${\cal O}\left(c^{-5}\right)$ as follows, 
\begin{eqnarray}
        g_{\alpha\beta}\left(t,\ve{x}\right) &=& \eta_{\alpha\beta} + h_{\alpha\beta}^{\left(2\right)}\left(t,\ve{x}\right) 
	+ h_{\alpha\beta}^{\left(3\right)}\left(t,\ve{x}\right) + h_{\alpha\beta}^{\left(4\right)}\left(t,\ve{x}\right) 
	+ {\cal O}\left(c^{-5}\right) ,    
\label{PN_Expansion_1}
\end{eqnarray}

\noindent
where $h_{\alpha\beta}^{\left(n\right)} = {\cal O}\left(c^{-n}\right)$ are small perturbations to the flat Minkowskian space-time metric. The expansion  
(\ref{PN_Expansion_1}) follows from (\ref{PM_Expansion_1}) in case of slow motions. 
The post-Newtonian expansion (\ref{PN_Expansion_1}) is actually a non-analytic series, because at higher orders $n \ge 8$  
non-analytic terms involving powers of logarithms occur \cite{Thorne, Blanchet_Damour1,Book_PN,Poujade_Blanchet,Blanchet_Damour3}.  
From the MPM formalism it follows that the most general PN solution is given in terms of only two mass-multipoles and spin-multipoles 
\cite{Zschocke_2PM_Metric}
\begin{eqnarray}
      && g_{\alpha\beta}\left(t,\ve{x}\right) = \eta_{\alpha\beta} + h_{\alpha\beta}^{\left(2\right)}\left(\hat{M}_L\left(s\right)\right)  
      + h_{\alpha\beta}^{\left(3\right)}\left(\hat{M}_L\left(s\right),\hat{S}_L\left(s\right)\right)   
      + h_{\alpha\beta}^{\left(4\right)}\left(\hat{M}_L\left(s\right),\hat{S}_L\left(s\right)\right)  
      + {\cal O}\left(c^{-5}\right) , 
\label{PN_Expansion_2}
\end{eqnarray}

\noindent 
where the metric perturbations $h_{\alpha\beta}^{\left(2\right)}$ and $h_{\alpha\beta}^{\left(3\right)}$ can be read off from Eqs.~(109) - (111) 
in \cite{Zschocke_2PM_Metric}, while $h_{\alpha\beta}^{\left(4\right)}$ is given by Eqs.~(115) - (117) in \cite{Zschocke_2PM_Metric}. 

In case of stationary source of matter the multipoles and the metric perturbations are time-independent and then the post-Newtonian expansion 
of the metric tensor reads
\begin{eqnarray}
        g_{\alpha\beta}\left(\ve{x}\right) &=& \eta_{\alpha\beta} + h_{\alpha\beta}^{\left(2\right)}\left(\hat{M}_L\right)  
        + h_{\alpha\beta}^{\left(3\right)}\left(\hat{S}_L\right) 
        + h_{\alpha\beta}^{\left(4\right)}\left(\hat{M}_L\right) 
	+ {\cal O}\left(c^{-5}\right) .  
\label{PN_Expansion_3}
\end{eqnarray}

\noindent
The terms $h_{\alpha\beta}^{\left(2\right)}$ are given by Eqs.~(127) and (129) in \cite{Zschocke_2PM_Metric}, 
the term $h_{\alpha\beta}^{\left(3\right)}$ is given by Eqs.~(128) in \cite{Zschocke_2PM_Metric}, 
and $h_{\alpha\beta}^{\left(4\right)}$ is given by Eqs.~(134) - (136) in \cite{Zschocke_2PM_Metric}.  
The STF multipoles are given by Eqs.~(125) and (126) in \cite{Zschocke_2PM_Metric} and read  
\begin{eqnarray}
	\hat{M}_L &=& \int d^3 x \; \hat{x}_L\;\frac{T^{00} + T^{kk}}{c^2} + {\cal O}\left(c^{-5}\right), 
\label{M_L}
\\
\nonumber\\ 
	\hat{S}_L &=& {\rm STF}_L \int d^3 x \; \hat{x}_{L-1}\;\epsilon_{i_l j k}\;x^j\;\frac{T^{0k}}{c} + {\cal O}\left(c^{-5}\right),  
\label{S_L}
\end{eqnarray}

\noindent
where $T^{\alpha\beta}$ is the stress-energy tensor of the body and the integration runs over the three-dimensional volume of the body; 
the STF notation is given in Appendix~\ref{Appendix_STF}. 

As described in the introductory Section, our knowledge about 2PN effects in the theory of light propagation in the Solar System is restricted   
to the case of light propagation in the gravitational field generated by bodies with mass-monopole structure. 
In this investigation we will go beyond the mass-monopole  
approximation and consider the problem of light propagation in the 2PN approximation in the field of one body at rest with time-independent monopole  
and quadrupole structure, while higher mass-multipoles are neglected. Furthermore, we do not consider any rotational motions or inner circulations 
of the massive body. Accordingly, there are no spin-multipoles and, therefore, no 1.5PN terms in the metric tensor.  
The mass-dipole term vanishes $M_a = 0$ because the origin of the coordinate system is assumed to be 
at the center of mass of the body. That means for our purposes the metric (\ref{PN_Expansion_3}) simplifies to the form 
\begin{eqnarray}
        g_{\alpha\beta}\left(\ve{x}\right) &=& \eta_{\alpha\beta} + h_{\alpha\beta}^{\left(2\right)}\left(M, \hat{M}_{ab}\right) 
	+ h_{\alpha\beta}^{\left(4\right)}\left(M, \hat{M}_{ab}\right) 
	+ {\cal O}\left(c^{-6}\right) .  
\label{PN_Expansion_4}
\end{eqnarray}

\noindent
There are no terms of the order ${\cal O}\left(c^{-3}\right)$ and ${\cal O}\left(c^{-5}\right)$ in the metric tensor 
(\ref{PN_Expansion_4}) because mass-multipoles are assumed to be time-independent and spin-multipoles are neglected. 
The metric perturbations in (\ref{PN_Expansion_4}) have been obtained in our article \cite{Zschocke_2PM_Metric} and they are given below:  
the 1PN terms $h_{\alpha\beta}^{\left(2\right)}$ are given by Eqs.~(\ref{metric_1PN_00}) and (\ref{metric_1PN_ij})
and the 2PN terms $h_{\alpha\beta}^{\left(4\right)}$ are given by Eqs.~(\ref{metric_2PN_00}) and (\ref{metric_2PN_ij}). 
The monopole and the STF quadrupole moment in (\ref{PN_Expansion_4}) are given by 
\begin{eqnarray}
	M &=& \int d^3 x \,\frac{T^{00} + T^{kk}}{c^2} + {\cal O}\left(c^{-5}\right) , 
\label{Mass}
\\
	\hat{M}_{ab} &=& \int d^3 x \,\hat{x}_{ab}\,\frac{T^{00} + T^{kk}}{c^2} + {\cal O}\left(c^{-5}\right) ,
\label{Quadrupole}
\end{eqnarray}

\noindent
where $\hat{x}_{ab} = \hat{x}^{ab} = x^a x^b - \delta^{ab}\,x^2/3$.   
The integrals in (\ref{Mass}) and (\ref{Quadrupole})   
run over the three-dimensional volume of the body, where the stress-energy tensor of the body is assumed to be time-independent;   
see also endnote \footnote{For matter we adopt the post-Newtonian statements $T^{\alpha\beta} = {\cal O}\left(c^2,c^1,c^0\right)$ 
\cite{DSX1}. The given order in Eqs.~(\ref{Mass}) - (\ref{Quadrupole}) follows from Eqs.~(6.2) - (6.3) in \cite{Blanchet4}; 
cf. Eqs.~(5.11a) and (5.11b) in \cite{Gauge_Transformation} or Eqs.~(99a) - (99b) in \cite{Blanchet6}.}. 
The explicit mathematical form of the quadrupole moment (\ref{Quadrupole}) is given below by Eq.~(\ref{Quadrupole_Tensor}) for the model 
of an axial symmetric body, which is a good approximation for the giant planets of the Solar System;   
see also endnote \footnote{In the literature the mass-quadrupole moment is sometimes introduced in the following form (e.g. Eq.~(3) 
in \cite{Klioner1991})  
$M_{ab} = \int d^3 x \,x^{ab}\,\left(T^{00} + T^{kk}\right)/c^2 + {\cal O}\left(c^{-5}\right)$ 
where $x^{ab} = x^a x^b$ so that $M_{ab}$ is not traceless: $M_{kk} = \delta^{ab}\,M_{ab} \neq 0$. 
However, the trace of the quadrupole moment (or any higher moments) does no contribute to the metric tensor 
due to STF relation (\ref{STF_Relation}). 
In general, trace terms of mass-moments or spin-moments cancel each other in the metric tensor and, therefore, the calculations are considerably 
simplified if the metric tensor is expressed in terms of STF multipoles.}.

\section{The exact geodesic equation}\label{Section0}  

The trajectory of a light signal propagating in curved space-time is determined by the geodesic equation (\ref{Geodetic_Equation}) and 
isotropic condition (\ref{Null_Condition}), which in terms of coordinate time read as follows \cite{Kopeikin_Efroimsky_Kaplan,Brumberg1991,MTW} 
(e.g. Eqs.~(1.2.48) - (1.2.49) in \cite{Brumberg1991} or Eqs.~(7.20) - (7.23) in \cite{Kopeikin_Efroimsky_Kaplan}): 
\begin{eqnarray}
	&& \frac{\ddot{x}^{i}\left(t\right)}{c^2} + \Gamma^{i}_{\mu\nu} \frac{\dot{x}^{\mu}\left(t\right)}{c} \frac{\dot{x}^{\nu}\left(t\right)}{c}
- \Gamma^{0}_{\mu\nu} \frac{\dot{x}^{\mu}\left(t\right)}{c} \frac{\dot{x}^{\nu}\left(t\right)}{c} \frac{\dot{x}^{i}\left(t\right)}{c} = 0\;, 
	\label{Geodetic_Equation2}
\\
	&& g_{\alpha\beta}\,\frac{\dot{x}^{\alpha}\left(t\right)}{c}\,\frac{\dot{x}^{\beta}\left(t\right)}{c} = 0 \;, 
\label{Null_Condition1}
\end{eqnarray}

\noindent
where $g_{\alpha\beta}$ are the covariant components of the metric tensor of space-time and a dot denotes total derivative  
with respect to coordinate time.  The isotropic condition (\ref{Null_Condition1}) states that light trajectories are null rays.  
This condition is satisfied at any point along the light trajectory; cf. text below Eq.~(\ref{First_Integration_1PN_Q}) 
as well as text below Eqs.~(\ref{First_Integration_2PN_Term_Q_Q}) and (\ref{unit_nu}). 
The Christoffel symbols in (\ref{Geodetic_Equation2}) are given by Eq.~(\ref{Christoffel_Symbols2}).  

The metric and the Christoffel symbols (\ref{Christoffel_Symbols2}) are functions of the entire space-time $(t,\ve{x})$, while if one inserts
these symbols in the geodesic equation (\ref{Geodetic_Equation2}) then the metric and the Christoffel symbols become relevant only  
at the coordinates of the photon $\ve{x}\left(t\right)$. Consequently, the derivatives of the metric tensor
contained in the Christoffel symbols (\ref{Christoffel_Symbols2}) of the geodesic equation (\ref{Geodetic_Equation2}) must be taken along the light trajectory,
\begin{eqnarray}
\frac{\partial g_{\alpha \beta}}{\partial x^{\mu}} = \frac{\partial g_{\alpha \beta}\left(t,\ve{x}\right)}{\partial x^{\mu}}
	\Bigg|_{\ve{x}=\ve{x}\mbox{$\left(t\right)$}}\;, 
\label{Derivatives_1}
\end{eqnarray}

\noindent
where $\ve{x}\left(t\right)$ is the exact light trajectory. The equation (\ref{Derivatives_1}) means that  
the differentiations of the metric tensor in (\ref{Christoffel_Symbols2}) have to be performed with respect to the space-time coordinates,
and afterwards the light trajectory has to be substituted.  

The geodesic equation (\ref{Geodetic_Equation2}) is a differential equation of second order of one variable, $t$, 
thus a well-defined initial-value problem 
(Cauchy problem) implies two initial conditions: the initial unit direction $\ve{\mu}$ of the light signal and the spatial position of the light source 
$\ve{x}_0$ at the moment of emission of the light signal:
\begin{eqnarray}
	\ve{\mu} &=& \frac{\dot{\ve{x}}\left(t\right)}{\left|\dot{\ve{x}}\left(t\right)\right|}\bigg|_{t = t_0} \quad {\rm with} \quad 
	\ve{\mu} \cdot \ve{\mu} = 1 \;,  
	\label{Initial_A}
	\\ 
	\ve{x}_0 \; &=& \; \ve{x}\left(t\right)\,\,\bigg|_{t = t_0}  \;.  
	\label{Initial_B}
\end{eqnarray}

\noindent
A graphical elucidation of these vectors is given in Figure~\ref{Diagram}.

\subsection{The first integration of the exact geodesic equation} 

As stated above, a unique solution of the first integration of geodesic equation (\ref{Geodetic_Equation2}) requires the initial condition 
(\ref{Initial_A}) for the direction $\ve{\mu}$ of the light ray which is emitted by the light source. The initial condition (\ref{Initial_A}) can be replaced 
by the initial condition of the direction $\ve{\sigma}$ of the light ray at past infinity 
\cite{Brumberg1991,Kopeikin_Efroimsky_Kaplan,KlionerKopeikin1992,Kopeikin1997,KSGE,Zschocke1,Zschocke2,Zschocke3,Zschocke4,Zschocke5},
\begin{eqnarray}
	\ve{\sigma} = \frac{\dot{\ve{x}}\left(t\right)}{c}\bigg|_{t = - \infty} \quad {\rm with} \quad \ve{\sigma} \cdot \ve{\sigma} = 1 \;,  
\label{Boundary_Condition}
\end{eqnarray}

\noindent
with $\ve{\sigma}$ being the unit-direction of the light ray at past infinity.  
For a graphical elucidation of these statements see also Figure~\ref{Diagram}. 
There is a one-to-one correspondence between condition (\ref{Boundary_Condition}) and (\ref{Initial_A}). In fact, 
one may find a unique relation between the tangent vectors $\ve{\sigma}$ and $\ve{\mu}$ (cf. Section $3.2.3$ in \cite{Brumberg1991}).  
The advantage of (\ref{Boundary_Condition}) is that the integration procedure and the expressions become simpler when using (\ref{Boundary_Condition}) 
instead of (\ref{Initial_A}). 
Then, the first integration of geodesic equation (\ref{Geodetic_Equation2}) yields the coordinate velocity of the light signal, 
\begin{eqnarray}
\frac{\dot{\ve{x}}\left(t\right)}{c} = \int\limits_{- \infty}^{t} d c {\rm t} \,\frac{\ddot{\ve{x}}\left({\rm t}\right)}{c^2}
	= \ve{\sigma} + \frac{\Delta \dot{\ve{x}}\left(t\right)}{c}\;, 
\label{First_Integration}
\end{eqnarray}

\noindent
where the integration variable ${\rm t}$ (roman style) runs from lower limit of integration $ - \infty$ to the upper limit of integration $t$ (mathnormal style) 
and where $\Delta \dot{\ve{x}}\left(t\right)/c$ denotes the correction to the unit-direction $\ve{\sigma}$ of the light ray at past infinity, so that 
\begin{eqnarray}
	\lim_{t \rightarrow - \infty}\;\frac{\Delta \dot{\ve{x}}\left(t\right)}{c} = 0\;. 
	\label{Boundary_Condition_1}
\end{eqnarray}

\noindent
Let us notice that (\ref{Boundary_Condition}) in combination with (\ref{PM_Expansion_1A}) implies that the null condition (\ref{Null_Condition1}) is satisfied 
at past infinity.  

\subsection{The second integration of the exact geodesic equation} 

As stated above, a unique solution of the second integration of geodesic equation (\ref{Geodetic_Equation2}) requires the initial condition (\ref{Initial_B})  
\cite{Brumberg1991,Kopeikin_Efroimsky_Kaplan,KlionerKopeikin1992,Kopeikin1997,KSGE,Zschocke1,Zschocke2,Zschocke3,Zschocke4,Zschocke5},
\begin{eqnarray}
	\ve{x}_0 = \ve{x}\left(t\right)\bigg|_{t=t_0}\;, 
\label{Initional_Condition}
\end{eqnarray}

\noindent 
with $\ve{x}_0$ being the initial position of the light source at the moment of the emission of the light signal.  
The second integration of geodesic equation (\ref{Geodetic_Equation2}) yields the trajectory of the light signal,
\begin{eqnarray}
\ve{x}\left(t\right) = \int\limits_{t_0}^{t} d c {\rm t} \,\frac{\dot{\ve{x}}\left({\rm t}\right)}{c}
	= \ve{x}_0 + c \left(t - t_0\right) \ve{\sigma} + \Delta \ve{x}\left(t,t_0\right) , 
\label{Second_Integration}
\end{eqnarray}

\noindent
where the integration variable ${\rm t}$ (roman style) runs from lower limit of integration $t_0$ to the upper limit of 
integration $t$ (mathnormal style) and where $\Delta \ve{x}\left(t,t_0\right)$ denotes the corrections to the unperturbed light trajectory, so that 
\begin{eqnarray}
        \lim_{t \rightarrow t_0}\;\Delta \ve{x}\left(t,t_0\right) = 0\;.
        \label{Boundary_Condition_2}
\end{eqnarray}

\section{The unperturbed light ray}\label{Section1}

The geodesic equation for a light signal, Eqs.~(\ref{Geodetic_Equation2}) and (\ref{Null_Condition1}), will be solved by iteration.
In the zeroth approximation, that is the Newtonian limit, the light signal propagates along a straight line.
We assume the light signal to be emitted at the position of the light source, $\ve{x}_0\,$, 
and travels along the unit-vector $\ve{\sigma}$. Thus, the coordinate velocity of this
so-called unperturbed light ray reads
\begin{eqnarray}
        \frac{\dot{\ve{x}}_{\rm N}}{c} = \ve{\sigma}\;,  
        \label{Unperturbed_Lightray_1}
\end{eqnarray}

\noindent
where the subindex ${\rm N}$ stands for Newtonian.
Then, the spatial position of the unperturbed light ray at coordinate time $t$ is given by
\begin{eqnarray}
	\ve{x}_{\rm N} &=& \ve{x}_0 + c \left(t - t_0\right) \ve{\sigma}\;. 
\label{Unperturbed_Lightray_2}
\end{eqnarray}

\noindent
Furthermore, we notice the absolute value of the spatial coordinate of the unperturbed light ray,
\begin{eqnarray}
	x_{\rm N} &=& \sqrt{x_0^2 + 2\,c\,\left(t-t_0\right) \left(\ve{\sigma} \cdot \ve{x}_0\right) + c^2 \left(t-t_0\right)^2}\;. 
\label{Unperturbed_Lightray_3}
\end{eqnarray}

\noindent
The iterative approach is based on the unperturbed light ray, that means the subsequent calculations are
finally expressed in terms of (\ref{Unperturbed_Lightray_1}) and (\ref{Unperturbed_Lightray_2}).
A further important quantity is the impact vector $\ve{d}_{\sigma}$ and its absolute value $d_{\sigma}$ of the
unperturbed light ray, which are defined as follows (cf. Eq.~(55) in \cite{Article_Zschocke1}) 
\begin{eqnarray}
	\ve{d}_{\sigma} &=& \ve{\sigma} \times \left(\ve{x}_0 \times \ve{\sigma}\right) = \ve{\sigma} \times \left(\ve{x}_{\rm N} \times \ve{\sigma}\right)
        \quad {\rm hence} \quad \ve{\sigma} \cdot \ve{d}_{\sigma} = 0\;, 
        \label{impact_vector_1}
        \\
	d_{\sigma} &=& \left| \ve{d}_{\sigma} \right| \;,\; \left(d_{\sigma}\right)^2 = x_0^2 - \left(\ve{\sigma} \cdot \ve{x}_0\right)^2
        = \left(x_{\rm N}\right)^2 - \left(\ve{\sigma} \cdot \ve{x}_{\rm N}\right)^2\;. 
	\label{impact_parameter_1}
\end{eqnarray}

\noindent
The impact parameter $d_{\sigma}$ is assumed to be larger or equal (grazing light ray) than the (equatorial) radius $P$ of the massive Solar System body, 
\begin{eqnarray}
        d_{\sigma} \ge P \;.  
	\label{impact_parameter_condition_A}
\end{eqnarray}

\begin{widetext} 
\begin{figure}[!ht]
\begin{center}
\includegraphics[scale=0.30]{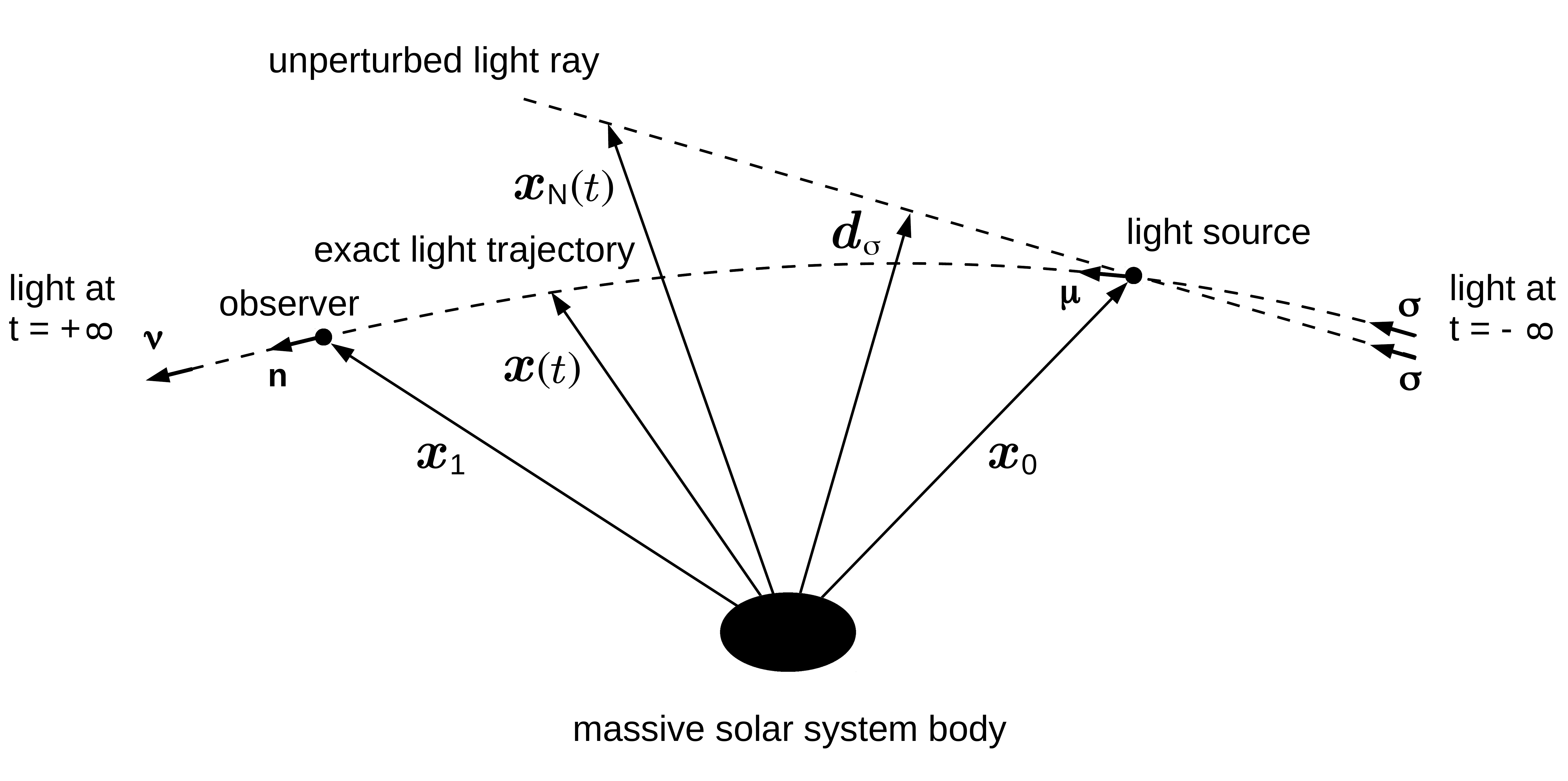}
\end{center}
\caption{ 
	A geometrical representation of the propagation of a light signal through the gravitational field of a massive Solar System body at rest. 
	The origin of the spatial coordinates $x^1,x^2,x^3$ is assumed to be located at the center of mass of the body. 
        The light signal is emitted by the light source at $\ve{x}_0$ in the direction of the unit vector $\ve{\mu}$ and propagates along the exact
        light trajectory $\ve{x}\left(t\right)$ in (\ref{Second_Integration}) which is determined by the geodesic equation (\ref{Geodetic_Equation2}).
        The unit tangent vector $\ve{\sigma}$ and $\ve{\nu}$ of the light trajectory at past infinity and future infinity are
        given by Eqs.~(\ref{Boundary_Condition}) and (\ref{limit_plus}), respectively.
	The unit tangent vector $\ve{n}$ of the light trajectory at the observer's position $\ve{x}_1$ is defined by Eq.~(\ref{vector_n}).
        The unperturbed light ray $\ve{x}_{\rm N}\left(t\right)$ is given by
        Eq.~(\ref{Unperturbed_Lightray_2}) and propagates in the direction of $\ve{\sigma}$ along a straight line through the position
        of the light source at $\ve{x}_0$. The impact vector $\ve{d}_{\sigma}$ of the unperturbed light ray is given by Eq.~(\ref{impact_vector_1}). 
	The impact vector $\hat{\ve{d}_{\sigma}}$ as given by Eq.~(\ref{impact_vector_x1}) is parallel to $\ve{d}_{\sigma}$ and a tiny bit smaller, 
	$\hat{d}_{\sigma} < d_{\sigma}$, but is not shown in the diagram.} 
\label{Diagram}
\end{figure}
\end{widetext}

\noindent 
For graphical illustration see Figure~\ref{Diagram}. From these definitions one obtains the relations
\begin{eqnarray}
	c \left(t-t_0\right) &=& \left(\ve{\sigma} \cdot \ve{x}_{\rm N}\right) - \left(\ve{\sigma}\cdot\ve{x}_0\right),  
\label{Relation_Integral1}
\\
	\ve{x}_{\rm N} &=& \ve{d}_{\sigma} + \left(\ve{\sigma} \cdot \ve{x}_{\rm N}\right) \ve{\sigma}\,, 
        \label{Equation_A}
        \\
        \left(\ve{\sigma} \cdot \ve{x}_{\rm N}\right)^2 &=& \left(x_{\rm N}\right)^2 - \left(d_{\sigma}\right)^2\,, 
        \label{Equation_B}
        \\
\left(\ve{\sigma} \cdot \ve{x}_{\rm N}\right)^4 &=& \left(x_{\rm N}\right)^4 - 2\,\left(x_{\rm N}\right)^2 \left(d_{\sigma}\right)^2 + \left(d_{\sigma}\right)^4\,,
        \label{Equation_C}
\end{eqnarray}

\noindent
which are frequently used in the subsequent considerations.

\section{Light propagation in 1PN approximation \label{Section2}} 

\subsection{The 1PN metric for a body with monopole and quadrupole term}

In our investigation, the body is assumed to be at rest and without rotational motion, hence the metric is time-independent.  
Then, the post-Newtonian series expansion of the metric tensor reads
\begin{eqnarray}
g_{\alpha\beta}\left(\ve{x}\right) = \eta_{\alpha\beta} + h^{(2)}_{\alpha\beta}\left(\ve{x}\right)
	+ {\cal O} \left(c^{-4}\right) , 
\label{metric_perturbation_1PN_10}
\end{eqnarray}

\noindent
where $h_{\alpha \beta}^{(2)} = {\cal O}\left(c^{-2}\right)$ and $\ve{x}$ is some field point.  
These metric coefficients are given in terms of so-called symmetric trace-free (STF) mass-multipoles $\hat{M}_L$ and spin-multipoles $\hat{S}_L$
as given by Eqs.~(5.33) and (5.35) in \cite{Multipole_Damour_2}.
If one takes into account only the mass-monopole and mass-quadrupole terms, then the metric perturbation in 1PN approximation reads
\cite{Klioner1991,Klioner2003b,Zschocke_2PM_Metric} 
\begin{eqnarray}
	h^{\left(2\right)}_{00}\left(\ve{x}\right) &=& \frac{2\,G}{c^2}\,\frac{M}{x}
	+ \frac{3\,G\,\hat{M}_{ab}}{c^2}\,\frac{x^a\,x^b}{x^5}\;, 
\label{metric_1PN_00}
\\
\nonumber\\
	h^{\left(2\right)}_{ij}\left(\ve{x}\right) &=& h^{\left(2\right)}_{00}\left(\ve{x}\right)\,\delta_{ij}\;,  
\label{metric_1PN_ij}
\end{eqnarray}

\noindent
while $h^{\left(2\right)}_{0i} = 0$ and the mass-monopole and mass-quadrupole are given by Eqs.~(\ref{Mass}) and (\ref{Quadrupole}), respectively.  
Let us notice that Eqs.~(\ref{metric_1PN_00}) and (\ref{metric_1PN_ij}) are identical  
with Eqs.~(1) and (2) in \cite{Klioner1991} owing to the STF relation (cf. Eq.~(1.158) in \cite{Poisson_Will})  
\begin{eqnarray}
	M_{ab}\,\hat{x}^{ab} = \hat{M}_{ab}\,x^{ab} = \hat{M}_{ab}\,\hat{x}^{ab}\;,  
	\label{STF_Relation}
\end{eqnarray}

\noindent 
where $M_{ab} = \hat{M}_{ab} + \delta_{ab}\,M_{kk}/3$.  
For the same reason Eqs.~(\ref{metric_1PN_00}) and (\ref{metric_1PN_ij}) agree with Eqs.~(145) and (147) in \cite{Zschocke_2PM_Metric}.

\subsection{The geodesic equation in 1PN approximation}  

Inserting the expansion (\ref{metric_perturbation_1PN_10}) into
the geodesic equation (\ref{Geodetic_Equation2}) yields the geodesic equation in 1PN approximation,  
\begin{eqnarray}
	\frac{\ddot{x}^i \left(t\right)}{c^2} &=& \frac{1}{2}\,h_{00,i}^{(2)}
        - h_{00,j}^{(2)} \frac{\dot{x}^i\left(t\right)}{c}\frac{\dot{x}^j\left(t\right)}{c}
	- h_{ij,k}^{(2)} \frac{\dot{x}^j\left(t\right)}{c}\frac{\dot{x}^k\left(t\right)}{c} 
	+ \frac{1}{2}\,h_{jk,i}^{(2)} \frac{\dot{x}^j\left(t\right)}{c}\frac{\dot{x}^k\left(t\right)}{c}
       + {\cal O}\left(c^{-4}\right) ,  
\label{Geodesic_Equation_1PN_A}
\end{eqnarray}

\noindent
where it has been taken into account that the space-time component of the metric perturbation $h^{\left(2\right)}_{0i} = 0$ vanishes for a body at rest. 
In (\ref{Geodesic_Equation_1PN_A}) the field point $\ve{x}$ as argument of the metric perturbations is taken at the position of the light signal, while 
$\dot{\ve{x}}\left(t\right)$ and $\ve{x}\left(t\right)$ are the coordinate velocity and spatial coordinate of the light signal at this field point. 
For the geodesic equation in 1PN approximation, one may take the unperturbed light ray ($\dot{\ve{x}}_{\rm N}$ and $\ve{x}_{\rm N}$)  
on the r.h.s. in (\ref{Geodesic_Equation_1PN_A}). 
In view of (\ref{metric_1PN_ij}), the 1PN geodesic equation in (\ref{Geodesic_Equation_1PN_A}) then simplifies considerably,  
\begin{eqnarray}
	\frac{\ddot{x}^i\left(t\right)}{c^2} = h_{00,i}^{(2)}\left(\ve{x}_{\rm N}\right) 
	- 2\,h_{00,j}^{(2)}\left(\ve{x}_{\rm N}\right) \; \sigma^i\,\sigma^j + {\cal O}\left(c^{-4}\right)\,.  
\label{Geodesic_Equation_1PN_B}
\end{eqnarray}

\noindent
Determining the spatial derivatives of $h_{00}^{(2)}\left(\ve{x}\right)$ in (\ref{metric_1PN_00}) and afterwards replacing $\ve{x}$ by $\ve{x}_{\rm N}$ 
(cf. (Eq.~\ref{Derivatives_1})) one gets 
\begin{eqnarray}
	h_{00,i}^{(2)}\left(\ve{x}_{\rm N}\right) &=& - 2\, \frac{G M}{c^2}\,\frac{x_{\rm N}^i}{\left(x_{\rm N}\right)^3} 
	+ 6\,\frac{G \hat{M}_{ai}}{c^2} \,\frac{x_{\rm N}^a}{\left(x_{\rm N}\right)^5}
	- 15\,\frac{G \hat{M}_{ab}}{c^2} \,\frac{x_{\rm N}^a x_{\rm N}^b x_{\rm N}^i}{\left(x_{\rm N}\right)^7}\;.  
\label{spatial_derivative_1PN}
\end{eqnarray}

\noindent
By inserting (\ref{spatial_derivative_1PN}) into (\ref{Geodesic_Equation_1PN_B}) one may separate terms proportional 
to the monopole and quadrupole, and obtains 
\begin{eqnarray}
	\frac{\ddot{\ve{x}}\left(t\right)}{c^2} = \frac{\ddot{\ve{x}}^{\rm M}_{\rm 1PN}\left(t\right)}{c^2} 
	+ \frac{\ddot{\ve{x}}^{\rm Q}_{\rm 1PN}\left(t\right)}{c^2} + {\cal O}\left(c^{-4}\right) ,  
\label{Geodesic_Equation_1PN_C}
\end{eqnarray}

\noindent
where the monopole and quadrupole terms are given by  
\begin{eqnarray}
	\frac{\ddot{x}^{i\,{\rm M}}_{\rm 1PN}\left(t\right)}{c^2} &=& \frac{G M}{c^2} \bigg[
	 -2 \,\frac{x_{\rm N}^i}{\left(x_{\rm N}\right)^3} 
	+ 4 \,\sigma^i\,\frac{\ve{\sigma} \cdot \ve{x}_{\rm N}}{\left(x_{\rm N}\right)^3}\bigg] ,  
\label{Geodesic_Equation_1PN_M}
\\
	 \frac{\ddot{x}^{i\,{\rm Q}}_{\rm 1PN}\left(t\right)}{c^2} &=& \frac{G \hat{M}_{ab}}{c^2}  
	\bigg[6\,\delta^{bi}\,\frac{x_{\rm N}^a}{\left(x_{\rm N}\right)^5} 
	- 12\,\sigma^i\,\sigma^b\,\frac{x_{\rm N}^a}{\left(x_{\rm N}\right)^5} 
	- 15\,\frac{x_{\rm N}^a\,x_{\rm N}^b\,x_{\rm N}^i}{\left(x_{\rm N}\right)^7}  
	+ 30\,\sigma^i\,\frac{\ve{\sigma} \cdot \ve{x}_{\rm N}}{\left(x_{\rm N}\right)^7}\,x_{\rm N}^a\,x_{\rm N}^b\bigg] , 
\label{Geodesic_Equation_1PN_Q}
\end{eqnarray}

\noindent
which agrees with \cite{Klioner1991};  
see also endnote \footnote{Let us note that the trace terms of the quadrupole moment 
are not zero but they cancel each other in Eq.~(26) in \cite{Klioner1991}, while in (\ref{Geodesic_Equation_1PN_Q}) there are no trace terms at all.}.
By means of relations (\ref{Equation_A}) and (\ref{Equation_B}) the r.h.s. of equations (\ref{Geodesic_Equation_1PN_M}) 
and (\ref{Geodesic_Equation_1PN_Q}) can be rewritten in terms of time-independent tensorial coefficients and time-dependent 
scalar functions,  
\begin{eqnarray}
	\frac{\ddot{x}^{i\,{\rm M}}_{\rm 1PN}\left(t\right)}{c^2} &=& \frac{G M}{c^2}\bigg[ 
		{\cal A}^i_{\left(3\right)} \,\ddot{\cal W}_{\left(3\right)}\left(t\right)   
	 + {\cal B}^i_{\left(3\right)}\, \ddot{\cal X}_{\left(3\right)}\left(t\right) \bigg] ,  
\label{Geodesic_Equation_1PN_M_Rewritten}
\\
	\frac{\ddot{x}^{i\,{\rm Q}}_{\rm 1PN}\left(t\right)}{c^2} &=& \frac{G \hat{M}_{ab}}{c^2}\!\!  
	\sum\limits_{n=5,7} \! \bigg[   
	{\cal C}^{i\,ab}_{\left(n\right)} \ddot{\cal W}_{\left(n\right)}\left(t\right) 
        + {\cal D}^{i\,ab}_{\left(n\right)} \ddot{\cal X}_{\left(n\right)}\left(t\right) \!\bigg] , 
\label{Geodesic_Equation_1PN_Q_Rewritten}
\end{eqnarray}

\noindent
where the time-independent coefficients read   
\begin{eqnarray}
	{\cal A}^i_{\left(3\right)} &=& + 2 \, \sigma^i \;, 
	\label{coefficient_A3}
	\\
	{\cal B}^i_{\left(3\right)} &=& - 2\, d_{\sigma}^i \;,
	\label{coefficient_B3}
	\\
	{\cal C}^{i\,ab}_{\left(5\right)} &=& + \,6\,\sigma^a\,\delta^{bi}  
        + 3\,\sigma^a\,\sigma^b\,\sigma^i \;, 
        \label{coefficient_C5}
        \\
	{\cal C}^{i\,ab}_{\left(7\right)} &=& - 15 \left(d_{\sigma}\right)^2 \sigma^a \sigma^b \sigma^i 
        + 15 \,d^a_{\sigma} d^b_{\sigma} \sigma^i 
        - 30 \,\sigma^a d^b_{\sigma}\,d^i_{\sigma} \;, 
        \label{coefficient_C7}
        \\
	{\cal D}^{i\,ab}_{\left(5\right)} 
	&=& +\, 6\,d^a_{\sigma}\,\delta^{bi} - 15\,\sigma^a\,\sigma^b\,d^i_{\sigma} 
        + 18\,\sigma^a\,d^b_{\sigma}\,\sigma^i \;,   
        \label{coefficient_D5}
        \\
	{\cal D}^{i\,ab}_{\left(7\right)} &=& - 15 \, d^a_{\sigma} d^b_{\sigma} d^i_{\sigma} 
        + 15 \left(d_{\sigma}\right)^2 \sigma^a \sigma^b d^i_{\sigma} 
        - 30 \left(d_{\sigma}\right)^2 \sigma^a d^b_{\sigma} \sigma^i ,  
        \label{coefficient_D7}
\end{eqnarray}

\noindent 
and the scalar functions are
\begin{eqnarray} 
        \ddot{\cal W}_{\left(n\right)}\left(t\right) &=& \frac{\ve{\sigma} \cdot \ve{x}_{\rm N}}{\left(x_{\rm N}\right)^n} \;, 
        \label{scalar_function_ddot_W}
        \\ 
        \ddot{\cal X}_{\left(n\right)}\left(t\right) &=& \frac{1}{\left(x_{\rm N}\right)^n}\;.   
        \label{scalar_function_ddot_X}
\end{eqnarray}

\noindent
In the following two subsections the first and second integration of differential equations (\ref{Geodesic_Equation_1PN_M_Rewritten})
and (\ref{Geodesic_Equation_1PN_Q_Rewritten}) will be performed.

\subsection{The first integration of geodesic equation in 1PN approximation}

The first integration of geodesic equation (\ref{Geodesic_Equation_1PN_C}) yields the coordinate velocity of the light signal in 1PN approximation,
\begin{eqnarray}
\frac{\dot{\ve{x}}\left(t\right)}{c} = \int\limits_{- \infty}^{t} d c {\rm t} \,\frac{\ddot{\ve{x}}\left({\rm t}\right)}{c^2}
= \ve{\sigma} + \frac{\Delta \dot{\ve{x}}_{\rm 1PN}\left(t\right)}{c} + {\cal O}\left(c^{-4}\right). 
\label{First_Integration_1PN}
\end{eqnarray}

\noindent
By inserting (\ref{Geodesic_Equation_1PN_M_Rewritten}) and (\ref{Geodesic_Equation_1PN_Q_Rewritten}) into the r.h.s. of 
(\ref{First_Integration_1PN}) one obtains the first integration of geodesic equation (coordinate velocity) for a light signal  
in the gravitational field of one body at rest  
\begin{eqnarray}
	\frac{\Delta \dot{\ve{x}}_{\rm 1PN}\left(t\right)}{c} &=& \frac{\Delta \dot{\ve{x}}^{\rm M}_{\rm 1PN}\left(t\right)}{c} 
+ \frac{\Delta \dot{\ve{x}}^{\rm Q}_{\rm 1PN}\left(t\right)}{c} \;,  
\label{First_Integration_1PN_M_Q}
\end{eqnarray}

\noindent
where their spatial components are
\begin{eqnarray}
	\frac{\Delta \dot{x}^{i\,{\rm M}}_{\rm 1PN}\left(t\right)}{c} &=& \frac{G M}{c^2} \bigg[
		{\cal A}^i_{\left(3\right)}\,\dot{\cal W}_{\left(3\right)}\left(t\right)   
	 + {\cal B}^i_{\left(3\right)}\,\dot{\cal X}_{\left(3\right)}\left(t\right) \bigg] , 
	\label{First_Integration_1PN_M} 
	\\
       \frac{\Delta \dot{x}^{i\,{\rm Q}}_{\rm 1PN}\left(t\right)}{c} &=& \frac{G \hat{M}_{ab}}{c^2}    
        \sum\limits_{n=5,7} \bigg[   
        {\cal C}^{i\,ab}_{\left(n\right)}\,\dot{\cal W}_{\left(n\right)}\left(t\right) 
        + {\cal D}^{i\,ab}_{\left(n\right)}\,\dot{\cal X}_{\left(n\right)}\left(t\right) \bigg] .  
\label{First_Integration_1PN_Q}
\end{eqnarray}

\noindent
The scalar master integrals $\dot{\cal W}_{\left(n\right)}$ and $\dot{\cal X}_{\left(n\right)}$ are defined by 
Eqs.~(\ref{Master_Integral_I1}) and (\ref{Master_Integral_J1}), respectively, while their explicit solutions for $n=3,5,7$  
are given by Eq.~(\ref{Master_Integral_A}) and Eqs.~(\ref{Master_Integral_B3}), (\ref{Master_Integral_B5}), (\ref{Master_Integral_B7}) 
in Appendix \ref{Appendix_Scalar_Master_Integrals}.  
One may show that Eq.~(\ref{First_Integration_1PN_M}) agrees with \cite{Brumberg1991} and  
Eq.~(\ref{First_Integration_1PN_Q}) agrees with \cite{Klioner1991};  
see also endnote \footnote{By means of $M_{ab} = \hat{M}_{ab} + \delta_{ab}\,M_{ii}/3$, one may recognize that the trace terms of the 
quadrupole
moment cancel each other in the coefficients given by Eqs.~(29) in \cite{Klioner1991}.}.
It is easy to check that the solutions in (\ref{First_Integration_1PN}) - (\ref{First_Integration_1PN_Q}) satisfy the 
null condition (\ref{Null_Condition1}) up to terms of the order ${\cal O}\left(c^{-4}\right)$.

\subsection{The second integration of the geodesic equation in 1PN approximation}

The second integration of geodesic equation (trajectory) for a light signal in the gravitational field of one body at rest endowed  
with the mass-monopole and mass-quadrupole in the 1PN approximation reads  
\begin{eqnarray}
	\ve{x}\left(t\right) &=& \int\limits_{t_0}^{t} d c {\rm t} \,\frac{\dot{\ve{x}}\left({\rm t}\right)}{c}
	 = \ve{x}_0 + \left(t - t_0\right) \ve{\sigma} + \Delta \ve{x}_{\rm 1PN}\left(t,t_0\right)
	+ {\cal O}\left(c^{-4}\right). 
\label{Second_Integration_1PN}
\end{eqnarray}

\noindent
Inserting the r.h.s. of (\ref{First_Integration_1PN}) with Eqs.~(\ref{First_Integration_1PN_M_Q}) - (\ref{First_Integration_1PN_Q}) 
into (\ref{Second_Integration_1PN}) yields  
\begin{eqnarray}
	 && \Delta \ve{x}_{\rm 1PN}\left(t,t_0\right) = 
	\Delta\ve{x}^{\rm M}_{\rm 1PN}\left(t,t_0\right) + \Delta\ve{x}^{\rm Q}_{\rm 1PN}\left(t,t_0\right) ,  
	\label{Second_Integration_1PN_M_Q}
\end{eqnarray}

\noindent
where their spatial components are 
\begin{eqnarray}
	 && \Delta x^{i\,{\rm M}}_{\rm 1PN}\left(t,t_0\right) = \frac{G M}{c^2} \bigg[ 
	 {\cal A}^i_{\left(3\right)}\,{\cal W}_{\left(3\right)}\left(t,t_0\right) 
	 + {\cal B}^i_{\left(3\right)}\,{\cal X}_{\left(3\right)}\left(t,t_0\right)\bigg]  ,
	\label{Second_Integration_1PN_M} 
	\\
	 && \Delta x^{i\,{\rm Q}}_{\rm 1PN}\left(t,t_0\right)  
	 = \frac{G \hat{M}_{ab}}{c^2} \sum\limits_{n=5,7} \bigg[ 
	 {\cal C}^{iab}_{\left(n\right)}\,{\cal W}_{\left(n\right)} \left(t,t_0\right)
	 + {\cal D}^{iab}_{\left(n\right)}\,{\cal X}_{\left(n\right)} \left(t,t_0\right)\bigg] . 
\label{Second_Integration_1PN_Q}
\end{eqnarray}

\noindent 
The scalar master integrals ${\cal W}_{\left(n\right)}$ and ${\cal X}_{\left(n\right)}$ are defined by
Eqs.~(\ref{Master_Integral_I2}) and (\ref{Master_Integral_J2}), respectively, while their explicit solutions for $n=3,5,7$ are given  
by Eqs.~(\ref{Master_Integral_E1}), (\ref{Master_Integral_E3}), (\ref{Master_Integral_E7}) and 
Eqs.~(\ref{Master_Integral_F}), (\ref{Master_Integral_G}), (\ref{Master_Integral_H}) in Appendix \ref{Appendix_Scalar_Master_Integrals}.
One may show that Eq.~(\ref{Second_Integration_1PN_M}) agrees with \cite{Brumberg1991} and
Eq.~(\ref{Second_Integration_1PN_Q}) agrees with \cite{Klioner1991};  
see also endnote \footnote{We note that $\displaystyle \ln \frac{x_{\rm N} - \ve{\sigma} \cdot \ve{x}_{\rm N}}{x_0 - \ve{\sigma} \cdot \ve{x}_0} 
= - \ln \frac{x_{\rm N} + \ve{\sigma} \cdot \ve{x}_{\rm N}}{x_0 + \ve{\sigma} \cdot \ve{x}_0}$.}.   

\section{Light propagation in 2PN approximation \label{Section3}} 

\subsection{The 2PM metric for a body with monopole and quadrupole term}

The post-Newtonian series expansion in 2PM approximation of the metric tensor reads
\begin{eqnarray}
g_{\alpha\beta}\left(\ve{x}\right) = \eta_{\alpha\beta} + h^{(2)}_{\alpha\beta}\left(\ve{x}\right)
+ h^{(4)}_{\alpha\beta}\left(\ve{x}\right)
+ {\cal O} \left(c^{-6}\right) . 
\label{metric_perturbation_2PN_10}
\end{eqnarray}

\noindent
The 1PN terms of the metric perturbation, $h^{(2)}_{\alpha\beta}$, were given by Eqs.~(\ref{metric_1PN_00}) - (\ref{metric_1PN_ij}). 
The 2PN terms of the metric perturbation, $h_{\alpha \beta}^{(4)}$, in canonical harmonic coordinates read \cite{Zschocke_2PM_Metric}  
\begin{eqnarray}
	h^{\left(4\right)}_{00}\left(\ve{x}\right) &=& - \frac{2\,G^2}{c^4}\,\frac{M^2}{x^2}
- \frac{6\,G^2}{c^4} \frac{M\,\hat{M}_{ab}}{x^4} \hat{n}_{ab}
 	 - \frac{G^2}{c^4}\frac{\hat{M}_{ab}\,\hat{M}_{cd}}{x^6} \left(\frac{3}{5} \delta_{ac} \delta_{bd}
           + \frac{18}{7} \delta_{ac} \hat{n}_{bd} + \frac{9}{2} \hat{n}_{abcd}\right) ,
\label{metric_2PN_00}
\\
\nonumber\\
	h^{\left(4\right)}_{ij}\left(\ve{x}\right) &=& \frac{G^2}{c^4}\,\frac{M^2}{x^2}\left(\frac{4}{3}\,\delta_{ij} + \hat{n}_{ij}\right)
	+ \frac{G^2\,M\,\hat{M}_{ab}}{c^4\;x^4}\left(\frac{15}{2}\,\hat{n}_{ijab} + \frac{32}{7}\,\delta_{ij}\,\hat{n}_{ab}
- \frac{12}{7}\,\delta_{a\,(i} \hat{n}_{j\,) b} \right)  
\nonumber\\
\nonumber\\
	&& \hspace{-0.75cm} +  \frac{G^2}{c^4}\frac{\hat{M}_{ab}\,\hat{M}_{cd}}{x^6}
\bigg(\frac{75}{4}\,\hat{n}_{ijabcd} - \frac{90}{11}\,\hat{n}_{ijac}\,\delta_{bd} + \frac{27}{11}\,\hat{n}_{abcd}\,\delta_{ij}
 - \frac{25}{84}\,\hat{n}_{ij}\,\delta_{ac}\,\delta_{bd} 
 + \frac{83}{42}\,\hat{n}_{ad}\,\delta_{bc}\,\delta_{ij}
 + \frac{16}{35}\,\delta_{ac}\,\delta_{bd}\,\delta_{ij} 
\nonumber\\
\nonumber\\
&& \hspace{-0.75cm} + \frac{18}{11}\,\hat{n}_{acd(i}\,\delta_{j)b}
                   - \frac{5}{21}\,\hat{n}_{a(i}\,\delta_{j)c}\,\delta_{bd} 
	+ \frac{10}{21}\,\delta_{ci}\,\delta_{dj}\,\hat{n}_{ab}
  - \frac{23}{42}\,\delta_{b(i}\,\delta_{j)c}\,\hat{n}_{ad} - \frac{6}{35}\,\delta_{ad}\,\delta_{b(i}\,\delta_{j)c} \bigg),  
\label{metric_2PN_ij}
\end{eqnarray}

\noindent
while $h^{\left(4\right)}_{0i} = 0$. 
The STF tensors $\hat{n}_L$ are given by Eq.~(\ref{Appendix_Cartesian_Tensor_General_Formula}); 
for the explicit form of $\hat{n}_{ab}$ and $\hat{n}_{abcd}$ as well as $\hat{n}_{abcde\!f}$ see Eqs.~(\ref{Appendix1_STF2}) and (\ref{Appendix1_STF4}) 
as well as (\ref{Appendix1_STF6}) in Appendix \ref{Appendix_STF}. 
The symmetrization operation is defined by Eq.~(\ref{Symmetrization}) in Appendix \ref{Appendix_STF} and $x = \left|\ve{x}\right|$ is the 
absolute of the spatial coordinates of some field point where the origin of the coordinate system is located at the center of mass of the body.

\subsection{The geodesic equation in 2PN approximation}  

The geodesic equation in 2PN approximation has been given, for instance, in \cite{Brumberg1991} as well as in several articles, 
e.g. \cite{Bruegmann2005,Zschocke3,Zschocke4}.  Inserting (\ref{metric_perturbation_2PN_10}) into the geodesic equation (\ref{Geodetic_Equation2}) yields 
\begin{eqnarray}
	\frac{\ddot{x}^i \left(t\right)}{c^2} &=& + \frac{1}{2}\,h_{00,i}^{(2)}
- h_{00,j}^{(2)} \frac{\dot{x}^i\left(t\right)}{c}\frac{\dot{x}^j\left(t\right)}{c}
- h_{ij,k}^{(2)} \frac{\dot{x}^j\left(t\right)}{c}\frac{\dot{x}^k\left(t\right)}{c}
 + \frac{1}{2}\,h_{jk,i}^{(2)} \frac{\dot{x}^j\left(t\right)}{c}\frac{\dot{x}^k\left(t\right)}{c}
 - \frac{1}{2}\,h_{ij}^{(2)} h_{00,j}^{(2)}
- h_{00}^{(2)} h_{00,j}^{(2)} \frac{\dot{x}^i\left(t\right)}{c} \frac{\dot{x}^j\left(t\right)}{c}
\nonumber\\
\nonumber\\
&& \hspace {-1.25cm} + h_{is}^{(2)} h_{sj,k}^{(2)} \frac{\dot{x}^j\left(t\right)}{c} \frac{\dot{x}^k\left(t\right)}{c}
- \frac{1}{2} h_{is}^{(2)} h_{jk,s}^{(2)} \frac{\dot{x}^j\left(t\right)}{c} \frac{\dot{x}^k\left(t\right)}{c}
 + \frac{1}{2} h_{00,i}^{(4)}
 - h_{00,j}^{(4)} \frac{\dot{x}^i\left(t\right)}{c} \frac{\dot{x}^j\left(t\right)}{c}
- h_{ij,k}^{(4)} \frac{\dot{x}^j\left(t\right)}{c} \frac{\dot{x}^k\left(t\right)}{c}
+ \frac{1}{2} h_{jk,i}^{(4)} \frac{\dot{x}^j\left(t\right)}{c} \frac{\dot{x}^k\left(t\right)}{c} 
\nonumber\\
	&&\hspace {-1.25cm} + {\cal O}\left(c^{-6}\right)\,. 
\label{Geodesic_Equation4}
\end{eqnarray}

\noindent
In (\ref{Geodesic_Equation4}) the field point $\ve{x}$ as argument of the metric perturbations is taken at the position of the light signal, while 
$\dot{\ve{x}}\left(t\right)$ and $\ve{x}\left(t\right)$ are the coordinate velocity and spatial coordinate of the light signal at this field point. 
For the geodesic equation in 2PN approximation, 
one has to take the 1PN light ray ($\dot{\ve{x}}_{\rm 1PN}$ and $\ve{x}_{\rm 1PN}$) for the first four terms on the r.h.s. in (\ref{Geodesic_Equation4}) 
and the unperturbed light ray ($\dot{\ve{x}}_{\rm N}$ and $\ve{x}_{\rm N}$) for the other terms on the r.h.s. in (\ref{Geodesic_Equation4}).  
Accordingly, in view of (\ref{metric_1PN_ij}) and (\ref{First_Integration_1PN}), the 2PM geodesic equation in (\ref{Geodesic_Equation4}) simplifies considerably, 
\begin{eqnarray}
	\frac{\ddot{x}^i\left(t\right)}{c^2} &=& + h_{00,i}^{(2)}\left(\ve{x}_{\rm 1PN}\right) 
	- 2\,h_{00,j}^{(2)}\left(\ve{x}_{\rm 1PN}\right) \,\sigma^i \sigma^j 
	 + h_{00,i}^{(2)}\left(\ve{x}_{\rm N}\right)\,\frac{\sigma^j \Delta \dot{x}^j_{\rm 1PN}\left(\ve{x}_{\rm N}\right)}{c}
	- 2\,h_{00,j}^{(2)}\left(\ve{x}_{\rm N}\right)\,\frac{\Delta \dot{x}^j_{\rm 1PN}\left(\ve{x}_{\rm N}\right)}{c}\,\sigma^i
	\nonumber\\
        \nonumber\\
	&& \hspace {-1.4cm} - 2\,h_{00,j}^{(2)}\left(\ve{x}_{\rm N}\right)\,\frac{\Delta \dot{x}^i_{\rm 1PN}\left(\ve{x}_{\rm N}\right)}{c}\,\sigma^j
        -\,h_{00}^{(2)}\left(\ve{x}_{\rm N}\right)\,h_{00,i}^{(2)}\left(\ve{x}_{\rm N}\right) 
	+ \frac{1}{2}\,h_{00,i}^{(4)}\left(\ve{x}_{\rm N}\right) 
	- h_{00,j}^{(4)}\left(\ve{x}_{\rm N}\right)\,\sigma^i \sigma^j 
	- h_{ij,k}^{(4)}\left(\ve{x}_{\rm N}\right)\,\sigma^j \sigma^k
	\nonumber\\
	&& \hspace {-1.4cm} + \frac{1}{2}\,h_{jk,i}^{(4)}\left(\ve{x}_{\rm N}\right)\,\sigma^j \sigma^k + {\cal O}\left(c^{-6}\right),   
\label{Geodesic_Equation_2PN_5}
\end{eqnarray}

\noindent
whereby the spatial derivatives are to be understood that one has to perform first these derivatives and afterwards 
one has to insert the light ray in the given approximation (cf. Eq.~(\ref{Derivatives_1})). This equation consists of $10$ individual terms, 
\begin{eqnarray}
	\frac{\ddot{x}}{c^2} &=& \frac{\ddot{x}_{A}^i}{c^2} 
        + \frac{\ddot{x}_{B}^i}{c^2} + \frac{\ddot{x}_{C}^i}{c^2} + \frac{\ddot{x}_{D}^i}{c^2} 
        + \frac{\ddot{x}_{E}^i}{c^2} + \frac{\ddot{x}_{F}^i}{c^2} + \frac{\ddot{x}_{G}^i}{c^2} + \frac{\ddot{x}_{H}^i}{c^2} 
	+ \frac{\ddot{x}_{I}^i}{c^2} 
        + \frac{\ddot{x}_{J}^i}{c^2} + {\cal O}\left(c^{-6}\right) , 
\label{Geodesic_Equation_2PN_10}
\end{eqnarray}

\noindent 
as defined by Eqs.~(\ref{Individual_A}) - (\ref{Individual_J}). These individual terms are given by Eqs.~(\ref{Appendix_Term_A_10}), (\ref{Appendix_Term_B_10}),  
(\ref{Appendix_Term_C_5}), (\ref{Appendix_Term_D_5}), (\ref{Appendix_Term_E_5}), (\ref{Appendix_Term_F_5}), (\ref{Appendix_Term_G_5}), (\ref{Appendix_Term_H_5}),  
(\ref{Appendix_Term_I_5}), (\ref{Appendix_Term_J_5}) in Appendix \ref{Appendix_2PN_Terms}. Similar to (\ref{Geodesic_Equation_1PN_C}), the sum of these terms in  
(\ref{Geodesic_Equation_2PN_10}) can be separated into monopole and quadrupole and mixed monopole-quadrupole terms,  
\begin{eqnarray}
	\frac{\ddot{\ve{x}}\left(t\right)}{c^2} &=& \frac{\ddot{\ve{x}}^{\rm M}_{\rm 1PN}\left(t\right)}{c^2} 
        + \frac{\ddot{\ve{x}}^{\rm Q}_{\rm 1PN}\left(t\right)}{c^2}
	+ \frac{\ddot{\ve{x}}^{{\rm M} \times {\rm M}}_{\rm 2PN}\left(t\right)}{c^2} 	
	+ \frac{\ddot{\ve{x}}^{{\rm M} \times {\rm Q}}_{\rm 2PN}\left(t\right)}{c^2} 	
	+ \frac{\ddot{\ve{x}}^{{\rm Q} \times {\rm Q}}_{\rm 2PN}\left(t\right)}{c^2} 	
	+ {\cal O}\left(c^{-6}\right) .
\label{Geodesic_Equation_2PN_15}
\end{eqnarray}

\noindent 
The 1PN terms $\ddot{\ve{x}}^{\rm M}_{\rm 1PN}$ and $\ddot{\ve{x}}^{\rm Q}_{\rm 1PN}$ in (\ref{Geodesic_Equation_2PN_15}) are proportional to $G^1$. 
These 1PN terms were given by Eq.~(\ref{Geodesic_Equation_1PN_C}) with (\ref{Geodesic_Equation_1PN_M}) and (\ref{Geodesic_Equation_1PN_Q}), 
and have already been considered in detail in the previous Section \ref{Section2}.  The 2PN terms $\ddot{\ve{x}}^{{\rm M} \times {\rm M}}_{\rm 2PN}$  
and $\ddot{\ve{x}}^{{\rm M} \times {\rm Q}}_{\rm 2PN}$ and $\ddot{\ve{x}}^{{\rm Q} \times {\rm Q}}_{\rm 2PN}$ in (\ref{Geodesic_Equation_2PN_15}) are  
proportional to $G^2$. These 2PN terms will be considered in the following.

\subsubsection{The terms $\ddot{\ve{x}}^{{\rm M} \times {\rm M}}_{\rm 2PN}$} 

The terms proportional to $M \times M$ in Eq.~(\ref{Appendix_Term_A_10}) and Eqs.~(\ref{Appendix_Term_B_10}) - (\ref{Appendix_Term_J_5}) in 
Appendix~\ref{Appendix_2PN_Terms} were summarized by Eq.~(\ref{Appendix_Term_M_M}).  
By inserting $\Delta \dot{\ve{x}}_{\rm 1PN}^{\rm M}$ and $\Delta \ve{x}_{\rm 1PN}^{\rm M}$ as given by Eqs.~(\ref{First_Integration_1PN_M})  
and (\ref{Second_Integration_1PN_M}), respectively, and using relations (\ref{Equation_A}) - (\ref{Equation_C})  
the terms in Eq.~(\ref{Appendix_Term_M_M}) can be written in terms of time-independent tensorial coefficients and time-dependent scalar functions, 
\begin{eqnarray}
	\frac{\ddot{x}^{i\,{\rm M} \times {\rm M}}_{\rm 2PN}\left(t\right)}{c^2} &=& \frac{G^2 M^2}{c^4}   
	\bigg[\sum\limits_{n=3}^{6} {\cal E}^{i}_{\left(n\right)} \ddot{\cal W}_{\left(n\right)}\left(t\right) 
	+ \sum\limits_{n=2}^{6} {\cal F}^{i}_{\left(n\right)}\, \ddot{\cal X}_{\left(n\right)}\left(t\right) 
	+ {\cal G}^{i}_{\left(5\right)}\,\ddot{\cal Y}_{\left(5\right)}\left(t\right)  
	+ \sum\limits_{n=3,5} {\cal H}^{i}_{\left(n\right)} \,\ddot{\cal Z}_{\left(n\right)}\left(t\right) \bigg] .  
\label{2PN_Term_M_M}
\end{eqnarray}

\noindent
The step from (\ref{Appendix_Term_M_M}) to (\ref{2PN_Term_M_M}) necessitates a similar procedure as the step from Eqs.~(\ref{Geodesic_Equation_1PN_M}) 
and (\ref{Geodesic_Equation_1PN_Q}) to Eqs.~(\ref{Geodesic_Equation_1PN_M_Rewritten}) and (\ref{Geodesic_Equation_1PN_Q_Rewritten}). 
The coefficients ${\cal E}^{i}_{\left(n\right)}$, ${\cal F}^{i}_{\left(n\right)}$, ${\cal G}^{i}_{\left(5\right)}$, ${\cal H}^{i}_{\left(n\right)}$   
are given by Eqs.~(\ref{coefficient_E3}) - (\ref{coefficient_H5}) in Appendix \ref{Appendix_2PN_Terms}. 
The scalar functions $\ddot{\cal W}_{\left(n\right)}$ and $\ddot{\cal X}_{\left(n\right)}$ were given by Eqs.~(\ref{scalar_function_ddot_W}) and 
(\ref{scalar_function_ddot_X}), respectively, while the scalar functions $\ddot{\cal Y}_{\left(n\right)}$ and $\ddot{\cal Z}_{\left(n\right)}$ read 
\begin{eqnarray}
	\ddot{\cal Y}_{\left(n\right)}\left(t\right) &=& \frac{\ve{\sigma} \cdot \ve{x}_{\rm N}}{\left(x_{\rm N}\right)^n}\,
        \ln \frac{x_{\rm N} - \ve{\sigma} \cdot \ve{x}_{\rm N}}{x_0 - \ve{\sigma} \cdot \ve{x}_0}\;,  
	\label{scalar_function_ddot_Y}
	\\
	\ddot{\cal Z}_{\left(n\right)}\left(t\right) &=& \frac{1}{\left(x_{\rm N}\right)^n}\,
	\ln \frac{x_{\rm N} - \ve{\sigma} \cdot \ve{x}_{\rm N}}{x_0 - \ve{\sigma} \cdot \ve{x}_0}\;. 
        \label{scalar_function_ddot_Z}
\end{eqnarray}

\noindent
Actually, in (\ref{2PN_Term_M_M}) one needs the scalar function $\ddot{\cal Y}_{\left(n\right)}$ only for $n=5$, while higher values of $n$  
are necessary for the mass-quadrupole terms which will be considered in (\ref{2PN_Term_M_Q}).

\subsubsection{The terms $\ddot{\ve{x}}^{{\rm M} \times {\rm Q}}_{\rm 2PN}$}

The terms proportional to $M \times \hat{M}_{ab}$ in Eq.~(\ref{Appendix_Term_A_10}) and Eqs.~(\ref{Appendix_Term_B_10}) - (\ref{Appendix_Term_J_5}) in 
Appendix~\ref{Appendix_2PN_Terms} were summarized by Eq.~(\ref{Appendix_Term_M_Q}). 
By inserting $\Delta \dot{\ve{x}}_{\rm 1PN}^{\rm M}$ and $\Delta \ve{x}_{\rm 1PN}^{\rm M}$ as given by Eqs.~(\ref{First_Integration_1PN_M})
and (\ref{Second_Integration_1PN_M}), respectively, as well as $\Delta \dot{\ve{x}}_{\rm 1PN}^{\rm Q}$ and $\Delta \ve{x}_{\rm 1PN}^{\rm Q}$ 
as given by Eqs.~(\ref{First_Integration_1PN_Q}) and (\ref{Second_Integration_1PN_Q}), and using relations (\ref{Equation_A}) - (\ref{Equation_C}), 
the terms in Eq.~(\ref{Appendix_Term_M_Q}) can be written in terms of time-independent tensorial coefficients and time-dependent scalar functions,
\begin{eqnarray}
	\frac{\ddot{x}^{i\,{\rm M} \times {\rm Q}}_{\rm 2PN}\left(t\right)}{c^2} &=& \frac{G M}{c^2}\,\frac{G \hat{M}_{ab}}{c^2}   
	\bigg[\sum\limits_{n=3}^{10} {\cal K}^{i\,ab}_{\left(n\right)}\;\ddot{\cal W}_{\left(n\right)}\left(t\right)
	+ \sum\limits_{n=2}^{10} {\cal L}^{i\,ab}_{\left(n\right)}\; \ddot{\cal X}_{\left(n\right)}\left(t\right) 
	 + \sum\limits_{n=7}^{9} {\cal M}^{i\,ab}_{\left(n\right)}\;\ddot{\cal Y}_{\left(n\right)}\left(t\right) 
	+ \sum\limits_{n=5}^{9} {\cal N}^{i\,ab}_{\left(n\right)}\;\ddot{\cal Z}_{\left(n\right)}\left(t\right) \bigg] .  
\label{2PN_Term_M_Q}
\end{eqnarray}

\noindent
The step from (\ref{Appendix_Term_M_Q}) to (\ref{2PN_Term_M_Q}) is achieved by a similar procedure as the step from Eqs.~(\ref{Geodesic_Equation_1PN_M})
and (\ref{Geodesic_Equation_1PN_Q}) to Eqs.~(\ref{Geodesic_Equation_1PN_M_Rewritten}) and (\ref{Geodesic_Equation_1PN_Q_Rewritten}), but the 
algebraic effort is more complex. 
The coefficients ${\cal K}^{i\,ab}_{\left(n\right)}$, ${\cal L}^{i\,ab}_{\left(n\right)}$, 
${\cal M}^{i\,ab}_{\left(n\right)}$, ${\cal N}^{i\,ab}_{\left(n\right)}$ are 
given by Eqs.~(\ref{coefficient_K3}) - (\ref{coefficient_N9}) in Appendix \ref{Appendix_2PN_Terms}. 
The scalar functions $\ddot{\cal W}_{\left(n\right)}$ and $\ddot{\cal X}_{\left(n\right)}$ were given by Eqs.~(\ref{scalar_function_ddot_W}) and
(\ref{scalar_function_ddot_X}), while the scalar functions $\ddot{\cal Y}_{\left(n\right)}$ and $\ddot{\cal Z}_{\left(n\right)}$ 
were given by Eqs.~(\ref{scalar_function_ddot_Y}) and (\ref{scalar_function_ddot_Z}).

\subsubsection{The terms $\ddot{\ve{x}}^{{\rm Q} \times {\rm Q}}_{\rm 2PN}$}

The terms proportional to $\hat{M}_{ab} \times \hat{M}_{cd}$ in Eq.~(\ref{Appendix_Term_A_10}) and Eqs.~(\ref{Appendix_Term_B_10}) - (\ref{Appendix_Term_J_5}) in  
Appendix~\ref{Appendix_2PN_Terms} were summarized by Eq.~(\ref{Appendix_Term_Q_Q}). 
By inserting $\Delta \dot{\ve{x}}_{\rm 1PN}^{\rm M}$ and $\Delta \ve{x}_{\rm 1PN}^{\rm M}$ as given by Eqs.~(\ref{First_Integration_1PN_M})
and (\ref{Second_Integration_1PN_M}), respectively, as well as $\Delta \dot{\ve{x}}_{\rm 1PN}^{\rm Q}$ and $\Delta \ve{x}_{\rm 1PN}^{\rm Q}$
as given by Eqs.~(\ref{First_Integration_1PN_Q}) and (\ref{Second_Integration_1PN_Q}), and using relations (\ref{Equation_A}) - (\ref{Equation_C}), 
the terms in Eq.~(\ref{Appendix_Term_Q_Q}) can be written in terms of time-independent tensorial coefficients and time-dependent scalar functions,
\begin{eqnarray}
	\frac{\ddot{x}^{i\,{\rm Q} \times {\rm Q}}_{\rm 2PN}\left(t\right)}{c^2} &=& \frac{G \hat{M}_{ab}}{c^2}\,\frac{G \hat{M}_{cd}}{c^2} \bigg[  
	\sum\limits_{n=5}^{14} {\cal P}^{i\,abcd}_{\left(n\right)}\,\ddot{\cal W}_{\left(n\right)}\left(t\right)
	+ \sum\limits_{n=4}^{14} {\cal Q}^{i\,abcd}_{\left(n\right)}\,\ddot{\cal X}_{\left(n\right)}\left(t\right) \bigg] . 
\label{2PN_Term_Q_Q}
\end{eqnarray}

\noindent
The step from (\ref{Appendix_Term_Q_Q}) to (\ref{2PN_Term_Q_Q}) requires a similar procedure as the step from Eqs.~(\ref{Geodesic_Equation_1PN_M})
and (\ref{Geodesic_Equation_1PN_Q}) to Eqs.~(\ref{Geodesic_Equation_1PN_M_Rewritten}) and (\ref{Geodesic_Equation_1PN_Q_Rewritten}), but the 
algebraic effort is more complex. 
The coefficients ${\cal P}^{i\,abcd}_{\left(n\right)}$ and ${\cal Q}^{i\,abcd}_{\left(n\right)}$ are  
given by Eqs.~(\ref{coefficient_P5}) - (\ref{coefficient_Q14}) in Appendix \ref{Appendix_2PN_Terms}.  
The scalar functions $\ddot{\cal W}_{\left(n\right)}$ and $\ddot{\cal X}_{\left(n\right)}$ were given by Eqs.~(\ref{scalar_function_ddot_W}) and
(\ref{scalar_function_ddot_X}).

\subsection{The first integration of geodesic equation in 2PN approximation}\label{Section2PN_1}  

The first integration of geodesic equation (\ref{Geodesic_Equation_2PN_10}) yields the coordinate velocity of the light signal in 2PN approximation, 
\begin{eqnarray}
	\frac{\dot{\ve{x}}\left(t\right)}{c} &=& \int\limits_{- \infty}^{t} d c {\rm t} \,\frac{\ddot{\ve{x}}\left({\rm t}\right)}{c^2}
	\nonumber\\ 
	&=& \ve{\sigma} + \frac{\Delta \dot{\ve{x}}_{\rm 1PN}\left(t\right)}{c} + \frac{\Delta \dot{\ve{x}}_{\rm 2PN}\left(t\right)}{c} 
	+ {\cal O}\left(c^{-6}\right). 
\label{First_Integration_2PN}
\end{eqnarray}
  
\noindent
By inserting (\ref{Geodesic_Equation_2PN_15}) into the r.h.s. of (\ref{First_Integration_2PN}) one gets
\begin{eqnarray}
	\frac{\Delta \dot{\ve{x}}_{\rm 1PN}\left(t\right)}{c} &=& \frac{\Delta \dot{\ve{x}}^{\rm M}_{\rm 1PN}\left(t\right)}{c} 
+ \frac{\Delta \dot{\ve{x}}^{\rm Q}_{\rm 1PN}\left(t\right)}{c}\;,   
\label{First_Integration_2PN_M_Q}
\\
	\frac{\Delta \dot{\ve{x}}_{\rm 2PN}\left(t\right)}{c} &=& \frac{\Delta \dot{\ve{x}}^{{\rm M} \times {\rm M}}_{\rm 2PN}\left(t\right)}{c}
	+ \frac{\Delta \dot{\ve{x}}^{{\rm M} \times {\rm Q}}_{\rm 2PN}\left(t\right)}{c}
	+ \frac{\Delta \dot{\ve{x}}^{{\rm Q} \times {\rm Q}}_{\rm 2PN}\left(t\right)}{c}\;. 
\label{First_Integration_2PN_MM_MQ_QQ}
\end{eqnarray}

\noindent 
The 1PN terms $\Delta \dot{\ve{x}}^{\rm M}_{\rm 1PN}$ and $\Delta \dot{\ve{x}}^{\rm Q}_{\rm 1PN}$ have been determined 
by Eqs.~(\ref{First_Integration_1PN_M}) and (\ref{First_Integration_1PN_Q}), respectively, and will not be represented here again. 

The first integration of the 2PN monopole-monopole term in (\ref{2PN_Term_M_M}) is given by
\begin{eqnarray}
	\frac{\Delta \dot{x}^{i\,{\rm M} \times {\rm M}}_{\rm 2PN}\left(t\right)}{c} &=& \frac{G^2 M^2}{c^4}   
	\bigg[\sum\limits_{n=3}^{6} {\cal E}^{i}_{\left(n\right)}\;\dot{\cal W}_{\left(n\right)}\left(t\right) 
	 + \sum\limits_{n=2}^{6} {\cal F}^{i}_{\left(n\right)} \; \dot{\cal X}_{\left(n\right)}\left(t\right)   
	+ {\cal G}^{i}_{\left(5\right)} \; \dot{\cal Y}_{\left(5\right)}\left(t\right)
	+ \sum\limits_{n=3,5} {\cal H}^{i}_{\left(n\right)} \; \dot{\cal Z}_{\left(n\right)}\left(t\right)\bigg] .
\label{First_Integration_2PN_Term_M_M}
\end{eqnarray}

\noindent 
The first integration of the 2PN monopole-quadrupole term in (\ref{2PN_Term_M_Q}) is given by   
\begin{eqnarray}
	\frac{\Delta \dot{x}^{i\,{\rm M} \times {\rm Q}}_{\rm 2PN}\left(t\right)}{c} &=& \frac{G M}{c^2}\,\frac{G \hat{M}_{ab}}{c^2} \bigg[  
	\sum\limits_{n=3}^{10} {\cal K}^{i\,ab}_{\left(n\right)}\,\dot{\cal W}_{\left(n\right)}\left(t\right)
	+ \sum\limits_{n=2}^{10} {\cal L}^{i\,ab}_{\left(n\right)}\,\dot{\cal X}_{\left(n\right)}\left(t\right)
	+\,\sum\limits_{n=7}^{9} {\cal M}^{i\,ab}_{\left(n\right)}\,\dot{\cal Y}_{\left(n\right)}\left(t\right)  
	+ \sum\limits_{n=5}^{9} {\cal N}^{i\,ab}_{\left(n\right)}\,
	\dot{\cal Z}_{\left(n\right)}\left(t\right) \bigg] . 
\label{First_Integration_2PN_Term_M_Q}
\end{eqnarray}

\noindent
The first integration of the 2PN quadrupole-quadrupole term in (\ref{2PN_Term_Q_Q}) is given by
\begin{eqnarray}
	\frac{\Delta \dot{x}^{i\,{\rm Q} \times {\rm Q}}_{\rm 2PN}\left(t\right)}{c} &=& \frac{G \hat{M}_{ab}}{c^2}\,\frac{G \hat{M}_{cd}}{c^2} \bigg[  
	\sum\limits_{n=5}^{14} {\cal P}^{i\,abcd}_{\left(n\right)}\,\dot{\cal W}_{\left(n\right)}\left(t\right)  
	+ \sum\limits_{n=4}^{14} {\cal Q}^{i\,abcd}_{\left(n\right)}\,\dot{\cal X}_{\left(n\right)}\left(t\right)\bigg] .  
\label{First_Integration_2PN_Term_Q_Q}
\end{eqnarray}

\noindent
The master integrals $\dot{\cal W}_{\left(n\right)}$, $\dot{\cal X}_{\left(n\right)}$, $\dot{\cal Y}_{\left(n\right)}$, $\dot{\cal Z}_{\left(n\right)}$ 
are defined by Eqs.~(\ref{Master_Integral_I1}) - (\ref{Master_Integral_L1}). 
The solution for the master integral $\dot{\cal W}_{\left(n\right)}$ is given explicitly by Eq.~(\ref{Master_Integral_A}),  
while the solutions for the master integrals $\dot{\cal X}_{\left(n\right)}$, 
$\dot{\cal Y}_{\left(n\right)}$, $\dot{\cal Z}_{\left(n\right)}$ can be given in closed form by the recurrence relations 
in Eqs.~(\ref{Master_Integral_B}) - (\ref{Master_Integral_D}) in Appendix \ref{Appendix_Scalar_Master_Integrals}.
The explicit form of $\dot{\cal X}_{\left(n\right)}$, $\dot{\cal Y}_{\left(n\right)}$, $\dot{\cal Z}_{\left(n\right)}$ 
for the first few orders in $n$ are given by Eqs.~(\ref{Master_Integral_B2}) - (\ref{Master_Integral_L5}) in Appendix \ref{Appendix_Scalar_Master_Integrals}. 

Two comments are in order: 
First of all, one may show, that 
$c\,\ve{\sigma} + \Delta \dot{\ve{x}}^{{\rm M}}_{\rm 1PN} + \Delta \dot{\ve{x}}^{{\rm M} \times {\rm M}}_{\rm 2PN}$ agrees with Eq.~(3.2.37)
in \cite{Brumberg1991} as well as with Eq.~(44) in our article \cite{Article_Zschocke1}; 
see also endnote \footnote{In order to demonstrate that agreement, one has to perform a series expansion of $\ve{x}_{\rm 1PN}$ which appears as 
argument in the
vectorial functions in \cite{Brumberg1991,Article_Zschocke1} (cf. Eqs.~(\ref{Appendix_x_A}) - (\ref{Appendix_x_C}) in Appendix \ref{Appendix_2PN_Terms}).}.
And second, it has been checked numerically by the computer algebra system {\it Maple} \cite{Maple} that
the solutions in (\ref{First_Integration_2PN}) - (\ref{First_Integration_2PN_Term_Q_Q})
satisfy the null condition (\ref{Null_Condition1}) along the entire light trajectory, which represents an independent
strict confirmation about the correctness of the time-independent coefficients given in Appendix \ref{Appendix_2PN_Terms}.

\subsection{The second integration of geodesic equation in 2PN approximation}\label{Section2PN_2}

The second integration of geodesic equation (\ref{Geodesic_Equation_2PN_10}) yields the trajectory of the light signal in 2PN approximation,
\begin{eqnarray}
	\hspace{0.0cm}  \ve{x}\left(t\right) &=& \int\limits_{- \infty}^{t} d c {\rm t} \,\frac{\dot{\ve{x}}\left({\rm t}\right)}{c}
	\nonumber\\ 
	&=& \ve{x}_0 + c \left( t- t_0\right) \ve{\sigma}  
	+ \Delta \ve{x}_{\rm 1PN}\left(t,t_0\right) + \Delta \ve{x}_{\rm 2PN}\left(t,t_0\right) 
	+ {\cal O}\left(c^{-6}\right). 
\label{Second_Integration_2PN}
\end{eqnarray}

\noindent
By inserting (\ref{First_Integration_2PN}) with (\ref{First_Integration_2PN_M_Q}) and (\ref{First_Integration_2PN_MM_MQ_QQ}) 
into the r.h.s. of (\ref{Second_Integration_2PN}) one gets
\begin{eqnarray}
        \Delta \ve{x}_{\rm 1PN}\left(t,t_0\right) &=& \Delta \ve{x}^{\rm M}_{\rm 1PN}\left(t,t_0\right) 
+ \Delta \ve{x}^{\rm Q}_{\rm 1PN}\left(t,t_0\right) , 
\label{Second_Integration_2PN_M_Q}
\\
        \Delta \ve{x}_{\rm 2PN}\left(t,t_0\right) &=& \Delta \ve{x}^{{\rm M} \times {\rm M}}_{\rm 2PN}\left(t,t_0\right)
        + \Delta \ve{x}^{{\rm M} \times {\rm Q}}_{\rm 2PN}\left(t,t_0\right)
	+ \Delta \ve{x}^{{\rm Q} \times {\rm Q}}_{\rm 2PN}\left(t,t_0\right) . 
\label{Second_Integration_2PN_MM_MQ_QQ}
\end{eqnarray}

\noindent
The 1PN terms $\Delta \ve{x}^{\rm M}_{\rm 1PN}$ and $\Delta \ve{x}^{\rm Q}_{\rm 1PN}$ have been determined
by Eqs.~(\ref{Second_Integration_1PN_M}) and (\ref{Second_Integration_1PN_Q}), respectively, and will not be re\-presented here again.

The second integration of the 2PN monopole-monopole term in (\ref{2PN_Term_M_M}) is given by
\begin{eqnarray}
	\Delta x^{i\,{\rm M} \times {\rm M}}_{\rm 2PN}\left(t,t_0\right) &=& \frac{G^2 M^2}{c^4}   
	\bigg[\sum\limits_{n=3}^{6} {\cal E}^{i}_{\left(n\right)}\,{\cal W}_{\left(n\right)}\left(t,t_0\right) 
	 + \sum\limits_{n=2}^{6} {\cal F}^{i}_{\left(n\right)}\,{\cal X}_{\left(n\right)}\left(t,t_0\right)   
	 + {\cal G}^{i}_{\left(5\right)}\,{\cal Y}_{\left(5\right)}\left(t,t_0\right)
	 + \sum\limits_{n=3,5} {\cal H}^{i}_{\left(n\right)}\,{\cal Z}_{\left(n\right)}\left(t,t_0\right)\bigg] .  
\label{Second_Integration_2PN_Term_M_M}
\end{eqnarray}

\noindent 
The second integration of the 2PN monopole-quadrupole term in (\ref{2PN_Term_M_Q}) is given by   
\begin{eqnarray}
	\Delta x^{i\,{\rm M} \times {\rm Q}}_{\rm 2PN}\left(t,t_0\right) &=& \frac{G M}{c^2}\,\frac{G \hat{M}_{ab}}{c^2} \bigg[    
	\sum\limits_{n=3}^{10} {\cal K}^{i\,ab}_{\left(n\right)}\,{\cal W}_{\left(n\right)}\left(t,t_0\right)
	 + \sum\limits_{n=2}^{10} {\cal L}^{i\,ab}_{\left(n\right)}\,{\cal X}_{\left(n\right)}\left(t,t_0\right) 
	 + \sum\limits_{n=7}^{9} {\cal M}^{i\,ab}_{\left(n\right)}\,{\cal Y}_{\left(n\right)}\left(t,t_0\right)  
	 + \sum\limits_{n=5}^{9} {\cal N}^{i\,ab}_{\left(n\right)}\, 
	{\cal Z}_{\left(n\right)}\left(t,t_0\right)\bigg] . 
	\nonumber\\ 
\label{Second_Integration_2PN_Term_M_Q}
\end{eqnarray}

\noindent
The second integration of the 2PN quadrupole-quadrupole term in (\ref{2PN_Term_Q_Q}) is given by
\begin{eqnarray}
	\Delta x^{i\,{\rm Q} \times {\rm Q}}_{\rm 2PN}\left(t,t_0\right) &=& \frac{G \hat{M}_{ab}}{c^2}\,\frac{G \hat{M}_{cd}}{c^2} \bigg[  
	\sum\limits_{n=5}^{14} {\cal P}^{i\,abcd}_{\left(n\right)}\,{\cal W}_{\left(n\right)}\left(t,t_0\right) 
	+ \sum\limits_{n=4}^{14} {\cal Q}^{i\,abcd}_{\left(n\right)}\,{\cal X}_{\left(n\right)}\left(t,t_0\right)\bigg] . 
	\label{Second_Integration_2PN_Term_Q_Q}
\end{eqnarray}

\noindent
The master integrals ${\cal W}_{\left(n\right)}$, ${\cal X}_{\left(n\right)}$, ${\cal Y}_{\left(n\right)}$, ${\cal Z}_{\left(n\right)}$ are defined by
Eqs.~(\ref{Master_Integral_I2}) - (\ref{Master_Integral_L2}).
The solutions for the master integrals ${\cal W}_{\left(n\right)}$, ${\cal X}_{\left(n\right)}$,
${\cal Y}_{\left(n\right)}$, ${\cal Z}_{\left(n\right)}$ can be given in closed form by the recurrence relations
in Eqs.~(\ref{Master_Integral_W}) - (\ref{Master_Integral_Z}) in Appendix \ref{Appendix_Scalar_Master_Integrals}.
The explicit form of ${\cal W}_{\left(n\right)}$, ${\cal X}_{\left(n\right)}$, ${\cal Y}_{\left(n\right)}$, ${\cal Z}_{\left(n\right)}$
for the first few orders in $n$ are given by Eqs.~(\ref{Master_Integral_E1}) - (\ref{Master_Integral_L5_2}) in Appendix \ref{Appendix_Scalar_Master_Integrals}.

One comment is in order: 
One may show, that
$\ve{x}_{\rm N} + \Delta \ve{x}^{{\rm M}}_{\rm 1PN} + \Delta \ve{x}^{{\rm M} \times {\rm M}}_{\rm 2PN}$  agrees with Eq.~(3.2.38) in \cite{Brumberg1991}
as well as with Eq.~(45) in our article \cite{Article_Zschocke1}.  
After some algebraic manipulations, mainly using relation (\ref{impact_parameter_1}), one may also demonstrate that (\ref{Second_Integration_2PN_Term_M_M})
agrees with Eq.~(J11) in \cite{Zschocke2}.

The presented solutions of the first integration (\ref{First_Integration_2PN}) and second integration (\ref{Second_Integration_2PN}) 
of geodesic equation are strictly valid for any astrometric configuration between light source, body and observer, 
that means for $\ve{\sigma} \cdot \ve{x}_0 \le 0$ and $\ve{\sigma} \cdot \ve{x}_0 > 0$ as well as for 
$\ve{\sigma} \cdot \ve{x}_1 \le 0$ and $\ve{\sigma} \cdot \ve{x}_1 > 0$.

\section{The total light deflection}\label{Total_Light_Deflection} 

The solution of the first integration of the initial value problem allows to determine the total light deflection.  
For that, one has to consider the unit tangent vectors along the light trajectory at $t \rightarrow \pm \infty$, 
\begin{eqnarray} 
        \ve{\sigma} &=& \frac{\dot{\ve x}\left(t\right)}{c}\bigg|_{t = - \infty} \;\;,  
        \label{unit_vector_sigma}
        \\
        \ve{\nu} &=& \frac{\dot{\ve x}\left(t\right)}{c}\bigg|_{t = + \infty}\;\;.  
        \label{unit_vector_nu}
\end{eqnarray}

\noindent
The total light deflection is defined as angle between these unit tangent vectors, $\delta\left(\ve{\sigma},\ve{\nu}\right)$,   
and the sine of total light deflection angle is given by  
\begin{eqnarray}
	\sin \delta\left(\ve{\sigma},\ve{\nu}\right)  &=& \left| \ve{\sigma} \times \ve{\nu} \right|\,.  
	\label{sine_total_light_deflection}
\end{eqnarray}

\noindent 
From (\ref{First_Integration_2PN}) - (\ref{First_Integration_2PN_MM_MQ_QQ}) with (\ref{First_Integration_1PN_M}) - (\ref{First_Integration_1PN_Q}) and 
(\ref{First_Integration_2PN_Term_M_M}) - (\ref{First_Integration_2PN_Term_Q_Q}) one obtains for the unit tangent vector $\ve{\nu}$ in (\ref{unit_vector_nu}):  
\begin{eqnarray} 
	\ve{\nu} &=& \ve{\sigma} + \ve{\nu}^{\rm M}_{\rm 1PN} + \ve{\nu}^{\rm Q}_{\rm 1PN} + \ve{\nu}^{{\rm M} \times {\rm M}}_{\rm 2PN} 
	+ \ve{\nu}^{{\rm M} \times {\rm Q}}_{\rm 2PN} + \ve{\nu}^{{\rm Q} \times {\rm Q}}_{\rm 2PN}\;, 
        \label{limit_plus}
        \\	
	\nu_{\rm 1PN}^{i\,{\rm M}} &=& \frac{G M}{c^2}\left[{\cal B}^i_{\left(3\right)}\,\dot{\cal X}^{\infty}_{\left(3\right)}  \right] , 
	\label{limit_plus_M}
	\\
	\nu_{\rm 1PN}^{i\,{\rm Q}} &=& \frac{G \hat{M}_{ab}}{c^2} 
	\left[\sum\limits_{n = 5,7} {\cal D}^{i\,ab}_{\left(n\right)}\;\dot{\cal X}^{\infty}_{\left(n\right)} \right] , 
        \label{limit_plus_Q}
	\\
	 \nu_{\rm 2PN}^{i\,{\rm M} \times {\rm M}} &=& 
	 \frac{G^2 M^2}{c^4} \left[\sum\limits_{n=2}^{6} {\cal F}^i_{\left(n\right)}\,\dot{\cal X}^{\infty}_{\left(n\right)}
	+ {\cal G}^i_{\left(5\right)}\,\dot{\cal Y}^{\infty}_{\left(5\right)}\right] ,  
        \label{limit_plus_M_M}
	\\ 
	\nu_{\rm 2PN}^{i\,{\rm M} \times {\rm Q}} &=& \frac{G M}{c^2}\,\frac{G \hat{M}_{ab}}{c^2} 
	\left[\sum\limits_{n=2}^{10} {\cal L}^{i\,ab}_{\left(n\right)}\;\dot{\cal X}^{\infty}_{\left(n\right)}  
	+ \sum\limits_{n=7}^{9} {\cal M}^{i\,ab}_{\left(n\right)}\;\dot{\cal Y}^{\infty}_{\left(n\right)} 
         \right] ,
	 \label{limit_plus_M_Q}
        \\
	\nu_{\rm 2PN}^{i\,{\rm Q} \times {\rm Q}} &=&  \frac{G \hat{M}_{ab}}{c^2}\,\frac{G \hat{M}_{cd}}{c^2} \left[
		\sum\limits_{n=4}^{14} {\cal Q}^{i\,abcd}_{\left(n\right)}\;\dot{\cal X}^{\infty}_{\left(n\right)} \right] , 
        \label{limit_plus_Q_Q}
\end{eqnarray}

\noindent
where the abbreviations 
\begin{eqnarray} 
        \dot{\cal X}^{\infty}_{\left(n\right)} = \dot{\cal X}_{\left(n\right)}\left(t\right) \bigg|_{t = + \infty} \; 
        {\rm and} \quad 
        \dot{\cal Y}^{\infty}_{\left(n\right)} = \dot{\cal Y}_{\left(n\right)}\left(t\right) \bigg|_{t = + \infty} 
        \label{Limit_X_Y} 
\end{eqnarray}

\noindent
have been introduced. It has been checked analytically that $\ve{\nu}$ in (\ref{limit_plus}) is in fact a unit vector, 
\begin{eqnarray}
\ve{\nu} \cdot \ve{\nu} = 1 + {\cal O}\left(c^{-6}\right).
	\label{unit_nu} 
\end{eqnarray}

\noindent
The result (\ref{unit_nu}) demonstrates that the null condition (\ref{Null_Condition1}) is satisfied along the light ray at 
future infinity up to terms of the order ${\cal O}\left(c^{-6}\right)$ (note that the metric perturbations $h^{\left(n\right)}_{\alpha\beta}$ vanish at 
infinite distances from the body according to Eq.~(\ref{PM_Expansion_1A})). Furthermore,  
the proof of (\ref{unit_nu}) implies an independent check about the correctness of those time-independent coefficients which arise in (\ref{limit_plus}). 

In (\ref{limit_plus_M}) - (\ref{limit_plus_Q_Q}) it has been taken into account that in the limit $t \rightarrow + \infty$ all terms 
with $\dot{\cal W}_{\left(n\right)}$ vanish, while all terms with $\dot{\cal Z}_{\left(n\right)}$ cancel each other. 
The limits in (\ref{Limit_X_Y}) are given by Eqs.~(\ref{Limit_X_Y_n}) in Appendix~\ref{Appendix_Scalar_Master_Integrals}.  

From (\ref{limit_plus}) one obtains for the  three-vector of total light deflection  
\begin{eqnarray}
	\ve{\sigma} \times \ve{\nu} &=& \ve{\sigma} \times \ve{\nu}^{\rm M}_{\rm 1PN} + \ve{\sigma} \times \ve{\nu}^{\rm Q}_{\rm 1PN}
	+ \ve{\sigma} \times \ve{\nu}^{{\rm M} \times {\rm M}}_{\rm 2PN} + \ve{\sigma} \times \ve{\nu}^{{\rm M} \times {\rm Q}}_{\rm 2PN} 
	+ \ve{\sigma} \times \ve{\nu}^{{\rm Q} \times {\rm Q}}_{\rm 2PN}\;, 
        \label{total_light_deflection}
\end{eqnarray}

\noindent
where the spatial components of the individual terms of the three-vector $\ve{\nu}$ are given by Eqs.~(\ref{limit_plus_M}) - (\ref{limit_plus_Q_Q}). 
The total light deflection (\ref{total_light_deflection}) is a coordinate-independent quantity and irrespective that it   
is calculated in a concrete coordinate system it is, in principle, an observable. The absolute value of the total light deflection 
(\ref{total_light_deflection}) can be estimated by 
\begin{eqnarray}
	&& \hspace{-0.75cm} \big| \ve{\sigma} \times \ve{\nu}\big| \le 
	\big| \ve{\sigma} \times \ve{\nu}^{\rm M}_{\rm 1PN}\big| 
	+ \big| \ve{\sigma} \times \ve{\nu}^{\rm Q}_{\rm 1PN}\big|
	+ \big| \ve{\sigma} \times \ve{\nu}^{{\rm M} \times {\rm M}}_{\rm 2PN} \big| 
	+ \big| \ve{\sigma} \times \ve{\nu}^{{\rm M} \times {\rm Q}}_{\rm 2PN} \big| 
	+ \big| \ve{\sigma} \times \ve{\nu}^{{\rm Q} \times {\rm Q}}_{\rm 2PN}\big| \;.
        \label{total_light_deflection_absolute_value}
\end{eqnarray}

\noindent 
In order to determine the upper limit of each individual term in (\ref{total_light_deflection_absolute_value}) the meaningful assumption is adopted,  
that to a good approximation the giant planets can be considered as axially symmetric bodies, where the STF quadrupole tensor (\ref{Quadrupole}) takes
the following form \cite{Klioner2003b,Zschocke6}:
\begin{eqnarray} 
        \hat{M}_{ab} = M\,J_2\,P^2\,\frac{1}{3}\; {\cal R} \;{\rm diag} \left(+1, +1, -2\right)\, {\cal R}^{\rm T} \;, 
        \label{Quadrupole_Tensor} 
\end{eqnarray}

\noindent
where ${\cal R}$ is the rotational matrix giving the orientation of the unit-vector ${\ve e}_3$ of the symmetry axis of the massive body in the coordinate system,  
$M$ is the mass of the body, $J_2$ is the actual second zonal harmonic coefficient, and $P$ is the equatorial radius of the body. 
The parameters in (\ref{Quadrupole_Tensor}) are given in Table~\ref{Table1} for the giant planets of the Solar System.
\begin{table}[h!]
	\caption{\label{Table1}The numerical parameters: Schwarzschild radius $m = GM/c^2$ (in meter), 
	equatorial radius $P$ (in meter), and the actual second zonal harmonic coefficient $J_2$  
	of giant planets of the Solar System \cite{JPL}. For later purposes in Section~\ref{Enhanced_Terms} the absolute value of the maximal distance 
	$x^{\rm max}_1$ between the giant planet and a Gaia-like observer located at Lagrange point $L_2$ is given in astronomical units (au).}
\footnotesize
\begin{tabular}{@{}|c|c|c|c|c|}
\hline
	Object & $m$ & $P$ & $J_2$ & $x^{\rm max}_1$\\
&&&&\\[-10pt]
\hline
Jupiter & $1.410$ & $ 71.49 \times 10^6$ & $14.697 \cdot 10^{-3}$ & $6$\\
Saturn & $ 0.422 $ & $ 60.27 \times 10^6$ & $16.331 \cdot 10^{-3}$  & $11$ \\
Uranus & $ 0.064 $ & $ 25.56 \times 10^6$ & $3.516 \cdot 10^{-3}$  & $21$ \\
Neptune & $ 0.076 $ & $ 24.76 \times 10^6$ & $3.538 \cdot 10^{-3}$  & $31$ \\
\hline
\end{tabular}\\
\end{table}
\normalsize

\noindent
By inserting (\ref{Quadrupole_Tensor}) into (\ref{limit_plus_M}) - (\ref{limit_plus_Q_Q}) one obtains (details are given in 
Appendices~\ref{Estimation_Q} - \ref{Estimation_Q_Q}):
\begin{eqnarray}
        \big| \ve{\sigma} \times \ve{\nu}^{\rm M}_{\rm 1PN}\big| &\le& 4\,\frac{G M}{c^2}\,\frac{1}{d_{\sigma}} \;,
\label{upper_limits_M}
\\
        \big| \ve{\sigma} \times \ve{\nu}^{\rm Q}_{\rm 1PN}\big| &\le& 4\,\frac{G M}{c^2}\,\left|J_2\right|\,\frac{P^2}{\left(d_{\sigma}\right)^3} \;,
\label{upper_limits_Q}
\\
        \left|\ve{\sigma} \times \ve{\nu}^{{\rm M} \times {\rm M}}_{\rm 2PN}\right| &=& \frac{G^2 M^2}{c^4} 
        \left|\frac{15}{4}\, \frac{\pi}{\left(d_{\sigma}\right)^2} 
        - 8 \,\frac{x_0 + \ve{\sigma} \cdot \ve{x}_0}{\left(d_{\sigma}\right)^3} \right| \,,  
\label{upper_limits_M_M}
\\ 
        \left| \ve{\sigma} \times \ve{\nu}^{{\rm M} \times {\rm Q}}_{\rm 2PN}\right| &\le& 
        4\,\frac{G^2 M^2}{c^4}\,\left|J_2\right|\,\frac{P^2}{\left(d_{\sigma}\right)^2}\,
        \left|\frac{15}{4}\, \frac{\pi}{\left(d_{\sigma}\right)^2} + 8\,\frac{x_0 + \ve{\sigma} \cdot \ve{x}_0}{\left(d_{\sigma}\right)^3} \right|\,, 
\label{upper_limits_M_Q}  
\\
        \left| \ve{\sigma} \times \ve{\nu}^{{\rm Q} \times {\rm Q}}_{\rm 2PN}\right| &\le&
        2\,\frac{G^2 M^2}{c^4}\,\left|J_2\right|^2\,\frac{P^4}{\left(d_{\sigma}\right)^4}\,
        \left|\frac{15}{4}\, \frac{\pi}{\left(d_{\sigma}\right)^2} + 12\,\frac{x_0 + \ve{\sigma} \cdot \ve{x}_0}{\left(d_{\sigma}\right)^3} \right| \,.
\label{upper_limits_Q_Q} 
\end{eqnarray}

\noindent
These upper limits are strictly valid for any astrometric configuration between source, body and observer, that means for 
$\ve{\sigma} \cdot \ve{x}_0 \le 0$ and $\ve{\sigma} \cdot \ve{x}_0 > 0$ as well as for $\ve{\sigma} \cdot \ve{x}_1 \le 0$ and $\ve{\sigma} \cdot \ve{x}_1 > 0$. 
The upper limits (\ref{upper_limits_M}) - (\ref{upper_limits_Q})
take their maximal value in case of a light ray grazing at the surface of the massive body ($d_{\sigma} = P_A$). These maximal numerical values
are presented in Table~\ref{Table2_1PN} for the giant planets of the Solar System.
\begin{table}[h!]
        \caption{\label{Table2_1PN}The upper limits of the individual terms in (\ref{upper_limits_M}) and (\ref{upper_limits_Q}) of the absolute value of total
        light deflection (\ref{total_light_deflection_absolute_value}) for grazing light rays ($d_{\sigma} = P$) at the giant planets of the Solar System.
        All values are given in micro-arcseconds.}
\footnotesize
\begin{tabular}{@{}|c|c|c|}
\hline
&&\\[-7pt]
Object & $\big|\ve{\sigma} \times \ve{\nu}^{\rm M}_{\rm 1PN}\big|$ & $\big|\ve{\sigma} \times \ve{\nu}^{\rm Q}_{\rm 1PN}\big|$ \\
&&\\[-7pt]
\hline
Jupiter & $16272.7$ & $239.2$ \\
Saturn & $5776.9$   & $94.3$  \\
Uranus & $2065.9$   & $7.3$   \\
Neptune & $2532.5$  & $9.0$   \\
\hline
\end{tabular}\\
\end{table}

\noindent 
If the massive body is, so to speak, located between the light source and the observer, then the following condition is valid, 
\begin{eqnarray}
        \ve{\sigma} \cdot \ve{x}_0 &\le& 0  \;,
        \label{configuration1}
\end{eqnarray}

\noindent
a condition that applies with certainty for any stellar light source. 
Then, the upper limits of the 2PN terms (\ref{upper_limits_M_M}) - (\ref{upper_limits_Q_Q}) simplify as follows  
\begin{eqnarray}
	\big| \ve{\sigma} \times \ve{\nu}^{{\rm M} \times {\rm M}}_{\rm 2PN}\big| &\le& 
        \frac{15}{4}\,\pi\,\frac{G^2 M^2}{c^4}\,\frac{1}{\left(d_{\sigma}\right)^2}\;, 
\label{upper_limits_M_M_configuration1}
\\ 
	\big| \ve{\sigma} \times \ve{\nu}^{{\rm M} \times {\rm Q}}_{\rm 2PN}\big| &\le& 
        15\,\pi\,\frac{G^2 M^2}{c^4}\,\left|J_2\right|\,\frac{P^2}{\left(d_{\sigma}\right)^4}\;,
\label{upper_limits_M_Q_configuration1}  
\\
	\big| \ve{\sigma} \times \ve{\nu}^{{\rm Q} \times {\rm Q}}_{\rm 2PN}\big| &\le& 
	\frac{15}{2}\,\pi\,\frac{G^2 M^2}{c^4}\,\left|J_2\right|^2\,\frac{P^4}{\left(d_{\sigma}\right)^6}\;. 
\label{upper_limits_Q_Q_configuration1} 
\end{eqnarray}

\noindent 
The upper limits (\ref{upper_limits_M_M_configuration1}) - (\ref{upper_limits_Q_Q_configuration1}) take their maximal value in case 
of a light ray grazing at the surface of the massive body ($d_{\sigma} = P_A$). These maximal numerical values
for such configurations are presented in Table~\ref{Table2} for the giant planets of the Solar System.
\begin{table}[h!]
	\caption{\label{Table2}The upper limits of the individual terms in (\ref{upper_limits_M_M_configuration1}) - (\ref{upper_limits_Q_Q_configuration1}) 
	of the absolute value of total light deflection (\ref{total_light_deflection_absolute_value}) for grazing light rays ($d_{\sigma} = P$) at the 
	giant planets of the Solar System. All values are given in nano-arcseconds.} 
\footnotesize
\begin{tabular}{@{}|c|c|c|c|}
\hline
&&&\\[-7pt]
	Object & $\left| \ve{\sigma} \times \ve{\nu}^{{\rm M} \times {\rm M}}_{\rm 2PN}\right|$ & $\left| \ve{\sigma} \times \ve{\nu}^{{\rm M} \times {\rm Q}}_{\rm 2PN}\right|$ & $\left| \ve{\sigma} \times \ve{\nu}^{{\rm Q} \times {\rm Q}}_{\rm 2PN}\right|$ \\ 
&&&\\[-7pt]
\hline
Jupiter & $9.4 \times 10^{-1}$ & $5.5 \times 10^{-2}$ & $4.1 \times 10^{-4}$ \\
Saturn  & $1.2 \times 10^{-1}$ & $7.8 \times 10^{-3}$ & $6.4 \times 10^{-5}$ \\
Uranus  & $1.5 \times 10^{-2}$ & $2.1 \times 10^{-4}$ & $3.7 \times 10^{-7}$ \\
Neptune & $2.3 \times 10^{-2}$ & $3.3 \times 10^{-4}$ & $5.8 \times 10^{-7}$ \\   
\hline
\end{tabular}\\ 
\end{table}

\noindent
In view of the small numerical values in Table~\ref{Table2} one might be inclined to believe that the light deflection caused by the quadrupole structure 
of any massive Solar System body is irrelevant in second post-Newtonian approximation for future astrometry missions on the sub-micro-arcsecond level and 
even on the nas-level of accuracy. This is, however, not true, because of the occurrence of so-called {\it enhanced 2PN terms}, which will be considered 
in the following Section.

\section{Enhanced terms}\label{Enhanced_Terms}  

In the previous Section it has been shown that in case of condition (\ref{configuration1}) the 2PN terms in the total light deflection
in Eqs.~(\ref{upper_limits_M_M}) - (\ref{upper_limits_Q_Q}) can be estimated as $\displaystyle {\rm const.} \times \frac{m^2}{d_{\sigma}^2}$. 
Therefore, they are so tiny that they can be neglected even
for high-precision astrometry on the nano-arcsecond level; their numeri\-cal values are given in Table~\ref{Table2} for grazing light rays ($d_{\sigma} = P$).
However, as worked out in several investigations \cite{Article_Zschocke1,Teyssandier,AshbyBertotti2010} there are so-called {\it enhanced terms} of
second post-Newtonian order in the theory of light propagation, whose upper limits are given by 
$\displaystyle {\rm const.} \times \frac{m^2}{d_{\sigma}^2}\,\frac{x_1}{d_{\sigma}}$, where $x_1$ is the distance between massive body and observer 
and $d_{\sigma}$ is the impact vector of the light ray. Hence $\displaystyle \frac{x_1}{d_{\sigma}} \gg 1$ for grazing rays where the impact vector equals 
the radius of the body. It is emphasized that these enhanced terms occur in case of condition (\ref{configuration1}), which is, for instance, always valid for 
light sources far away from the Solar System, like stellar light sources. 

The occurrence of {\it enhanced terms} becomes apparent when the boundary value problem of light propagation is considered instead the initial value problem. 
Let $\left(ct_0,\ve{x}_0\right)$ be the four-coordinates of the light source at the moment of emission of the light signal, and $\left(ct_1, \ve{x}_1\right)$ be 
the four-coordinates of the observer at the moment of reception of the light signal. Then, the boundary value problem of light propagation is determined as 
solution of geodesic equation (\ref{Geodetic_Equation2}) with two boundary conditions 
\begin{eqnarray}
	\ve{x}_0 &=& \ve{x}\left(t\right)\,\,\bigg|_{t = t_0}  \;,
	\label{boundary_0}
	\\
	\ve{x}_1 &=& \ve{x}\left(t\right)\,\,\bigg|_{t = t_1}  \;. 
	\label{boundary_1}
\end{eqnarray}

\noindent
In reality an observer will not be located at infinity but at some finite distance from the massive Solar System body. Therefore, for relativistic modeling of 
real astrometric observations it is not sufficient to determine the initial value problem, defined by Eqs.~(\ref{Initial_B}) and (\ref{Boundary_Condition}), 
but the solution of the boundary value problem, defined by Eqs.~(\ref{boundary_0}) and (\ref{boundary_1}), is required.   
This fact implies that for real astrometric measurements not the unit tangent-vector 
$\ve{\nu}$ at plus infinity is relevant, but the unit tangent-vector $\ve{n}$ at the time of reception, $t_1$, that means at the exact position $\ve{x}_1$ 
of the observer (cf. Fig.~\ref{Diagram}). This unit tangent-vector is defined by 
\begin{eqnarray} 
        \ve{n} = \frac{\dot{\ve{x}}\left(t_1\right)}{\left| \dot{\ve{x}}\left(t_1\right)\right|}\;,  
        \label{vector_n}
\end{eqnarray}

\noindent 
and the sine of the light deflection angle $\delta\left(\ve{\sigma},\ve{n}\right)$ at the observer's position is given by
\begin{eqnarray} 
        \sin \delta\left(\ve{\sigma},\ve{n}\right) &=& \left| \ve{\sigma} \times \ve{n} \right| .  
        \label{light_deflection_vector_n}
\end{eqnarray}

\noindent 
The unit tangent-vector (\ref{vector_n}) is of fundamental importance for astrometry and implies that one has to introduce the impact vector 
$\hat{\ve{d}_{\sigma}}$ defined below by Eq.~(\ref{impact_vector_x1}). Therefore, this impact vector becomes indispensable for modeling of real 
astrometric measurements and we will see that $\hat{\ve{d}_{\sigma}}$ is the reason for the occurrence of {\it enhanced terms}   
in the 2PN light deflection. 

Now we consider the unit vector in (\ref{vector_n}). Inserting (\ref{First_Integration_2PN}) into (\ref{vector_n}) yields for this vector in the 2PN approximation 
\begin{eqnarray} 
        \ve{n} &=& \ve{\sigma} + \ve{n}_{\rm 1PN}\left(\ve{x}_{\rm N}\right) + \ve{n}_{\rm 2PN}\left(\ve{x}_{\rm N}\right) + {\cal O}\left(c^{-6}\right) , 
        \label{vector_n_2PN_A}
\end{eqnarray} 

\noindent 
where $\ve{n}_{\rm 1PN} = {\cal O}\left(c^{-2}\right)$ and $\ve{n}_{\rm 2PN} = {\cal O}\left(c^{-4}\right)$, which read  
\begin{eqnarray} 
        \ve{n}_{\rm 1PN} &=& \frac{\Delta \dot{\ve x}_{\rm 1PN}}{c} - 
        \frac{\left(\ve{\sigma} \cdot \Delta \dot{\ve x}_{\rm 1PN}\right)}{c}\,\ve{\sigma} \;, 
        \label{Appendix_n_1PN_Enhanced}
        \\
        \ve{n}_{\rm 2PN} &=& \frac{\Delta \dot{\ve x}_{\rm 2PN}}{c} - \frac{\left(\ve{\sigma} \cdot \Delta \dot{\ve x}_{\rm 2PN}\right)}{c}\,\ve{\sigma}
	- \frac{\Delta \dot{\ve x}_{\rm 1PN}}{c}\!\left(\frac{\ve{\sigma} \cdot \Delta \dot{\ve x}_{\rm 1PN}}{c}\right) 
	- \frac{1}{2} \!\left(\frac{\Delta \dot{\ve x}_{\rm 1PN}}{c} \cdot \frac{\Delta \dot{\ve x}_{\rm 1PN}}{c}\right) \ve{\sigma}
        + \frac{3}{2} \!\left(\frac{\ve{\sigma} \cdot \Delta \dot{\ve x}_{\rm 1PN}}{c}\right)^2 \ve{\sigma} ,  
        \label{Appendix_n_2PN_Enhanced}
\end{eqnarray}

\noindent 
where $\Delta \dot{\ve x}_{\rm 1PN}$ is given by Eqs.~(\ref{First_Integration_1PN_M_Q}) - (\ref{First_Integration_1PN_Q}), while 
$\Delta \dot{\ve x}_{\rm 2PN}$ is given by Eqs.~(\ref{First_Integration_2PN_MM_MQ_QQ}) - (\ref{First_Integration_2PN_Term_Q_Q}). 

The 1PN terms $\ve{n}_{\rm 1PN}$ and the 2PN terms $\ve{n}_{\rm 2PN}$ in (\ref{vector_n_2PN_A}) depend on the spatial coordinates of the light source $\ve{x}_0$ 
as well as on the spatial coordinates of the unperturbed light signal at the moment of reception: $\ve{x}_{\rm N}\left(t_1\right)$. However, for the boundary 
value problem the solution of the unit tangent-vector $\ve{n}$ must be given in terms of the position of the observer $\ve{x}_1$. But only in the 2PN terms 
$\ve{n}_{\rm 2PN}$ in (\ref{vector_n_2PN_A}) the argument $\ve{x}_{\rm N}\left(t_1\right)$ can be replaced by the spatial position of the observer $\ve{x}_1$ 
because of 
\begin{eqnarray}
        \ve{x}_1 = \ve{x}_{\rm N}\left(t_1\right) + {\cal O} \left(c^{-2}\right) , 
        \label{Replacement_1PN}
\end{eqnarray}

\noindent 
hence such a replacement in $\ve{n}_{\rm 2PN}$ would cause an error of the order ${\cal O}\left(c^{-6}\right)$ in Eq.~(\ref{vector_n_2PN_A}) 
which is in line with the 2PN approximation. However, the argument $\ve{x}_{\rm N}\left(t_1\right)$ of the 1PN terms $\ve{n}_{\rm 1PN}$ 
in (\ref{vector_n_2PN_A}) can not be replaced by $\ve{x}_1$, because such a replacement  
would cause an error of the order ${\cal O}\left(c^{-4}\right)$ in (\ref{vector_n_2PN_A}) which would spoil the 2PN approximation. 
Therefore, one has to rewrite $\ve{n}_{\rm 1PN}$ in Eq.~(\ref{vector_n_2PN_A}) in such a form that it's argument is not 
the spatial coordinate of the unperturbed light signal, $\ve{x}_{\rm N}\left(t_1\right)$, but given by the spatial coordinate of the 
light signal in 1PN approximation (\ref{Second_Integration_1PN}), 
that means $\ve{x}_{\rm 1PN}\left(t_1\right) = \ve{x}_{\rm N}\left(t_1\right) + \Delta \ve{x}_{\rm 1PN}\left(t_1\right)$, 
where $\Delta \ve{x}_{\rm 1PN}$ is given by Eqs.~(\ref{Second_Integration_1PN_M_Q}) - (\ref{Second_Integration_1PN_Q}).
Then Eq.~(\ref{vector_n_2PN_A}) reads 
\begin{eqnarray} 
        \ve{n} &=& \ve{\sigma} + \ve{n}_{\rm 1PN}\left(\ve{x}_{\rm 1PN}\right) + \ve{n}_{\rm 2PN}\left(\ve{x}_{\rm N}\right) 
	+ \Delta \ve{n}_{\rm 2PN}\left(\ve{x}_{\rm N}\right) 
	+ {\cal O}\left(c^{-6}\right), 
        \label{vector_n_2PN_B} 
\end{eqnarray}

\noindent
where $\Delta \ve{n}_{\rm 2PN}$ is defined by
\begin{eqnarray}
        \Delta \ve{n}_{\rm 2PN}\left(\ve{x}_{\rm N}\right) = \ve{n}_{\rm 1PN}\left(\ve{x}_{\rm N}\right) - \ve{n}_{\rm 1PN}\left(\ve{x}_{\rm 1PN}\right) 
	+ {\cal O}\left(c^{-6}\right). 
        \label{Delta_n_2PN}
\end{eqnarray}

\noindent 
The expression for $\ve{n}_{\rm 1PN}\left(\ve{x}_{\rm N}\right)$ and $\ve{n}_{\rm 1PN}\left(\ve{x}_{\rm 1PN}\right)$ are given  
by Eqs.~(\ref{Appendix_n_N}) and (\ref{Appendix_n_1PN}) in Appendix \ref{Appendix_Enhanced_Terms} in its explicit form. 
Now, in Eq.~(\ref{vector_n_2PN_B}) it is allowed to replace the argument in $\ve{n}_{\rm 1PN}\left(\ve{x}_{\rm 1PN}\right)$ by the exact spatial position of the 
observer $\ve{x}_1$, because of 
\begin{eqnarray}
        \ve{x}_1 = \ve{x}_{\rm 1PN}\left(t_1\right) + {\cal O} \left(c^{-4}\right) , 
        \label{Replacement_2PN}
\end{eqnarray}

\noindent 
which would cause an error of the order ${\cal O} \left(c^{-6}\right)$ in (\ref{vector_n_2PN_B}) and would be in line with the 2PN approximation. 
To be consistent with the 2PN approximation, the additional term $\Delta \ve{n}_{\rm 2PN}$ in (\ref{Delta_n_2PN}) has to be series expanded up to terms of the 
order ${\cal O}\left(c^{-6}\right)$. The additional term $\Delta \ve{n}_{\rm 2PN}$ is of second post-Newtonian order ${\cal O}\left(c^{-4}\right)$ 
and, in view of relation (\ref{Replacement_1PN}), in $\Delta \ve{n}_{\rm 2PN}$ one may replace $\ve{x}_{\rm N}\left(t_1\right)$ by $\ve{x}_1$, that means 
the spatial position of the unperturbed light ray at the time of reception by the exact spatial position of the observer. 
Then, the unit tangent vector at the observers position follows from (\ref{vector_n_2PN_B})  
\begin{eqnarray}
        \ve{n} &=& \ve{\sigma} + \ve{n}_{\rm 1PN}\left(\ve{x}_1\right) + \ve{n}_{\rm 2PN}\left(\ve{x}_1\right)  
        + \Delta \ve{n}_{\rm 2PN}\left(\ve{x}_1\right) + {\cal O}\left(c^{-6}\right) . 
        \label{unit_vector_n_2PN}
\end{eqnarray}

\noindent 
Let us consider each individual term in (\ref{unit_vector_n_2PN}).   
The expression $\ve{n}_{\rm 1PN}\left(\ve{x}_1\right)$ in (\ref{unit_vector_n_2PN}) follows from (\ref{Appendix_n_1PN}) when 
$\ve{x}_{\rm 1PN}\left(t_1\right)$ is replaced by $\ve{x}_1$. Then, by inserting the multipole tensor (\ref{Quadrupole_Tensor_in_z_axis}) into 
(\ref{Appendix_n_1PN}) one obtains  
\begin{eqnarray}
	\ve{n}_{\rm 1PN}\left(\ve{x}_1\right) &=& \ve{n}^{\rm M}_{\rm 1PN}\left(\ve{x}_1\right) + \ve{n}^{\rm Q}_{\rm 1PN}\left(\ve{x}_1\right), 
\label{unit_vector_n_1PN}
\\
	\ve{n}^{\rm M}_{\rm 1PN}\left(\ve{x}_1\right) &=& - 2\,\frac{G M}{c^2} \left(1 + \frac{\ve{\sigma} \cdot \ve{x}_1}{x_1}\right) 
	\frac{\hat{\ve{d}_{\sigma}}}{\left(\hat{d}_{\sigma}\right)^2} \;,
	\label{unit_vector_n_1PN_M} 
	\\
	\ve{n}^{\rm Q}_{\rm 1PN}\left(\ve{x}_1\right) &=& - 2\,\frac{G M}{c^2}\,J_2\,\frac{P^2}{\left(\hat{d}_{\sigma}\right)^3}
	\left(1 + \frac{\ve{\sigma} \cdot \ve{x}_1}{x_1}\right) 
	\left[\!\left(1 - \left(\ve{\sigma} \cdot \ve{e}_3\right)^2  
	- 4 \frac{\left(\hat{\ve{d}_{\sigma}} \cdot \ve{e}_3\right)^2}{\left(\hat{d}_{\sigma}\right)^2} \right)\frac{\hat{\ve{d}_{\sigma}}}{\hat{d}_{\sigma}} 
	+ 2 \, \frac{\hat{\ve{d}_{\sigma}} \cdot \ve{e}_3}{\hat{d}_{\sigma}}\,\ve{e}_3 \right] 
	\nonumber\\ 
	&& +\,{\cal O} \left(\frac{G M}{c^2}\,\frac{J_2}{\hat{d}_{\sigma}}\,\frac{P^2}{\left(x_1\right)^2}\right). 
	\label{unit_vector_n_1PN_Q} 
\end{eqnarray}

\noindent
In (\ref{unit_vector_n_1PN_Q}) terms proportional to $\ve{\sigma}$ have been omitted because they do not contribute to the light deflection angle in 
(\ref{light_deflection_vector_n}), while terms proportional to 
$\displaystyle \frac{P^2}{\left(x_1\right)^2} \ll 1$ have been neglected because they turn out to be much 
smaller than $1\,{\rm nas}$. 
In the impact vector (\ref{Appendix_impact_vector_x1}) the spatial coordinate of the light signal in 1PN approximation $\ve{x}_{\rm 1PN}\left(t_1\right)$ 
can be replaced by the exact position of the observer $\ve{x}_1$ because of relation (\ref{Replacement_2PN}). 
Thus, the impact vector in (\ref{unit_vector_n_1PN_M}) - (\ref{unit_vector_n_1PN_Q}) can be written in the form  
\begin{eqnarray}
        \hat{\ve{d}_{\sigma}} &=& \left(\ve{\sigma} \times \left( \ve{x}_1 \times \ve{\sigma} \right) \right)  
	\quad {\rm and} \quad \hat{d}_{\sigma} = \left| \hat{\ve{d}_{\sigma}}\right|
        \label{impact_vector_x1}
\end{eqnarray}

\noindent 
up to terms of the order ${\cal O}\left(c^{-4}\right)$ which would cause an error of the order ${\cal O}\left(c^{-6}\right)$ in (\ref{unit_vector_n_1PN_M}) 
and (\ref{unit_vector_n_1PN_Q}) in line with the 2PN approximation; note that $\ve{\sigma} \cdot \hat{\ve{d}_{\sigma}} = 0$. 
The impact parameter $\hat{d}_{\sigma}$ is also larger than or equal to the equatorial radius $P$ of the massive Solar System body 
\begin{eqnarray}
	\hat{d}_{\sigma} \ge P \;.  
        \label{impact_parameter_condition_B}
\end{eqnarray}

\noindent
The expression of the additional terms $\Delta \ve{n}_{\rm 2PN}\left(\ve{x}_1\right)$ in (\ref{unit_vector_n_2PN}) is given by  
(see Eqs.~(\ref{Appendix_Enhanced_Terms_5_MM}) - (\ref{Appendix_Enhanced_Terms_5_QQ}) in Appendix \ref{Appendix_Enhanced_Terms})  
\begin{eqnarray}
	\Delta \ve{n}_{\rm 2PN} &=& \Delta \ve{n}_{\rm 2PN}^{{\rm M} \times {\rm M}} 
	+ \Delta \ve{n}_{\rm 2PN}^{{\rm M} \times {\rm Q}} + \Delta \ve{n}_{\rm 2PN}^{{\rm Q} \times {\rm Q}} \;,  
        \label{Enhanced_Terms_5}
        \\
	 \Delta \ve{n}_{\rm 2PN}^{{\rm M} \times {\rm M}}\left(\ve{x}_1\right)  &=& 
	 4\,\frac{G^2 M^2}{c^4}\,\frac{1}{\left(\hat{d}_{\sigma}\right)^2} \left(1 + \frac{\ve{\sigma} \cdot \ve{x}_1}{x_1}\right)  
	 \frac{x_1 + \ve{\sigma} \cdot \ve{x}_1 - x_0 - \ve{\sigma} \cdot \ve{x}_0}{\hat{d}_{\sigma}} 
	 \frac{\hat{\ve{d}_{\sigma}}}{\hat{d}_{\sigma}}\,,
         \label{Enhanced_Terms_5_MM}
	 \\
	 \nonumber\\ 
         \Delta \ve{n}_{\rm 2PN}^{{\rm M} \times {\rm Q}}\left(\ve{x}_1\right)  &=& 
	 16\,\frac{G^2 M^2}{c^4}\,J_2\,\frac{P^2}{\left(\hat{d}_{\sigma}\right)^4} 
	 \left(1 + \frac{\ve{\sigma} \cdot \ve{x}_1}{x_1}\right) 
	 \frac{x_1 + \ve{\sigma} \cdot \ve{x}_1 - x_0 - \ve{\sigma} \cdot \ve{x}_0}{\hat{d}_{\sigma}} 
	 \nonumber\\ 
	&& \times 
	\Bigg[ \left(1 -  \left(\ve{\sigma} \cdot \ve{e}_3\right)^2 - 3\,\bigg(\frac{\hat{\ve{d}_{\sigma}} \cdot \ve{e}_3}{\hat{d}_{\sigma}}\bigg)^2 \right)  
	\frac{\hat{\ve{d}_{\sigma}}}{\hat{d}_{\sigma}} 
	+ \frac{\hat{\ve{d}_{\sigma}} \cdot \ve{e}_3}{\hat{d}_{\sigma}} \; \ve{e}_3  
        \Bigg] ,  
        \label{Enhanced_Terms_5_MQ}
        \\
	 \Delta \ve{n}_{\rm 2PN}^{{\rm Q} \times {\rm Q}}\left(\ve{x}_1\right) &=&
	 12 \,\frac{G^2 M^2}{c^4} \,\left(J_2\right)^2\,\frac{P^4}{\left(\hat{d}_{\sigma}\right)^6} 
        \left(1 + \frac{\ve{\sigma} \cdot \ve{x}_1}{x_1}\right)    
	 \frac{x_1 + \ve{\sigma} \cdot \ve{x}_1 - x_0 - \ve{\sigma} \cdot \ve{x}_0}{\hat{d}_{\sigma}} 
	\left(1 - \left(\ve{\sigma} \cdot \ve{e}_3 \right)^2\right)^2 \,\frac{\hat{\ve{d}_{\sigma}}}{\hat{d}_{\sigma}}  \;,   
        \label{Enhanced_Terms_5_QQ}
\end{eqnarray}

\noindent
where in (\ref{Enhanced_Terms_5_MQ}) a term proportional to $\ve{\sigma}$ has been omitted because it does not contribute to the light deflection 
(\ref{light_deflection_vector_n}). In what follows it is shown that these terms are the so-called {\it enhanced 2PN terms}, because they are 
proportional to the large factor $\displaystyle \frac{x_1}{\hat{d}_{\sigma}}$, where $x_1$ is the distance between body and observer and 
$\hat{d}_{\sigma}$ equals the radius of the body for grazing light rays. 

In Appendix~\ref{Appendix_Enhanced_Terms} it is shown that these {\it enhanced 2PN terms} in Eqs.~(\ref{Enhanced_Terms_5_MM}) - (\ref{Enhanced_Terms_5_QQ}) 
are solely caused by the new impact vector (\ref{impact_vector_x1}) which is indispensable for modeling of real 
astrometric observations; cf. text below Eq.~(\ref{Appendix_Enhanced_Terms_5_QQ}) in Appendix~\ref{Appendix_Enhanced_Terms}.
These expressions are obtained from Eqs.~(\ref{Appendix_Enhanced_Terms_5_MM}) - (\ref{Appendix_Enhanced_Terms_5_QQ}) by 
replacing the spatial coordinate of the unperturbed light ray $\ve{x}_{\rm N}\left(t_1\right)$ by the spatial coordinates of the observer $\ve{x}_1$. Such 
replacement causes an error of the order ${\cal O}\left(c^{-6}\right)$, hence is in line with the 2PN approximation. 
We notice that in (\ref{Enhanced_Terms_5_MM}) - (\ref{Enhanced_Terms_5_QQ}) the impact vector $\ve{d}_{\sigma}$ has been replaced by $\hat{\ve{d}_{\sigma}}$ 
which is in line with the 2PN approximation, because of 
\begin{eqnarray} 
	\hat{\ve{d}_{\sigma}}  &=& \ve{d}_{\sigma} + {\cal O}\left(c^{-2}\right)
	\label{Relation_20} 
\end{eqnarray} 

\noindent
which results from (\ref{Replacement_1PN}). 
Finally, the term $\ve{n}_{\rm 2PN}\left(\ve{x}_1\right)$ in (\ref{unit_vector_n_2PN}) is formally given by Eq.~(\ref{Appendix_n_2PN_Enhanced}), 
when $\ve{x}_{\rm N}\left(t_1\right)$ is replaced by $\ve{x}_1$.  

By inspection of the unit vector in (\ref{unit_vector_n_2PN}), one obtains for the upper limit of the light deflection (\ref{light_deflection_vector_n}) 
\begin{eqnarray}
        \big| \ve{\sigma} \times \ve{n} \big| &\le& \big| \ve{\sigma} \times \ve{n}_{\rm 1PN}\left(\ve{x}_1\right) \big|
        + \big| \ve{\sigma} \times \ve{n}_{\rm 2PN}\left(\ve{x}_1\right) \big| 
	+ \big| \ve{\sigma} \times \Delta \ve{n}_{\rm 2PN}\left(\ve{x}_1\right) \big| 
	+ {\cal O}\left(c^{-6}\right) . 
        \label{light_deflection_sigma_n_5}
\end{eqnarray}

\noindent
Then, by using (\ref{unit_vector_n_1PN}) - (\ref{unit_vector_n_1PN_Q}) and (\ref{Enhanced_Terms_5}) - (\ref{Enhanced_Terms_5_QQ}), 
as well as (\ref{Appendix_n_2PN_Enhanced}) with (\ref{First_Integration_2PN_M_Q}) - (\ref{First_Integration_2PN_MM_MQ_QQ}) one obtains 
for the upper limit (\ref{light_deflection_sigma_n_5}) of the light deflection the following expression:   
\begin{eqnarray}
	\big| \ve{\sigma} \times \ve{n} \big| &\le& \big| \ve{\sigma} \times \ve{n}^{\rm M}_{\rm 1PN}\left(\ve{x}_1\right) \big| 
	+ \big| \ve{\sigma} \times \ve{n}^{\rm Q}_{\rm 1PN}\left(\ve{x}_1\right) \big|
	+ |\ve{\sigma}\times\ve{n}^{{\rm M}\times{\rm M}}_{\rm 2PN}\!\left(\ve{x}_1\right)\!\big|
	+ |\ve{\sigma}\times\ve{n}^{{\rm M}\times{\rm Q}}_{\rm 2PN}\!\left(\ve{x}_1\right)\!\big|
	+ |\ve{\sigma}\times\ve{n}^{{\rm Q}\times{\rm Q}}_{\rm 2PN}\!\left(\ve{x}_1\right)\!\big|
	\nonumber\\ 
	&&  
	+ \big| \ve{\sigma} \times \Delta \ve{n}^{{\rm M} \times {\rm M}}_{\rm 2PN}\left(\ve{x}_1\right) \big|
	+ \big| \ve{\sigma} \times \Delta \ve{n}^{{\rm M} \times {\rm Q}}_{\rm 2PN}\left(\ve{x}_1\right) \big|
	+ \big| \ve{\sigma} \times \Delta \ve{n}^{{\rm Q} \times {\rm Q}}_{\rm 2PN}\left(\ve{x}_1\right) \big| 
	+ {\cal O}\left(c^{-6}\right) .  
	\label{light_deflection_sigma_n}
\end{eqnarray}

\noindent
The 2PN terms in the second line of (\ref{light_deflection_sigma_n}) are given by Eqs.~(\ref{Appendix_Proof_n2_MM}) - (\ref{Appendix_Proof_n2_QQ}) 
in Appendix~\ref{Proof}. The upper limits of each individual term in Eq.~(\ref{light_deflection_sigma_n}) read 
\begin{eqnarray}
	\left| \ve{\sigma} \times \ve{n}^{\rm M}_{\rm 1PN}\right| &\le& 4\,\frac{G M}{c^2}\,\frac{1}{\hat{d}_{\sigma}} \;, 
\label{upper_limits_1PN_M}
\\
\left| \ve{\sigma} \times \ve{n}^{\rm Q}_{\rm 1PN}\right| &\le& 4\,\frac{G M}{c^2}\,\left|J_2\right|\,\frac{P^2}{\left(\hat{d}_{\sigma}\right)^3} \;,  
\label{upper_limits_1PN_Q}
\\
\bigg| \ve{\sigma} \times \ve{n}^{{\rm M} \times {\rm M}}_{\rm 2PN} \bigg| 
	&\le& \frac{G^2 M^2}{c^4}  
        \left| \frac{15}{4}\, \frac{\pi}{\left(\hat{d}_{\sigma}\right)^2} 
	- 8 \,\frac{x_0 + \ve{\sigma} \cdot \ve{x}_0}{\left(\hat{d}_{\sigma}\right)^3} \right| , 
        \label{upper_limit_n2_MM}
	\\
        \left| \ve{\sigma} \times \ve{n}^{{\rm M} \times {\rm Q}}_{\rm 2PN}\right| &\le& 
	4\,\frac{G^2 M^2}{c^4}\,\left|J_2\right|\,\frac{P^2}{\left(\hat{d}_{\sigma}\right)^2}\,
	\left| \frac{15}{4}\, \frac{\pi}{\left(\hat{d}_{\sigma}\right)^2} + 8\,\frac{x_0 + \ve{\sigma} \cdot \ve{x}_0}{\left(\hat{d}_{\sigma}\right)^3} \right| , 
\\
        \label{upper_limit_n2_MQ}
        \left| \ve{\sigma} \times \ve{n}^{{\rm Q} \times {\rm Q}}_{\rm 2PN}\right| &\le&
	2\,\frac{G^2 M^2}{c^4}\,\left|J_2\right|^2\,\frac{P^4}{\left(\hat{d}_{\sigma}\right)^4}\,
	\left| \frac{15}{4}\, \frac{\pi}{\left(\hat{d}_{\sigma}\right)^2} + 12\,\frac{x_0 + \ve{\sigma} \cdot \ve{x}_0}{\left(\hat{d}_{\sigma}\right)^3} \right| , 
        \label{upper_limit_n2_QQ}
\\
	\left| \ve{\sigma} \times \Delta \ve{n}_{\rm 2PN}^{{\rm M} \times {\rm M}}\right| &\le& 16\,\frac{G^2 M^2}{c^4}\,
	\frac{1}{\left(\hat{d}_{\sigma}\right)^2} \, 
	\frac{x_1}{\hat{d}_{\sigma}} \;, 
\label{Enhanced_Terms_upper_limits_M_M}
\\ 
	\left| \ve{\sigma} \times \Delta \ve{n}_{\rm 2PN}^{{\rm M} \times {\rm Q}}\right| &\le& 64\,\frac{G^2 M^2}{c^4}\,\left|J_2\right|\,
	\frac{P^2}{\left(\hat{d}_{\sigma}\right)^4} \, 
	\frac{x_1}{\hat{d}_{\sigma}} \;, 
\label{Enhanced_Terms_upper_limits_M_Q}  
\\
	\left| \ve{\sigma} \times \Delta \ve{n}_{\rm 2PN}^{{\rm Q} \times {\rm Q}} \right| &\le& 48\,\frac{G^2 M^2}{c^4}\,\left|J_2\right|^2\,
	\frac{P^4}{\left(\hat{d}_{\sigma}\right)^6}\,\frac{x_1}{\hat{d}_{\sigma}} \;.  
\label{Enhanced_Terms_upper_limits_Q_Q} 
\end{eqnarray}

\noindent
These upper limits are strictly valid for any astrometric configuration, that means for $\ve{\sigma} \cdot \ve{x}_0 \le 0$ and
$\ve{\sigma} \cdot \ve{x}_0 > 0$ as well as for $\ve{\sigma} \cdot \ve{x}_1 \le 0$ and $\ve{\sigma} \cdot \ve{x}_1 > 0$.

The estimation for the 1PN terms in (\ref{upper_limits_1PN_M}) - (\ref{upper_limits_1PN_Q}) are not complicated; for instance 
(\ref{upper_limits_1PN_Q}) is shown by a very similar procedure as considered in Appendix \ref{Estimation_Q}, while the upper limits for the 2PN terms 
(\ref{upper_limit_n2_MM}) - (\ref{upper_limit_n2_QQ}) are considered in Appendix~\ref{Proof}.  
The upper limits for the {\it enhanced terms} in (\ref{Enhanced_Terms_upper_limits_M_M}) - (\ref{Enhanced_Terms_upper_limits_Q_Q})
follow straightforward from their expressions in (\ref{Enhanced_Terms_5_MM}) - (\ref{Enhanced_Terms_5_QQ}) by using the relation (see endnote  
\footnote{If $\ve{\sigma} \cdot \ve{x}_0 \le 0$ then $\displaystyle x_0 + \ve{\sigma} \cdot \ve{x}_0 \le \frac{\left(d_{\sigma}\right)^2}{x_0} \le d_{\sigma}$ 
which follows from  
(\ref{Relation_5}) with the minus-sign; note that $1 - \sqrt{1 - a^2} \le a^2$ for $0 \le a \le 1$. 
If $\ve{\sigma} \cdot \ve{x}_0 > 0$ then $x_0 \le x_1$ 
which follows from geometrical considerations (light signal must reach the observer's position). These facts allow us to show  
(\ref{Relation_25}).})  
\begin{eqnarray}
0 \le 	x_1 + \ve{\sigma} \cdot \ve{x}_1 - x_0 - \ve{\sigma} \cdot \ve{x}_0 \le x_1 + \ve{\sigma} \cdot \ve{x}_1\;, 
\label{Relation_25}
\end{eqnarray}

\noindent 
which is strictly valid for any astrometric configuration, that means for $\ve{\sigma} \cdot \ve{x}_0 \le 0$ and $\ve{\sigma} \cdot \ve{x}_0 > 0$ 
as well as $\ve{\sigma} \cdot \ve{x}_1 \le 0$ and $\ve{\sigma} \cdot \ve{x}_1 > 0$. 
Furthermore, for the upper limit (\ref{Enhanced_Terms_upper_limits_M_Q}) relation (\ref{Relation_10}) has been applied.  

Now let us consider the {\it enhanced terms} in (\ref{Enhanced_Terms_upper_limits_M_M}) - (\ref{Enhanced_Terms_upper_limits_Q_Q}). These terms are proportional to 
the large factor $\displaystyle \frac{x_1}{\hat{d_{\sigma}}}$, but this fact does not imply that these terms are divergent because their validity is restricted 
to the case of realistic observers located in the Solar System or at least having a distance of less than several hundred ${\rm au}$ from the massive Solar System 
body (see text below Eq.~(\ref{Appendix_x_F})). 

The {\it enhanced term} in (\ref{Enhanced_Terms_upper_limits_M_M}) has already been obtained 
in case of 2PN light propagation in the monopole field some time ago \cite{Article_Zschocke1,Teyssandier,AshbyBertotti2010};  
e.g. relation (100) in \cite{Article_Zschocke1} or relation(66) in \cite{Teyssandier} as well as Section $9$ in \cite{AshbyBertotti2010}.  
The estimations in (\ref{Enhanced_Terms_upper_limits_M_Q}) and (\ref{Enhanced_Terms_upper_limits_Q_Q}) are new results which  
demonstrate that {\it enhanced terms} do also arise in the case of 2PN light propagation in the quadrupole field.  
As mentioned, these {\it enhanced terms} originate 
because the light deflection in (\ref{light_deflection_vector_n}) is expressed in terms of the new impact vector (\ref{impact_vector_x1}). 

There is an important difference between the 2PN terms in (\ref{upper_limit_n2_MM}) - (\ref{upper_limit_n2_QQ}) and the 
{\it enhanced 2PN terms} in (\ref{Enhanced_Terms_upper_limits_M_M}) - (\ref{Enhanced_Terms_upper_limits_Q_Q}). 
Namely, if one considers again the most relevant astrometric configurations, where the body is, so to speak, located between the light source and the observer, 
that means where the condition (\ref{configuration1}) holds, then the 2PN terms in (\ref{upper_limit_n2_MM}) - (\ref{upper_limit_n2_QQ}) simplify as follows   
\begin{eqnarray}
        \left| \ve{\sigma} \times \ve{n}^{{\rm M} \times {\rm M}}_{\rm 2PN}\right| &\le& 
	\frac{15}{4}\,\pi\,\frac{G^2 M^2}{c^4}\,\frac{1}{\left(\hat{d}_{\sigma}\right)^2}\;, 
\label{upper_limit_n2_MM_configuration1}
\\ 
        \left| \ve{\sigma} \times \ve{n}^{{\rm M} \times {\rm Q}}_{\rm 2PN}\right| &\le& 
	15\,\pi\,\frac{G^2 M^2}{c^4}\,\left|J_2\right|\,\frac{P^2}{\left(\hat{d}_{\sigma}\right)^4}\;,
\label{upper_limit_n2_MQ_configuration1}  
\\
        \left| \ve{\sigma} \times \ve{n}^{{\rm Q} \times {\rm Q}}_{\rm 2PN}\right| &\le& 
	\frac{15}{2}\,\pi\,\frac{G^2 M^2}{c^4}\,\left|J_2\right|^2\,\frac{P^4}{\left(\hat{d}_{\sigma}\right)^6}\;. 
\label{upper_limit_n2_QQ_configuration1} 
\end{eqnarray}

\noindent 
These terms are smaller than $1\,{\rm nas}$ for grazing light rays at the giant planets, hence they are negligible for astrometry on the sub-micro-arcsecond 
level and nano-arcsecond level of accuracy. In contrast, the upper limits for the 1PN terms (\ref{upper_limits_1PN_M}) - (\ref{upper_limits_1PN_Q}) and the 
enhanced 2PN terms (\ref{Enhanced_Terms_upper_limits_M_M}) - (\ref{Enhanced_Terms_upper_limits_Q_Q}) remain unaffected and keep their validity under condition 
$\ve{\sigma} \cdot \ve{x}_0 \le 0$.  

For grazing light rays, $\hat{d}_{\sigma} = P_A$, the numerical magnitude of the upper limits (\ref{upper_limits_1PN_M}) and (\ref{upper_limits_1PN_Q}) 
coincide with (\ref{upper_limits_M}) and (\ref{upper_limits_Q}), respectively. Their numerical values were already given in Table~\ref{Table2_1PN}.
Similarly, in order to quantify the numerical magnitude of the upper limits (\ref{Enhanced_Terms_upper_limits_M_M}) - (\ref{Enhanced_Terms_upper_limits_Q_Q}) 
we consider a light ray grazing at the surface of the massive body: $\hat{d}_{\sigma} = P_A$. 
The numerical values of these upper limits for such configurations are presented in Table~\ref{Table3} for the giant planets of the Solar System.
\begin{table}[h!]
	\caption{\label{Table3}The upper limits of the {\it enhanced 2PN terms} of light deflection given by
        Eqs.~(\ref{Enhanced_Terms_upper_limits_M_M}) - (\ref{Enhanced_Terms_upper_limits_Q_Q}) for grazing light rays
        ($\hat{d}_{\sigma} = P$) at the giant planets of the Solar System. For the distance $x_1$ a Gaia-like observer at Lagrange point $L_2$
        is assumed and the maximal distances $x^{\rm max}_1$ between the giant planets and observer are used (see parameters given in Table~\ref{Table1}).
	All values are given in micro-arcseconds.}
\footnotesize
\begin{tabular}{@{}|c|c|c|c|}
\hline
&&&\\[-7pt]
	Object & $\left| \ve{\sigma} \times \Delta \ve{n}_{\rm 2PN}^{{\rm M} \times {\rm M}} \right|$ & $\left| \ve{\sigma} \times \Delta \ve{n}_{\rm 2PN}^{{\rm M} \times {\rm Q}}\right|$ & $\left| \ve{\sigma} \times \Delta \ve{n}_{\rm 2PN}^{{\rm Q} \times {\rm Q}}\right|$ \\
&&&\\[-7pt]
\hline
	Jupiter & $16.11$ & $0.95$ & $0.010$ \\
	Saturn  & $4.42$ &  $0.29$ & $0.003$ \\
	Uranus  & $2.58$ &  $0.04$ & $0.001$ \\
	Neptune & $5.83$ &  $0.08$ & $0.002$ \\
\hline
\end{tabular}\\  
\end{table}

\noindent 
The upper limit of the {\it enhanced 2PN term} in (\ref{Enhanced_Terms_upper_limits_M_M}) caused by the monopole structure is considerably larger 
(for numerical values see Table~\ref{Table3}) than the upper limit of the 2PN monopole term of the total light deflection in (\ref{upper_limits_M_M_configuration1}) 
(for numerical values see Table~\ref{Table2}) and 
is already relevant on the \muas-level of accuracy \cite{Article_Zschocke1,Teyssandier,AshbyBertotti2010}. 
This fact has been taken into account in the refined model GREM \cite{Klioner2003b,Article_Zschocke1} which is in use for data reduction of the ESA 
astrometry mission Gaia \cite{GAIA,GAIA_ESA,GAIA1}. 

Similarly, the upper limits of the {\it enhanced 2PN terms} in (\ref{Enhanced_Terms_upper_limits_M_Q}) and (\ref{Enhanced_Terms_upper_limits_Q_Q}) caused 
by the quadrupole structure are considerably larger (for numerical values see Table~\ref{Table3}) than the upper limits of the 2PN quadrupole terms in 
(\ref{upper_limits_M_Q_configuration1}) and (\ref{upper_limits_Q_Q_configuration1}) of the total light deflection (for numerical values see Table~\ref{Table2}). 

The {\it enhanced 2PN terms} of the quadrupole amount up to $0.95\,\muas$ and $0.29\,\muas$ for grazing rays at Jupiter and Saturn, hence they are still below 
the present state-of-the-art in astrometric measurements. For instance, the aimed accuracy of the ESA astrometry mission {\it Gaia} 
is about $5\,\muas$ in the final data release getting published after the end of mission expected around $2025$ \cite{Gaia_Archive}. 
But in view of the impressive advancements of astrometric science during recent decades it is certain that future ground-based or space-based astrometric 
measurements aiming at the sub-micro-arcsecond level will be able to detect the quadrupole light deflection effect in 2PN approximation.

\section{Summary and Outlook}\label{Summary}  

The impressive progress in astrometric science makes it necessary to consider several subtle effects in the 
theory of light propagation. In particular, for high precision astrometry on the sub-micro-arcsecond level of accuracy 
2PN effects need to be investigated. Thus far, however, only the impact of the monopole structure of massive bodies on 
2PN light propagation has been considered in the literature. These 2PN effects of the monopole case have thoroughly been studied 
during recent years and in meanwhile a detailed understanding of these 2PN effects has been achieved. 

On the other side, actually nothing is known about 2PN light deflection effects beyond the monopole structure of the massive bodies. 
For determining the geodesic equation in 2PN approximation one needs the metric of one body at rest in 2PN approximation. 
For that, the Multipole Post-Minkowskian (MPM) formalism has been used, which  
allows to determine the gravitational field of some massive body being of arbitrary shape, inner structure 
and rotational motion. According to the MPM approach, the gravitational field of some massive body is 
described by a multipole decomposition in terms of symmetric and trace-free mass-multipoles $\hat{M}_L$ (shape and inner structure of the body) 
and spin-multipoles $\hat{S}_L$ (rotational motions and inner circulations of the body). 

In this investigation we have considered the problem of light propagation in 2PN approximation in the field of one body at rest with 
time-independent monopole and quadrupole structure, $\hat{M}_L = \{M, \hat{M}_{ab}\}$, while higher mass-multipoles as well as spin-multipoles 
of the body have been neglected. The primary results of this investigation are the following:  

${\bf 1.}$  
The coordinate velocity (first integration of geodesic equation) 
and the trajectory (second integration of geodesic equation) 
of a light signal has been determined in 2PN approximation, which propagates in the gravitational field 
of one body at rest, where the monopole structure and the quadrupole structure of the body has been taken into account. 
The first integration and the second integration can be reduced in each case to only four master integrals, 
Eqs.~(\ref{Master_Integral_I1}) - (\ref{Master_Integral_L1}) and Eqs.~(\ref{Master_Integral_I2}) - (\ref{Master_Integral_L2}).  
These integrals can be solved in closed form to any order by recurrence relations as given by Eqs.~(\ref{Master_Integral_A}) - (\ref{Master_Integral_D}) and 
Eqs.~(\ref{Master_Integral_W}) - (\ref{Master_Integral_Z}).    

${\bf 2.}$  
The solutions for the first and second integration of geodesic equation are given by 
\begin{eqnarray}
	\frac{\dot{\ve{x}}\left(t\right)}{c} &=& \ve{\sigma} + \frac{\Delta\dot{\ve{x}}_{\rm 1PN}}{c} 
	+  \frac{\Delta \dot{\ve{x}}_{\rm 2PN}}{c} + {\cal O}\left(c^{-6}\right) , 
        \label{Summary_5}
\\
	\ve{x}\left(t\right) &=& \ve{x}_{\rm N} + \Delta\ve{x}_{\rm 1PN}  
        + \Delta\ve{x}_{\rm 2PN} 
	+ {\cal O}\left(c^{-6}\right) ,  
        \label{Summary_10} 
\end{eqnarray}

\noindent
where the 1PN and 2PN corrections $\Delta\dot{\ve{x}}_{\rm 1PN}$ and $\Delta\dot{\ve{x}}_{\rm 2PN}$ are given by 
Eqs.~(\ref{First_Integration_1PN_M_Q}) - (\ref{First_Integration_1PN_Q}) and Eqs.~(\ref{First_Integration_2PN_MM_MQ_QQ}) - (\ref{First_Integration_2PN_Term_Q_Q}), 
respectively, while the 1PN and 2PN corrections $\Delta\ve{x}_{\rm 1PN}$ and $\Delta\ve{x}_{\rm 2PN}$ are given by 
Eqs.~(\ref{Second_Integration_1PN_M_Q}) - (\ref{Second_Integration_1PN_Q}) 
and Eqs.~(\ref{Second_Integration_2PN_MM_MQ_QQ}) - (\ref{Second_Integration_2PN_Term_Q_Q}), respectively. 
These solutions are valid for any astrometric configuration between source, body, and observer, that means for 
$\ve{\sigma} \cdot \ve{x}_0 \le 0$ and $\ve{\sigma} \cdot \ve{x}_0 > 0$ as well as $\ve{\sigma} \cdot \ve{x}_1 \le 1$ and $\ve{\sigma} \cdot \ve{x}_1 > 0$. 

${\bf 3.}$  
Using the given solution of the geodesic equation in 2PN approximation, the total light deflection has been determined in Eq.~(\ref{total_light_deflection}). 
The upper limits of the 1PN terms of the total light deflection are given by Eqs.~(\ref{upper_limits_M}) - (\ref{upper_limits_Q}) and their numerical 
values are given in Table~\ref{Table2_1PN}. The upper limits of the 2PN terms of the total light deflection are given by 
Eqs.~(\ref{upper_limits_M_M}) - (\ref{upper_limits_Q_Q}). These upper limits are valid for any astrometric configuration between light source, massive body, 
and observer. For configurations with $\ve{\sigma} \cdot \ve{x}_0 \le 0$ (body is between source and observer) the upper limits of the 2PN terms of 
total light deflection are given by the inequalities (\ref{upper_limits_M_M_configuration1}) - (\ref{upper_limits_Q_Q_configuration1}) 
and turn out to be less than $1$ nano-arcsecond as presented in Table~\ref{Table2}.

${\bf 4.}$  
In view of the smallness of the 2PN terms in the total light deflection  
one might believe that the 2PN quadrupole corrections are negligible even for high-precision astrometry on the sub-micro-arcsecond level of accuracy. 
This is, however, not true. It is a well-known fact that there are {\it enhanced terms} for the case of 2PN light deflection in the gravitational field 
of one monopole at rest, which are proportional to $x_1/\hat{d}_{\sigma}$ and given by the inequality (\ref{Enhanced_Terms_upper_limits_M_M}),  
\begin{eqnarray} 
        \left| \ve{\sigma} \times \Delta \ve{n}_{\rm 2PN}^{{\rm M} \times {\rm M}}\right| &\le& 16\,\frac{G^2 M^2}{c^4}\,
        \frac{1}{\left(\hat{d}_{\sigma}\right)^2} \,
        \frac{x_1}{\hat{d}_{\sigma}} \;,
\label{Summary_Enhanced_Terms_M_M}
\end{eqnarray}

\noindent 
where $x_1$ is the distance between massive body and observer and $\hat{d}_{\sigma}$ is the impact vector of the light ray. Hence $x_1/\hat{d}_{\sigma} \gg 1$ for 
grazing rays where the impact vector equals the radius of the body. These {\it enhanced 2PN terms} amount up to $16$ micro-arcseconds for a grazing light ray at 
Jupiter and $4$ micro-arcseconds for a grazing light ray at Saturn (cf. Table~\ref{Table3}), besides that the total 2PN light deflection 
for one monopole at rest is less than $1$ nano-arcsecond for grazing light rays at Jupiter \cite{Article_Zschocke1,Teyssandier,AshbyBertotti2010}. 

Our considerations have revealed that there exist similar {\it enhanced 2PN terms} in the light deflection caused by the quadrupole structure of the body, 
which are also proportional to the large factor $x_1/\hat{d}_{\sigma}$. The upper limits of these {\it enhanced 2PN terms} are given by the inequalities 
(\ref{Enhanced_Terms_upper_limits_M_Q}) - (\ref{Enhanced_Terms_upper_limits_Q_Q}),  
\begin{eqnarray} 
        \left| \ve{\sigma} \times \Delta \ve{n}_{\rm 2PN}^{{\rm M} \times {\rm Q}}\right| &\le& 64\,\frac{G^2 M^2}{c^4}\,\left|J_2\right|\,
        \frac{P^2}{\left(\hat{d}_{\sigma}\right)^4} \,
        \frac{x_1}{\hat{d}_{\sigma}} \;,
\label{Summary_Enhanced_Terms_M_Q}
\\
        \left| \ve{\sigma} \times \Delta \ve{n}_{\rm 2PN}^{{\rm Q} \times {\rm Q}} \right| &\le& 48\,\frac{G^2 M^2}{c^4}\,\left|J_2\right|^2\,
        \frac{P^4}{\left(\hat{d}_{\sigma}\right)^6}\,\frac{x_1}{\hat{d}_{\sigma}} \;.
\label{Summary_Enhanced_Terms_Q_Q}
\end{eqnarray} 

\noindent 
The enhanced 2PN quadrupole terms in (\ref{Summary_Enhanced_Terms_M_Q}) amount up to $0.95$ micro-arcseconds for a grazing light ray at Jupiter and $0.29$ 
micro-arcseconds for a grazing light ray at Saturn (cf. Table~\ref{Table3}), which shows that they are relevant on the sub-micro-arcsecond level of astrometric 
accuracy. It has been emphasized that these {\it enhanced 2PN terms} in (\ref{Enhanced_Terms_upper_limits_M_M}) - (\ref{Enhanced_Terms_upper_limits_Q_Q}) are 
solely caused by the fact that the light deflection at observer's position in (\ref{light_deflection_vector_n}) is expressed in terms of the impact vector 
given by Eq.~(\ref{impact_vector_x1}) which is indispensable for modeling of real astrometric measurements. 

Finally, it is noticed that a set of several checks has been performed: (a) the calculations which lead to the time-independent coefficients 
(\ref{coefficient_K3}) - (\ref{coefficient_N9}) and (\ref{coefficient_P5}) - (\ref{coefficient_Q14}) have been performed twice, namely 
analytically and by means of the computer algebra system {\it Maple} \cite{Maple}. 
(b) it is a non-trivial fact that their prefactors are either integers or half-integral numbers.  
(c) it has been checked analytically that the tangent vector $\ve{\nu}$ in (\ref{limit_plus}) is a unit vector.   
(d) it has been checked numerically by {\it Maple} that the null-condition (\ref{Null_Condition1}) is satisfied along the entire light ray. 
(e) our results in case of 1PN and 2PN monopole as well as in case of 1PN quadrupole agree with known achievements in the literature.  

Actually, one has to realize that at the sub-micro-arcsecond level of accuracy celestial light sources like stars and quasars, cannot be treated as pointlike 
but extended objects. Accordingly, one may separate the surface of celestial objects into sufficiently small surface elements such, that each individual surface 
element can then be treated as a pointlike light source. This treatment becomes more relevant the higher the aimed astrometric accuracies are.

Furthermore, in the limit of geometrical optics, light propagates along null-geodesics, governed by the geodesic equation (\ref{Geodetic_Equation}) 
and the null-condition (\ref{Null_Condition}). In reality, however, light propagates as electromagnetic wave, and one has to investigate at which scale of 
accuracy the wave nature of light must be taken into account. Such an investigation has been performed in \cite{Geometrical_Optics}, where is was found that 
deviations from geometric optics are negligible even on the nano-arcsecond level, because the wavelength of electromagnetic waves in the optical band is much 
smaller than the curvature of space-time in the Solar System. In this respect we also refer to the recent investigations \cite{Turyshev1,Turyshev2,Turyshev3} 
where by using the eikonal approach the issue of image formation has been considered, which could be observed with telescopes located far out of the solar system.

The investigations presented are a basis for further considerations in the area of high-precision astrometry.   
In particular, in a subsequent investigation the boundary value problem of geodesic equation will be solved in 2PN approximation for the case of one body 
at rest with monopole and quadrupole structure, where also the Shapiro time delay will comprehensively be treated. 
Furthermore, the approach presented is also applicable to solve the problem of 2PN light propagation in the gravitational field of massive bodies with 
higher multipoles beyond the monopole and quadrupole structure of the bodies.

\section*{Acknowledgment}

This work was funded by the German Research Foundation (Deutsche Forschungsgemeinschaft DFG) under grant number 447922800. Sincere gratitude 
is expressed to Prof. Sergei A. Klioner for kind encouragement and enduring support. Prof. Michael Soffel, Priv.-Doz. G\"unter Plunien, 
Dr. Alexey Butkevich, Prof. Ralf Sch\"utzhold, Prof. Laszlo Csernai, and Prof. Burkhard K\"ampfer are greatly acknowledged 
for inspiring discussions about astrometry and general theory of relativity. 
Dr. Sebastian Bablok and Dipl.-Inf. Robin Geyer are kindly acknowledged for computer assistance. 

\appendix

\section{Notation}\label{Appendix0}  

Throughout the investigation the following notation is in use:  

\begin{itemize}
\item $G$ is the Newtonian constant of gravitation.  
\item $c$ is the vacuum speed of light in Minkowskian space-time.  
\item $M$ is the rest mass of the body defined by Eq.~(\ref{Mass}).   
\item $\hat{M}_{ab}$ is the quadrupole moment of the body defined by Eq.~(\ref{Quadrupole}).   
\item $m = G\,M/c^2$ is the Schwarzschild radius of the body.  
\item $P$ denotes the equatorial radius of the body.   
\item Lower case Latin indices $i$, $j$, \dots take values $1,2,3$.
\item $\dot{f}$ denotes total time derivative of $f$. 
\item $f_{\,,\,i} = \partial f / \partial x^{i}$ denotes partial derivative of $f$ with respect to $x^i$. 
\item $\delta_{ij} = \delta^{ij} = {\rm diag} \left(+1,+1,+1\right)$ is Kronecker delta.
\item $n! = n \left(n-1\right)\left(n-2\right)\cdot\cdot\cdot 2 \cdot 1$ is the factorial for positive integer $\left(0! = 1\right)$. 
\item $n!! = n \left(n-2\right) \left(n-4\right)\cdot\cdot\cdot \left(2\;{\rm or}\;1\right)$ is the double factorial for positive integer $\left(0!! = 1\right)$. 
\item $\varepsilon_{ijk} = \varepsilon^{ijk}$ with $\varepsilon_{123} = + 1$ is the fully anti-symmetric Levi-Civita symbol.  
\item Triplet of three-vectors are in boldface: e.g. $\ve{a}$, $\ve{b}$, $\ve{\sigma}$, $\ve{x}$. 
\item Contravariant components of three-vectors: $a^{i} = \left(a^{\,1},a^2,a^3\right)$.
\item Absolute value of a three-vector: $a = |\ve{a}| = \sqrt{a^{\,1}\,a^{\,1}+a^2\,a^2+a^3\,a^3}$.
\item Scalar product of three-vectors: $\ve{a}\,\cdot\,\ve{b}=\delta_{ij}\,a^i\,b^j$.  
\item Vector product of two three-vectors: $\left(\ve{a}\times\ve{b}\right)^i=\varepsilon_{ijk}\,a^j\,b^k$.  
\item Angle between three-vectors $\ve{a}$ and $\ve{b}$ is denoted by $\delta\left(\ve{a},\ve{b}\right)$.  
\item Lower case Greek indices take values 0,1,2,3. 
\item $f_{\,,\,\mu} = \partial f / \partial x^{\mu}$ denotes partial derivative of $f$ with respect to $x^{\mu}$. 
\item $\eta_{\alpha\beta} = \eta^{\alpha \beta} = {\rm diag}\left(-1,+1,+1,+1\right)$ is the metric tensor of flat space-time.
\item $g_{\alpha\beta}$ and $g^{\alpha\beta}$ are the covariant and contravariant components of the metric tensor.  
\item Contravariant components of four-vectors: $a^{\mu} = \left(a^{\,0},a^{\,1},a^2,a^3\right)$.
\item milli-arcsecond (mas): $1\,{\rm mas} \simeq 4.85 \times 10^{-9}\,{\rm rad}$.
\item micro-arcsecond (\muas): $1\,\muas \simeq 4.85 \times 10^{-12}\,{\rm rad}$.
\item nano-arcsecond (nas): $1\,{\rm nas} \simeq 4.85 \times 10^{-15}\,{\rm rad}$.
\item ${\rm au} = 1.496 \times 10^{11}\,{\rm m}$ is the astronomical unit 
\item repeated indices are implicitly summed over ({\it Einstein's} sum convention).  
\end{itemize}

\section{The STF tensors $\hat{n}_L$}\label{Appendix_STF}

In this appendix we summarize some formulae of symmetric trace-free (STF) tensors which are relevant for our investigation.  
Further standard STF notation and some STF relations can be found in \cite{Thorne,Blanchet_Damour1,Coope1,Coope2,Coope3,Soffel_Hartmann};  
	see also appendix B in \cite{Zschocke_2PM_Metric}. The symmetric part of some tensor ${\cal J}^L$ with indices $L= i_1, \dots i_l$ is given by  
\begin{eqnarray}
	{\cal J}^{\left(L\right)} = {\cal J}^{\left(i_1 \dots i_l\right)} = \frac{1}{l!} \sum\limits_{\sigma}  
	{\cal J}^{i_{\sigma\left(1\right)} \dots i_{\sigma\left(l\right)}} \;, 
	\label{Symmetrization} 
\end{eqnarray}

\noindent
where $\sigma$ is running over all permutations of the indices $1, \dots, l$. If one subtracts all traces from (\ref{Symmetrization}) then  
one gets the symmetric trace-free (STF) part of tensor ${\cal J}^L$. A general formula for the STF part of some tensor is provided by Eq.~(2.2) in \cite{Thorne}  
or Eqs.~(A 2a) - (A 2c) in \cite{Blanchet_Damour1}. Here, we will not repeat that formula, but just apply for a specific case of it,  
as given by Eq.~(\ref{Appendix_Cartesian_Tensor_General_Formula}) below.  

In particular, in our investigation the following Cartesian tensors are of primary importance,  
\begin{eqnarray}
	x^L = x^{i_1}\,\dots\,x^{i_l} \quad {\rm and} \quad  
        n^L = \frac{x^{i_1}}{x}\,\dots\,\frac{x^{i_l}}{x}  
\label{Appendix_Cartesian_Tensor}
\end{eqnarray}

\noindent 
where $L = i_1 , \dots , i_l$ is a multi-index, $x^i$ are the spatial coordinates of some arbitrary field point in space and $x = \left|\ve{x}\right|$. 
Note that $x_L = x^L$ and $n_L = n^L$. 
For instance, 
\begin{eqnarray}
	n^{ab} &=& \frac{x^a x^b}{x^2}\;,
\label{n_ab}
\\
	n^{abcd} &=& \frac{x^a x^b x^c x^d}{x^4}\;. 
\label{n_abcd}
\end{eqnarray}

\noindent
The symmetric trace-free (STF) part of these Cartesian tensors in (\ref{Appendix_Cartesian_Tensor}) are denoted by
\begin{eqnarray}
\hat{x}^L = {\rm STF}_{i_1 \dots i_l}\;x^{i_1}\,\dots\,x^{i_l}  \; {\rm and} \;  
\hat{n}^L = {\rm STF}_{i_1 \dots i_l}\;\frac{x^{i_1}}{x}\,\dots\,\frac{x^{i_l}}{x} 
\label{Appendix_Cartesian_Tensor_STF}
\end{eqnarray}

\noindent
where ${\rm STF}_{i_1 \dots i_l}$ is the symmetrization and trace-free operation with respect to the spatial indices $i_1 \dots i_l$. 
Note that $\hat{x}_L = \hat{x}^L$ and $\hat{n}_L = \hat{n}^L$. 
A general formula to express $\hat{n}^L$ in terms of $n^L$ is provided by (cf. Eq.~(1.154) in \cite{Poisson_Will}) 
\begin{eqnarray}
	\hat{n}^L &=& \sum\limits_{k=0}^{[l/2]} \left(-1\right)^k \frac{\left(2l-2k-1\right)!!}{\left(2l-1\right)!!}\;
	\frac{l!}{\left(l-2k\right)! \left(2k\right)!!}\;
	 \delta^{( i_1 i_2} \dots \delta^{i_{2k-1} i_{2k}}\;n^{i_{2k+1 \dots i_l} )}  
\label{Appendix_Cartesian_Tensor_General_Formula}
\end{eqnarray}

\noindent
where round brackets of the tensorial term indicate symmetrization according to Eq.~(\ref{Symmetrization}), double factorial is denoted by $m!!$ 
and $[l/2] = l/2$ for even $l$, while $[l/2] = (l-1)/2$ for odd $l$.  
For instance,  
\begin{eqnarray}
	\hat{n}^{ab} &=& n^{ab} - \frac{1}{3}\,\delta^{ab}\;,
\label{Appendix1_STF2}
\\
\nonumber\\
\hat{n}^{abcd} &=& n^{abcd}
 - \frac{1}{7} \left(\delta^{a b} n^{c d} + \delta^{a c} n^{b d} + \delta^{a d} n^{b c} + \delta^{b c} n^{a d}
+ \delta^{b d} n^{a c} + \delta^{c d} n^{a b} \right)
 + \frac{1}{35}\left(\delta^{a b}\,\delta^{c d} + \delta^{a c}\,\delta^{b d} + \delta^{a d}\,\delta^{b c}\right)\,,
\label{Appendix1_STF4}
\\
\nonumber\\
\hat{n}^{abcde\!f} &=& n^{abcde\!f}
	- \frac{1}{11} \left(\delta^{ab}\,n^{cde\!f} + \emph{sym.}\right) 
	+ \frac{1}{99} \left(\delta^{ab}\,\delta^{cd}\, n^{e\!f} + \emph{sym.}\right) 
	- \frac{1}{693} \left(\delta^{ab}\,\delta^{cd}\,\delta^{e\!f} + \emph{sym.}\right).   
\label{Appendix1_STF6}
\end{eqnarray}

\noindent
These STF tensors appear in the 2PN metric coefficients in Eqs.~(\ref{metric_2PN_00}) - (\ref{metric_2PN_ij}). 
The expressions (\ref{Appendix1_STF2}) and (\ref{Appendix1_STF4}) were also given by Eqs.~(1.153a) and (1.153c) in \cite{Poisson_Will}; 
see also Eqs.~(1.8.2) and (1.8.4) in \cite{Poisson_Lecture_Notes}; the abbreviation \emph{sym.} in (\ref{Appendix1_STF6}) 
means symmetrization with respect to the indices. 

The geodesic equation in 1PN and 2PN approximation, Eqs.~(\ref{Geodesic_Equation_1PN_A}) and (\ref{Geodesic_Equation4}), contains spatial derivatives  
of the metric coefficients in (\ref{metric_2PN_00}) - (\ref{metric_2PN_ij}).  
Therefore, one needs the spatial derivative of the STF tensor in (\ref{Appendix_Cartesian_Tensor_General_Formula}). 
The general formula is given by (cf. Eq.~(A 24) in \cite{Blanchet_Damour1})  
\begin{eqnarray}
	\left(\hat{n}^L\right)_{,\,k} = \frac{l + 1}{x}\,n^k\,\hat{n}^{L} - \frac{2l + 1}{x}\,\hat{n}^{kL} \;, 
\label{Appendix_Cartesian_Tensor_General_Formula_Derivative}
\end{eqnarray}

\noindent 
which can finally be expressed in terms of $n^L$ by means of relation (\ref{Appendix_Cartesian_Tensor_General_Formula}). 

As explained in the text below Eq.~(\ref{Geodesic_Equation4}), that field point becomes the spatial coordinate of the light ray, which is expanded in terms  
of the unperturbed light ray; see Eqs.~(\ref{Appendix_x_A}) - (\ref{Appendix_x_C}) in Appendix \ref{Appendix_2PN_Terms}. 
Therefore, one finally needs to consider the tensor  
$n_L$ in (\ref{Appendix_Cartesian_Tensor}) and the STF tensor $\hat{n}_L$ in (\ref{Appendix_Cartesian_Tensor_STF}) where 
the field point is replaced by the unperturbed light ray, that means  
\begin{eqnarray}
	n^L_{\rm N} = \frac{x_{\rm N}^{i_1}}{x_{\rm N}}\,\dots\,\frac{x_{\rm N}^{i_l}}{x_{\rm N}} \;, 
\label{Appendix_Cartesian_Tensor_N}
\end{eqnarray}

\noindent
and 
\begin{eqnarray}
\hat{n}^L_{\rm N} = {\rm STF}_{i_1 \dots i_l}\;\frac{x_{\rm N}^{i_1}}{x_{\rm N}}\,\dots\,\frac{x_{\rm N}^{i_l}}{x_{\rm N}} \;, 
\label{Appendix_Cartesian_Tensor_N_STF}
\end{eqnarray}

\noindent
where $x^i_{\rm N}$ are the spatial coordinates of the unperturbed light ray as given by Eq.~(\ref{Unperturbed_Lightray_2}) 
and $x_{\rm N} = \left|\ve{x}_{\rm N}\right|$. 
From (\ref{Appendix_Cartesian_Tensor_General_Formula}) it is obvious that the general formula to express $\hat{n}_{\rm N}^L$ in terms of $n_{\rm N}^L$ reads  
\begin{eqnarray}
	\hat{n}_{\rm N}^L &=& \sum\limits_{k=0}^{[l/2]} \left(-1\right)^k \frac{\left(2l-2k-1\right)!!}{\left(2l-1\right)!!}\;
	\frac{l!}{\left(l-2k\right)! \left(2k\right)!!}\;
	\delta^{( i_1 i_2} \dots \delta^{i_{2k-1} i_{2k}}\;n_{\rm N}^{i_{2k+1 \dots i_l} )} \;. 
\label{Appendix_Cartesian_Tensor_General_Formula_N}
\end{eqnarray}

\noindent 
Similarly, we need the spatial derivative of $\hat{n}^L$ in (\ref{Appendix_Cartesian_Tensor_General_Formula_Derivative}), 
where (after derivation) the field point $\ve{x}$ is replaced by the unperturbed light ray $\ve{x}_{\rm N}$, that means   
\begin{eqnarray}
	\left(\hat{n}_{\rm N}^L\right)_{,\,k} = \frac{l + 1}{x}\,n_{\rm N}^k\,\hat{n}_{\rm N}^{L} - \frac{2l + 1}{x}\,\hat{n}_{\rm N}^{kL}\;. 
\label{Appendix_Cartesian_Tensor_General_Formula_Derivative_N}
\end{eqnarray}

\noindent
By means of (\ref{Appendix_Cartesian_Tensor_General_Formula_N}) one may express such quantities in terms of $n_{\rm N}^L$. 
In what follows, such terms are slightly imprecise called  
spatial derivative of $\hat{n}_{\rm N}^L$, while these terms are actually spatial derivatives of 
$\hat{n}^L$ where after the differentiation the field point is replaced by the unperturbed light ray.

\section{The auxiliary tensors $T^{ijab}_{\rm N}$ and $T^{iijabcd}_{\rm N}$}\label{Appendix_Auxiliary_Tensors}

These tensors $T^{ijab}_{\rm N}$ and $T^{iijabcd}_{\rm N}$ are composed of STF tensors $\hat{n}^L_{\rm N}$, nevertheless they themselves   
are neither symmetric in their indices nor trace-free. They are just abbreviations to simplify the notations.  
The auxiliary tensor $T^{ijab}_{\rm N}$ is defined by 
\begin{eqnarray}
 T_{\rm N}^{ijab} = \frac{15}{2}\,\hat{n}_{\rm N}^{ijab} + \frac{32}{7}\,\delta^{ij}\,\hat{n}_{\rm N}^{ab} 
        - \frac{6}{7}\,\delta^{ai}\,\hat{n}_{\rm N}^{jb} - \frac{6}{7}\,\delta^{aj}\,\hat{n}_{\rm N}^{ib}\;,
\label{Appendix_Tensor_S}
\end{eqnarray}

\noindent
where the STF tensors $\hat{n}_{\rm N}^{ab}$ and $\hat{n}_{\rm N}^{abcd}$ were given by Eq.~(\ref{Appendix_Cartesian_Tensor_General_Formula_N}).  
The spatial derivative of $T^{ijab}_{\rm N}$ reads 
\begin{eqnarray}
	\left(T_{\rm N}^{ijab}\right)_{,\,k} &=& \frac{15}{2} \left(\hat{n}_{\rm N}^{ijab}\right)_{,\,k} 
	+ \frac{32}{7}\,\delta^{ij}\left(\hat{n}_{\rm N}^{ab}\right)_{,\,k} 
	- \frac{6}{7}\,\delta^{ai}\left(\hat{n}_{\rm N}^{jb}\right)_{,\,k} 
	 - \frac{6}{7}\,\delta^{aj}\left(\hat{n}_{\rm N}^{ib}\right)_{,\,k} \;, 
\label{Appendix_Tensor_S_diff}
\end{eqnarray}

\noindent
where the terms on the r.h.s. are spatial derivatives of STF tensors and were given by Eq.(\ref{Appendix_Cartesian_Tensor_General_Formula_Derivative_N}). 
The auxiliary tensor $T^{ijabcd}_{\rm N}$ is defined by  
\begin{eqnarray}
	T_{\rm N}^{ijabcd} &=& \frac{75}{4}\,\hat{n}^{ijabcd}_{\rm N} - \frac{90}{11}\,\delta^{bd}\,\hat{n}^{ijac}_{\rm N}  
	+ \frac{27}{11}\,\delta^{ij}\,\hat{n}^{abcd}_{\rm N} 
	- \frac{25}{84}\,\delta^{ac}\,\delta^{bd}\,\hat{n}^{ij}_{\rm N} 
	+ \frac{83}{42}\,\delta^{bc}\,\delta^{ij}\,\hat{n}^{ad}_{\rm N}  
	 + \frac{9}{11}\,\delta^{jb}\,\hat{n}^{acdi}_{\rm N}   
	 \nonumber\\ 
        && \hspace{-1.5cm} + \frac{9}{11}\,\delta^{ib}\,\hat{n}^{acdj}_{\rm N} 
          - \frac{5}{42}\,\delta^{jc}\,\delta^{bd}\,\hat{n}^{ai}_{\rm N}  
        - \frac{5}{42}\,\delta^{ic}\,\delta^{bd}\,\hat{n}^{aj}_{\rm N}
	+ \frac{10}{21}\,\delta^{ci}\,\delta^{dj}\,\hat{n}^{ab}_{\rm N} 
	   - \frac{23}{84}\,\delta^{bi}\,\delta^{jc}\,\hat{n}^{ad}_{\rm N}  
           - \frac{23}{84}\,\delta^{bj}\,\delta^{ic}\,\hat{n}^{ad}_{\rm N}  \;,  
\label{Appendix_Tensor_T}
\end{eqnarray}

\noindent
where the STF tensors $\hat{n}_{\rm N}^{ab}$, $\hat{n}_{\rm N}^{abcd}$, and $\hat{n}_{\rm N}^{abcde\!f}$ were given 
by Eq.~(\ref{Appendix_Cartesian_Tensor_General_Formula_N}).  
The spatial derivative of $T^{ijabcd}_{\rm N}$ reads
\begin{eqnarray}
	\left(T_{\rm N}^{ijabcd}\right)_{,\,k} &=& \frac{75}{4} \left(\hat{n}^{ijabcd}_{\rm N}\right)_{,\,k} 
	- \frac{90}{11}\,\delta^{bd} \left(\hat{n}^{ijac}_{\rm N}\right)_{,\,k}
	+ \frac{27}{11}\,\delta^{ij}\left(\hat{n}^{abcd}_{\rm N}\right)_{,\,k} 
	- \frac{25}{84}\,\delta^{ac}\,\delta^{bd}\left(\hat{n}^{ij}_{\rm N}\right)_{,\,k}  
        \nonumber\\
	&& \hspace{-1.5cm} + \frac{83}{42}\,\delta^{bc}\,\delta^{ij}\left(\hat{n}^{ad}_{\rm N}\right)_{,\,k}  
	+ \frac{9}{11}\,\delta^{jb}\left(\hat{n}^{acdi}_{\rm N}\right)_{,\,k} 
	+ \frac{9}{11}\,\delta^{ib}\left(\hat{n}^{acdj}_{\rm N}\right)_{,\,k}  
	- \frac{5}{42}\,\delta^{jc}\,\delta^{bd}\left(\hat{n}^{ai}_{\rm N}\right)_{,\,k} 
        \nonumber\\
	&& \hspace{-1.5cm} - \frac{5}{42}\,\delta^{ic}\,\delta^{bd}\left(\hat{n}^{aj}_{\rm N}\right)_{,\,k} 
	+ \frac{10}{21}\,\delta^{ci}\,\delta^{dj}\left(\hat{n}^{ab}_{\rm N}\right)_{,\,k} 
	- \frac{23}{84}\,\delta^{bi}\,\delta^{jc}\left(\hat{n}^{ad}_{\rm N}\right)_{,\,k} 
	- \frac{23}{84}\,\delta^{bj}\,\delta^{ic}\left(\hat{n}^{ad}_{\rm N}\right)_{,\,k}  \;, 
\label{Appendix_Tensor_T_diff}
\end{eqnarray}

\noindent
where the terms on the r.h.s. are spatial derivatives of STF tensors and were given by Eq.~(\ref{Appendix_Cartesian_Tensor_General_Formula_Derivative_N}).

\section{Scalar master integrals}\label{Appendix_Scalar_Master_Integrals}

\subsection{Scalar master integrals for the first integration} 

The first integration of geodesic equation in 1PN and 2PN approximation can be reduced to finally only four kind of scalar master integrals,  
\begin{eqnarray}
	\dot{\cal W}_{\left(n\right)}\left(t\right) &=& \int\limits_{- \infty}^{t} d c {\rm t}\;
	\frac{\ve{\sigma} \cdot \ve{x}_{\rm N}}{\left(x_{\rm N}\right)^n}\;,
\label{Master_Integral_I1}
\\
	\dot{\cal X}_{\left(n\right)}\left(t\right) &=& \int\limits_{- \infty}^{t} d c {\rm t}\;\frac{1}{\left(x_{\rm N}\right)^n}\;, 
\label{Master_Integral_J1}
\\
        \dot{\cal Y}_{\left(n\right)}\left(t\right) &=& \int\limits_{- \infty}^{t} d c {\rm t}\;
	\frac{\ve{\sigma} \cdot \ve{x}_{\rm N}}{\left(x_{\rm N}\right)^n} \; 
	\ln \frac{x_{\rm N} - \ve{\sigma} \cdot \ve{x}_{\rm N}}{x_0 - \ve{\sigma} \cdot \ve{x}_0} \;, 
\label{Master_Integral_K1}
\\
	\dot{\cal Z}_{\left(n\right)}\left(t\right) &=& \int\limits_{- \infty}^{t} d c {\rm t}\;\frac{1}{\left(x_{\rm N}\right)^n}\; 
	\ln \frac{x_{\rm N} - \ve{\sigma} \cdot \ve{x}_{\rm N}}{x_0 - \ve{\sigma} \cdot \ve{x}_0} \;,   
\label{Master_Integral_L1}
\end{eqnarray}

\noindent
where $\ve{x}_{\rm N} = \ve{x}_{\rm N}\left({\rm t}\right)$ and $x_{\rm N} = x_{\rm N}\left({\rm t}\right)$ are functions of the 
integration variable ${\rm t}$ according to (\ref{Unperturbed_Lightray_2}) 
and (\ref{Unperturbed_Lightray_3}). 
Clearly, the limits  
\begin{eqnarray}
        \lim_{t \rightarrow - \infty} \dot{\cal W}_{\left(n\right)}\left(t\right) &=& 0 \;, \quad 
        \lim_{t \rightarrow - \infty} \dot{\cal X}_{\left(n\right)}\left(t\right) = 0 \;, \quad 
        \lim_{t \rightarrow - \infty} \dot{\cal Y}_{\left(n\right)}\left(t\right) = 0 \;, \quad 
        \lim_{t \rightarrow - \infty} \dot{\cal Z}_{\left(n\right)}\left(t\right) = 0 \;,
\label{Limit_Master_Integral_1}
\end{eqnarray}

\noindent
are in accordance with relation (\ref{Boundary_Condition_1}).
The scalar master integrals for the first integration can be given in closed form,   
\begin{eqnarray}
	        \dot{\cal W}_{\left(n\right)}\left(t\right) 
	&=& - \frac{1}{n-2}\,\frac{1}{\left(x_{\rm N}\right)^{n-2}}  \;,  
\label{Master_Integral_A}
\\
        \dot{\cal X}_{\left(n\right)}\left(t\right)   
	&=& \frac{1}{n-2}\, \frac{1}{\left(d_{\sigma}\right)^2} 
	\left(\left(n-3\right)\dot{\cal X}_{\left(n-2\right)}\left(t\right) + \frac{\ve{\sigma} \cdot \ve{x}_{\rm N}}{\left(x_{\rm N}\right)^{n-2}}\right) ,
\label{Master_Integral_B}
\\
        \dot{\cal Y}_{\left(n\right)}\left(t\right)   
	&=& - \frac{1}{n-2}\left(\dot{\cal X}_{\left(n-1\right)}\left(t\right) + \frac{1}{\left(x_{\rm N}\right)^{n-2}}\, 
	\ln \frac{x_{\rm N} - \ve{\sigma} \cdot \ve{x}_{\rm N}}{x_0 - \ve{\sigma} \cdot \ve{x}_0}\right) , 
\label{Master_Integral_C}
\\
        \dot{\cal Z}_{\left(n\right)}\left(t\right)   
	&=& \frac{1}{n-2} \frac{1}{\left(d_{\sigma}\right)^2}\!\Bigg(\!  
	\!\left(n-3\right)\!\dot{\cal Z}_{\left(n-2\right)}\left(t\right)  
	- \frac{1}{n-3} \frac{1}{\left(x_{\rm N}\right)^{n-3}} 
	+  \frac{\ve{\sigma} \cdot \ve{x}_{\rm N}}{\left(x_{\rm N}\right)^{n-2}}  
	\ln \frac{x_{\rm N} - \ve{\sigma} \cdot \ve{x}_{\rm N}}{x_0 - \ve{\sigma} \cdot \ve{x}_0}\!\Bigg) ,
\label{Master_Integral_D}
\end{eqnarray}
	
\noindent 
where $n \ge 3$ in Eqs.~(\ref{Master_Integral_A}) and (\ref{Master_Integral_C}) and 
$n \ge 4$ in Eqs.~(\ref{Master_Integral_B}) and (\ref{Master_Integral_D}). While (\ref{Master_Integral_A}) represents the 
explicit solution of the integrals (\ref{Master_Integral_I1}) to any order, the implicit  
recurrence relations (\ref{Master_Integral_B}) - (\ref{Master_Integral_D}) allow us to determine the explicit solutions of 
the integrals (\ref{Master_Integral_J1}) - (\ref{Master_Integral_L1}) to any order of the inverse powers in $n$ as soon as the first orders of them are known, 
which have to be determined directly by using standard integrals. Furthermore, they are very useful for estimating their upper limits of impact on 
light deflection and they can be used to make some cross checks of the explicit solutions for the integrals. These integrals 
(\ref{Master_Integral_A}) - (\ref{Master_Integral_D}) are stable, that means they are convergent with respect to the limit $n \rightarrow \infty$. 
The explicit solutions for the first few orders of the inverse powers in $n$ are given by  
\begin{eqnarray}
	        \dot{\cal X}_{\left(2\right)}\left(t\right)  
	&=& \frac{1}{d_{\sigma}}\left(\arctan \frac{\ve{\sigma} \cdot \ve{x}_{\rm N}}{d_{\sigma}} + \frac{\pi}{2}\right) , 
\label{Master_Integral_B2}
\end{eqnarray}

\begin{eqnarray}
	        \dot{\cal X}_{\left(3\right)}\left(t\right)  
	&=& \frac{1}{\left(d_{\sigma}\right)^2} \left(\frac{\ve{\sigma} \cdot \ve{x}_{\rm N}}{x_{\rm N}} + 1\right) ,
	\label{Master_Integral_B3}
\end{eqnarray}
 
\begin{eqnarray}
        \dot{\cal X}_{\left(4\right)}\left(t\right)    
	&=& \frac{1}{2}\,\frac{1}{\left(d_{\sigma}\right)^2} 
	\left(\frac{\ve{\sigma} \cdot \ve{x}_{\rm N}}{\left(x_{\rm N}\right)^2} 
	+ \frac{1}{d_{\sigma}} \left(\arctan \frac{\ve{\sigma} \cdot \ve{x}_{\rm N}}{d_{\sigma}} + \frac{\pi}{2}\right) \right) ,
\label{Master_Integral_B4}
\end{eqnarray}

\begin{eqnarray}
	        \dot{\cal X}_{\left(5\right)}\left(t\right)  
        &=& \frac{2}{3}\,\frac{1}{\left(d_{\sigma}\right)^2} \left(\frac{\ve{\sigma} \cdot \ve{x}_{\rm N}}{x_{\rm N}}\,
	\frac{1}{\left(d_{\sigma}\right)^2} + \frac{1}{2}\,\frac{\ve{\sigma} \cdot \ve{x}_{\rm N}}{\left(x_{\rm N}\right)^3} 
	+ \frac{1}{\left(d_{\sigma}\right)^2}\right) ,  
	\label{Master_Integral_B5}
\end{eqnarray}

\begin{eqnarray}
        \dot{\cal X}_{\left(6\right)}\left(t\right) 
	&=& \frac{1}{4} \frac{1}{\left(d_{\sigma}\right)^2} 
        \left(\frac{\ve{\sigma} \cdot \ve{x}_{\rm N}}{\left(x_{\rm N}\right)^4} 
	+ \frac{3}{2}\,\frac{1}{\left(d_{\sigma}\right)^2} \frac{\ve{\sigma} \cdot \ve{x}_{\rm N}}{\left(x_{\rm N}\right)^2}   
	+ \frac{3}{2}\,\frac{1} 
	{\left(d_{\sigma}\right)^3} \left(\arctan \frac{\ve{\sigma} \cdot \ve{x}_{\rm N}}{d_{\sigma}} + \frac{\pi}{2}\right) \right) ,
	\label{Master_Integral_B6}
\end{eqnarray}

\begin{eqnarray}
	        \dot{\cal X}_{\left(7\right)}\left(t\right)  
	&=& \frac{8}{15}\,\frac{1}{\left(d_{\sigma}\right)^2} 
	\left(\frac{\ve{\sigma} \cdot \ve{x}_{\rm N}}{x_{\rm N}}\,\frac{1}{\left(d_{\sigma}\right)^4} 
	+ \frac{1}{2}\,\frac{\ve{\sigma} \cdot \ve{x}_{\rm N}}{\left(x_{\rm N}\right)^3}\,\frac{1}{\left(d_{\sigma}\right)^2} 
	+ \frac{3}{8}\,\frac{\ve{\sigma} \cdot \ve{x}_{\rm N}}{\left(x_{\rm N}\right)^5} + \frac{1}{\left(d_{\sigma}\right)^4}\right) , 
\label{Master_Integral_B7}
\end{eqnarray}

\begin{eqnarray}
	        \dot{\cal Y}_{\left(5\right)}\left(t\right)  
	&=& - \frac{1}{3}\,\frac{1}{\left(x_{\rm N}\right)^3}\,
	\ln \frac{x_{\rm N} - \ve{\sigma} \cdot \ve{x}_{\rm N}}{x_0 - \ve{\sigma} \cdot \ve{x}_0} 
	- \frac{1}{6}\frac{1}{\left(d_{\sigma}\right)^2} 
        \left(\frac{\ve{\sigma} \cdot \ve{x}_{\rm N}}{\left(x_{\rm N}\right)^2} 
        + \frac{1}{d_{\sigma}} \left(\arctan \frac{\ve{\sigma} \cdot \ve{x}_{\rm N}}{d_{\sigma}} + \frac{\pi}{2}\right) \right) ,
	\label{Master_Integral_K5}
\end{eqnarray}

\begin{eqnarray}
	        \dot{\cal Z}_{\left(3\right)}\left(t\right)  
	&=& \frac{1}{\left(d_{\sigma}\right)^2} \left(\frac{\ve{\sigma} \cdot \ve{x}_{\rm N}}{x_{\rm N}} + 1 \right) 
	\ln \frac{x_{\rm N} - \ve{\sigma} \cdot \ve{x}_{\rm N}}{x_0 - \ve{\sigma} \cdot \ve{x}_0} 
	- \frac{1}{\left(d_{\sigma}\right)^2} \left( \ln \frac{x_{\rm N} - \ve{\sigma} \cdot \ve{x}_{\rm N}}{2\,x_{\rm N}}\right) , 
	\label{Master_Integral_L3}
\end{eqnarray}

\begin{eqnarray}
	        \dot{\cal Z}_{\left(5\right)}\left(t\right) 
	&=& \frac{2}{3}\frac{1}{\left(d_{\sigma}\right)^2} \left(\!\frac{\ve{\sigma} \cdot \ve{x}_{\rm N}}{x_{\rm N}} 
	\frac{1}{\left(d_{\sigma}\right)^2} + \frac{1}{2} \frac{\ve{\sigma} \cdot \ve{x}_{\rm N}}{\left(x_{\rm N}\right)^3} 
	+ \frac{1}{\left(d_{\sigma}\right)^2}\!\right) 
        \ln \frac{x_{\rm N} - \ve{\sigma} \cdot \ve{x}_{\rm N}}{x_0 - \ve{\sigma} \cdot \ve{x}_0} 
	- \frac{1}{6}\,\frac{1}{\left(d_{\sigma}\right)^2} \frac{1}{\left(x_{\rm N}\right)^2} 
	- \frac{2}{3}\frac{1}{\left(d_{\sigma}\right)^4}
        \left(\ln \frac{x_{\rm N} - \ve{\sigma} \cdot \ve{x}_{\rm N}}{2\,x_{\rm N}} \right) ,
\nonumber\\ 
        \label{Master_Integral_L5}
\end{eqnarray}

\noindent
where the time arguments have been omitted in the terms on the r.h.s. in order to simplify the notation. 

For the expression of the total light deflection one needs the limits in (\ref{Limit_X_Y}). 
Using $\lim_{t \rightarrow + \infty} \ve{\sigma} \cdot \ve{x}_{\rm N} / x_{\rm N} = 1$, which follows from (\ref{Unperturbed_Lightray_2}) 
and (\ref{Unperturbed_Lightray_3}), one obtains: 
\begin{eqnarray}
	\dot{\cal X}^{\infty}_{\left(2\right)} &=& \frac{\pi}{d_{\sigma}} \;,\; 
	\dot{\cal X}^{\infty}_{\left(3\right)} = \frac{2}{\left(d_{\sigma}\right)^2}\;,\;  
	\dot{\cal X}^{\infty}_{\left(4\right)} = \frac{\pi}{2}\,\frac{1}{\left(d_{\sigma}\right)^3}\;,\; 
	\dot{\cal X}^{\infty}_{\left(5\right)} = \frac{4}{3}\,\frac{1}{\left(d_{\sigma}\right)^4} \;,\; 
	\dot{\cal X}^{\infty}_{\left(6\right)} = \frac{3}{8}\,\frac{\pi}{\left(d_{\sigma}\right)^5} \;,\;
	\dot{\cal X}^{\infty}_{\left(7\right)} = \frac{16}{15}\,\frac{1}{\left(d_{\sigma}\right)^6} \;,\;
	\nonumber\\
	\dot{\cal X}^{\infty}_{\left(8\right)} &=& \frac{5}{16}\,\frac{\pi}{\left(d_{\sigma}\right)^7}\;,\;
        \dot{\cal X}^{\infty}_{\left(9\right)} = \frac{32}{35}\,\frac{1}{\left(d_{\sigma}\right)^8} \;,\;
	\dot{\cal X}^{\infty}_{\left(10\right)} = \frac{35}{128}\,\frac{\pi}{\left(d_{\sigma}\right)^9} \;,\;
        \dot{\cal X}^{\infty}_{\left(11\right)} = \frac{256}{315}\,\frac{1}{\left(d_{\sigma}\right)^{10}} \;,\;
	\dot{\cal X}^{\infty}_{\left(12\right)} = \frac{63}{256}\,\frac{\pi}{\left(d_{\sigma}\right)^{11}}\;,\;
	\nonumber\\
	\dot{\cal X}^{\infty}_{\left(13\right)} &=& \frac{512}{693}\,\frac{1}{\left(d_{\sigma}\right)^{12}} \;,\;
	\dot{\cal X}^{\infty}_{\left(14\right)} = \frac{231}{1024}\,\frac{\pi}{\left(d_{\sigma}\right)^{13}}\;,\;
	\dot{\cal Y}^{\infty}_{\left(5\right)} = - \frac{1}{6}\,\frac{\pi}{\left(d_{\sigma}\right)^{3}} \;,\;
	\dot{\cal Y}^{\infty}_{\left(7\right)} = - \frac{3}{40}\,\frac{\pi}{\left(d_{\sigma}\right)^{5}}\;,\;
	\dot{\cal Y}^{\infty}_{\left(9\right)} = - \frac{5}{112}\,\frac{\pi}{\left(d_{\sigma}\right)^{7}}\;. 
        \label{Limit_X_Y_n}
\end{eqnarray}

\subsection{Scalar master integrals for the second integration}

The second integration of geodesic equation in 1PN and 2PN approximation can be reduced to finally only four scalar master integrals,
\begin{eqnarray}
        {\cal W}_{\left(n\right)}\left(t,t_0\right) &=& 
	\int\limits_{t_0}^{t} d c {\rm t} \; \dot{\cal W}_{\left(n\right)}\left({\rm t}\right) 
	= {\cal W}_{\left(n\right)}\left(t\right) - {\cal W}_{\left(n\right)}\left(t_0\right) , 
\label{Master_Integral_I2}
\end{eqnarray}
\begin{eqnarray}
        {\cal X}_{\left(n\right)}\left(t,t_0\right) &=& 
	\int\limits_{t_0}^{t} d c {\rm t} \; \dot{\cal X}_{\left(n\right)}\left({\rm t}\right) 
	= {\cal X}_{\left(n\right)}\left(t\right) - {\cal X}_{\left(n\right)}\left(t_0\right) ,
\label{Master_Integral_J2}
\end{eqnarray}
\begin{eqnarray}
        {\cal Y}_{\left(n\right)}\left(t,t_0\right) &=& 
	\int\limits_{t_0}^{t} d c {\rm t} \; \dot{\cal Y}_{\left(n\right)}\left({\rm t}\right) 
	= {\cal Y}_{\left(n\right)}\left(t\right) - {\cal Y}_{\left(n\right)}\left(t_0\right) ,
\label{Master_Integral_K2}
\end{eqnarray}
\begin{eqnarray}
        {\cal Z}_{\left(n\right)}\left(t,t_0\right) &=& 
	\int\limits_{t_0}^{t} d c {\rm t} \; \dot{\cal Z}_{\left(n\right)}\left({\rm t}\right) 
	= {\cal Z}_{\left(n\right)}\left(t\right) - {\cal Z}_{\left(n\right)}\left(t_0\right) .
\label{Master_Integral_L2}
\end{eqnarray}

\noindent 
Clearly, the limits  
\begin{eqnarray}
        \lim_{t \rightarrow t_0} {\cal W}_{\left(n\right)}\left(t,t_0\right) &=& 0 \;, \quad 
        \lim_{t \rightarrow t_0} {\cal X}_{\left(n\right)}\left(t,t_0\right) = 0 \;, \quad 
        \lim_{t \rightarrow t_0} {\cal Y}_{\left(n\right)}\left(t,t_0\right) = 0 \;, \quad 
        \lim_{t \rightarrow t_0} {\cal Z}_{\left(n\right)}\left(t,t_0\right) = 0 \;,
\label{Limit_Master_Integral_2}
\end{eqnarray}

\noindent
are in accordance with relation (\ref{Boundary_Condition_2}).
The scalar master integrals for the second integration can be given in closed form, 
\begin{eqnarray}
{\cal W}_{\left(n\right)}\left(t\right) &=& - \frac{1}{n-2}\,\dot{\cal X}_{\left(n-2\right)}\left(t\right) ,
\label{Master_Integral_W} 
\\
{\cal X}_{\left(n\right)}\left(t\right) &=& \frac{1}{n-2}\,\frac{1}{\left(d_{\sigma}\right)^2} 
	\left(\dot{\cal W}_{\left(n-2\right)}\left(t\right) 
	+ \left(n-3\right) {\cal X}_{\left(n-2\right)}\left(t\right)\right) , 
\label{Master_Integral_X} 
\\
{\cal Y}_{\left(n\right)}\left(t\right) &=& - \frac{1}{n-2}\left(\dot{\cal Z}_{\left(n-2\right)}\left(t\right) 
	+ {\cal X}_{\left(n-1\right)}\left(t\right)\right) , 
\label{Master_Integral_Y} 
\\
{\cal Z}_{\left(n\right)}\left(t\right) &=& \frac{1}{n-2}\,\frac{1}{\left(d_{\sigma}\right)^2}\Bigg(  
	\left(n-3\right){\cal Z}_{\left(n-2\right)}\left(t\right) 
	- \frac{1}{n-3}\dot{\cal X}_{\left(n-3\right)}\left(t\right) 
	+ \dot{\cal Y}_{\left(n-2\right)}\left(t\right)\Bigg) , 
\label{Master_Integral_Z} 
\end{eqnarray}

\noindent
where $n \ge 4$ in Eq.~(\ref{Master_Integral_W}) and
$n \ge 5$ in Eqs.~(\ref{Master_Integral_X}) - (\ref{Master_Integral_Z}).
The recurrence relations (\ref{Master_Integral_W}) - (\ref{Master_Integral_Z}) allow us to determine the solutions of 
the integrals (\ref{Master_Integral_I2}) - (\ref{Master_Integral_L2}) to any order of the power $n$ as soon as the very few orders of them are known, 
which have to be determined directly by using standard integrals. 
Furthermore, they are very useful for estimating
their upper limits of impact on time delay and light deflection and they can be used to cross check some of the explicit solutions for the integrals. 
These integrals (\ref{Master_Integral_W}) - (\ref{Master_Integral_Z}) are stable, that means they are convergent in the limit $n \rightarrow \infty$.
The explicit solutions for the first few orders of the power in $n$   
are given by (constants which cancel each other due to (\ref{Master_Integral_I2}) - (\ref{Master_Integral_L2}) have been omitted)  
\begin{eqnarray}
         {\cal W}_{\left(3\right)}\left(t\right) &=&  
	\ln \left(x_{\rm N} - \ve{\sigma} \cdot \ve{x}_{\rm N}\right) , 
\label{Master_Integral_E1}  
\end{eqnarray}

\begin{eqnarray}
	 {\cal W}_{\left(4\right)}\left(t\right) &=& - \frac{1}{2}\,\frac{1}{d_{\sigma}} \, 
	\arctan \frac{\ve{\sigma} \cdot \ve{x}_{\rm N}}{d_{\sigma}} \;,
        \label{Master_Integral_E2}
\end{eqnarray}

\begin{eqnarray}
	 {\cal W}_{\left(5\right)}\left(t\right) &=& - \frac{1}{3}\,
	\frac{1}{\left(d_{\sigma}\right)^2}\,\frac{\ve{\sigma} \cdot \ve{x}_{\rm N}}{x_{\rm N}} \;, 
\label{Master_Integral_E3}  
\end{eqnarray}

\begin{eqnarray}
	 {\cal W}_{\left(6\right)}\left(t\right) &=& - \frac{1}{8}\,\frac{1}{\left(d_{\sigma}\right)^2} 
	\left(\frac{\ve{\sigma} \cdot \ve{x}_{\rm N}}{\left(x_{\rm N}\right)^2} 
	+ \frac{1}{d_{\sigma}}\, \arctan \frac{\ve{\sigma} \cdot \ve{x}_{\rm N}}{d_{\sigma}} \right) , 
\label{Master_Integral_E6}
\end{eqnarray}
 
\begin{eqnarray}
	 {\cal W}_{\left(7\right)}\left(t\right) &=& - \frac{2}{15}\,\frac{1}{\left(d_{\sigma}\right)^2}  
	\left(\frac{\ve{\sigma} \cdot \ve{x}_{\rm N}}{x_{\rm N}}\,\frac{1}{\left(d_{\sigma}\right)^2} 
	+ \frac{1}{2}\,\frac{\ve{\sigma} \cdot \ve{x}_{\rm N}}{\left(x_{\rm N}\right)^3} \right) , 
	\label{Master_Integral_E7}
\end{eqnarray}

\begin{eqnarray}
       {\cal X}_{\left(2\right)}\left(t\right) &=& + \frac{\ve{\sigma} \cdot \ve{x}_{\rm N}}{d_{\sigma}} \, 
	\left(\arctan \frac{\ve{\sigma} \cdot \ve{x}_{\rm N}}{d_{\sigma}} + \frac{\pi}{2}\right)  
        - \ln x_{\rm N} \;,  
\label{Master_Integral_F1}  
\end{eqnarray}

\begin{eqnarray}
	 {\cal X}_{\left(3\right)}\left(t\right) &=& + \frac{1}{\left(d_{\sigma}\right)^2}  
	 \left(x_{\rm N} + \ve{\sigma} \cdot \ve{x}_{\rm N} \right) ,  
	\label{Master_Integral_F}
\end{eqnarray}

\begin{eqnarray}
	 {\cal X}_{\left(4\right)}\left(t\right) &=& + \frac{1}{2}\,\frac{\ve{\sigma} \cdot \ve{x}_{\rm N}}{\left(d_{\sigma}\right)^3} 
        \left(\arctan \frac{\ve{\sigma} \cdot \ve{x}_{\rm N}}{d_{\sigma}} + \frac{\pi}{2}\right) , 
\label{Master_Integral_F2}  
\end{eqnarray}

\begin{eqnarray}
	 {\cal X}_{\left(5\right)}\left(t\right)  
	&=& + \frac{2}{3}\,\frac{1}{\left(d_{\sigma}\right)^2} 
	\left(\frac{x_{\rm N} + \ve{\sigma} \cdot \ve{x}_{\rm N}}{\left(d_{\sigma}\right)^2} - \frac{1}{2}\,\frac{1}{x_{\rm N}}\right) , 
	\label{Master_Integral_G}
\end{eqnarray}
 
\begin{eqnarray}
        {\cal X}_{\left(6\right)}\left(t\right)  
	&=& - \frac{1}{8}\,\frac{1}{\left(d_{\sigma}\right)^2} 
	\left(\frac{1}{\left(x_{\rm N}\right)^2} - 3\,\frac{\ve{\sigma} \cdot \ve{x}_{\rm N}}{\left(d_{\sigma}\right)^3} 
	\left(\arctan \frac{\ve{\sigma} \cdot \ve{x}_{\rm N}}{d_{\sigma}} + \frac{\pi}{2}\right) \right) ,
	\label{Master_Integral_G1}
\end{eqnarray}

\begin{eqnarray}
	 {\cal X}_{\left(7\right)}\left(t\right) &=& + \frac{8}{15}\,\frac{1}{\left(d_{\sigma}\right)^2} 
        \left(\frac{x_{\rm N}}{\left(d_{\sigma}\right)^4} - \frac{1}{2}\,\frac{1}{x_{\rm N}}\,\frac{1}{\left(d_{\sigma}\right)^2} 
	- \frac{1}{8}\,\frac{1}{\left(x_{\rm N}\right)^3} 
	+ \frac{\ve{\sigma} \cdot \ve{x}_{\rm N}}{\left(d_{\sigma}\right)^4}\right) , 
	\label{Master_Integral_H}
\end{eqnarray}
	
\begin{eqnarray}
	 {\cal Y}_{\left(5\right)}\left(t\right) &=& - \frac{1}{3}\,\frac{1}{\left(d_{\sigma}\right)^2} 
	\left(\left(\frac{\ve{\sigma} \cdot \ve{x}_{\rm N}}{x_{\rm N}} + 1\right) 
	\ln \frac{x_{\rm N} - \ve{\sigma} \cdot \ve{x}_{\rm N}}{x_0 - \ve{\sigma} \cdot \ve{x}_0} 
	+ \frac{1}{2}\,\frac{\ve{\sigma} \cdot \ve{x}_{\rm N}}{d_{\sigma}}
	\left(\arctan \frac{\ve{\sigma}\cdot\ve{x}_{\rm N}}{d_{\sigma}} + \frac{\pi}{2}\right)\right)
	+ \,\frac{1}{3}\,\frac{1}{\left(d_{\sigma}\right)^2}\,  
	\ln \frac{x_{\rm N} - \ve{\sigma} \cdot \ve{x}_{\rm N}}{2\,x_{\rm N}} \;,
	\nonumber\\ 
	\label{Master_Integral_K5_2}
\end{eqnarray}
 
\begin{eqnarray}
	{\cal Z}_{\left(3\right)}\left(t\right) &=& \frac{x_{\rm N} + \ve{\sigma} \cdot \ve{x}_{\rm N}}{\left(d_{\sigma}\right)^2} \,
	\ln \frac{x_{\rm N} - \ve{\sigma} \cdot \ve{x}_{\rm N}}{x_0 - \ve{\sigma} \cdot \ve{x}_0} 
	+ \frac{1}{d_{\sigma}}\,\arctan \frac{\ve{\sigma} \cdot \ve{x}_{\rm N}}{d_{\sigma}} 
	- \frac{\ve{\sigma} \cdot \ve{x}_{\rm N}}{\left(d_{\sigma}\right)^2}\,\ln \frac{x_{\rm N} - \ve{\sigma} \cdot \ve{x}_{\rm N}}{2\,x_{\rm N}} \;, 
        \label{Master_Integral_L3_2}
\end{eqnarray}

\begin{eqnarray}
	 {\cal Z}_{\left(5\right)}\left(t\right) &=& \frac{2}{3}\,\frac{1}{\left(d_{\sigma}\right)^2} 
	\left(\frac{x_{\rm N}}{\left(d_{\sigma}\right)^2} + \frac{\ve{\sigma} \cdot \ve{x}_{\rm N}}{\left(d_{\sigma}\right)^2}  
	- \frac{1}{2}\,\frac{1}{x_{\rm N}}\right) \ln \frac{x_{\rm N} - \ve{\sigma} \cdot \ve{x}_{\rm N}}{x_0 - \ve{\sigma} \cdot \ve{x}_0} 
	+ \frac{1}{6}\,\frac{1}{\left(d_{\sigma}\right)^3}\,\arctan \frac{\ve{\sigma} \cdot \ve{x}_{\rm N}}{d_{\sigma}} 
	- \frac{2}{3}\,\frac{\ve{\sigma} \cdot \ve{x}_{\rm N}}{\left(d_{\sigma}\right)^4}\,
	\ln \frac{x_{\rm N} - \ve{\sigma} \cdot \ve{x}_{\rm N}}{2\,x_{\rm N}} \;,
	\nonumber\\ 
	\label{Master_Integral_L5_2}
\end{eqnarray}

\noindent
where the time arguments have been omitted in order to simplify the notation. Note that terms like 
$\ln \left(x_{\rm N} - \ve{\sigma} \cdot \ve{x}_{\rm N}\right)$ in (\ref{Master_Integral_E1}) and 
$\ln x_{\rm N}$ in (\ref{Master_Integral_F1}) are well-defined in view of (\ref{Master_Integral_I2}) and (\ref{Master_Integral_J2}).

\section{The terms of 2PN geodesic equation}\label{Appendix_2PN_Terms}

In this appendix each term of the 2PN geodesic equation (\ref{Geodesic_Equation_2PN_5}) is given, which can be 
written in the following form: 
\begin{eqnarray}
	\frac{\ddot{x}^i}{c^2} &=& \frac{\ddot{x}_{A}^i}{c^2} 
	+ \frac{\ddot{x}_{B}^i}{c^2} + \frac{\ddot{x}_{C}^i}{c^2} + \frac{\ddot{x}_{D}^i}{c^2} 
	+ \frac{\ddot{x}_{E}^i}{c^2} + \frac{\ddot{x}_{F}^i}{c^2} + \frac{\ddot{x}_{G}^i}{c^2} 
	+ \frac{\ddot{x}_{H}^i}{c^2} + \frac{\ddot{x}_{I}^i}{c^2} 
	+ \frac{\ddot{x}_{J}^i}{c^2} + {\cal O}\left(c^{-6}\right) ,  
\label{Geodesic_Equation_Appendix_5}
\end{eqnarray}

\noindent
where the individual terms are   
\begin{eqnarray}
	     \frac{\ddot{x}_{A}^i}{c^2} &=&  h_{00,i}^{(2)}\left(\ve{x}_{\rm 1PN}\right) \;,
	     \label{Individual_A}
\end{eqnarray}

\begin{eqnarray}
	\frac{\ddot{x}_{B}^i}{c^2} &=& - 2\,h_{00,j}^{(2)}\left(\ve{x}_{\rm 1PN}\right) \,\sigma^i \sigma^j \;, 
	     \label{Individual_B}
\end{eqnarray}

\begin{eqnarray}
	\frac{\ddot{x}_{C}^i}{c^2} &=& h_{00,i}^{(2)}\left(\ve{x}_{\rm N}\right)\,\frac{\sigma^j \Delta \dot{x}^j_{\rm 1PN}}{c}\;,
	     \label{Individual_C}
\end{eqnarray}

\begin{eqnarray}
	\frac{\ddot{x}_{D}^i}{c^2} &=& -2\,h_{00,j}^{(2)}\left(\ve{x}_{\rm N}\right)\,\frac{\Delta \dot{x}^j_{\rm 1PN}}{c}\,\sigma^i\;,
	     \label{Individual_D}
\end{eqnarray}

\begin{eqnarray}
 	\frac{\ddot{x}_{E}^i}{c^2} &=& - 2\,h_{00,j}^{(2)}\left(\ve{x}_{\rm N}\right)\,\frac{\Delta \dot{x}^i_{\rm 1PN}}{c}\,\sigma^j\;, 
	     \label{Individual_E}
\end{eqnarray}

\begin{eqnarray}
	\frac{\ddot{x}_{F}^i}{c^2} &=& - h_{00}^{(2)}\left(\ve{x}_{\rm N}\right)\,h_{00,i}^{(2)}\left(\ve{x}_{\rm N}\right), 
	     \label{Individual_F}
\end{eqnarray}
   
\begin{eqnarray}
 	\frac{\ddot{x}_{G}^i}{c^2} &=& \frac{1}{2}\,h_{00,i}^{(4)}\left(\ve{x}_{\rm N}\right),   
	     \label{Individual_G}
\end{eqnarray}

\begin{eqnarray}
	\frac{\ddot{x}_{H}^i}{c^2} &=& - h_{00,j}^{(4)}\left(\ve{x}_{\rm N}\right)\,\sigma^i \sigma^j\;, 
	     \label{Individual_H}
\end{eqnarray}

\begin{eqnarray}
 	\frac{\ddot{x}_{I}^i}{c^2} &=& - h_{ij,k}^{(4)}\left(\ve{x}_{\rm N}\right)\,\sigma^j \sigma^k\;,   
	     \label{Individual_I}
\end{eqnarray}

\begin{eqnarray}
	\frac{\ddot{x}_{J}^i}{c^2} &=& \frac{1}{2}\,h_{jk,i}^{(4)}\left(\ve{x}_{\rm N}\right)\,\sigma^j \sigma^k\;.  
	     \label{Individual_J}
\end{eqnarray}

\noindent 
The following subsections are devoted to each of these $10$ individual terms. In what follows, one needs the spatial coordinates  
and the inverse powers of their absolute values in 1PN approximation, that is, up to terms of the order ${\cal O}\left(c^{-4}\right)$:  
\begin{eqnarray}
	\ve{x}_{\rm 1PN} &=& \ve{x}_{\rm N} + \Delta \ve{x}_{\rm 1PN}\;,  
        \label{Appendix_x_A}
        \\
        x_{\rm 1PN} &=& x_{\rm N} + \frac{\ve{x}_{\rm N} \cdot \Delta \ve{x}_{\rm 1PN}}{x_{\rm N}}\;, 
        \label{Appendix_x_B}
        \\
        \frac{1}{\left(x_{\rm 1PN}\right)^n} &=& \frac{1}{\left(x_{\rm N}\right)^n} - \frac{n}{\left(x_{\rm N}\right)^n} \, 
        \frac{\ve{x}_{\rm N} \cdot \Delta \ve{x}_{\rm 1PN}}{\left(x_{\rm N}\right)^2}\;,    
        \label{Appendix_x_C}
\end{eqnarray}

\noindent
where Eq.~(\ref{Appendix_x_A}) was given by Eq.~(\ref{Second_Integration_1PN}) and expression $\Delta \ve{x}_{\rm 1PN}$ is given 
by Eqs.~(\ref{Second_Integration_1PN_M_Q}) - (\ref{Second_Integration_1PN_Q}).

\subsection{The term $\ddot{x}_{A}^i$}\label{Appendix_Term_A}  

Determining the spatial derivatives of $h_{00}^{(2)}\left(\ve{x}\right)$ in (\ref{metric_1PN_00}) and taking 
$\ve{x} = \ve{x}_{\rm 1PN} + {\cal O}\left(c^{-4}\right)$ one gets 
\begin{eqnarray}
	 \frac{\ddot{x}_{A}^i}{c^2} &=& - 2\, \frac{G M}{c^2}\,\frac{x_{\rm 1PN}^i}{\left(x_{\rm 1PN}\right)^3} 
	+ 6\,\frac{G \hat{M}_{ai}}{c^2} \,\frac{x_{\rm 1PN}^a}{\left(x_{\rm 1PN}\right)^5}
	- 15\,\frac{G \hat{M}_{ab}}{c^2} \,\frac{x_{\rm 1PN}^a x_{\rm 1PN}^b x_{\rm 1PN}^i}{\left(x_{\rm 1PN}\right)^7}\,. 
\label{Appendix_Term_A_5}
\end{eqnarray}

\noindent 
By means of (\ref{Appendix_x_A}) and (\ref{Appendix_x_C}) one obtains   
\begin{eqnarray}
	 \frac{\ddot{x}_{A}^i}{c^2} &=& - 2\, \frac{G M}{c^2}\,\frac{x_{\rm N}^i}{\left(x_{\rm N}\right)^3} 
         + \,6\,\frac{G \hat{M}_{ai}}{c^2} \,\frac{x_{\rm N}^a}{\left(x_{\rm N}\right)^5} 
	- \,15\,\frac{G \hat{M}_{ab}}{c^2} \,\frac{x_{\rm N}^a x_{\rm N}^b x_{\rm N}^i}{\left(x_{\rm N}\right)^7} 
	- 2\, \frac{G M}{c^2}\,\frac{\Delta x_{\rm 1PN}^i}{\left(x_{\rm N}\right)^3}
	+ 6\,\frac{G M}{c^2}\,\frac{x_{\rm N}^i}{\left(x_{\rm N}\right)^5}\,\left(\ve{x}_{\rm N} \cdot \Delta \ve{x}_{\rm 1PN}\right)  
\nonumber\\
	&&  
        + \,6\,\frac{G \hat{M}_{ai}}{c^2} \,\frac{\Delta x_{\rm 1PN}^a}{\left(x_{\rm N}\right)^5}
	- 30 \,\frac{G \hat{M}_{ai}}{c^2} \,\frac{x_{\rm N}^a}{\left(x_{\rm N}\right)^7} \,\left(\ve{x}_{\rm N} \cdot \Delta \ve{x}_{\rm 1PN}\right)
	- \,15\,\frac{G \hat{M}_{ab}}{c^2} \,\frac{x_{\rm N}^a x_{\rm N}^b}{\left(x_{\rm N}\right)^7}\,\Delta x_{\rm 1PN}^i  
	\nonumber\\ 
	&& 
	- \,30\,\frac{G \hat{M}_{ab}}{c^2} \,\frac{x_{\rm N}^a x_{\rm N}^i}{\left(x_{\rm N}\right)^7}\,\Delta x_{\rm 1PN}^b
        + 105\,\frac{G \hat{M}_{ab}}{c^2} \,\frac{x_{\rm N}^a x_{\rm N}^b x_{\rm N}^i}{\left(x_{\rm N}\right)^9}\,
	\left(\ve{x}_{\rm N} \cdot \Delta \ve{x}_{\rm 1PN}\right) ,
\label{Appendix_Term_A_10}
\end{eqnarray}

\noindent
where the symmetry $\hat{M}_{ab} = \hat{M}_{ba}$ has been used in the next-to-last term.   
According to Eq.~(\ref{Second_Integration_1PN_M_Q}), the term $\Delta \ve{x}_{\rm 1PN}$ is separated into  
a monopole and quadrupole term, $\Delta \ve{x}_{\rm 1PN} = \Delta \ve{x}^{\rm M}_{\rm 1PN} + \Delta \ve{x}^{\rm Q}_{\rm 1PN}$,  
given by Eq.~(\ref{Second_Integration_1PN_M}) and (\ref{Second_Integration_1PN_Q}), respectively.

\subsection{The term $\ddot{x}_{B}^i$}\label{Appendix_Term_B} 

Determining the spatial derivatives of $h_{00}^{(2)}\left(\ve{x}\right)$ in (\ref{metric_1PN_00}) and taking $\ve{x} = \ve{x}_{\rm 1PN}$ one gets 
\begin{eqnarray}
	\frac{\ddot{x}_{B}^i}{c^2} &=& 4\,\frac{G M}{c^2}\,\sigma^a \sigma^i \frac{x_{\rm 1PN}^a}{\left(x_{\rm 1PN}\right)^3}   
	- 12\,\frac{G \hat{M}_{ab}}{c^2}\,\sigma^b \sigma^i \frac{x_{\rm 1PN}^a}{\left(x_{\rm 1PN}\right)^5} 
	+ 30\,\frac{G \hat{M}_{ab}}{c^2}\,\sigma^c \sigma^i \frac{x_{\rm 1PN}^a x_{\rm 1PN}^b x_{\rm 1PN}^c}{\left(x_{\rm 1PN}\right)^7}\,.  
	\label{Appendix_Term_B_5}
\end{eqnarray}

\noindent
By inserting the spatial coordinates and their absolute values in 1PN approximation (\ref{Appendix_x_A}) - (\ref{Appendix_x_C}) one obtains  
\begin{eqnarray}
	\frac{\ddot{x}_{B}^i}{c^2} &=& 4\,\frac{G M}{c^2}\,\sigma^i\,\frac{\ve{\sigma} \cdot \ve{x}_{\rm N}}{\left(x_{\rm N}\right)^3}  
        - 12\,\frac{G \hat{M}_{ab}}{c^2}\,\sigma^b \sigma^i\,\frac{x_{\rm N}^a}{\left(x_{\rm N}\right)^5} 
	+ 30\,\frac{G \hat{M}_{ab}}{c^2}\,\sigma^i\,x_{\rm N}^a x_{\rm N}^b \,\frac{\ve{\sigma} \cdot \ve{x}_{\rm N}}{\left(x_{\rm N}\right)^7}
	+ 4\, \frac{G M}{c^2}\,\sigma^i\,\frac{\ve{\sigma} \cdot \Delta \ve{x}_{\rm 1PN}}{\left(x_{\rm N}\right)^3} 
	\nonumber\\
	&&  
	- 12\,\frac{G M}{c^2}\,\sigma^i\,\frac{\ve{\sigma} \cdot \ve{x}_{\rm N}}{\left(x_{\rm N}\right)^5}\,
	\left(\ve{x}_{\rm N} \cdot \Delta \ve{x}_{\rm 1PN}\right)    
	- 12\,\frac{G \hat{M}_{ab}}{c^2}\,\sigma^b \sigma^i\,\frac{\Delta x_{\rm 1PN}^a}{\left(x_{\rm N}\right)^5} 
	+ 60\,\frac{G \hat{M}_{ab}}{c^2}\,\sigma^b \sigma^i\,\frac{x_{\rm N}^a}{\left(x_{\rm N}\right)^7}\, 
	\left(\ve{x}_{\rm N} \cdot \Delta \ve{x}_{\rm 1PN}\right)      
	\nonumber\\
	&&  
	+ 30 \frac{G \hat{M}_{ab}}{c^2}\,\sigma^i \frac{x_{\rm N}^a x_{\rm N}^b}{\left(x_{\rm N}\right)^7}\,\left(\ve{\sigma}\cdot\Delta \ve{x}_{\rm 1PN}\right) 
	+ 60 \frac{G \hat{M}_{ab}}{c^2}\,\sigma^i x_{\rm N}^a \frac{\ve{\sigma} \cdot \ve{x}_{\rm N}}{\left(x_{\rm N}\right)^7}\,\Delta x_{\rm 1PN}^b   
	- 210 \frac{G \hat{M}_{ab}}{c^2}\,\sigma^i x_{\rm N}^a x_{\rm N}^b \frac{\ve{\sigma} \cdot \ve{x}_{\rm N}}{\left(x_{\rm N}\right)^9} 
	\left(\ve{x}_{\rm N} \cdot \Delta \ve{x}_{\rm 1PN}\right) , 
	\nonumber\\ 
        \label{Appendix_Term_B_10}
\end{eqnarray}

\noindent 
where the symmetry $\hat{M}_{ab} = \hat{M}_{ba}$ has been used in the next-to-last term.  
According to Eq.~(\ref{Second_Integration_1PN_M_Q}), the term $\Delta \ve{x}_{\rm 1PN}$ is separated into  
a monopole and quadrupole term, $\Delta \ve{x}_{\rm 1PN} = \Delta \ve{x}^{\rm M}_{\rm 1PN} + \Delta \ve{x}^{\rm Q}_{\rm 1PN}$,            
given by Eq.~(\ref{Second_Integration_1PN_M}) and (\ref{Second_Integration_1PN_Q}), respectively.

\subsection{The term $\ddot{x}_{C}^i$}\label{Appendix_Term_C}  

Using (\ref{spatial_derivative_1PN}) one obtains  
\begin{eqnarray}
	\frac{\ddot{x}_{C}^i}{c^2} &=& - 2\, \frac{G M}{c^2}\,\frac{x_{\rm N}^i}{\left(x_{\rm N}\right)^3}\,
	\frac{\ve{\sigma} \cdot \Delta \dot{\ve{x}}_{\rm 1PN}}{c}  
	+ 6\,\frac{G \hat{M}_{ai}}{c^2} \,\frac{x_{\rm N}^a}{\left(x_{\rm N}\right)^5}\,\frac{\ve{\sigma} \cdot \Delta \dot{\ve{x}}_{\rm 1PN}}{c}  
	- 15\,\frac{G \hat{M}_{ab}}{c^2} \,\frac{x_{\rm N}^a x_{\rm N}^b x_{\rm N}^i}{\left(x_{\rm N}\right)^7}\,
	\frac{\ve{\sigma} \cdot \Delta \dot{\ve{x}}_{\rm 1PN}}{c}\;. 
	\label{Appendix_Term_C_5}
\end{eqnarray}

\noindent
According to Eq.~(\ref{First_Integration_1PN_M_Q}), the term $\Delta \dot{\ve{x}}_{\rm 1PN}$ is separated into  
a monopole and quadrupole term, $\Delta \dot{\ve{x}}_{\rm 1PN} = \Delta \dot{\ve{x}}^{\rm M}_{\rm 1PN} + \Delta \dot{\ve{x}}^{\rm Q}_{\rm 1PN}$,            
given by Eq.~(\ref{First_Integration_1PN_M}) and (\ref{First_Integration_1PN_Q}), respectively.

\subsection{The term $\ddot{x}_{D}^i$}\label{Appendix_Term_D}  

Using (\ref{spatial_derivative_1PN}) one obtains
\begin{eqnarray}
        \frac{\ddot{x}_{D}^i}{c^2} &=& + 4\, \frac{G M}{c^2}\,\sigma^i\,\frac{1}{\left(x_{\rm N}\right)^3}\,
	\frac{\ve{x}_{\rm N} \cdot \Delta \dot{\ve{x}}_{\rm 1PN}}{c}  
	- 12 \,\frac{G \hat{M}_{aj}}{c^2}\,\sigma^i\,\frac{x_{\rm N}^a}{\left(x_{\rm N}\right)^5}\,\frac{\Delta \dot{x}^j_{\rm 1PN}}{c}   
	+ 30\,\frac{G \hat{M}_{ab}}{c^2}\,\sigma^i\,\frac{x_{\rm N}^a x_{\rm N}^b}{\left(x_{\rm N}\right)^7}\,
	\frac{\ve{x}_{\rm N} \cdot \Delta \dot{\ve{x}}_{\rm 1PN}}{c}\;. 
	\label{Appendix_Term_D_5}
\end{eqnarray}

\noindent
According to Eq.~(\ref{First_Integration_1PN_M_Q}), the term $\Delta \dot{\ve{x}}_{\rm 1PN}$ is separated into 
a monopole and quadrupole term, $\Delta \dot{\ve{x}}_{\rm 1PN} = \Delta \dot{\ve{x}}^{\rm M}_{\rm 1PN} + \Delta \dot{\ve{x}}^{\rm Q}_{\rm 1PN}$,
given by Eq.~(\ref{First_Integration_1PN_M}) and (\ref{First_Integration_1PN_Q}), respectively.

\subsection{The term $\ddot{x}_{E}^i$}\label{Appendix_Term_E} 

Using (\ref{spatial_derivative_1PN}) one obtains
\begin{eqnarray}
	\frac{\ddot{x}_{E}^i}{c^2} &=& + 4\,\frac{G M}{c^2}\,\frac{\ve{\sigma} \cdot \ve{x}_{\rm N}}{\left(x_{\rm N}\right)^3}\,
        \frac{\Delta \dot{x}^i_{\rm 1PN}}{c}  
        - 12 \,\frac{G \hat{M}_{ab}}{c^2}\,\sigma^b\,\frac{x_{\rm N}^a}{\left(x_{\rm N}\right)^5}\,\frac{\Delta \dot{x}^i_{\rm 1PN}}{c} 
	+ 30 \,\frac{G \hat{M}_{ab}}{c^2}\,x_{\rm N}^a x_{\rm N}^b \,\frac{\ve{\sigma} \cdot \ve{x}_{\rm N}}{\left(x_{\rm N}\right)^7}\,
	\frac{\Delta \dot{x}^i_{\rm 1PN}}{c}\;. 
        \label{Appendix_Term_E_5}
\end{eqnarray}

\noindent
According to Eq.~(\ref{First_Integration_1PN_M_Q}), the term $\Delta \dot{\ve{x}}_{\rm 1PN}$ is separated into 
a monopole and quadrupole term, $\Delta \dot{\ve{x}}_{\rm 1PN} = \Delta \dot{\ve{x}}^{\rm M}_{\rm 1PN} + \Delta \dot{\ve{x}}^{\rm Q}_{\rm 1PN}$,
given by Eq.~(\ref{First_Integration_1PN_M}) and (\ref{First_Integration_1PN_Q}), respectively.

\subsection{The term $\ddot{x}_{F}^i$}\label{Appendix_Term_F}

Using (\ref{metric_1PN_00}) and (\ref{spatial_derivative_1PN}) one obtains
\begin{eqnarray}
	\frac{\ddot{x}_{F}^i}{c^2} &=& + 4\, \frac{G^2 M^2}{c^4}\,\frac{x^i_{\rm N}}{\left(x_{\rm N}\right)^4} 
	-12\,\frac{G M}{c^2}\,\frac{G \hat{M}_{ai}}{c^2}\,\frac{x^a_{\rm N}}{\left(x_{\rm N}\right)^6} 
	+ 36\, \frac{G M}{c^2}\,\frac{G \hat{M}_{ab}}{c^2}\,\frac{x^a_{\rm N} x^b_{\rm N} x^i_{\rm N}}{\left(x_{\rm N}\right)^8}
	 - 18\,\frac{G \hat{M}_{ab}}{c^2}\,\frac{G \hat{M}_{ci}}{c^2}\,\frac{x^a_{\rm N} x^b_{\rm N} x^c_{\rm N}}{\left(x_{\rm N}\right)^{10}}  
	\nonumber\\
	&& 
   + 45\,\frac{G \hat{M}_{ab}}{c^2}\,\frac{G \hat{M}_{cd}}{c^2}\,\frac{x^a_{\rm N} x^b_{\rm N} x^c_{\rm N} x^d_{\rm N} x^i_{\rm N}}{\left(x_{\rm N}\right)^{12}}\;.  
        \label{Appendix_Term_F_5}
\end{eqnarray}

\subsection{The term $\ddot{x}_{G}^i$}\label{Appendix_Term_G}  

Using (\ref{metric_2PN_00}) by performing the spatial derivatives and then using  
$x^i = x_{\rm N}^i + {\cal O}\left(c^{-2}\right)$, one obtains 
\begin{eqnarray}
        \frac{\ddot{x}_{G}^i}{c^2} &=& 
	+ 2\,\frac{G^2 M^2}{c^4}\,\frac{x^i_{\rm N}}{\left(x_{\rm N}\right)^4}
	+ 12 \frac{G M}{c^2}\,\frac{G \hat{M}_{ab}}{c^2}\,\frac{x^i_{\rm N}}{\left(x_{\rm N}\right)^6}\,\hat{n}_{\rm N}^{ab}
	- 3\,\frac{G M}{c^2}\,\frac{G \hat{M}_{ab}}{c^2}\,\frac{\left(\hat{n}_{\rm N}^{ab}\right)_{,\,i}}{\left(x_{\rm N}\right)^4} 
	+ \frac{9}{5}\,\frac{G \hat{M}_{ab}}{c^2}\,\frac{G \hat{M}_{ab}}{c^2}\,\frac{x^i_{\rm N}}{\left(x_{\rm N}\right)^8} 
\nonumber\\
	&& 
	+ \frac{54}{7}\,\frac{G \hat{M}_{ab}}{c^2}\,\frac{G \hat{M}_{ac}}{c^2}\,\frac{x^i_{\rm N}}{\left(x_{\rm N}\right)^8}\,\hat{n}_{\rm N}^{bc} 
	- \frac{9}{7}\,\frac{G \hat{M}_{ab}}{c^2}\,\frac{G \hat{M}_{ac}}{c^2}\,\frac{\left(\hat{n}_{\rm N}^{bc}\right)_{,\,i}}{\left(x_{\rm N}\right)^6}
	+ \frac{27}{2}\,\frac{G \hat{M}_{ab}}{c^2}\,\frac{G \hat{M}_{cd}}{c^2}\,\frac{x^i_{\rm N}}{\left(x_{\rm N}\right)^8}\,\hat{n}_{\rm N}^{abcd} 
\nonumber\\
	&& 
	- \frac{9}{4}\,\frac{G \hat{M}_{ab}}{c^2}\,\frac{G \hat{M}_{cd}}{c^2}\,\frac{\left(\hat{n}_{\rm N}^{abcd}\right)_{,\,i}}{\left(x_{\rm N}\right)^6}\;, 
                \label{Appendix_Term_G_5}
\end{eqnarray}

\noindent
where the STF tensors $\hat{n}^{ab}_{\rm N}$ and $\hat{n}^{abcd}_{\rm N}$ are given by were by Eq.~(\ref{Appendix_Cartesian_Tensor_General_Formula_N})  
and their spatial derivatives are given by were by Eq.~(\ref{Appendix_Cartesian_Tensor_General_Formula_Derivative_N}) in \ref{Appendix_STF}. 
By using $\hat{M}_{ab}\,\delta^{ab} = 0$ and summarizing some terms, the expression (\ref{Appendix_Term_G_5}) can be simplified.  
Such simplifications will not be performed at this stage but will be postponed for awhile.

\subsection{The term $\ddot{x}_{H}^i$}\label{Appendix_Term_H}  

Using (\ref{metric_2PN_00}) by performing the spatial derivatives and then using     
$x^i = x_{\rm N}^i + {\cal O}\left(c^{-2}\right)$, one obtains  
\begin{eqnarray}
        \frac{\ddot{x}_{H}^i}{c^2} &=&  
	- 4\,\frac{G^2 M^2}{c^4}\,\sigma^i\,\frac{\ve{\sigma} \cdot \ve{x}_{\rm N}}{\left(x_{\rm N}\right)^4}  
	- 24\,\frac{G M}{c^2}\,\frac{G \hat{M}_{ab}}{c^2}\,\sigma^i\,\frac{\ve{\sigma} \cdot \ve{x}_{\rm N}}{\left(x_{\rm N}\right)^6}\,\hat{n}_{\rm N}^{ab}   
	+ 6\,\frac{G M}{c^2}\,\frac{G \hat{M}_{ab}}{c^2}\,\sigma^i \sigma^j\,\frac{\left(\hat{n}_{\rm N}^{ab}\right)_{,\,j}}{\left(x_{\rm N}\right)^4}   
	- \frac{18}{5}\,\frac{G \hat{M}_{ab}}{c^2}\,\frac{G \hat{M}_{ab}}{c^2}\,\sigma^i\, 
	\,\frac{\ve{\sigma} \cdot \ve{x}_{\rm N}}{\left(x_{\rm N}\right)^8}  
\nonumber\\
	&&  
	- \frac{108}{7}\,\frac{G \hat{M}_{ab}}{c^2}\,\frac{G \hat{M}_{ac}}{c^2}\,\sigma^i\, 
	\,\frac{\ve{\sigma} \cdot \ve{x}_{\rm N}}{\left(x_{\rm N}\right)^8}\,\hat{n}_{\rm N}^{bc}    
	+ \frac{18}{7}\,\frac{G \hat{M}_{ab}}{c^2}\,\frac{G \hat{M}_{ac}}{c^2}\,\sigma^i \sigma^j\, 
	\,\frac{\left(\hat{n}_{\rm N}^{bc}\right)_{,\,j}}{\left(x_{\rm N}\right)^6}   
	- 27\,\frac{G \hat{M}_{ab}}{c^2}\,\frac{G \hat{M}_{cd}}{c^2}
	\,\sigma^i\,\frac{\ve{\sigma} \cdot \ve{x}_{\rm N}}{\left(x_{\rm N}\right)^8}\,\hat{n}_{\rm N}^{abcd}  
\nonumber\\
	&&  
	+ \frac{9}{2}\,\frac{G \hat{M}_{ab}}{c^2}\,\frac{G \hat{M}_{cd}}{c^2}\,\sigma^i \sigma^j\,  
	\,\frac{\left(\hat{n}_{\rm N}^{abcd}\right)_{,\,j}}{\left(x_{\rm N}\right)^6}  \;,  
                \label{Appendix_Term_H_5}
\end{eqnarray}

\noindent 
where $\hat{n}_{\rm N}^{ab}$ and $\hat{n}_{\rm N}^{abcd}$ are given by Eq.~(\ref{Appendix_Cartesian_Tensor_General_Formula_N}). 
By using $\hat{M}_{ab}\,\delta^{ab} = 0$ and summarizing some terms, the expression (\ref{Appendix_Term_H_5}) can be simplified.
Such simplifications will not be performed at this stage but will be postponed for awhile.

\subsection{The term $\ddot{x}_{I}^i$}\label{Appendix_Term_I}  

Using (\ref{metric_2PN_ij}) by performing the spatial derivatives and then using
$x^i = x_{\rm N}^i + {\cal O}\left(c^{-2}\right)$, one obtains
\begin{eqnarray}
        \frac{\ddot{x}_{I}^i}{c^2} &=& 
	+ \frac{G^2 M^2}{c^4}\,\sigma^i\,\frac{\ve{\sigma} \cdot \ve{x}_{\rm N}}{\left(x_{\rm N}\right)^4}  
	+ 3\,\frac{G^2 M^2}{c^4}\,\frac{x^i_{\rm N}}{\left(x_{\rm N}\right)^4}
	- 4\,\frac{G^2 M^2}{c^4}\,\left(d_{\sigma}\right)^2\,\frac{x^i_{\rm N}}{\left(x_{\rm N}\right)^6}
	+ 4\,\frac{G M}{c^2}\,\frac{G \hat{M}_{ab}}{c^2}\,\sigma^j\,\frac{\ve{\sigma} \cdot \ve{x}_{\rm N}}{\left(x_{\rm N}\right)^6}\,T_{\rm N}^{ijab}  
	\nonumber\\
	&&  
	 - \frac{G M}{c^2}\,\frac{G \hat{M}_{ab}}{c^2}\,\sigma^j \sigma^k\,\frac{1}{\left(x_{\rm N}\right)^4}\,\left(T_{\rm N}^{ijab}\right)_{,\,k}   
	+ 6\,\frac{G \hat{M}_{ab}}{c^2}\,\frac{G \hat{M}_{cd}}{c^2}\,\sigma^j\,\frac{\ve{\sigma} \cdot \ve{x}_{\rm N}}{\left(x_{\rm N}\right)^8}\,
	T_{\rm N}^{ijabcd}
        \nonumber\\
	&&  
	- \frac{G \hat{M}_{ab}}{c^2}\,\frac{G \hat{M}_{cd}}{c^2}\,\sigma^j \sigma^k\,\frac{1}{\left(x_{\rm N}\right)^6} 
	\left(T_{\rm N}^{ijabcd}\right)_{,\,k}
	+ \frac{96}{35}\,\frac{G \hat{M}_{ab}}{c^2}\,\frac{G \hat{M}_{ab}}{c^2}\,\sigma^i\,\frac{\ve{\sigma} \cdot \ve{x}_{\rm N}}{\left(x_{\rm N}\right)^8} 
	- \frac{36}{35}\,\frac{G \hat{M}_{ab}}{c^2}\,\frac{G \hat{M}_{ai}}{c^2}\,\sigma^b\,\frac{\ve{\sigma} \cdot \ve{x}_{\rm N}}{\left(x_{\rm N}\right)^8}\,,  
\label{Appendix_Term_I_5}
\end{eqnarray}

\noindent
where in the first line relation (\ref{Equation_B}) has been used.  
The tensors $T_{\rm N}^{ijab}$ and $T_{\rm N}^{ijabcd}$ are abbreviations to simplify equation (\ref{Appendix_Term_I_5}). 
These auxiliary quantities are given by Eqs.~(\ref{Appendix_Tensor_S}) and (\ref{Appendix_Tensor_T}) and their spatial 
derivatives by Eqs.~(\ref{Appendix_Tensor_S_diff}) and (\ref{Appendix_Tensor_T_diff}) in \ref{Appendix_Auxiliary_Tensors}.

\subsection{The term $\ddot{x}_{J}^i$}\label{Appendix_Term_J}  

Using (\ref{metric_2PN_ij}) by performing the spatial derivatives and then using
$x^i = x_{\rm N}^i + {\cal O}\left(c^{-2}\right)$, one obtains

\begin{eqnarray}
        \frac{\ddot{x}_{J}^i}{c^2} &=&  
	 + \frac{G^2 M^2}{c^4}\,\sigma^i\,\frac{\ve{\sigma} \cdot \ve{x}_{\rm N}}{\left(x_{\rm N}\right)^4}  
        - 3\,\frac{G^2 M^2}{c^4}\,\frac{x^i_{\rm N}}{\left(x_{\rm N}\right)^4}
        + 2\,\frac{G^2 M^2}{c^4}\,\left(d_{\sigma}\right)^2\,\frac{x^i_{\rm N}}{\left(x_{\rm N}\right)^6}
	- 2\,\frac{G M}{c^2}\,\frac{G \hat{M}_{ab}}{c^2}\,\sigma^j \sigma^k\,\frac{x^i_{\rm N}}{\left(x_{\rm N}\right)^6}\,T_{\rm N}^{jkab} 
        \nonumber\\
	&& \hspace{-1.0cm} 
	 + \frac{1}{2}\,\frac{G M}{c^2}\,\frac{G \hat{M}_{ab}}{c^2}\,\sigma^j \sigma^k\,\frac{1}{\left(x_{\rm N}\right)^4}\,\left(T_{\rm N}^{jkab}\right)_{,\,i}
	- 3\,\frac{G \hat{M}_{ab}}{c^2}\,\frac{G \hat{M}_{cd}}{c^2}\,\sigma^j \sigma^k\,\frac{x^i_{\rm N}}{\left(x_{\rm N}\right)^8}\,
        T_{\rm N}^{jkabcd}   
        \nonumber\\
	&& \hspace{-1.0cm} 
	+ \frac{1}{2}\,\frac{G \hat{M}_{ab}}{c^2}\,\frac{G \hat{M}_{cd}}{c^2}\,\sigma^j \sigma^k\,\frac{1}{\left(x_{\rm N}\right)^6}\,
        \left(T_{\rm N}^{jkabcd}\right)_{,\,i}  
	- \frac{48}{35}\,\frac{G \hat{M}_{ab}}{c^2}\,\frac{G \hat{M}_{ab}}{c^2}\,\frac{x^i_{\rm N}}{\left(x_{\rm N}\right)^8}  
	+ \frac{18}{35}\,\frac{G \hat{M}_{aj}}{c^2}\,\frac{G \hat{M}_{ak}}{c^2}\,\sigma^j \sigma^k\,\frac{x^i_{\rm N}}{\left(x_{\rm N}\right)^8} \;,   
\label{Appendix_Term_J_5}
\end{eqnarray}

\noindent
where the tensors $T_{\rm N}^{jkab}$ and $T_{\rm N}^{jkabcd}$ are given by Eqs.~(\ref{Appendix_Tensor_S}) and (\ref{Appendix_Tensor_T}), while 
their spatial derivatives are given by Eqs.~(\ref{Appendix_Tensor_S_diff}) and (\ref{Appendix_Tensor_T_diff}).

\subsection{The terms proportional to $M \times M$}  

Summarizing all those terms in Eq.~(\ref{Appendix_Term_A_10}) and Eqs.~(\ref{Appendix_Term_B_10}) - (\ref{Appendix_Term_J_5}) which are 
proportional to $M \times M$ one obtains:  
\begin{eqnarray}
        \frac{\ddot{x}^{i\;{\rm M} \times {\rm M}}_{\rm 2PN}}{c^2} &=&  
	- 2\,\frac{G M}{c^2}\,\frac{\Delta x_{\rm 1PN}^{i\,{\rm M}}}{\left(x_{\rm N}\right)^3}
	+ 6\,\frac{G M}{c^2}\,\frac{x_{\rm N}^i}{\left(x_{\rm N}\right)^5}\left(\ve{x}_{\rm N} \cdot \Delta \ve{x}^{\rm M}_{\rm 1PN}\right) 
	+\,4\,\frac{G M}{c^2}\,\sigma^i\,\frac{\ve{\sigma} \cdot \Delta \ve{x}^{\rm M}_{\rm 1PN}}{\left(x_{\rm N}\right)^3}   
	- 12\,\frac{G M}{c^2}\,\sigma^i\,\frac{\ve{\sigma} \cdot \ve{x}_{\rm N}}{\left(x_{\rm N}\right)^5}\,
	\left(\ve{x}_{\rm N} \cdot \Delta \ve{x}^{\rm M}_{\rm 1PN}\right)  
	\nonumber\\ 
	&& -\,2\, \frac{G M}{c^2}\,\frac{x_{\rm N}^i}{\left(x_{\rm N}\right)^3}\,
        \frac{\ve{\sigma} \cdot \Delta \dot{\ve{x}}^{\rm M}_{\rm 1PN}}{c}  
         + 4\,\frac{G M}{c^2}\,\sigma^i\,\frac{1}{\left(x_{\rm N}\right)^3}\,
        \frac{\ve{x}_{\rm N} \cdot \Delta \dot{\ve{x}}^{\rm M}_{\rm 1PN}}{c}  
	+\,4\,\frac{G M}{c^2}\,\frac{\ve{\sigma} \cdot \ve{x}_{\rm N}}{\left(x_{\rm N}\right)^3}\,
	\frac{\Delta \dot{x}^{i\,{\rm M}}_{\rm 1PN}}{c}  
        + \,6\, \frac{G^2 M^2}{c^4}\,\frac{x^i_{\rm N}}{\left(x_{\rm N}\right)^4}
	\nonumber\\ 
	&& - \,4\,\frac{G^2 M^2}{c^4}\,\sigma^i\,\frac{\ve{\sigma} \cdot \ve{x}_{\rm N}}{\left(x_{\rm N}\right)^4}
        + 2\,\frac{G^2 M^2}{c^4}\,\sigma^i\,\frac{\ve{\sigma} \cdot \ve{x}_{\rm N}}{\left(x_{\rm N}\right)^4}
        - 2\,\frac{G^2 M^2}{c^4}\,\left(d_{\sigma}\right)^2\,\frac{x^i_{\rm N}}{\left(x_{\rm N}\right)^6}\;, 
	\label{Appendix_Term_M_M}
\end{eqnarray}

\noindent
where $\Delta\dot{\ve{x}}^{\rm M}_{\rm 1PN}$ and $\Delta\ve{x}^{\rm M}_{\rm 1PN}$ are given by Eqs.~(\ref{First_Integration_1PN_M}) 
and (\ref{Second_Integration_1PN_M}), respectively.  
The equation (\ref{Appendix_Term_M_M}) can be written in the form given by (\ref{2PN_Term_M_M}) by means of relation (\ref{Equation_A}).  
Then, the coefficients in (\ref{2PN_Term_M_M}) read as follows:
\begin{eqnarray} 
	{\cal E}^{i}_{\left(3\right)} &=& - \frac{4}{\left(d_{\sigma}\right)^2}\,d_{\sigma}^i \;,   
\label{coefficient_E3}
\\
	{\cal E}^{i}_{\left(4\right)} &=& - 4\,\sigma^i  \;, 
\label{coefficient_E4}
\\
	{\cal E}^{i}_{\left(5\right)} &=& - 12\,d_{\sigma}^i - 12\,\left(x_0 + \ve{\sigma} \cdot \ve{x}_0 \right)\,\sigma^i  \;, 
\label{coefficient_E5}
\\
	{\cal E}^{i}_{\left(6\right)} &=& - 2\, \left(d_{\sigma}\right)^2\, \sigma^i  \;, 
\label{coefficient_E6}
\\
	{\cal F}^{i}_{\left(2\right)} &=& - \frac{4}{\left(d_{\sigma}\right)^2}\,d_{\sigma}^i  \;, 
\label{coefficient_F2}
\\
	{\cal F}^{i}_{\left(3\right)} &=& + 4\,\sigma^i - \frac{4}{\left(d_{\sigma}\right)^2}\left(x_0 + \ve{\sigma} \cdot \ve{x}_0\right) d_{\sigma}^i \;, 
\label{coefficient_F3}
\end{eqnarray}

\begin{eqnarray} 
	{\cal F}^{i}_{\left(4\right)} &=& + 6\,d_{\sigma}^i  \;, 
\label{coefficient_F4}
\\
	{\cal F}^{i}_{\left(5\right)} &=& - 12 \left(d_{\sigma}\right)^2\,\sigma^i + 12 \left(x_0 + \ve{\sigma} \cdot \ve{x}_0 \right) d_{\sigma}^i \;,  
\label{coefficient_F5}
\\
	{\cal F}^{i}_{\left(6\right)} &=& -2\,\left(d_{\sigma}\right)^2\,d_{\sigma}^i  \;, 
\label{coefficient_F6}
\\
	{\cal G}^{i}_{\left(5\right)} &=& + 12\,d_{\sigma}^i  \;, 
\label{coefficient_G5}
\\
	{\cal H}^{i}_{\left(3\right)} &=& - 8\,\sigma^i  \;, 
\label{coefficient_H3}
\\
	{\cal H}^{i}_{\left(5\right)} &=& + 12\,\left(d_{\sigma}\right)^2\,\sigma^i  \;.
\label{coefficient_H5}
\end{eqnarray}

\subsection{The terms proportional to $M \times \hat{M}_{ab}$} 

Summarizing all those terms in Eq.~(\ref{Appendix_Term_A_10}) and Eqs.~(\ref{Appendix_Term_B_10}) - (\ref{Appendix_Term_J_5}) which are 
proportional to $M \times \hat{M}_{ab}$ one obtains 
\begin{eqnarray}
        \frac{\ddot{x}^{i\;{\rm M} \times {\rm Q}}_{\rm 2PN}}{c^2} &=&  
	- 2\,\frac{G M}{c^2}\,\frac{\Delta x_{\rm 1PN}^{i\,{\rm Q}}}{\left(x_{\rm N}\right)^3}  
	+ 6\,\frac{G M}{c^2}\,\frac{x_{\rm N}^i}{\left(x_{\rm N}\right)^5} \left(\ve{x}_{\rm N} \cdot \Delta \ve{x}^{\rm Q}_{\rm 1PN}\right) 
        + \,6\,\frac{G \hat{M}_{ai}}{c^2} \,\frac{\Delta x^{a\,{\rm M}}_{\rm 1PN}}{\left(x_{\rm N}\right)^5}
\nonumber\\
	&& -30\,\frac{G \hat{M}_{ai}}{c^2}\,\frac{x_{\rm N}^a}{\left(x_{\rm N}\right)^7}
	\left(\ve{x}_{\rm N} \cdot \Delta \ve{x}^{\rm M}_{\rm 1PN}\right) 
	- \,15\,\frac{G \hat{M}_{ab}}{c^2} \,\frac{x_{\rm N}^a x_{\rm N}^b}{\left(x_{\rm N}\right)^7}\;\Delta x_{\rm 1PN}^{i\,{\rm M}}
	- \,30\,\frac{G \hat{M}_{ab}}{c^2} \,\frac{x_{\rm N}^a x_{\rm N}^i}{\left(x_{\rm N}\right)^7}\;\Delta x_{\rm 1PN}^{b\,{\rm M}}
        \nonumber\\
	&& + 105\,\frac{G \hat{M}_{ab}}{c^2} \,\frac{x_{\rm N}^a x_{\rm N}^b x_{\rm N}^i}{\left(x_{\rm N}\right)^9}
	\left(\ve{x}_{\rm N} \cdot \Delta \ve{x}^{\rm M}_{\rm 1PN}\right) 
	+ 4\,\frac{G M}{c^2}\,\sigma^i \frac{\ve{\sigma} \cdot \Delta \ve{x}_{\rm 1PN}^{\rm Q}}{\left(x_{\rm N}\right)^3} 
        \nonumber\\
	&& - 12\,\frac{G \hat{M}_{ab}}{c^2}\,\sigma^b \sigma^i \frac{\Delta x_{\rm 1PN}^{a\,{\rm M}}}{\left(x_{\rm N}\right)^5}
        - 12\,\frac{G M}{c^2}\,\sigma^i\,\frac{\ve{\sigma} \cdot \ve{x}_{\rm N}}{\left(x_{\rm N}\right)^5}\,
	\left(\ve{x}_{\rm N} \cdot \Delta \ve{x}^{\rm Q}_{\rm 1PN}\right) 
        \nonumber\\
	&& + 60\,\frac{G \hat{M}_{ab}}{c^2}\,\sigma^b \sigma^i \frac{x_{\rm N}^a}{\left(x_{\rm N}\right)^7}
	\left(\ve{x}_{\rm N} \cdot \Delta \ve{x}^{\rm M}_{\rm 1PN}\right)\,  
	+ 30\,\frac{G \hat{M}_{ab}}{c^2}\,\sigma^i \frac{x_{\rm N}^a x_{\rm N}^b}{\left(x_{\rm N}\right)^7}
	\left(\ve{\sigma} \cdot \Delta \ve{x}^{\rm M}_{\rm 1PN}\right) 
        \nonumber\\
	&& + 60\,\frac{G \hat{M}_{ab}}{c^2}\,\sigma^i\,x_{\rm N}^a \frac{\ve{\sigma} \cdot \ve{x}_{\rm N}}{\left(x_{\rm N}\right)^7}  
	\,\Delta x_{\rm 1PN}^{b\,{\rm M}}
	- 210\,\frac{G \hat{M}_{ab}}{c^2}\,\sigma^i x_{\rm N}^a x_{\rm N}^b\,  
	\frac{\ve{\sigma} \cdot \ve{x}_{\rm N}}{\left(x_{\rm N}\right)^9}   
	\left(\ve{x}_{\rm N} \cdot \Delta \ve{x}^{\rm M}_{\rm 1PN}\right) 
        \nonumber\\
	 && - 2\, \frac{G M}{c^2}\,\frac{x_{\rm N}^i}{\left(x_{\rm N}\right)^3}\,
        \frac{\ve{\sigma} \cdot \Delta \dot{\ve{x}}^{\rm Q}_{\rm 1PN}}{c}
        + 6\,\frac{G \hat{M}_{ai}}{c^2} \,\frac{x_{\rm N}^a}{\left(x_{\rm N}\right)^5}\,\frac{\ve{\sigma} \cdot \Delta \dot{\ve{x}}^{\rm M}_{\rm 1PN}}{c}
        - 15\,\frac{G \hat{M}_{ab}}{c^2} \,\frac{x_{\rm N}^a x_{\rm N}^b x_{\rm N}^i}{\left(x_{\rm N}\right)^7}\,
        \frac{\ve{\sigma} \cdot \Delta \dot{\ve{x}}^{\rm M}_{\rm 1PN}}{c}
\nonumber\\
	&& + 4\,\frac{G M}{c^2}\,\sigma^i\,\frac{1}{\left(x_{\rm N}\right)^3}\,
        \frac{\ve{x}_{\rm N} \cdot \Delta \dot{\ve{x}}^{\rm Q}_{\rm 1PN}}{c}
        - 12 \,\frac{G \hat{M}_{ab}}{c^2}\,\sigma^i\,\frac{x_{\rm N}^a}{\left(x_{\rm N}\right)^5}\,\frac{\Delta \dot{x}^{b\,{\rm M}}_{\rm 1PN}}{c}
	+ 4\,\frac{G M}{c^2}\,\frac{\ve{\sigma} \cdot \ve{x}_{\rm N}}{\left(x_{\rm N}\right)^3}\,
        \frac{\Delta \dot{x}^{i\,{\rm Q}}_{\rm 1PN}}{c}
\nonumber\\
	&& + 30\,\frac{G \hat{M}_{ab}}{c^2}\,\sigma^i\,\frac{x_{\rm N}^a x_{\rm N}^b}{\left(x_{\rm N}\right)^7}\,
        \frac{\ve{x}_{\rm N} \cdot \Delta \dot{\ve{x}}^{\rm M}_{\rm 1PN}}{c}
        - 12\,\frac{G \hat{M}_{ab}}{c^2}\,\sigma^b\,\frac{x_{\rm N}^a}{\left(x_{\rm N}\right)^5}\,\frac{\Delta \dot{x}^{i\,{\rm M}}_{\rm 1PN}}{c} 
	-12\, \frac{G M}{c^2} \frac{G \hat{M}_{ai}}{c^2}\,\frac{x^a_{\rm N}}{\left(x_{\rm N}\right)^6}
\nonumber\\
	&& + 30 \,\frac{G \hat{M}_{ab}}{c^2}\,x_{\rm N}^a x_{\rm N}^b\, 
	\frac{\ve{\sigma} \cdot \ve{x}_{\rm N}}{\left(x_{\rm N}\right)^7}\,  
        \frac{\Delta \dot{x}^{i\,{\rm M}}_{\rm 1PN}}{c}
	+ 36\, \frac{G M}{c^2} \frac{G \hat{M}_{ab}}{c^2}\,\frac{x^a_{\rm N} x^b_{\rm N} x^i_{\rm N}}{\left(x_{\rm N}\right)^8}
	 + 12 \frac{G M}{c^2}\frac{G \hat{M}_{ab}}{c^2}\,\frac{x^i_{\rm N}}{\left(x_{\rm N}\right)^6}\,\hat{n}_{\rm N}^{ab}
\nonumber\\
	 && -3\,\frac{G M}{c^2} \frac{G \hat{M}_{ab}}{c^2}\,\frac{\left(\hat{n}_{\rm N}^{ab}\right)_{,\,i}}{\left(x_{\rm N}\right)^4}
	- 24 \frac{G M}{c^2}\,\frac{G \hat{M}_{ab}}{c^2}\,\sigma^i \frac{\ve{\sigma} \cdot \ve{x}_{\rm N}}{\left(x_{\rm N}\right)^6}\,\hat{n}_{\rm N}^{ab}
	+ 6\,\frac{G M}{c^2} \frac{G \hat{M}_{ab}}{c^2}\,\sigma^i \sigma^j \frac{\left(\hat{n}_{\rm N}^{ab}\right)_{,\,j}}{\left(x_{\rm N}\right)^4}
\nonumber\\
	 && + 4\,\frac{G M}{c^2}\,\frac{G \hat{M}_{ab}}{c^2}\,\sigma^j \frac{\ve{\sigma} \cdot \ve{x}_{\rm N}}{\left(x_{\rm N}\right)^6}\,T_{\rm N}^{ijab}
	 - \frac{G M}{c^2}\,\frac{G \hat{M}_{ab}}{c^2}\,\sigma^j \sigma^k \frac{1}{\left(x_{\rm N}\right)^4}\,\left(T_{\rm N}^{ijab}\right)_{,\,k}
\nonumber\\
	 &&  - 2\,\frac{G M}{c^2}\,\frac{G \hat{M}_{ab}}{c^2}\,\sigma^j \sigma^k \frac{x^i_{\rm N}}{\left(x_{\rm N}\right)^6}\,T_{\rm N}^{jkab}
	 + \frac{1}{2}\,\frac{G M}{c^2}\,\frac{G \hat{M}_{ab}}{c^2}\,\sigma^j \sigma^k \frac{1}{\left(x_{\rm N}\right)^4}\,\left(T_{\rm N}^{jkab}\right)_{,\,i}\;. 
	\label{Appendix_Term_M_Q}
\end{eqnarray}

\noindent
One may rewrite Eq.~(\ref{Appendix_Term_M_Q}) in the form as given by Eq.~(\ref{2PN_Term_M_Q}) by inserting the explicit form of   
$\Delta\dot{\ve{x}}^{\rm M}_{\rm 1PN}$, $\Delta\dot{\ve{x}}^{\rm Q}_{\rm 1PN}$ and $\Delta\ve{x}^{\rm M}_{\rm 1PN}$, $\Delta\ve{x}^{\rm Q}_{\rm 1PN}$,  
as given by Eqs.~(\ref{First_Integration_1PN_M}), (\ref{First_Integration_1PN_Q}) and Eqs.~(\ref{Second_Integration_1PN_M}), (\ref{Second_Integration_1PN_Q}), 
respectively. 
Furthermore, one has to insert the explicit expressions for $T_{\rm N}^{ijab}$ and its spatial derivative as given by 
Eqs.~(\ref{Appendix_Tensor_S}) and (\ref{Appendix_Tensor_S_diff}). In addition, relations (\ref{Equation_A}) - (\ref{Equation_C}) are used several times.  
Then, one finds the following time-independent tensorial coefficients in Eq.~(\ref{2PN_Term_M_Q}):  
\begin{eqnarray} 
	{\cal K}^{i\,ab}_{\left(3\right)} &=&  
	+ \frac{8}{\left(d_{\sigma}\right)^4}\,d_{\sigma}^a\,\delta^{bi} 
	- \frac{4}{\left(d_{\sigma}\right)^4}\sigma^a\,\sigma^b\,d_{\sigma}^i 
        - \frac{8}{\left(d_{\sigma}\right)^4}\,\sigma^a\,d_{\sigma}^b\,\sigma^i  
	- \frac{16}{\left(d_{\sigma}\right)^6}\,d_{\sigma}^a\,d_{\sigma}^b\,d_{\sigma}^i  
	\;.  
\label{coefficient_K3}
\end{eqnarray}
\begin{eqnarray} 
	{\cal K}^{i\,ab}_{\left(4\right)} &=&  
	+ \frac{4}{\left(d_{\sigma}\right)^2}\,\sigma^a\,\delta^{bi} 
	+ \frac{4}{\left(d_{\sigma}\right)^2}\,\sigma^a\,\sigma^b\,\sigma^i 
	- \frac{8}{\left(d_{\sigma}\right)^4}\,\sigma^a d_{\sigma}^b\,d_{\sigma}^i 
	+ \frac{16}{\left(d_{\sigma}\right)^4}\,d_{\sigma}^a\,d_{\sigma}^b\,\sigma^i 
	\;.  
\label{coefficient_K4}
\end{eqnarray}
\begin{eqnarray} 
	{\cal K}^{i\,ab}_{\left(5\right)} &=&  
	- \frac{12}{\left(d_{\sigma}\right)^2}\,d_{\sigma}^a\,\delta^{bi} 
	- \frac{18}{\left(d_{\sigma}\right)^2}\,\sigma^a\,\sigma^b\,d_{\sigma}^i 
	- \frac{12}{\left(d_{\sigma}\right)^2}\,\sigma^a\,d_{\sigma}^b\,\sigma^i 
        - \frac{24}{\left(d_{\sigma}\right)^4}\,d_{\sigma}^a\,d_{\sigma}^b\,d_{\sigma}^i 
	+ \frac{6}{\left(d_{\sigma}\right)^2}\,\frac{\ve{\sigma} \cdot \ve{x}_0}{x_0}\,\sigma^a\,\sigma^b\,d_{\sigma}^i 
	\nonumber\\
	&& + \frac{12}{\left(d_{\sigma}\right)^2}\,\frac{\ve{\sigma} \cdot \ve{x}_0}{x_0}\,\sigma^a\,d_{\sigma}^b\,\sigma^i 
	+ \frac{12}{\left(d_{\sigma}\right)^4}\,\frac{\ve{\sigma} \cdot \ve{x}_0}{x_0}\,d_{\sigma}^a\,d_{\sigma}^b\,d_{\sigma}^i 
	- 6\,\frac{\ve{\sigma} \cdot \ve{x}_0}{\left(x_0\right)^3}\,\sigma^a\,\sigma^b\,d_{\sigma}^i
	+ 12\,\frac{\ve{\sigma} \cdot \ve{x}_0}{\left(x_0\right)^3}\,\sigma^a\,d_{\sigma}^b\,\sigma^i   
	\nonumber\\ 
	&& + \frac{6}{\left(d_{\sigma}\right)^2}\,\frac{\ve{\sigma}\cdot\ve{x}_0}{\left(x_0\right)^3}\,d_{\sigma}^a\,d_{\sigma}^b\,d_{\sigma}^i
	+ \frac{6}{x_0}\,\sigma^a\,\sigma^b\,\sigma^i 
	+ \frac{12}{\left(d_{\sigma}\right)^2}\,\frac{1}{x_0}\,d_{\sigma}^a\,d_{\sigma}^b\,\sigma^i
	- \frac{6}{\left(x_0\right)^3} \left(d_{\sigma}\right)^2 \sigma^a\,\sigma^b\,\sigma^i 
	- \frac{12}{\left(x_0\right)^3}\,\sigma^a\,d_{\sigma}^b\,d_{\sigma}^i 
	\nonumber\\
	&& + \frac{6}{\left(x_0\right)^3}\,d_{\sigma}^a\,d_{\sigma}^b\,\sigma^i 
	- \frac{12}{\left(d_{\sigma}\right)^2}\left(x_0 + \ve{\sigma} \cdot \ve{x}_0\right)\sigma^a\,\sigma^b\,\sigma^i 
	- \frac{24}{\left(d_{\sigma}\right)^4}\left(x_0 + \ve{\sigma} \cdot \ve{x}_0\right) d_{\sigma}^a\,d_{\sigma}^b\,\sigma^i 
	\;. 
\label{coefficient_K5}
\end{eqnarray}
\begin{eqnarray} 
        {\cal K}^{i\,ab}_{\left(6\right)} &=&  
	+ 22\,\sigma^a\,\delta^{bi} 
	+ \frac{32}{\left(d_{\sigma}\right)^2}\,\sigma^a\,d_{\sigma}^b\,d_{\sigma}^i 
	- \frac{64}{\left(d_{\sigma}\right)^2}\,d_{\sigma}^a\,d_{\sigma}^b\,\sigma^i 
	\;. 
\label{coefficient_K6}
\end{eqnarray}
\begin{eqnarray}
        {\cal K}^{i\,ab}_{\left(7\right)} &=&  
	+ 60\,d_{\sigma}^a\,\delta^{bi}   
	- 180\,\sigma^a\sigma^b\,d_{\sigma}^i
	+ 240\,\sigma^a\,d_{\sigma}^b\,\sigma^i
	+ \frac{30}{\left(d_{\sigma}\right)^2}\,d_{\sigma}^a\,d_{\sigma}^b\,d_{\sigma}^i 
	- 60 \left(x_0 + \ve{\sigma} \cdot \ve{x}_0\right) \sigma^a\,\delta^{bi} 
	\nonumber\\
	&& - 90 \left(x_0 + \ve{\sigma} \cdot \ve{x}_0\right) \sigma^a\,\sigma^b\,\sigma^i 
	- \frac{120}{\left(d_{\sigma}\right)^2} \left(x_0 + \ve{\sigma} \cdot \ve{x}_0\right) \sigma^a\,d_{\sigma}^b\,d_{\sigma}^i
	+ \frac{60}{\left(d_{\sigma}\right)^2} \left(x_0 + \ve{\sigma} \cdot \ve{x}_0\right) d_{\sigma}^a\,d_{\sigma}^b\,\sigma^i  
	\;. 
\label{coefficient_K7}
\end{eqnarray}
\begin{eqnarray}
         {\cal K}^{i\,ab}_{\left(8\right)} &=&   
	 + \frac{21}{2} \left(d_{\sigma}\right)^2 \sigma^a\,\delta^{bi} 
	 - 84 \left(d_{\sigma}\right)^2 \sigma^a\,\sigma^b\,\sigma^i 
	 - 63\,\sigma^a\,d_{\sigma}^b\,d_{\sigma}^i  
	 + 69\,d_{\sigma}^a\,d_{\sigma}^b\,\sigma^i 
	 \;.  
\label{coefficient_K8}
\end{eqnarray}
\begin{eqnarray}
        {\cal K}^{i\,ab}_{\left(9\right)} &=&  
	+ 210 \left(d_{\sigma}\right)^2 \sigma^a\,\sigma^b\,d_{\sigma}^i 
	- 420 \left(d_{\sigma}\right)^2 \sigma^a d_{\sigma}^b\,\sigma^i
	- 210\,d_{\sigma}^a\,d_{\sigma}^b\,d_{\sigma}^i 
	+ 210 \left(d_{\sigma}\right)^2 \left(x_0 + \ve{\sigma} \cdot \ve{x}_0\right) \sigma^a\,\sigma^b\,\sigma^i 
	\nonumber\\
	&&  
	+ 420 \left(x_0 + \ve{\sigma} \cdot \ve{x}_0\right) \sigma^a\,d_{\sigma}^b\,d_{\sigma}^i  
	- 210 \left(x_0 + \ve{\sigma} \cdot \ve{x}_0\right) d_{\sigma}^a\,d_{\sigma}^b\,\sigma^i   
	\;.  
\label{coefficient_K9}
\end{eqnarray}
\begin{eqnarray}
          {\cal K}^{i\,ab}_{\left(10\right)} &=&
	  + 30 \left(d_{\sigma}\right)^4 \sigma^a\,\sigma^b\,\sigma^i 
	  - 60 \left(d_{\sigma}\right)^2 \sigma^a\,d_{\sigma}^b\,d_{\sigma}^i
	  - 30 \left(d_{\sigma}\right)^2 d_{\sigma}^a\,d_{\sigma}^b\,\sigma^i 
	  \;.  
\label{coefficient_K10}
\end{eqnarray}
\begin{eqnarray} 
	{\cal L}^{i\,ab}_{\left(2\right)} &=&  
   	+ \frac{8}{\left(d_{\sigma}\right)^4}\,d_{\sigma}^a\,\delta^{bi} 
	- \frac{4}{\left(d_{\sigma}\right)^4}\,\sigma^a\,\sigma^b\,d_{\sigma}^i 
	- \frac{8}{\left(d_{\sigma}\right)^4}\,\sigma^a\,d_{\sigma}^b\,\sigma^i
	- \frac{16}{\left(d_{\sigma}\right)^6}\,d_{\sigma}^a\,d_{\sigma}^b\,d_{\sigma}^i 
	\;.  
\label{coefficient_L2}
\end{eqnarray}
\begin{eqnarray}
	{\cal L}^{i\,ab}_{\left(3\right)} &=&  
	+ \frac{4}{\left(d_{\sigma}\right)^2}\,\sigma^a\,\sigma^b\,\sigma^i 
	+ \frac{8}{\left(d_{\sigma}\right)^4}\,d_{\sigma}^a\,d_{\sigma}^b\,\sigma^i 
	- \frac{4}{\left(d_{\sigma}\right)^2}\,\frac{\ve{\sigma} \cdot \ve{x}_0}{x_0}\,\sigma^a\,\delta^{bi}  
	+ \frac{8}{\left(d_{\sigma}\right)^4}\,\frac{\ve{\sigma} \cdot \ve{x}_0}{x_0}\,\sigma^a\,d_{\sigma}^b\,d_{\sigma}^i  
	\nonumber\\
	&& - \frac{8}{\left(d_{\sigma}\right)^4}\,\frac{\ve{\sigma} \cdot \ve{x}_0}{x_0}\,d_{\sigma}^a\,d_{\sigma}^b\,\sigma^i  
	+ 4\,\frac{\ve{\sigma} \cdot \ve{x}_0}{\left(x_0\right)^3}\,\sigma^a\,\sigma^b\,\sigma^i 
	+ \frac{4}{\left(d_{\sigma}\right)^2}\,\frac{\ve{\sigma} \cdot \ve{x}_0}{\left(x_0\right)^3}\,\sigma^a\,d_{\sigma}^b\,d_{\sigma}^i  
	\nonumber\\
	&& - \frac{4}{\left(d_{\sigma}\right)^2}\,\frac{\ve{\sigma} \cdot \ve{x}_0}{\left(x_0\right)^3}\,d_{\sigma}^a\,d_{\sigma}^b\,\sigma^i
	- \frac{4}{\left(d_{\sigma}\right)^2}\,\frac{1}{x_0}\,d_{\sigma}^a\,\delta^{bi}    
	+ \frac{2}{\left(d_{\sigma}\right)^2}\,\frac{1}{x_0}\,\sigma^a\,\sigma^b\,d_{\sigma}^i   
	+ \frac{4}{\left(d_{\sigma}\right)^2}\,\frac{1}{x_0}\,\sigma^a\,d_{\sigma}^b\,\sigma^i 
	\nonumber\\
	&& + \frac{8}{\left(d_{\sigma}\right)^4}\,\frac{1}{x_0}\,  d_{\sigma}^a\,d_{\sigma}^b\,d_{\sigma}^i   
        - \frac{2}{\left(x_0\right)^3}\,\sigma^a\,\sigma^b\,d_{\sigma}^i 
	+ \frac{8}{\left(x_0\right)^3}\,\sigma^a\,d_{\sigma}^b\,\sigma^i 
	+ \frac{2}{\left(d_{\sigma}\right)^2}\,\frac{1}{\left(x_0\right)^3}\,  d_{\sigma}^a\,d_{\sigma}^b\,d_{\sigma}^i   
	\nonumber\\
	&& + \frac{8}{\left(d_{\sigma}\right)^4} \left(x_0 + \ve{\sigma} \cdot \ve{x}_0\right) d_{\sigma}^a\,\delta^{bi}   
	- \frac{4}{\left(d_{\sigma}\right)^4} \left(x_0 + \ve{\sigma} \cdot \ve{x}_0\right) \sigma^a\,\sigma^b\,d_{\sigma}^i 
	\nonumber\\
	&& - \frac{8}{\left(d_{\sigma}\right)^4}\left(x_0 + \ve{\sigma} \cdot \ve{x}_0\right) \sigma^a\,d_{\sigma}^b\,\sigma^i 
	- \frac{16}{\left(d_{\sigma}\right)^6} \left(x_0 + \ve{\sigma} \cdot \ve{x}_0\right) d_{\sigma}^a\,d_{\sigma}^b\,d_{\sigma}^i   
	\;.  
\label{coefficient_L3}
\end{eqnarray}
\begin{eqnarray}
	{\cal L}^{i\,ab}_{\left(4\right)} &=& 
	- \frac{16}{\left(d_{\sigma}\right)^2}\,d_{\sigma}^a\,\delta^{bi}  
	- \frac{22}{\left(d_{\sigma}\right)^2}\,\sigma^a\,\sigma^b\,d_{\sigma}^i 
	- \frac{20}{\left(d_{\sigma}\right)^2}\,\sigma^a\,d_{\sigma}^b\,\sigma^i
	- \frac{28}{\left(d_{\sigma}\right)^4}\,d_{\sigma}^a\,d_{\sigma}^b\,d_{\sigma}^i 
	\;. 
\label{coefficient_L4}
\end{eqnarray}
\begin{eqnarray}
	{\cal L}^{i\,ab}_{\left(5\right)} &=& 
	+ 60\,\sigma^a\,\delta^{bi}   
        + 18\,\sigma^a\,\sigma^b\,\sigma^i 
	+ \frac{24}{\left(d_{\sigma}\right)^2}\,\sigma^a\,d_{\sigma}^b\,d_{\sigma}^i  
	- \frac{60}{\left(d_{\sigma}\right)^2}\,d_{\sigma}^a\,d_{\sigma}^b\,\sigma^i 
        + 6\,\frac{\ve{\sigma} \cdot \ve{x}_0}{x_0}\,\sigma^a\,\sigma^b\,\sigma^i
	\nonumber\\
	&& - \frac{12}{\left(d_{\sigma}\right)^2}\,\frac{\ve{\sigma} \cdot \ve{x}_0}{x_0}\,\sigma^a\,d_{\sigma}^b\,d_{\sigma}^i 
	+ \frac{12}{\left(d_{\sigma}\right)^2}\,\frac{\ve{\sigma} \cdot \ve{x}_0}{x_0}\,d_{\sigma}^a\,d_{\sigma}^b\,\sigma^i
	- 6\,\frac{\ve{\sigma} \cdot \ve{x}_0}{\left(x_0\right)^3}\,\left(d_{\sigma}\right)^2\,\sigma^a\,\sigma^b\,\sigma^i
	\nonumber\\
	&& - 12\,\frac{\ve{\sigma} \cdot \ve{x}_0}{\left(x_0\right)^3}\,\sigma^a\,d_{\sigma}^b\,d_{\sigma}^i 
        + 6\,\frac{\ve{\sigma} \cdot \ve{x}_0}{\left(x_0\right)^3}\,d_{\sigma}^a\,d_{\sigma}^b\,\sigma^i
	- \frac{12}{\left(d_{\sigma}\right)^2}\,\frac{1}{x_0}\,d_{\sigma}^a\,d_{\sigma}^b\,d_{\sigma}^i 
	- \frac{6}{x_0}\,\sigma^a\,\sigma^b\,d_{\sigma}^i 
	\nonumber\\
	&& + 6\,\frac{\left(d_{\sigma}\right)^2}{\left(x_0\right)^3}\,\sigma^a\,\sigma^b\,d_{\sigma}^i 
	- 12\,\frac{\left(d_{\sigma}\right)^2}{\left(x_0\right)^3}\,\sigma^a\,d_{\sigma}^b\,\sigma^i
	- \frac{6}{\left(x_0\right)^3}\,d_{\sigma}^a\,d_{\sigma}^b\,d_{\sigma}^i  
        + \frac{12}{\left(d_{\sigma}\right)^2} \left(x_0 + \ve{\sigma} \cdot \ve{x}_0\right) d_{\sigma}^a\,\delta^{bi}  
	\nonumber\\
	&& - \frac{18}{\left(d_{\sigma}\right)^2} \left(x_0 + \ve{\sigma} \cdot \ve{x}_0\right) \sigma^a\,\sigma^b\,d_{\sigma}^i   
	+ \frac{36}{\left(d_{\sigma}\right)^2} \left(x_0 + \ve{\sigma} \cdot \ve{x}_0 \right) \sigma^a\,d_{\sigma}^b\,\sigma^i 
	+ \frac{24}{\left(d_{\sigma}\right)^4} \left(x_0 + \ve{\sigma} \cdot \ve{x}_0\right)\,d_{\sigma}^a\,d_{\sigma}^b\,d_{\sigma}^i 
	\;.  
\label{coefficient_L5}
\end{eqnarray}
\begin{eqnarray}
	{\cal L}^{i\,ab}_{\left(6\right)} &=&  
		18\,d_{\sigma}^a\,\delta^{bi} 
		- 20\,\sigma^a\,\sigma^b\,d_{\sigma}^i 
		+ 132\,\sigma^a\,d_{\sigma}^b\,\sigma^i
		+ \frac{50}{\left(d_{\sigma}\right)^2}\,d_{\sigma}^a\,d_{\sigma}^b\,d_{\sigma}^i   
		\;.  
\label{coefficient_L6}
\end{eqnarray}
\begin{eqnarray}
	{\cal L}^{i\,ab}_{\left(7\right)} &=& 
	- 60 \left(d_{\sigma}\right)^2 \sigma^a\,\delta^{bi} 
	- 240 \left(d_{\sigma}\right)^2\,\sigma^a\,\sigma^b\,\sigma^i 
	 - 420\,\sigma^a d_{\sigma}^b\,d_{\sigma}^i  
	+ 210\,d_{\sigma}^a\,d_{\sigma}^b\,\sigma^i  
	- 60 \left(x_0 + \ve{\sigma} \cdot \ve{x}_0\right) d_{\sigma}^a\,\delta^{bi} 
	\nonumber\\
	&& 
	+ 240 \left(x_0 + \ve{\sigma} \cdot \ve{x}_0\right) \sigma^a\,\sigma^b\,d_{\sigma}^i 
	- 360 \left(x_0 + \ve{\sigma} \cdot \ve{x}_0\right) \sigma^a\,d_{\sigma}^b\,\sigma^i 
	- \frac{90}{\left(d_{\sigma}\right)^2} \left(x_0 + \ve{\sigma} \cdot \ve{x}_0\right) d_{\sigma}^a\,d_{\sigma}^b\,d_{\sigma}^i  
	\;.  
\label{coefficient_L7}
\end{eqnarray}
\begin{eqnarray}
	{\cal L}^{i\,ab}_{\left(8\right)} &=&  
	+ \frac{21}{2} \left(d_{\sigma}\right)^2 d_{\sigma}^a\,\delta^{bi} 
        + 9 \left(d_{\sigma}\right)^2 \sigma^a\,\sigma^b\,d_{\sigma}^i 
	- 183 \left(d_{\sigma}\right)^2 \sigma^a\,d_{\sigma}^b\,\sigma^i 
	- 24\,d_{\sigma}^a\,d_{\sigma}^b\,d_{\sigma}^i 
	\;.  
\label{coefficient_L8}
\end{eqnarray}
\begin{eqnarray}
	{\cal L}^{i\,ab}_{\left(9\right)} &=& 
	+ 210 \left(d_{\sigma}\right)^4 \sigma^a\,\sigma^b\,\sigma^i 
	+ 420 \left(d_{\sigma}\right)^2 \sigma^a d_{\sigma}^b\,d_{\sigma}^i 
	- 210 \left(d_{\sigma}\right)^2 d_{\sigma}^a\, d_{\sigma}^b\,\sigma^i 
	- 210 \left(d_{\sigma}\right)^2 \left(x_0 + \ve{\sigma} \cdot \ve{x}_0\right) \sigma^a \sigma^b\,d_{\sigma}^i   
	\nonumber\\
        &&  
	+ 420 \left(d_{\sigma}\right)^2 \left(x_0 + \ve{\sigma} \cdot \ve{x}_0\right) \sigma^a\,d_{\sigma}^b\,\sigma^i   
	+ 210 \left(x_0 + \ve{\sigma} \cdot \ve{x}_0\right) d_{\sigma}^a\,d_{\sigma}^b\,d_{\sigma}^i 
	\;.  
\label{coefficient_L9}
\end{eqnarray}
\begin{eqnarray}
	{\cal L}^{i\,ab}_{\left(10\right)} &=&  
	+ 30 \left(d_{\sigma}\right)^4 \sigma^a\,\sigma^b\,d_{\sigma}^i 
	+ 60 \left(d_{\sigma}\right)^4 \sigma^a\,d_{\sigma}^b\,\sigma^i 
	- 30 \left(d_{\sigma}\right)^2 d_{\sigma}^a\,d_{\sigma}^b\,d_{\sigma}^i 
	\;.  
\label{coefficient_L10}
\end{eqnarray}
\begin{eqnarray}
	{\cal M}^{i\,ab}_{\left(7\right)} &=& 
	- 60\,d_{\sigma}^a\,\delta^{bi}  
	+ 150\,\sigma^a \sigma^b\,d_{\sigma}^i 
	- 180\,\sigma^a d_{\sigma}^b\,\sigma^i 
	\;.  
\label{coefficient_M7}
\end{eqnarray}
\begin{eqnarray}
	{\cal M}^{i\,ab}_{\left(8\right)} &=& 0  
	\;. 
\label{coefficient_M8}
\end{eqnarray}
\begin{eqnarray}
	{\cal M}^{i\,ab}_{\left(9\right)} &=& 
	- 210 \left(d_{\sigma}\right)^2 \sigma^a\,\sigma^b\,d_{\sigma}^i  
	+ 420 \left(d_{\sigma}\right)^2 \sigma^a\,d_{\sigma}^b\,\sigma^i
	+ 210\,d_{\sigma}^a\,d_{\sigma}^b\,d_{\sigma}^i  
	\;.  
\label{coefficient_M9}
\end{eqnarray}
\begin{eqnarray}
	{\cal N}^{i\,ab}_{\left(5\right)} &=& - 48\,\sigma^a\,\delta^{bi} - 24\,\sigma^a \sigma^b\,\sigma^i 
	\;.  
\label{coefficient_N5}
\end{eqnarray}
\begin{eqnarray}
	{\cal N}^{i\,ab}_{\left(6\right)} &=& 0  
       \;.
\label{coefficient_N6}
\end{eqnarray}
\begin{eqnarray}
	{\cal N}^{i\,ab}_{\left(7\right)} &=&   
	+ 60 \left(d_{\sigma}\right)^2 \sigma^a\,\delta^{bi}   
	+ 210 \left(d_{\sigma}\right)^2 \sigma^a\,\sigma^b\,\sigma^i
	+ 360\,\sigma^a\,d_{\sigma}^b\,d_{\sigma}^i   
	- 180\,d_{\sigma}^a\,d_{\sigma}^b\,\sigma^i  
	\;.  
\label{coefficient_N7}
\end{eqnarray}
\begin{eqnarray}
	{\cal N}^{i\,ab}_{\left(8\right)} &=& 0 
        \;. 
\label{coefficient_N8}
\end{eqnarray}
\begin{eqnarray}
	{\cal N}^{i\,ab}_{\left(9\right)} &=& 
        - 210 \left(d_{\sigma}\right)^4 \sigma^a\,\sigma^b\,\sigma^i 
	- 420 \left(d_{\sigma}\right)^2 \sigma^a\,d_{\sigma}^b\,d_{\sigma}^i
	+ 210 \left(d_{\sigma}\right)^2 d_{\sigma}^a\,d_{\sigma}^b\,\sigma^i 
	\;.  
\label{coefficient_N9}
\end{eqnarray}

\subsection{The terms proportional to $\hat{M}_{ab} \times \hat{M}_{cd}$}

Summarizing all those terms in Eq.~(\ref{Appendix_Term_A_10}) and Eqs.~(\ref{Appendix_Term_B_10}) - (\ref{Appendix_Term_J_5}) which are 
proportional to $\hat{M}_{ab} \times \hat{M}_{cd}$ one obtains:
\begin{eqnarray}
        \frac{\ddot{x}^{i\;{\rm Q} \times {\rm Q}}_{\rm 2PN}}{c^2} &=& 
	6\,\frac{G \hat{M}_{ai}}{c^2}\,\frac{\Delta x_{\rm 1PN}^{a\,{\rm Q}}}{\left(x_{\rm N}\right)^5}
	- 30\,\frac{G \hat{M}_{ai}}{c^2}\,\frac{x_{\rm N}^a}{\left(x_{\rm N}\right)^7}
	\left(\ve{x}_{\rm N} \cdot \Delta \ve{x}^{\rm Q}_{\rm 1PN}\right)  
	- \,15\,\frac{G \hat{M}_{ab}}{c^2}\,\frac{x_{\rm N}^a x_{\rm N}^b}{\left(x_{\rm N}\right)^7}\,\Delta x_{\rm 1PN}^{i\,{\rm Q}} 
        \nonumber\\
	&& - \,30\,\frac{G \hat{M}_{ab}}{c^2}\,\frac{x_{\rm N}^a x_{\rm N}^i}{\left(x_{\rm N}\right)^7}\,\Delta x_{\rm 1PN}^{b\,{\rm Q}}  
	+ 105\,\frac{G \hat{M}_{ab}}{c^2}\,\frac{x_{\rm N}^a x_{\rm N}^b x_{\rm N}^i}{\left(x_{\rm N}\right)^9}\,
	\left(\ve{x}_{\rm N} \cdot \Delta \ve{x}^{\rm Q}_{\rm 1PN}\right) 
	- 12\,\frac{G \hat{M}_{ab}}{c^2}\,\sigma^b \sigma^i \frac{\Delta x_{\rm 1PN}^{a\,{\rm Q}}}{\left(x_{\rm N}\right)^5}\,
        \nonumber\\
	&& + 60\,\frac{G \hat{M}_{ab}}{c^2}\,\sigma^b \sigma^i\, \frac{x_{\rm N}^a}{\left(x_{\rm N}\right)^7}\, 
	\left(\ve{x}_{\rm N} \cdot \Delta \ve{x}^{\rm Q}_{\rm 1PN}\right)   
	+ 30\,\frac{G \hat{M}_{ab}}{c^2}\,\sigma^i\,
	\frac{x_{\rm N}^a x_{\rm N}^b}{\left(x_{\rm N}\right)^7}\,\left(\ve{\sigma} \cdot \Delta \ve{x}_{\rm 1PN}^{c\,{\rm Q}}\right)
       \nonumber\\
	&& + 60\,\frac{G \hat{M}_{ab}}{c^2}\,\sigma^i\,\frac{\ve{\sigma} \cdot \ve{x}_{\rm N}}{\left(x_{\rm N}\right)^7}\,
	\,x_{\rm N}^a\,\Delta x_{\rm 1PN}^{b\,{\rm Q}}      
	- 210\,\frac{G \hat{M}_{ab}}{c^2}\,\sigma^i\,\frac{\ve{\sigma} \cdot \ve{x}_{\rm N}}{\left(x_{\rm N}\right)^9}\,
	x_{\rm N}^a x_{\rm N}^b\,\left(\ve{x}_{\rm N} \cdot \Delta \ve{x}^{\rm Q}_{\rm 1PN}\right) 
       \nonumber\\
	&& + 6\,\frac{G \hat{M}_{ai}}{c^2} \,\frac{x_{\rm N}^a}{\left(x_{\rm N}\right)^5}\,\frac{\ve{\sigma} \cdot \Delta \dot{\ve{x}}^{\rm Q}_{\rm 1PN}}{c}  
        - 15\,\frac{G \hat{M}_{ab}}{c^2} \,\frac{x_{\rm N}^a x_{\rm N}^b x_{\rm N}^i}{\left(x_{\rm N}\right)^7}\,
        \frac{\ve{\sigma} \cdot \Delta \dot{\ve{x}}^{\rm Q}_{\rm 1PN}}{c}
        \nonumber\\
	&& - 12\,\frac{G \hat{M}_{aj}}{c^2}\,\sigma^i\,\frac{x_{\rm N}^a}{\left(x_{\rm N}\right)^5}\,\frac{\Delta \dot{x}^{j\,{\rm Q}}_{\rm 1PN}}{c} 
        + 30\,\frac{G \hat{M}_{ab}}{c^2}\,\sigma^i\,\frac{x_{\rm N}^a x_{\rm N}^b}{\left(x_{\rm N}\right)^7}\,
        \frac{\ve{x}_{\rm N} \cdot \Delta \dot{\ve{x}}^{\rm Q}_{\rm 1PN}}{c}
        \nonumber\\
	&& - 12 \,\frac{G \hat{M}_{ab}}{c^2}\,\sigma^b\,\frac{x_{\rm N}^a}{\left(x_{\rm N}\right)^5}\,\frac{\Delta \dot{x}^{i\,{\rm Q}}_{\rm 1PN}}{c}
	+ 30 \,\frac{G \hat{M}_{ab}}{c^2} \,x_{\rm N}^a x_{\rm N}^b \,\frac{\ve{\sigma} \cdot \ve{x}_{\rm N}}{\left(x_{\rm N}\right)^7}\,
        \frac{\Delta \dot{x}^{i\,{\rm Q}}_{\rm 1PN}}{c}
        \nonumber\\
	&& - 18\,\frac{G \hat{M}_{ab}}{c^2}\,\frac{G \hat{M}_{ci}}{c^2}\,\frac{x^a_{\rm N} x^b_{\rm N} x^c_{\rm N}}{\left(x_{\rm N}\right)^{10}}
	+ 45\,\frac{G \hat{M}_{ab}}{c^2}\,\frac{G \hat{M}_{cd}}{c^2}\,\frac{x^a_{\rm N} x^b_{\rm N} x^c_{\rm N} x^d_{\rm N} x^i_{\rm N}}{\left(x_{\rm N}\right)^{12}}
       \nonumber\\
	&& + \frac{9}{5}\,\frac{G \hat{M}_{ab}}{c^2}\,\frac{G \hat{M}_{ab}}{c^2}\,\frac{x^i_{\rm N}}{\left(x_{\rm N}\right)^8}
	+ \frac{54}{7}\,\frac{G \hat{M}_{ab}}{c^2}\,\frac{G \hat{M}_{ac}}{c^2}\,\frac{x^i_{\rm N}}{\left(x_{\rm N}\right)^8}\,\hat{n}_{\rm N}^{bc}
	- \frac{9}{7}\,\frac{G \hat{M}_{ab}}{c^2}\,\frac{G \hat{M}_{ac}}{c^2}\,\frac{\left(\hat{n}_{\rm N}^{bc}\right)_{,\,i}}{\left(x_{\rm N}\right)^6}
\nonumber\\
\nonumber\\
	 && + \frac{27}{2}\,\frac{G \hat{M}_{ab}}{c^2}\,\frac{G \hat{M}_{cd}}{c^2}\,\frac{x^i_{\rm N}}{\left(x_{\rm N}\right)^8}\,\hat{n}_{\rm N}^{abcd}
	- \frac{9}{4}\,\frac{G \hat{M}_{ab}}{c^2}\,\frac{G \hat{M}_{cd}}{c^2}\,\frac{\left(\hat{n}_{\rm N}^{abcd}\right)_{,\,i}}{\left(x_{\rm N}\right)^6}
       \nonumber\\
	&& - \frac{18}{5}\,\frac{G \hat{M}_{ab}}{c^2}\,\frac{G \hat{M}_{ab}}{c^2}\,\sigma^i\,\frac{\ve{\sigma} \cdot \ve{x}_{\rm N}}{\left(x_{\rm N}\right)^8}
	- \frac{108}{7}\,\frac{G \hat{M}_{ab}}{c^2}\,\frac{G \hat{M}_{ac}}{c^2}\,\sigma^i\,
	\frac{\ve{\sigma} \cdot \ve{x}_{\rm N}}{\left(x_{\rm N}\right)^8}\,\hat{n}_{\rm N}^{bc}  
\nonumber\\
\nonumber\\
	&& + \frac{18}{7}\,\frac{G \hat{M}_{ab}}{c^2}\,\frac{G \hat{M}_{ac}}{c^2}\,\sigma^i \sigma^j  
	\frac{\left(\hat{n}_{\rm N}^{bc}\right)_{,\,j}}{\left(x_{\rm N}\right)^6}
        - 27\,\frac{G \hat{M}_{ab}}{c^2}\,\frac{G \hat{M}_{cd}}{c^2}\,\sigma^i\,
	\frac{\ve{\sigma} \cdot \ve{x}_{\rm N}}{\left(x_{\rm N}\right)^8}\,\hat{n}_{\rm N}^{abcd}
        \nonumber\\
	&& + \frac{9}{2}\,\frac{G \hat{M}_{ab}}{c^2}\,\frac{G \hat{M}_{cd}}{c^2}\,\sigma^i \sigma^j  
	\,\frac{\left(\hat{n}_{\rm N}^{abcd}\right)_{,\,j}}{\left(x_{\rm N}\right)^6}
        \nonumber\\
	&& + 6\,\frac{G \hat{M}_{ab}}{c^2}\,\frac{G \hat{M}_{cd}}{c^2}\,\sigma^j\,\frac{\ve{\sigma} \cdot \ve{x}_{\rm N}}{\left(x_{\rm N}\right)^8}\,
        T_{\rm N}^{ijabcd}
	- \frac{G \hat{M}_{ab}}{c^2}\,\frac{G \hat{M}_{cd}}{c^2}\,\sigma^j \sigma^k\,\frac{1}{\left(x_{\rm N}\right)^6}\,
        \left(T_{\rm N}^{ijabcd}\right)_{,\,k}  
        \nonumber\\
 && + \frac{96}{35}\,\frac{G \hat{M}_{ab}}{c^2}\,\frac{G \hat{M}_{ab}}{c^2}\,\sigma^i\,\frac{\ve{\sigma} \cdot \ve{x}_{\rm N}}{\left(x_{\rm N}\right)^8} 
	- \frac{36}{35}\,\frac{G \hat{M}_{ai}}{c^2}\,\frac{G \hat{M}_{ab}}{c^2}\,\sigma^b\,\frac{\ve{\sigma} \cdot \ve{x}_{\rm N}}{\left(x_{\rm N}\right)^8} 
       \nonumber\\
 &&  - 3\,\frac{G \hat{M}_{ab}}{c^2}\, \frac{G \hat{M}_{cd}}{c^2}\,\sigma^j \sigma^k\,\frac{x^i_{\rm N}}{\left(x_{\rm N}\right)^8}\,
        T_{\rm N}^{jkabcd}
	+ \frac{1}{2}\,\frac{G \hat{M}_{ab}}{c^2}\,\frac{G \hat{M}_{cd}}{c^2}\,\sigma^j \sigma^k\,\frac{1}{\left(x_{\rm N}\right)^6}\,
	\left(T_{\rm N}^{jkabcd}\right)_{,\,i}   
        \nonumber\\
	&& - \frac{48}{35}\,\frac{G \hat{M}_{ab}}{c^2}\,\frac{G \hat{M}_{ab}}{c^2}\,\frac{x^i_{\rm N}}{\left(x_{\rm N}\right)^8}
	+ \frac{18}{35}\,\frac{G \hat{M}_{ab}}{c^2}\,\frac{G \hat{M}_{ac}}{c^2}\,\sigma^b \sigma^c\,\frac{x^i_{\rm N}}{\left(x_{\rm N}\right)^8} 
	\;. 
        \label{Appendix_Term_Q_Q}
\end{eqnarray}

\noindent 
One may rewrite Eq.~(\ref{Appendix_Term_Q_Q}) in the form as given by Eq.~(\ref{2PN_Term_Q_Q}) by inserting the explicit form of
$\Delta\dot{\ve{x}}^{\rm M}_{\rm 1PN}$, $\Delta\dot{\ve{x}}^{\rm Q}_{\rm 1PN}$ and $\Delta\ve{x}^{\rm M}_{\rm 1PN}$, $\Delta\ve{x}^{\rm Q}_{\rm 1PN}$,
as given by Eqs.~(\ref{First_Integration_1PN_M}), (\ref{First_Integration_1PN_Q}) and Eqs.~(\ref{Second_Integration_1PN_M}), (\ref{Second_Integration_1PN_Q}),
respectively.  
Furthermore, one has to insert the expression for $T_{\rm N}^{ijabcd}$ and its spatial derivative as given by
Eqs.~(\ref{Appendix_Tensor_T}) and (\ref{Appendix_Tensor_T_diff}). Furthermore, the STF structure of the indices of the quadrupole tensors $\hat{M}_{ab}$ and $\hat{M}_{cd}$ 
implies that the dummy variables $a,b,c,d$ are symmetric under exchange $\left(a \leftrightarrow b\right)$ and $\left(c \leftrightarrow d\right)$  
as well as $\left(a \leftrightarrow c\; \land \;b \leftrightarrow d\right)$ and $\left(a \leftrightarrow d\; \land \;b \leftrightarrow c\right)$,  
which simplify the calculations significantly.   
Also the relations (\ref{Equation_A}) - (\ref{Equation_C}) are frequently been used.
Then, one finds the following time-independent tensorial coefficients in Eq.~(\ref{2PN_Term_Q_Q}):
\begin{eqnarray}
	{\cal P}^{i\,abcd}_{\left(5\right)} &=& + \frac{24}{\left(d_{\sigma}\right)^4}\,d_{\sigma}^a \delta^{bc}\,\delta^{di}\, 
	- \frac{24}{\left(d_{\sigma}\right)^4}\,\sigma^a \sigma^b d_{\sigma}^c\,\delta^{di}   
	- \frac{6}{\left(d_{\sigma}\right)^4}\,\sigma^a \sigma^b \sigma^c \sigma^d d_{\sigma}^i 
	 - \frac{48}{\left(d_{\sigma}\right)^4}\,\sigma^a \sigma^b \sigma^c d_{\sigma}^d \sigma^i 
	+ \frac{24}{\left(d_{\sigma}\right)^4}\, \sigma^a \delta^{bc}\,d_{\sigma}^d \sigma^i 
	\nonumber\\ 
	&& - \frac{48}{\left(d_{\sigma}\right)^6}\,d_{\sigma}^a d_{\sigma}^b d_{\sigma}^c\,\delta^{di}  
	- \frac{24}{\left(d_{\sigma}\right)^6}\, \sigma^a \sigma^b d_{\sigma}^c d_{\sigma}^d d_{\sigma}^i 
	- \frac{48}{\left(d_{\sigma}\right)^6}\,\sigma^a d_{\sigma}^b d_{\sigma}^c d_{\sigma}^d \sigma^i 
	\;.
	\nonumber\\ 
\label{coefficient_P5}
\\ 
	{\cal P}^{i\,abcd}_{\left(6\right)} &=& + \frac{126}{\left(d_{\sigma}\right)^2}\,\sigma^a \sigma^b \sigma^c\,\delta^{di}   
	   - \frac{12}{\left(d_{\sigma}\right)^2}\,\sigma^a \delta^{bc}\,\delta^{di}  
	   - \frac{36}{\left(d_{\sigma}\right)^2}\,\sigma^a\,\delta^{bc}\,\sigma^d \sigma^i 
	   + \frac{48}{\left(d_{\sigma}\right)^2}\,\sigma^a \sigma^b \sigma^c \sigma^d \sigma^i 
	   + \frac{96}{\left(d_{\sigma}\right)^4}\,\sigma^a d_{\sigma}^b d_{\sigma}^c\,\delta^{di}  
	   \nonumber\\
	   && + \frac{168}{\left(d_{\sigma}\right)^4}\,d_{\sigma}^a d_{\sigma}^b \sigma^c\,\delta^{di}     
	   + \frac{84}{\left(d_{\sigma}\right)^4}\,\sigma^a \sigma^b \sigma^c d_{\sigma}^d d_{\sigma}^i  
	   - \frac{120}{\left(d_{\sigma}\right)^4}\,\sigma^a\,\delta^{bc}\,d_{\sigma}^d d_{\sigma}^i 
	   + \frac{72}{\left(d_{\sigma}\right)^4}\,d_{\sigma}^a\,\delta^{bc}\,d_{\sigma}^d \sigma^i 
	   + \frac{48}{\left(d_{\sigma}\right)^4}\,d_{\sigma}^a d_{\sigma}^b \sigma^c \sigma^d \sigma^i 
	   \nonumber\\ 
	   && - \frac{72}{\left(d_{\sigma}\right)^4}\,\sigma^a d_{\sigma}^b \sigma^c d_{\sigma}^d \sigma^i 
	   + \frac{96}{\left(d_{\sigma}\right)^6}\,\sigma^a d_{\sigma}^b d_{\sigma}^c d_{\sigma}^d d_{\sigma}^i  
	   - \frac{144}{\left(d_{\sigma}\right)^6}\,d_{\sigma}^a d_{\sigma}^b d_{\sigma}^c d_{\sigma}^d \sigma^i  
           \;. 
\label{coefficient_P6}
\end{eqnarray}

\begin{eqnarray}
	{\cal P}^{i\,abcd}_{\left(7\right)} &=& - \frac{180}{\left(d_{\sigma}\right)^2}\,\sigma^a \sigma^b \sigma^c \sigma^d d_{\sigma}^i  
	   + \frac{420}{\left(d_{\sigma}\right)^2}\,\sigma^a \sigma^b \sigma^c d_{\sigma}^d \sigma^i 
	   - \frac{120}{\left(d_{\sigma}\right)^2}\,\sigma^a\,\delta^{bc}\,d_{\sigma}^d \sigma^i   
	  + \frac{180}{\left(d_{\sigma}\right)^4}\,d_{\sigma}^a d_{\sigma}^b d_{\sigma}^c\,\delta^{di}     
	   - \frac{270}{\left(d_{\sigma}\right)^4}\,\sigma^a \sigma^b d_{\sigma}^c d_{\sigma}^d d_{\sigma}^i  
	   \nonumber\\
	   && - \frac{120}{\left(d_{\sigma}\right)^4}\,d_{\sigma}^a\,\delta^{bc}\,d_{\sigma}^d d_{\sigma}^i  
	   + \frac{120}{\left(d_{\sigma}\right)^4}\,\sigma^a d_{\sigma}^b \sigma^c d_{\sigma}^d d_{\sigma}^i  
	   + \frac{540}{\left(d_{\sigma}\right)^4}\,\sigma^a d_{\sigma}^b\,d_{\sigma}^c\,d_{\sigma}^d\,\sigma^i  
           + \frac{120}{\left(d_{\sigma}\right)^6}\,d_{\sigma}^a d_{\sigma}^b d_{\sigma}^c d_{\sigma}^d d_{\sigma}^i 
           \nonumber\\
	   && - \frac{30}{\left(d_{\sigma}\right)^2} \frac{\ve{\sigma} \cdot \ve{x}_0}{x_0}\,\sigma^a \sigma^b d_{\sigma}^c\,\delta^{di}  
	      + \frac{135}{\left(d_{\sigma}\right)^2} \frac{\ve{\sigma} \cdot \ve{x}_0}{x_0}\, \sigma^a \sigma^b \sigma^c \sigma^d d_{\sigma}^i  
           - \frac{60}{\left(d_{\sigma}\right)^2} \frac{\ve{\sigma} \cdot \ve{x}_0}{x_0}\,\sigma^a\,\delta^{bc}\,\sigma^d d_{\sigma}^i 
           + \frac{60}{\left(d_{\sigma}\right)^2} \frac{\ve{\sigma} \cdot \ve{x}_0}{x_0}\, d_{\sigma}^a\,\delta^{bc}\,\sigma^d \sigma^i  
           \nonumber\\
	&& - \frac{60}{\left(d_{\sigma}\right)^4} \frac{\ve{\sigma} \cdot \ve{x}_0}{x_0}\,d_{\sigma}^a d_{\sigma}^b d_{\sigma}^c\,\delta^{di}   
           + \frac{150}{\left(d_{\sigma}\right)^4} \frac{\ve{\sigma} \cdot \ve{x}_0}{x_0}\, \sigma^a \sigma^b d_{\sigma}^c d_{\sigma}^d d_{\sigma}^i  
           + \frac{240}{\left(d_{\sigma}\right)^4} \frac{\ve{\sigma} \cdot \ve{x}_0}{x_0}\, \sigma^a d_{\sigma}^b \sigma^c d_{\sigma}^d d_{\sigma}^i  
           \nonumber\\
	&& - \frac{300}{\left(d_{\sigma}\right)^4} \frac{\ve{\sigma} \cdot \ve{x}_0}{x_0}\, \sigma^a d_{\sigma}^b d_{\sigma}^c d_{\sigma}^d \sigma^i  
           + 60\,\frac{\ve{\sigma} \cdot \ve{x}_0}{\left(x_0\right)^3}\,\sigma^a d_{\sigma}^b \sigma^c\,\delta^{di}    
           + 30\,\frac{\ve{\sigma} \cdot \ve{x}_0}{\left(x_0\right)^3}\,\sigma^a \sigma^b d_{\sigma}^c\,\delta^{di}  
	   - 75\,\frac{\ve{\sigma} \cdot \ve{x}_0}{\left(x_0\right)^3}\, \sigma^a \sigma^b \sigma^c \sigma^d d_{\sigma}^i  
           \nonumber\\
	  && + 180\,\frac{\ve{\sigma} \cdot \ve{x}_0}{\left(x_0\right)^3}\,\sigma^a \sigma^b \sigma^c d_{\sigma}^d \sigma^i  
           - \frac{30}{\left(d_{\sigma}\right)^2} \frac{\ve{\sigma} \cdot \ve{x}_0}{\left(x_0\right)^3}\,d_{\sigma}^a d_{\sigma}^b d_{\sigma}^c \,\delta^{di}  
	   + \frac{120}{\left(d_{\sigma}\right)^2}\,\frac{\ve{\sigma} \cdot \ve{x}_0}{\left(x_0\right)^3}\,\sigma^a d_{\sigma}^b \sigma^c d_{\sigma}^d d_{\sigma}^i
           + \frac{75}{\left(d_{\sigma}\right)^2}\,\frac{\ve{\sigma} \cdot \ve{x}_0}{\left(x_0\right)^3}\,\sigma^a \sigma^b d_{\sigma}^c d_{\sigma}^d d_{\sigma}^i 
           \nonumber\\
	 && - \frac{150}{\left(d_{\sigma}\right)^2}\,\frac{\ve{\sigma} \cdot \ve{x}_0}{\left(x_0\right)^3}\,\sigma^a d_{\sigma}^b d_{\sigma}^c d_{\sigma}^d \sigma^i
	 + \frac{30}{x_0}\,\sigma^a \sigma^b \sigma^c \delta^{di}     
           + \frac{45}{x_0}\,\sigma^a \sigma^b \sigma^c \sigma^d \sigma^i 
           + \frac{60}{\left(d_{\sigma}\right)^2}\frac{1}{x_0}\,d_{\sigma}^a d_{\sigma}^b \sigma^c\,\delta^{di}   
           - \frac{60}{\left(d_{\sigma}\right)^2} \frac{1}{x_0}\,\sigma^a d_{\sigma}^b d_{\sigma}^c \delta^{di}     
           \nonumber\\
	&& + \frac{120}{\left(d_{\sigma}\right)^2} \frac{1}{x_0}\,\sigma^a d_{\sigma}^b \sigma^c \sigma^d d_{\sigma}^i 
           - \frac{60}{\left(d_{\sigma}\right)^2} \frac{1}{x_0}\,\sigma^a \, \delta^{bc}\,d_{\sigma}^d d_{\sigma}^i        
           + \frac{60}{\left(d_{\sigma}\right)^2} \frac{1}{x_0}\,d_{\sigma}^a \, \delta^{bc}\,d_{\sigma}^d \sigma^i 
	   + \frac{60}{\left(d_{\sigma}\right)^2} \frac{1}{x_0}\,d_{\sigma}^a d_{\sigma}^b \sigma^c \sigma^d \sigma^i 
           \nonumber\\
	   && + \frac{240}{\left(d_{\sigma}\right)^4} \frac{1}{x_0}\,\sigma^a d_{\sigma}^b d_{\sigma}^c d_{\sigma}^d d_{\sigma}^i 
           - \frac{120}{\left(d_{\sigma}\right)^4} \frac{1}{x_0}\,d_{\sigma}^a d_{\sigma}^b d_{\sigma}^c d_{\sigma}^d \sigma^i     
	   - \frac{30}{\left(x_0\right)^3}\left(d_{\sigma}\right)^2 \sigma^a \sigma^b \sigma^c\,\delta^{di}    
	   - \frac{45}{\left(x_0\right)^3}\left(d_{\sigma}\right)^2 \sigma^a \sigma^b \sigma^c \sigma^d \sigma^i 
           \nonumber\\
	   && + \frac{60}{\left(x_0\right)^3}\,\sigma^a d_{\sigma}^b d_{\sigma}^c\,\delta^{di} 
	   + \frac{30}{\left(x_0\right)^3}\,d_{\sigma}^a d_{\sigma}^b \sigma^c\,\delta^{di} 
	   - \frac{210}{\left(x_0\right)^3}\, \sigma^a \sigma^b \sigma^c d_{\sigma}^d d_{\sigma}^i  
	   + \frac{180}{\left(x_0\right)^3}\,\sigma^a d_{\sigma}^b \sigma^c d_{\sigma}^d \sigma^i  
	   \nonumber\\ 
	 &&  + \frac{75}{\left(x_0\right)^3}\,d_{\sigma}^a d_{\sigma}^b \sigma^c \sigma^d \sigma^i 
	   + \frac{90}{\left(d_{\sigma}\right)^2} \frac{1}{\left(x_0\right)^3}\,\sigma^a d_{\sigma}^b d_{\sigma}^c d_{\sigma}^d d_{\sigma}^i  
      	   - \frac{60}{\left(d_{\sigma}\right)^2} \frac{1}{\left(x_0\right)^3}\,d_{\sigma}^a d_{\sigma}^b d_{\sigma}^c d_{\sigma}^d \sigma^i 
	   \nonumber\\
	&& - \frac{60}{\left(d_{\sigma}\right)^2}\left(x_0 + \ve{\sigma} \cdot \ve{x}_0\right) \sigma^a \sigma^b \sigma^c\,\delta^{di}     
           - \frac{90}{\left(d_{\sigma}\right)^2}\left(x_0 + \ve{\sigma} \cdot \ve{x}_0\right) \sigma^a \sigma^b \sigma^c \sigma^d \sigma^i  
	   - \frac{120}{\left(d_{\sigma}\right)^4}\left(x_0 + \ve{\sigma} \cdot \ve{x}_0\right) d_{\sigma}^a d_{\sigma}^b \sigma^c\,\delta^{di} 
           \nonumber\\
	&& + \frac{120}{\left(d_{\sigma}\right)^4}\left(x_0 + \ve{\sigma} \cdot \ve{x}_0\right) \sigma^a d_{\sigma}^b d_{\sigma}^c\,\delta^{di}
	   - \frac{240}{\left(d_{\sigma}\right)^4}\left(x_0 + \ve{\sigma} \cdot \ve{x}_0\right) \sigma^a d_{\sigma}^b \sigma^c \sigma^d d_{\sigma}^i 
           + \frac{120}{\left(d_{\sigma}\right)^4}\left(x_0 + \ve{\sigma} \cdot \ve{x}_0\right) \sigma^a\,\delta^{bc}\,d_{\sigma}^d d_{\sigma}^i 
           \nonumber\\
        && - \frac{120}{\left(d_{\sigma}\right)^4}\left(x_0 + \ve{\sigma} \cdot \ve{x}_0\right) d_{\sigma}^a\,\delta^{bc}\,d_{\sigma}^d \sigma^i 
           - \frac{120}{\left(d_{\sigma}\right)^4}\left(x_0 + \ve{\sigma} \cdot \ve{x}_0\right) d_{\sigma}^a\,d_{\sigma}^b \sigma^c \sigma^d \sigma^i 
	   - \frac{480}{\left(d_{\sigma}\right)^6}\left(x_0+\ve{\sigma}\cdot\ve{x}_0\right)\sigma^a d_{\sigma}^b d_{\sigma}^c d_{\sigma}^d d_{\sigma}^i 
           \nonumber\\
	 &&  + \frac{240}{\left(d_{\sigma}\right)^6}\left(x_0+\ve{\sigma}\cdot\ve{x}_0\right) d_{\sigma}^a d_{\sigma}^b d_{\sigma}^c d_{\sigma}^d \sigma^i 
	   \;. 
\label{coefficient_P7}
\end{eqnarray}

\begin{eqnarray}
	{\cal P}^{i\,abcd}_{\left(8\right)} &=& + 9\,\sigma^a\,\delta^{bc}\,\delta^{di} 
	- \frac{399}{2}\,\sigma^a \sigma^b \sigma^c \delta^{di} 
	+ 3\,\delta^{ac}\,\delta^{bd}\,\sigma^i
	- 351\,\sigma^a \sigma^b \sigma^c \sigma^d \sigma^i 
	+ 57\,\sigma^a\,\delta^{bc}\,\sigma^d \sigma^i  
	- \frac{66}{\left(d_{\sigma}\right)^2}\,d_{\sigma}^a d_{\sigma}^b \sigma^c\,\delta^{di} 
	\nonumber\\
	&& - \frac{132}{\left(d_{\sigma}\right)^2}\,\sigma^a d_{\sigma}^b d_{\sigma}^c\,\delta^{di} 
	+ \frac{120}{\left(d_{\sigma}\right)^2}\,d_{\sigma}^a\,\delta^{bc}\,\sigma^d d_{\sigma}^i 
	- \frac{468}{\left(d_{\sigma}\right)^2}\,d_{\sigma}^a \sigma^b \sigma^c \sigma^d d_{\sigma}^i 
	- \frac{396}{\left(d_{\sigma}\right)^2}\,\sigma^a \sigma^b d_{\sigma}^c d_{\sigma}^d \sigma^i
	- \frac{216}{\left(d_{\sigma}\right)^2}\,\sigma^a d_{\sigma}^b \sigma^c d_{\sigma}^d \sigma^i 
	\nonumber\\
	&& - \frac{84}{\left(d_{\sigma}\right)^2}\,d_{\sigma}^a\,\delta^{bc}\,d_{\sigma}^d \sigma^i
	- \frac{1332}{\left(d_{\sigma}\right)^4}\,\sigma^a d_{\sigma}^b d_{\sigma}^c d_{\sigma}^d d_{\sigma}^i 
	+ \frac{648}{\left(d_{\sigma}\right)^4}\,d_{\sigma}^a d_{\sigma}^b d_{\sigma}^c d_{\sigma}^d \sigma^i 
	\;.
	\label{coefficient_P8}
\end{eqnarray}

\begin{eqnarray}
	{\cal P}^{i\,abcd}_{\left(9\right)} &=&  
          + 210\,\sigma^a \sigma^b \sigma^c \sigma^d d_{\sigma}^i 
	 - 420\, \sigma^a \sigma^b \sigma^c d_{\sigma}^d \sigma^i 
	 + \frac{210}{\left(d_{\sigma}\right)^2}\, d_{\sigma}^a d_{\sigma}^b \sigma^c \sigma^d d_{\sigma}^i 
	 - \frac{840}{\left(d_{\sigma}\right)^2}\, \sigma^a d_{\sigma}^b d_{\sigma}^c d_{\sigma}^d \sigma^i  
         \nonumber\\ 
	 && - \frac{420}{\left(d_{\sigma}\right)^4}\, d_{\sigma}^a d_{\sigma}^b d_{\sigma}^c d_{\sigma}^d d_{\sigma}^i
         - 105\,\frac{\ve{\sigma} \cdot \ve{x}_0}{x_0} \, \sigma^a \sigma^b \sigma^c \sigma^d d_{\sigma}^i 
         - \frac{105}{\left(d_{\sigma}\right)^2}\,\frac{\ve{\sigma} \cdot \ve{x}_0}{x_0} \,d_{\sigma}^a d_{\sigma}^b \sigma^c \sigma^d d_{\sigma}^i  
         \nonumber\\ 
	&& - \frac{420}{\left(d_{\sigma}\right)^2}\,\frac{\ve{\sigma} \cdot \ve{x}_0}{x_0} \,\sigma^a d_{\sigma}^b \sigma^c d_{\sigma}^d d_{\sigma}^i  
         + \frac{210}{\left(d_{\sigma}\right)^2}\,\frac{\ve{\sigma} \cdot \ve{x}_0}{x_0} \,\sigma^a d_{\sigma}^b d_{\sigma}^c d_{\sigma}^d \sigma^i  
         + \frac{210}{\left(d_{\sigma}\right)^4}\,\frac{\ve{\sigma} \cdot \ve{x}_0}{x_0} \,d_{\sigma}^a d_{\sigma}^b d_{\sigma}^c d_{\sigma}^d d_{\sigma}^i  
         \nonumber\\ 
	&& + 105 \left(d_{\sigma}\right)^2 \frac{\ve{\sigma} \cdot \ve{x}_0}{\left(x_0\right)^3}\,
        \sigma^a \sigma^b \sigma^c \sigma^d d_{\sigma}^i
        - 420 \left(d_{\sigma}\right)^2 \frac{\ve{\sigma} \cdot \ve{x}_0}{\left(x_0\right)^3}\,
        \sigma^a \sigma^b \sigma^c d_{\sigma}^d \sigma^i 
        - 210\,\frac{\ve{\sigma} \cdot \ve{x}_0}{\left(x_0\right)^3}\, \sigma^a \sigma^b d_{\sigma}^c d_{\sigma}^d d_{\sigma}^i  
        \nonumber\\
	&& - 420\,\frac{\ve{\sigma} \cdot \ve{x}_0}{\left(x_0\right)^3}\,\sigma^a d_{\sigma}^b \sigma^c d_{\sigma}^d d_{\sigma}^i  
        + 420\,\frac{\ve{\sigma} \cdot \ve{x}_0}{\left(x_0\right)^3}\, \sigma^a d_{\sigma}^b d_{\sigma}^c d_{\sigma}^d \sigma^i
    + \frac{105}{\left(d_{\sigma}\right)^2}\,\frac{\ve{\sigma}\cdot\ve{x}_0}{\left(x_0\right)^3}\,d_{\sigma}^a d_{\sigma}^b d_{\sigma}^c d_{\sigma}^d d_{\sigma}^i  
         \nonumber\\ 
	 && - 105\,\frac{\left(d_{\sigma}\right)^2}{x_0}\,\sigma^a \sigma^b \sigma^c \sigma^d \sigma^i  
         - \frac{210}{x_0}\, \sigma^a \sigma^b \sigma^c d_{\sigma}^d d_{\sigma}^i
         - \frac{105}{x_0}\,d_{\sigma}^a d_{\sigma}^b \sigma^c \sigma^d \sigma^i 
         - \frac{420}{\left(d_{\sigma}\right)^2}\,\frac{1}{x_0}\,\sigma^a d_{\sigma}^b d_{\sigma}^c d_{\sigma}^d d_{\sigma}^i 
         \nonumber\\ 
	&& + \frac{210}{\left(d_{\sigma}\right)^2}\,\frac{1}{x_0}\,d_{\sigma}^a d_{\sigma}^b d_{\sigma}^c d_{\sigma}^d \sigma^i 
	 + 105\,\frac{\left(d_{\sigma}\right)^4}{\left(x_0\right)^3}\,\sigma^a \sigma^b \sigma^c \sigma^d \sigma^i  
	 + 420\,\frac{\left(d_{\sigma}\right)^2}{\left(x_0\right)^3}\,\sigma^a \sigma^b \sigma^c d_{\sigma}^d d_{\sigma}^i 
         \nonumber\\ 
	 && - 210\,\frac{\left(d_{\sigma}\right)^2}{\left(x_0\right)^3}\,\sigma^a \sigma^b d_{\sigma}^c d_{\sigma}^d \sigma^i 
	 - 420\,\frac{\left(d_{\sigma}\right)^2}{\left(x_0\right)^3}\,\sigma^a d_{\sigma}^b \sigma^c d_{\sigma}^d \sigma^i 
	 - \frac{420}{\left(x_0\right)^3}\, d_{\sigma}^a d_{\sigma}^b d_{\sigma}^c \sigma^d d_{\sigma}^i
	 \nonumber\\ 
	&& + \frac{105}{\left(x_0\right)^3}\, d_{\sigma}^a d_{\sigma}^b d_{\sigma}^c d_{\sigma}^d \sigma^i 
         + 210\,\left(x_0 + \ve{\sigma} \cdot \ve{x}_0\right) \sigma^a \sigma^b \sigma^c \sigma^d \sigma^i 
	 \nonumber\\
	&& + \frac{420}{\left(d_{\sigma}\right)^2} \left(x_0 + \ve{\sigma} \cdot \ve{x}_0\right)\sigma^a \sigma^b \sigma^c d_{\sigma}^d d_{\sigma}^i  
        + \frac{210}{\left(d_{\sigma}\right)^2}  \left(x_0 + \ve{\sigma} \cdot \ve{x}_0\right) d_{\sigma}^a d_{\sigma}^b \sigma^c \sigma^d \sigma^i  
	\nonumber\\
	&& + \frac{840}{\left(d_{\sigma}\right)^4} \left(x_0 + \ve{\sigma} \cdot \ve{x}_0\right) \sigma^a d_{\sigma}^b d_{\sigma}^c d_{\sigma}^d d_{\sigma}^i  
	- \frac{420}{\left(d_{\sigma}\right)^4} \left(x_0 + \ve{\sigma} \cdot \ve{x}_0\right) d_{\sigma}^a d_{\sigma}^b d_{\sigma}^c d_{\sigma}^d \sigma^i 
	\;. 
\label{coefficient_P9}
\end{eqnarray}
\begin{eqnarray}
	{\cal P}^{i\,abcd}_{\left(10\right)} &=& 
        + \frac{381}{2} \left(d_{\sigma}\right)^2 \sigma^a \sigma^b \sigma^c\,\delta^{di}    
	- 13 \left(d_{\sigma}\right)^2 \sigma^a \,\delta^{bc}\, \,\delta^{di}   
	+ \frac{777}{2} \left(d_{\sigma}\right)^2 \sigma^a \sigma^b \sigma^c \sigma^d \sigma^i 
	\nonumber\\
	&& + 95 \left(d_{\sigma}\right)^2 \sigma^a \delta^{bc}\,\sigma^d \sigma^i 
	- 5 \left(d_{\sigma}\right)^2 \delta^{ac}\,\delta^{bd}\,\sigma^i  
	- 356\,\sigma^a d_{\sigma}^b d_{\sigma}^c\,\delta^{di}   
        - 172\,d_{\sigma}^a d_{\sigma}^b \sigma^c\,\delta^{di} 
	\nonumber\\
	&& + 1761\,\sigma^a \sigma^b \sigma^c d_{\sigma}^d d_{\sigma}^i
        - 110\,\sigma^a\,\delta^{bc}\,d_{\sigma}^d d_{\sigma}^i 
	- 112\,d_{\sigma}^a d_{\sigma}^b \sigma^c \sigma^d \sigma^i
	- 218\,\sigma^a d_{\sigma}^b \sigma^c d_{\sigma}^d \sigma^i 
	\nonumber\\
	&& - 50\,d_{\sigma}^a\,\delta^{bc}\,d_{\sigma}^d \sigma^i 
	+ \frac{816}{\left(d_{\sigma}\right)^2}\,d_{\sigma}^a d_{\sigma}^b d_{\sigma}^c \sigma^d d_{\sigma}^i   
	- \frac{324}{\left(d_{\sigma}\right)^2}\,d_{\sigma}^a d_{\sigma}^b d_{\sigma}^c d_{\sigma}^d \sigma^i  
	\;. 
\label{coefficient_P10}
\end{eqnarray}
\begin{eqnarray}
	{\cal P}^{i\,abcd}_{\left(11\right)} &=& 0 \;. 
\label{coefficient_P11}
\end{eqnarray}
\begin{eqnarray}
	{\cal P}^{i\,abcd}_{\left(12\right)} &=& 
	- \frac{45}{2} \left(d_{\sigma}\right)^4 \sigma^a \sigma^b \sigma^c \,\delta^{di} 
	- 165 \left(d_{\sigma}\right)^4 \sigma^a \sigma^b \sigma^c \sigma^d \sigma^i 
	- 75 \left(d_{\sigma}\right)^4 \sigma^a\, \delta^{bc}\,\sigma^d \sigma^i  
        \nonumber\\
	&& + 45 \left(d_{\sigma}\right)^2 \sigma^a d_{\sigma}^b d_{\sigma}^c\,\delta^{di}   
	+ \frac{45}{2} \left(d_{\sigma}\right)^2 d_{\sigma}^a d_{\sigma}^b \sigma^c \,\delta^{di}   
	- 2235 \left(d_{\sigma}\right)^2 \sigma^a \sigma^b \sigma^c d_{\sigma}^d d_{\sigma}^i  
        \nonumber\\
	&& + 150 \left(d_{\sigma}\right)^2 \sigma^a \, \delta^{bc}\,d_{\sigma}^d d_{\sigma}^i 
	+ 480 \left(d_{\sigma}\right)^2 \sigma^a \sigma^b d_{\sigma}^c d_{\sigma}^d \sigma^i  
	+ 960 \left(d_{\sigma}\right)^2 \sigma^a d_{\sigma}^b \sigma^c d_{\sigma}^d \sigma^i  
	\nonumber\\
	&& + 75 \left(d_{\sigma}\right)^2 d_{\sigma}^a \, \delta^{bc}\,d_{\sigma}^d \sigma^i 
	+ 1935\,d_{\sigma}^a d_{\sigma}^b d_{\sigma}^c \sigma^d d_{\sigma}^i 
	- 315\,d_{\sigma}^a d_{\sigma}^b d_{\sigma}^c d_{\sigma}^d \sigma^i 
	\;. 
\label{coefficient_P12}
\end{eqnarray}
\begin{eqnarray}
	{\cal P}^{i\,abcd}_{\left(13\right)} &=& 0 \;. 
\label{coefficient_P13}
\end{eqnarray}
\begin{eqnarray}
	{\cal P}^{i\,abcd}_{\left(14\right)} &=& 
        - \frac{225}{2} \left(d_{\sigma}\right)^6  \sigma^a \sigma^b \sigma^c \sigma^d \sigma^i  
	+  450 \left(d_{\sigma}\right)^4 \sigma^a \sigma^b \sigma^c d_{\sigma}^d d_{\sigma}^i  
        + 225 \left(d_{\sigma}\right)^4 \sigma^a \sigma^b d_{\sigma}^c d_{\sigma}^d \sigma^i 
	\nonumber\\
	&& + 450 \left(d_{\sigma}\right)^4 \sigma^a d_{\sigma}^b \sigma^c d_{\sigma}^d \sigma^i  
	- 450 \left(d_{\sigma}\right)^2 \sigma^a d_{\sigma}^b d_{\sigma}^c d_{\sigma}^d d_{\sigma}^i  
	- \frac{225}{2} \left(d_{\sigma}\right)^2 d_{\sigma}^a d_{\sigma}^b d_{\sigma}^c d_{\sigma}^d \sigma^i  
	\;. 
\label{coefficient_P14}
\end{eqnarray}
\begin{eqnarray}
	{\cal Q}^{i\,abcd}_{\left(4\right)} &=& + \frac{24}{\left(d_{\sigma}\right)^4}\,d_{\sigma}^a \delta^{bc}\,\delta^{di}   
	- \frac{24}{\left(d_{\sigma}\right)^4}\,\sigma^a d_{\sigma}^b \sigma^c\,\delta^{di} 
	- \frac{6}{\left(d_{\sigma}\right)^4}\,\sigma^a \sigma^b \sigma^c \sigma^d d_{\sigma}^i 
	- \frac{48}{\left(d_{\sigma}\right)^4}\,\sigma^a \sigma^b \sigma^c d_{\sigma}^d \sigma^i 
	\nonumber\\
	&& + \frac{24}{\left(d_{\sigma}\right)^4}\,\sigma^a \,\delta^{bc}\,d_{\sigma}^d \sigma^i 
       - \frac{48}{\left(d_{\sigma}\right)^6}\,d_{\sigma}^a d_{\sigma}^b d_{\sigma}^c \,\delta^{di}   
       - \frac{24}{\left(d_{\sigma}\right)^6}\,\sigma^a \sigma^b d_{\sigma}^c d_{\sigma}^d d_{\sigma}^i 
       - \frac{48}{\left(d_{\sigma}\right)^6}\,\sigma^a d_{\sigma}^b d_{\sigma}^c d_{\sigma}^d \sigma^i\;.  
	\label{coefficient_Q4}
\end{eqnarray}
\begin{eqnarray}
	{\cal Q}^{i\,abcd}_{\left(5\right)} &=& 
	+ \frac{60}{\left(d_{\sigma}\right)^2}\,\sigma^a \sigma^b \sigma^c\,\delta^{di}   
	+ \frac{30}{\left(d_{\sigma}\right)^2}\,\sigma^a \sigma^b \sigma^c \sigma^d \sigma^i 
	+ \frac{120}{\left(d_{\sigma}\right)^4}\,d_{\sigma}^a d_{\sigma}^b \sigma^c\,\delta^{di}    
	+ \frac{72}{\left(d_{\sigma}\right)^4}\,\sigma^a d_{\sigma}^b d_{\sigma}^c\,\delta^{di} 
	\nonumber\\
	&& + \frac{144}{\left(d_{\sigma}\right)^4}\,\sigma^a \sigma^b \sigma^c d_{\sigma}^d d_{\sigma}^i 
	- \frac{120}{\left(d_{\sigma}\right)^4}\,\sigma^a \,\delta^{bc}\,d_{\sigma}^d d_{\sigma}^i
        + \frac{24}{\left(d_{\sigma}\right)^4}\,\sigma^a \sigma^b d_{\sigma}^c d_{\sigma}^d \sigma^i 
 	+ \frac{72}{\left(d_{\sigma}\right)^4}\, d_{\sigma}^a \,\delta^{bc}\,d_{\sigma}^d \sigma^i
	\nonumber\\
	&& - \frac{144}{\left(d_{\sigma}\right)^4}\,\sigma^a d_{\sigma}^b \sigma^c d_{\sigma}^d \sigma^i 
        + \frac{96}{\left(d_{\sigma}\right)^6}\, \sigma^a d_{\sigma}^b d_{\sigma}^c d_{\sigma}^d d_{\sigma}^i
	- \frac{144}{\left(d_{\sigma}\right)^6}\, d_{\sigma}^a d_{\sigma}^b d_{\sigma}^c d_{\sigma}^d \sigma^i 
	\nonumber\\ 
	&& + \frac{12}{\left(d_{\sigma}\right)^2}\,\frac{\ve{\sigma} \cdot \ve{x}_0}{x_0}\,\sigma^a \delta^{bc}\,\delta^{di}    
	- \frac{66}{\left(d_{\sigma}\right)^2}\,\frac{\ve{\sigma} \cdot \ve{x}_0}{x_0}\, \sigma^a \sigma^b \sigma^c\,\delta^{di}   
	+ \frac{36}{\left(d_{\sigma}\right)^2}\,\frac{\ve{\sigma} \cdot \ve{x}_0}{x_0}\,\sigma^a\,\delta^{bc}\,\sigma^d \sigma^i 
	\nonumber\\
	&& - \frac{18}{\left(d_{\sigma}\right)^2}\,\frac{\ve{\sigma} \cdot \ve{x}_0}{x_0}\,\sigma^a \sigma^b \sigma^c \sigma^d \sigma^i 
 	- \frac{48}{\left(d_{\sigma}\right)^4}\,\frac{\ve{\sigma} \cdot \ve{x}_0}{x_0}\,d_{\sigma}^a d_{\sigma}^b \sigma^c\,\delta^{di}   
	- \frac{24}{\left(d_{\sigma}\right)^4}\,\frac{\ve{\sigma} \cdot \ve{x}_0}{x_0}\,\sigma^a d_{\sigma}^b d_{\sigma}^c \,\delta^{di} 
	\nonumber\\ 
	&& + \frac{60}{\left(d_{\sigma}\right)^4}\,\frac{\ve{\sigma} \cdot \ve{x}_0}{x_0}\,\sigma^a \sigma^b \sigma^c d_{\sigma}^d d_{\sigma}^i 
 	- \frac{24}{\left(d_{\sigma}\right)^4}\,\frac{\ve{\sigma} \cdot \ve{x}_0}{x_0}\, \sigma^a \sigma^b d_{\sigma}^c d_{\sigma}^d \sigma^i 
        - \frac{72}{\left(d_{\sigma}\right)^4}\,\frac{\ve{\sigma} \cdot \ve{x}_0}{x_0}\,\sigma^a d_{\sigma}^b \sigma^c d_{\sigma}^d \sigma^i 
	\nonumber\\ 
	&& + 24\,\frac{\ve{\sigma} \cdot \ve{x}_0}{\left(x_0\right)^3}\,\sigma^a \sigma^b \sigma^c\,\delta^{di}     
        + 12\,\frac{\ve{\sigma} \cdot \ve{x}_0}{\left(x_0\right)^3}\,\sigma^a \sigma^b \sigma^c \sigma^d \sigma^i  
        - \frac{24}{\left(d_{\sigma}\right)^2}\,\frac{\ve{\sigma} \cdot \ve{x}_0}{\left(x_0\right)^3}\,
        d_{\sigma}^a d_{\sigma}^b \sigma^c\,\delta^{di}    
        \nonumber\\
	&& - \frac{12}{\left(d_{\sigma}\right)^2}\,\frac{\ve{\sigma} \cdot \ve{x}_0}{\left(x_0\right)^3}\,\sigma^a d_{\sigma}^b d_{\sigma}^c \,\delta^{di}   
        + \frac{30}{\left(d_{\sigma}\right)^2}\,\frac{\ve{\sigma} \cdot \ve{x}_0}{\left(x_0\right)^3}\,\sigma^a \sigma^b \sigma^c d_{\sigma}^d d_{\sigma}^i  
        - \frac{12}{\left(d_{\sigma}\right)^2}\,\frac{\ve{\sigma} \cdot \ve{x}_0}{\left(x_0\right)^3}\, \sigma^a \sigma^b d_{\sigma}^c d_{\sigma}^d \sigma^i  
        \nonumber\\
	&& - \frac{36}{\left(d_{\sigma}\right)^2}\,\frac{\ve{\sigma} \cdot \ve{x}_0}{\left(x_0\right)^3}\,\sigma^a d_{\sigma}^b \sigma^c d_{\sigma}^d \sigma^i  
        - \frac{12}{\left(d_{\sigma}\right)^2}\,\frac{1}{x_0}\,\sigma^a d_{\sigma}^b \sigma^c\,\delta^{di}   
        + \frac{12}{\left(d_{\sigma}\right)^2}\,\frac{1}{x_0}\,d_{\sigma}^a \delta^{bc}\,\delta^{di}   
\nonumber\\
	&& - \frac{36}{\left(d_{\sigma}\right)^2}\,\frac{1}{x_0}\,\sigma^a \sigma^b d_{\sigma}^c \,\delta^{di}   
 	+ \frac{15}{\left(d_{\sigma}\right)^2}\,\frac{1}{x_0}\,\sigma^a \sigma^b \sigma^c \sigma^d d_{\sigma}^i 
	- \frac{24}{\left(d_{\sigma}\right)^2}\,\frac{1}{x_0}\,\sigma^a \sigma^b \sigma^c d_{\sigma}^d \sigma^i  
	\nonumber\\
	&& + \frac{36}{\left(d_{\sigma}\right)^2}\,\frac{1}{x_0}\,\sigma^a \delta^{bc}\, d_{\sigma}^d \sigma^i 
        - \frac{24}{\left(d_{\sigma}\right)^4}\,\frac{1}{x_0}\,d_{\sigma}^a d_{\sigma}^b d_{\sigma}^c \,\delta^{di}   
        + \frac{60}{\left(d_{\sigma}\right)^4}\,\frac{1}{x_0}\,\sigma^a \sigma^b d_{\sigma}^c d_{\sigma}^d d_{\sigma}^i 
	\nonumber\\
	&& - \frac{72}{\left(d_{\sigma}\right)^4}\,\frac{1}{x_0}\,\sigma^a d_{\sigma}^b d_{\sigma}^c d_{\sigma}^d \sigma^i 
        + \frac{6}{\left(x_0\right)^3}\,\sigma^a \sigma^b d_{\sigma}^c\,\delta^{di}   
	+ \frac{48}{\left(x_0\right)^3}\,\sigma^a d_{\sigma}^b \sigma^c \,\delta^{di}  
	- \frac{15}{\left(x_0\right)^3}\,\sigma^a \sigma^b \sigma^c \sigma^d d_{\sigma}^i 
	\nonumber\\
	&& + \frac{42}{\left(x_0\right)^3}\,\sigma^a \sigma^b \sigma^c d_{\sigma}^d \sigma^i 
        - \frac{6}{\left(d_{\sigma}\right)^2}\,\frac{1}{\left(x_0\right)^3}\,d_{\sigma}^a d_{\sigma}^b d_{\sigma}^c\,\delta^{di}  
	+ \frac{15}{\left(d_{\sigma}\right)^2}\,\frac{1}{\left(x_0\right)^3}\,\sigma^a \sigma^b d_{\sigma}^c d_{\sigma}^d d_{\sigma}^i 
	\nonumber\\
	&& - \frac{18}{\left(d_{\sigma}\right)^2}\,\frac{1}{\left(x_0\right)^3}\,\sigma^a d_{\sigma}^b d_{\sigma}^c d_{\sigma}^d \sigma^i 
        - \frac{24}{\left(d_{\sigma}\right)^4} \left(x_0 + \ve{\sigma} \cdot \ve{x}_0\right) d_{\sigma}^a \delta^{bc}\,\delta^{di}    
	\nonumber\\
     && + \frac{72}{\left(d_{\sigma}\right)^4} \left(x_0 + \ve{\sigma} \cdot \ve{x}_0\right) \sigma^a \sigma^b d_{\sigma}^c\,\delta^{di}  
        + \frac{24}{\left(d_{\sigma}\right)^4} \left(x_0 + \ve{\sigma} \cdot \ve{x}_0\right) \sigma^a d_{ \sigma}^b \sigma^c\,\delta^{di}   
        \nonumber\\
     && - \frac{30}{\left(d_{\sigma}\right)^4} \left(x_0 + \ve{\sigma} \cdot \ve{x}_0\right) \sigma^a \sigma^b \sigma^c \sigma^d d_{ \sigma}^i 
        + \frac{48}{\left(d_{\sigma}\right)^4} \left(x_0 + \ve{\sigma} \cdot \ve{x}_0\right) \sigma^a \sigma^b \sigma^c d_{ \sigma}^d \sigma^i  
        \nonumber\\
     && - \frac{72}{\left(d_{\sigma}\right)^4} \left(x_0 + \ve{\sigma} \cdot \ve{x}_0\right) \sigma^a \delta^{bc}\,d_{ \sigma}^d \sigma^i 
        + \frac{48}{\left(d_{\sigma}\right)^6} \left(x_0 + \ve{\sigma} \cdot \ve{x}_0\right) d_{ \sigma}^a d_{ \sigma}^b d_{\sigma}^c\,\delta^{di} 
        \nonumber\\
  && - \frac{120}{\left(d_{\sigma}\right)^6} \left(x_0 + \ve{\sigma}\cdot\ve{x}_0\right) \sigma^a \sigma^b d_{ \sigma}^c d_{ \sigma}^d d_{ \sigma}^i  
        + \frac{144}{\left(d_{\sigma}\right)^6} \left(x_0 + \ve{\sigma} \cdot \ve{x}_0\right) \sigma^a d_{ \sigma}^b d_{ \sigma}^c d_{ \sigma}^d 
        \sigma^i 
	\;. 
\label{coefficient_Q5}
\end{eqnarray}
\begin{eqnarray}
	{\cal Q}^{i\,abcd}_{\left(6\right)} &=&  
	    + \frac{30}{\left(d_{\sigma}\right)^2}\,\sigma^a \sigma^b d_{\sigma}^c\,\delta^{di}  
           - \frac{12}{\left(d_{\sigma}\right)^2}\,d_{\sigma}^a \delta^{bc}\,\delta^{di}  
	   + \frac{12}{\left(d_{\sigma}\right)^2}\,\sigma^a d_{\sigma}^b \sigma^c\,\delta^{di} 
	   - \frac{312}{\left(d_{\sigma}\right)^2}\,\sigma^a \sigma^b \sigma^c \sigma^d d_{\sigma}^i 
	   + \frac{60}{\left(d_{\sigma}\right)^2}\,\sigma^a\,\delta^{bc}\,\sigma^d d_{\sigma}^i 
	   \nonumber\\
	  && + \frac{444}{\left(d_{\sigma}\right)^2}\,\sigma^a \sigma^b \sigma^c d_{\sigma}^d \sigma^i 
	   - \frac{192}{\left(d_{\sigma}\right)^2}\,\sigma^a\,\delta^{bc}\,d_{\sigma}^d \sigma^i 
           + \frac{264}{\left(d_{\sigma}\right)^4}\,d_{\sigma}^a d_{\sigma}^b d_{\sigma}^c\,\delta^{di}  
	   - \frac{120}{\left(d_{\sigma}\right)^4}\, d_{\sigma}^a \,\delta^{bc}\,d_{\sigma}^d d_{\sigma}^i 
	   \nonumber\\
	   && - \frac{408}{\left(d_{\sigma}\right)^4}\, \sigma^a \sigma^b d_{\sigma}^c d_{\sigma}^d d_{\sigma}^i 
	   - \frac{120}{\left(d_{\sigma}\right)^4}\,\sigma^a d_{\sigma}^b \sigma^c d_{\sigma}^d d_{\sigma}^i 
	   + \frac{864}{\left(d_{\sigma}\right)^4}\,\sigma^a d_{\sigma}^b d_{\sigma}^c d_{\sigma}^d \sigma^i 
	   + \frac{120}{\left(d_{\sigma}\right)^6}\,d_{\sigma}^a d_{\sigma}^b d_{\sigma}^c d_{\sigma}^d d_{\sigma}^i
           \;. 
	\label{coefficient_Q6}
\end{eqnarray}
\begin{eqnarray}
	{\cal Q}^{i\,abcd}_{\left(7\right)} &=&  
	   - 60\,\sigma^a \sigma^b \sigma^c\,\delta^{di}    
	   - 240\,\sigma^a \sigma^b \sigma^c \sigma^d \sigma^i
 	   - \frac{120}{\left(d_{\sigma}\right)^2}\,\sigma^a d_{\sigma}^b d_{\sigma}^c\,\delta^{di}  
	   - \frac{120}{\left(d_{\sigma}\right)^2}\,d_{\sigma}^a d_{\sigma}^b \sigma^c \,\delta^{di}  
	   \nonumber\\
	   && - \frac{540}{\left(d_{\sigma}\right)^2}\,\sigma^a \sigma^b \sigma^c d_{\sigma}^d d_{\sigma}^i 
	   + \frac{120}{\left(d_{\sigma}\right)^2}\,\sigma^a \,\delta^{bc}\,d_{\sigma}^d d_{\sigma}^i 
	   - \frac{120}{\left(d_{\sigma}\right)^2}\,d_{\sigma}^a \,\delta^{bc}\,d_{\sigma}^d \sigma^i 
	   + \frac{240}{\left(d_{\sigma}\right)^2}\,\sigma^a d_{\sigma}^b \sigma^c d_{\sigma}^d \sigma^i  
	   \nonumber\\ 
	   && - \frac{270}{\left(d_{\sigma}\right)^2}\,\sigma^a \sigma^b d_{\sigma}^c d_{\sigma}^d \sigma^i 
	   - \frac{840}{\left(d_{\sigma}\right)^4}\,\sigma^a d_{\sigma}^b d_{\sigma}^c d_{\sigma}^d d_{\sigma}^i  
  	   + \frac{540}{\left(d_{\sigma}\right)^4}\,d_{\sigma}^a d_{\sigma}^b d_{\sigma}^c d_{\sigma}^d \sigma^i
           + 60\,\frac{\ve{\sigma} \cdot \ve{x}_0}{x_0}\,\sigma^a \sigma^b \sigma^c \, \delta^{di} 
	   \nonumber\\ 
	&& + 135\,\frac{\ve{\sigma} \cdot \ve{x}_0}{x_0}\,\sigma^a \sigma^b \sigma^c \sigma^d \sigma^i
           - 60\,\frac{\ve{\sigma} \cdot \ve{x}_0}{x_0}\,\sigma^a \, \delta^{bc}\,\sigma^d \sigma^i 
           + \frac{60}{\left(d_{\sigma}\right)^2}\,\frac{\ve{\sigma} \cdot \ve{x}_0}{x_0}\,\sigma^a d_{\sigma}^b d_{\sigma}^c\,\delta^{di} 
           \nonumber\\
	&& + \frac{30}{\left(d_{\sigma}\right)^2}\,\frac{\ve{\sigma} \cdot \ve{x}_0}{x_0}\,d_{\sigma}^a d_{\sigma}^b \sigma^c\,\delta^{di}  
        - \frac{30}{\left(d_{\sigma}\right)^2}\,\frac{\ve{\sigma} \cdot \ve{x}_0}{x_0}\,\sigma^a \sigma^b \sigma^c d_{\sigma}^d d_{\sigma}^i
        - \frac{60}{\left(d_{\sigma}\right)^2}\,\frac{\ve{\sigma} \cdot \ve{x}_0}{x_0}\, d_{\sigma}^a \, \delta^{bc}\, \sigma^d d_{\sigma}^i 
        \nonumber\\
	&& + \frac{150}{\left(d_{\sigma}\right)^2}\,\frac{\ve{\sigma} \cdot \ve{x}_0}{x_0}\, d_{\sigma}^a d_{\sigma}^b \sigma^c \sigma^d \sigma^i 
        + \frac{420}{\left(d_{\sigma}\right)^2}\,\frac{\ve{\sigma} \cdot \ve{x}_0}{x_0}\, \sigma^ad_{\sigma}^b \sigma^c d_{\sigma}^d \sigma^i
        + \frac{540}{\left(d_{\sigma}\right)^4}\,\frac{\ve{\sigma}\cdot\ve{x}_0}{x_0}\,\sigma^a d_{\sigma}^b d_{\sigma}^c d_{\sigma}^d d_{\sigma}^i 
        \nonumber\\
	&& - \frac{180}{\left(d_{\sigma}\right)^4}\,\frac{\ve{\sigma} \cdot \ve{x}_0}{x_0}\,d_{\sigma}^a d_{\sigma}^b d_{\sigma}^c d_{\sigma}^d \sigma^i 
        - 30 \left(d_{\sigma}\right)^2 \frac{\ve{\sigma} \cdot \ve{x}_0}{\left(x_0\right)^3} \sigma^a \sigma^b \sigma^c \,\delta^{di} 
        - 105 \left(d_{\sigma}\right)^2 \frac{\ve{\sigma} \cdot \ve{x}_0}{\left(x_0\right)^3} \sigma^a \sigma^b \sigma^c \sigma^d \sigma^i
        \nonumber\\
	&& + 60\,\frac{\ve{\sigma} \cdot \ve{x}_0}{\left(x_0\right)^3}\,\sigma^a d_{\sigma}^b d_{\sigma}^c \, \delta^{di} 
        + 30\,\frac{\ve{\sigma} \cdot \ve{x}_0}{\left(x_0\right)^3}\,d_{\sigma}^a d_{\sigma}^b \sigma^c\,\delta^{di}  
        - 420\,\frac{\ve{\sigma} \cdot \ve{x}_0}{\left(x_0\right)^3}\,\sigma^a \sigma^b \sigma^c d_{\sigma}^d d_{\sigma}^i 
        \nonumber\\ 
	&& + 195\,\frac{\ve{\sigma} \cdot \ve{x}_0}{\left(x_0\right)^3}\,\sigma^a \sigma^b d_{\sigma}^c d_{\sigma}^d \sigma^i
        + 360\,\frac{\ve{\sigma} \cdot \ve{x}_0}{\left(x_0\right)^3} \sigma^a d_{\sigma}^b \sigma^c d_{\sigma}^d \sigma^i 
	+ \frac{270}{\left(d_{\sigma}\right)^2}\, \frac{\ve{\sigma} \cdot \ve{x}_0}{\left(x_0\right)^3}\,\sigma^a d_{\sigma}^b
        d_{\sigma}^c d_{\sigma}^d d_{\sigma}^i
	\nonumber\\
	&& - \frac{90}{\left(d_{\sigma}\right)^2}\, \frac{\ve{\sigma} \cdot \ve{x}_0}{\left(x_0\right)^3}\,
        d_{\sigma}^a d_{\sigma}^b d_{\sigma}^c d_{\sigma}^d \sigma^i 
        + \frac{60}{x_0}\,\sigma^a \sigma^b d_{\sigma}^c\,\delta^{di}  
        - \frac{120}{x_0}\, \sigma^a \sigma^b \sigma^c \sigma^d d_{\sigma}^i 
        + \frac{210}{x_0}\, \sigma^a \sigma^b \sigma^c d_{\sigma}^d \sigma^i
	\nonumber\\
	&& - \frac{60}{x_0}\, \sigma^a\,\delta^{bc}\,d_{\sigma}^d \sigma^i 
        + \frac{30}{\left(d_{\sigma}\right)^2}\,\frac{1}{x_0}\,d_{\sigma}^a d_{\sigma}^b d_{\sigma}^c \,\delta^{di}  
        - \frac{225}{\left(d_{\sigma}\right)^2}\,\frac{1}{x_0}\,\sigma^a \sigma^b d_{\sigma}^c d_{\sigma}^d d_{\sigma}^i 
        - \frac{60}{\left(d_{\sigma}\right)^2}\,\frac{1}{x_0}\, d_{\sigma}^a\,\delta^{bc}\,d_{\sigma}^d d_{\sigma}^i 
       \nonumber\\
	&& + \frac{60}{\left(d_{\sigma}\right)^2}\,\frac{1}{x_0}\, \sigma^a d_{\sigma}^b \sigma^c d_{\sigma}^d d_{\sigma}^i 
        + \frac{450}{\left(d_{\sigma}\right)^2}\,\frac{1}{x_0}\,\sigma^a d_{\sigma}^b d_{\sigma}^c d_{\sigma}^d \sigma^i 
        + \frac{180}{\left(d_{\sigma}\right)^4}\,\frac{1}{x_0}\,d_{\sigma}^a d_{\sigma}^b d_{\sigma}^c d_{\sigma}^d d_{\sigma}^i 
        \nonumber\\
	&& - \frac{30}{\left(x_0\right)^3} \left(d_{\sigma}\right)^2 \sigma^a \sigma^b d_{\sigma}^c\,\delta^{di}   
        - \frac{60}{\left(x_0\right)^3} \left(d_{\sigma}\right)^2 \sigma^a d_{\sigma}^b \sigma^c\,\delta^{di}   
        + \frac{120}{\left(x_0\right)^3} \left(d_{\sigma}\right)^2 \sigma^a \sigma^b \sigma^c \sigma^d d_{\sigma}^i 
 	\nonumber\\
        && - \frac{390}{\left(x_0\right)^3} \left(d_{\sigma}\right)^2 \sigma^a \sigma^b \sigma^c d_{\sigma}^d \sigma^i 
        + \frac{30}{\left(x_0\right)^3}\,d_{\sigma}^a d_{\sigma}^b d_{\sigma}^c\,\delta^{di} 
	- \frac{165}{\left(x_0\right)^3} \, \sigma^a \sigma^b d_{\sigma}^c d_{\sigma}^d d_{\sigma}^i
  	\nonumber\\
	&& - \frac{360}{\left(x_0\right)^3}\, \sigma^a d_{\sigma}^b \sigma^c d_{\sigma}^d d_{\sigma}^i 
        + \frac{360}{\left(x_0\right)^3} \, \sigma^a d_{\sigma}^b d_{\sigma}^c d_{\sigma}^d \sigma^i 
  	+ \frac{45}{\left(d_{\sigma}\right)^2}\,\frac{1}{\left(x_0\right)^3}\,d_{\sigma}^a d_{\sigma}^b d_{\sigma}^c d_{\sigma}^d d_{\sigma}^i 
	\nonumber\\
        && - \frac{120}{\left(d_{\sigma}\right)^2} \left(x_0 + \ve{\sigma} \cdot \ve{x}_0\right) \sigma^a \sigma^b d_{\sigma}^c\,\delta^{di} 
        + \frac{240}{\left(d_{\sigma}\right)^2} \left(x_0 + \ve{\sigma} \cdot \ve{x}_0\right) \sigma^a \sigma^b \sigma^c \sigma^d d_{\sigma}^i 
        \nonumber\\
	&& + \frac{120}{\left(d_{\sigma}\right)^4} \left(x_0 + \ve{\sigma} \cdot \ve{x}_0\right) d_{\sigma}^a\,\delta^{bc}\,d_{\sigma}^d d_{\sigma}^i 
        - \frac{420}{\left(d_{\sigma}\right)^2} \left(x_0 + \ve{\sigma} \cdot \ve{x}_0\right) \sigma^a \sigma^b \sigma^c d_{\sigma}^d \sigma^i 
        \nonumber\\
        && + \frac{120}{\left(d_{\sigma}\right)^2} \left(x_0 + \ve{\sigma}\cdot\ve{x}_0\right) \sigma^a \,\delta^{bc}\, d_{\sigma}^d \sigma^i  
        - \frac{60}{\left(d_{\sigma}\right)^4} \left(x_0 + \ve{\sigma} \cdot \ve{x}_0\right) d_{\sigma}^a d_{\sigma}^b d_{\sigma}^c\,\delta^{di}
        \nonumber\\
        && + \frac{450}{\left(d_{\sigma}\right)^4} \left(x_0 + \ve{\sigma}\cdot\ve{x}_0\right) \sigma^a \sigma^b d_{\sigma}^c d_{\sigma}^d d_{\sigma}^i
        - \frac{120}{\left(d_{\sigma}\right)^4} \left(x_0 + \ve{\sigma}\cdot\ve{x}_0\right) \sigma^a d_{\sigma}^b \sigma^c d_{\sigma}^d d_{\sigma}^i
	\nonumber\\ 
	&& - \frac{900}{\left(d_{\sigma}\right)^4} \left(x_0 + \ve{\sigma} \cdot \ve{x}_0\right)
        d_{\sigma}^a d_{\sigma}^b d_{\sigma}^c \sigma^d \sigma^i
	- \frac{360}{\left(d_{\sigma}\right)^6} \left(x_0 + \ve{\sigma} \cdot \ve{x}_0\right)
        d_{\sigma}^a d_{\sigma}^b d_{\sigma}^c d_{\sigma}^d d_{\sigma}^i
	\;. 
	\label{coefficient_Q7}
\end{eqnarray}
\begin{eqnarray}
	{\cal Q}^{i\,abcd}_{\left(8\right)} &=&  
 	+ 12\,d_{\sigma}^a \delta^{bc}\,\delta^{di}  
        - 306\,\sigma^a d_{\sigma}^b \sigma^c \,\delta^{di}  
        - \frac{315}{2}\,\sigma^a \sigma^b d_{\sigma}^c \,\delta^{di}   
        - 102\,\sigma^a\,\delta^{bc}\,\sigma^d d_{\sigma}^i 
        + \frac{9}{2}\,\delta^{ac}\,\delta^{bd}\, d_{\sigma}^i 
	\nonumber\\
	&& + \frac{1683}{2}\, \sigma^a \sigma^b \sigma^c \sigma^d d_{\sigma}^i 
	- 876\,\sigma^a \sigma^b \sigma^c d_{\sigma}^d \sigma^i 
        + 93\,d_{\sigma}^a\,\delta^{bc}\, \sigma^d \sigma^i 
        - \frac{114}{\left(d_{\sigma}\right)^2}\,d_{\sigma}^a d_{\sigma}^b d_{\sigma}^c\,\delta^{di}  
	\nonumber\\
	&& + \frac{588}{\left(d_{\sigma}\right)^2}\,\sigma^a \sigma^b d_{\sigma}^c d_{\sigma}^d d_{\sigma}^i
        + \frac{420}{\left(d_{\sigma}\right)^2}\,\sigma^a d_{\sigma}^b \sigma^c d_{\sigma}^d d_{\sigma}^i
 	+ \frac{60}{\left(d_{\sigma}\right)^2}\, d_{\sigma}^a \,\delta^{bc}\,d_{\sigma}^d d_{\sigma}^i 
        - \frac{1884}{\left(d_{\sigma}\right)^2}\, \sigma^a d_{\sigma}^b d_{\sigma}^c d_{\sigma}^d \sigma^i
	\nonumber\\
	&& - \frac{690}{\left(d_{\sigma}\right)^4}\, d_{\sigma}^a d_{\sigma}^b d_{\sigma}^c d_{\sigma}^d d_{\sigma}^i 
        \;. 
\label{coefficient_Q8}
\end{eqnarray}
\begin{eqnarray}
	{\cal Q}^{i\,abcd}_{\left(9\right)} &=&  
	 + 210 \left(d_{\sigma}\right)^2 \sigma^a \sigma^b \sigma^c \sigma^d \sigma^i
	+ 420\,\sigma^a \sigma^b \sigma^c d_{\sigma}^d d_{\sigma}^i 
        + 210 \, d_{\sigma}^a d_{\sigma}^b \sigma^c \sigma^d \sigma^i 
 	+ \frac{840}{\left(d_{\sigma}\right)^2}\, \sigma^a d_{\sigma}^b d_{\sigma}^c d_{\sigma}^d d_{\sigma}^i 
	\nonumber\\
	&& -  \frac{420}{\left(d_{\sigma}\right)^2}\, d_{\sigma}^a d_{\sigma}^b d_{\sigma}^c d_{\sigma}^d \sigma^i 
        - 105 \left(d_{\sigma}\right)^2 \frac{\ve{\sigma}\cdot\ve{x}_0}{x_0} \sigma^a \sigma^b \sigma^c \sigma^d \sigma^i
        - 105\,\frac{\ve{\sigma}\cdot\ve{x}_0}{x_0}\,\sigma^a \sigma^b d_{\sigma}^c d_{\sigma}^d \sigma^i 
        \nonumber\\
	&& - 420\,\frac{\ve{\sigma}\cdot\ve{x}_0}{x_0}\,\sigma^a d_{\sigma}^b \sigma^c d_{\sigma}^d \sigma^i 
	- \frac{630}{\left(d_{\sigma}\right)^2}\,\frac{\ve{\sigma}\cdot\ve{x}_0}{x_0}\,\sigma^a d_{\sigma}^b d_{\sigma}^c d_{\sigma}^d d_{\sigma}^i
 	+ \frac{210}{\left(d_{\sigma}\right)^2}\,\frac{\ve{\sigma}\cdot\ve{x}_0}{x_0}\,d_{\sigma}^a d_{\sigma}^b d_{\sigma}^c d_{\sigma}^d \sigma^i
	\nonumber\\
	&& + 105 \left(d_{\sigma}\right)^4 \frac{\ve{\sigma}\cdot\ve{x}_0}{\left(x_0\right)^3} \sigma^a \sigma^b \sigma^c \sigma^d \sigma^i 
	+ 420 \left(d_{\sigma}\right)^2 \frac{\ve{\sigma}\cdot\ve{x}_0}{\left(x_0\right)^3} \sigma^a \sigma^b \sigma^c d_{\sigma}^d d_{\sigma}^i 
	- 210 \left(d_{\sigma}\right)^2 \frac{\ve{\sigma}\cdot\ve{x}_0}{\left(x_0\right)^3} \sigma^a \sigma^b d_{\sigma}^c d_{\sigma}^d \sigma^i 
	\nonumber\\ 
	&& - 420 \left(d_{\sigma}\right)^2 \frac{\ve{\sigma}\cdot\ve{x}_0}{\left(x_0\right)^3} \sigma^a d_{\sigma}^b \sigma^c d_{\sigma}^d \sigma^i 
	- 420\,\frac{\ve{\sigma}\cdot\ve{x}_0}{\left(x_0\right)^3}\, d_{\sigma}^a d_{\sigma}^b d_{\sigma}^c \sigma^d d_{\sigma}^i 
 	+ 105\,\frac{\ve{\sigma}\cdot\ve{x}_0}{\left(x_0\right)^3}\,d_{\sigma}^a d_{\sigma}^b d_{\sigma}^c d_{\sigma}^d \sigma^i 
	\nonumber\\
	&& + \frac{105}{x_0} \left(d_{\sigma}\right)^2  \sigma^a \sigma^b \sigma^c \sigma^d d_{\sigma}^i 
        - \frac{210}{x_0} \left(d_{\sigma}\right)^2 \sigma^a \sigma^b \sigma^c d_{\sigma}^d \sigma^i 
	+ \frac{105}{x_0}\, d_{\sigma}^a d_{\sigma}^b \sigma^c \sigma^d d_{\sigma}^i 
        - \frac{420}{x_0}\, \sigma^a d_{\sigma}^b d_{\sigma}^c d_{\sigma}^d \sigma^i 
        \nonumber\\
	&& - \frac{210}{\left(d_{\sigma}\right)^2}\, \frac{1}{x_0}\, d_{\sigma}^a d_{\sigma}^b d_{\sigma}^c d_{\sigma}^d d_{\sigma}^i 
        - \frac{105}{\left(x_0\right)^3}\left(d_{\sigma}\right)^4 \sigma^a \sigma^b \sigma^c \sigma^d d_{\sigma}^i
        + \frac{420}{\left(x_0\right)^3}\left(d_{\sigma}\right)^4 \sigma^a \sigma^b \sigma^c d_{\sigma}^d \sigma^i
        \nonumber\\
	&& + \frac{210}{\left(x_0\right)^3}\left(d_{\sigma}\right)^2 d_{\sigma}^a d_{\sigma}^b \sigma^c \sigma^d d_{\sigma}^i 
	+ \frac{420}{\left(x_0\right)^3}\left(d_{\sigma}\right)^2 \sigma^a d_{\sigma}^b \sigma^c d_{\sigma}^d d_{\sigma}^i 
	- \frac{420}{\left(x_0\right)^3}\left(d_{\sigma}\right)^2 \sigma^a d_{\sigma}^b d_{\sigma}^c d_{\sigma}^d \sigma^i 
	\nonumber\\
	&& - \frac{105}{\left(x_0\right)^3}\, d_{\sigma}^a d_{\sigma}^b d_{\sigma}^c d_{\sigma}^d d_{\sigma}^i 
        - 210 \left(x_0 + \ve{\sigma} \cdot \ve{x}_0 \right) \sigma^a \sigma^b \sigma^c \sigma^d d_{\sigma}^i 
        + 420 \left(x_0 + \ve{\sigma} \cdot \ve{x}_0 \right) \sigma^a \sigma^b \sigma^c d_{\sigma}^d \sigma^i 
	\nonumber\\ 
	&& - \frac{210}{\left(d_{\sigma}\right)^2} \left(x_0 + \ve{\sigma}\cdot\ve{x}_0\right) d_{\sigma}^a d_{\sigma}^b \sigma^c \sigma^d d_{\sigma}^i
        + \frac{840}{\left(d_{\sigma}\right)^2} \left(x_0 + \ve{\sigma}\cdot\ve{x}_0\right) \sigma^a d_{\sigma}^b d_{\sigma}^c d_{\sigma}^d \sigma^i 
	\nonumber\\
	&& + \frac{420}{\left(d_{\sigma}\right)^4} \left(x_0 + \ve{\sigma}\cdot\ve{x}_0\right) d_{\sigma}^a d_{\sigma}^b d_{\sigma}^c d_{\sigma}^d 
        d_{\sigma}^i 
	\;.
\label{coefficient_Q9}
\end{eqnarray}
\begin{eqnarray}
	{\cal Q}^{i\,abcd}_{\left(10\right)} &=&  
	+ 385 \left(d_{\sigma}\right)^2 \sigma^a d_{\sigma}^b \sigma^c \,\delta^{di}  
 	- 13 \left(d_{\sigma}\right)^2 d_{\sigma}^a\,\delta^{bc}\,\delta^{di}  
	+ \frac{397}{2} \left(d_{\sigma}\right)^2 \sigma^a \sigma^b d_{\sigma}^c \,\delta^{di} 
	\nonumber\\
	&& - 5 \left(d_{\sigma}\right)^2 \delta^{ac}\,\delta^{bd}\,d_{\sigma}^i 
	+ 115 \left(d_{\sigma}\right)^2 \sigma^a\,\delta^{bc}\,\sigma^d d_{\sigma}^i 
	- \frac{2127}{2} \left(d_{\sigma}\right)^2 \sigma^a \sigma^b \sigma^c \sigma^d d_{\sigma}^i 
	\nonumber\\
	&& + 220 \left(d_{\sigma}\right)^2 \sigma^a\,\delta^{bc}\,d_{\sigma}^d \sigma^i 
 	+ 1371 \left(d_{\sigma}\right)^2 \sigma^a \sigma^b \sigma^c d_{\sigma}^d \sigma^i 
	- 180\,d_{\sigma}^a d_{\sigma}^b d_{\sigma}^c\,\delta^{di}  
        + 716\,\sigma^a \sigma^b d_{\sigma}^c d_{\sigma}^d d_{\sigma}^i 
	\nonumber\\
	&& + 1438\,\sigma^a d_{\sigma}^b \sigma^c d_{\sigma}^d d_{\sigma}^i 
	- 70\,d_{\sigma}^a \,\delta^{bc}\,d_{\sigma}^d d_{\sigma}^i 
        - 96\,d_{\sigma}^a d_{\sigma}^b d_{\sigma}^c \sigma^d \sigma^i 
	+ \frac{300}{\left(d_{\sigma}\right)^2}\, d_{\sigma}^a d_{\sigma}^b d_{\sigma}^c d_{\sigma}^d d_{\sigma}^i 
	\;.
\label{coefficient_Q10}
\end{eqnarray}
\begin{eqnarray}
	{\cal Q}^{i\,abcd}_{\left(11\right)} &=& 0 \;.
\label{coefficient_Q11}
\end{eqnarray}
\begin{eqnarray}
	{\cal Q}^{i\,abcd}_{\left(12\right)} &=& 
	- 45 \left(d_{\sigma}\right)^4 \sigma^a d_{\sigma}^b \sigma^c \,\delta^{di}   
        - \frac{45}{2} \left(d_{\sigma}\right)^4 \sigma^a \sigma^b d_{\sigma}^c \,\delta^{di}   
	+ \frac{1305}{2} \left(d_{\sigma}\right)^4 \sigma^a \sigma^b \sigma^c \sigma^d d_{\sigma}^i  
	\nonumber\\
	&& - 75 \left(d_{\sigma}\right)^4 \sigma^a \,\delta^{bc}\,\sigma^d d_{\sigma}^i 
        - 585 \left(d_{\sigma}\right)^4 \sigma^a \sigma^b \sigma^c d_{\sigma}^d \sigma^i 
        - 150 \left(d_{\sigma}\right)^4 \sigma^a \,\delta^{bc}\, d_{\sigma}^d \sigma^i 
	\nonumber\\
	&& + \frac{45}{2} \left(d_{\sigma}\right)^2 d_{\sigma}^a d_{\sigma}^b d_{\sigma}^c \,\delta^{di}  
        - 1155 \left(d_{\sigma}\right)^2 \sigma^a \sigma^b d_{\sigma}^c d_{\sigma}^d d_{\sigma}^i 
        - 2310 \left(d_{\sigma}\right)^2 \sigma^a d_{\sigma}^b \sigma^c d_{\sigma}^d d_{\sigma}^i 
	\nonumber\\
	&& + 75 \left(d_{\sigma}\right)^2 d_{\sigma}^a \, \delta^{bc}\, d_{\sigma}^d d_{\sigma}^i 
	+ 885 \left(d_{\sigma}\right)^2 \sigma^a d_{\sigma}^b d_{\sigma}^c d_{\sigma}^d \sigma^i 
        + \frac{1005}{2}\, d_{\sigma}^a d_{\sigma}^b d_{\sigma}^c d_{\sigma}^d d_{\sigma}^i 
	\;.
\label{coefficient_Q12}
\end{eqnarray}
\begin{eqnarray}
	{\cal Q}^{i\,abcd}_{\left(13\right)} &=& 0 \;.
\label{coefficient_Q13}
\end{eqnarray}
\begin{eqnarray}
	{\cal Q}^{i\,abcd}_{\left(14\right)} &=&  
        - \frac{225}{2} \left(d_{\sigma}\right)^6 \sigma^a \sigma^b \sigma^c \sigma^d d_{\sigma}^i 
	- 450 \left(d_{\sigma}\right)^6 \sigma^a \sigma^b \sigma^c d_{\sigma}^d \sigma^i 
        + 225 \left(d_{\sigma}\right)^4 \sigma^a \sigma^b d_{\sigma}^c d_{\sigma}^d d_{\sigma}^i 
	\nonumber\\
	&& + 450 \left(d_{\sigma}\right)^4 \sigma^a d_{\sigma}^b \sigma^c d_{\sigma}^d d_{\sigma}^i 
	+ 450 \left(d_{\sigma}\right)^4 \sigma^a d_{\sigma}^b d_{\sigma}^c d_{\sigma}^d \sigma^i 
        - \frac{225}{2} \left(d_{\sigma}\right)^2 d_{\sigma}^a d_{\sigma}^b d_{\sigma}^c d_{\sigma}^d d_{\sigma}^i
	\;. 
\label{coefficient_Q14}
\end{eqnarray}

\section{Estimation of the expression in (\ref{upper_limits_Q})}\label{Estimation_Q}

In this appendix we want to proof the inequality in (\ref{upper_limits_Q}) as instructive example. 
That means we consider the upper limit of the absolute value $\big| \ve{\sigma} \times \ve{\nu}_{\rm 1PN}^{\rm Q} \big|$   
where $\ve{\nu}_{\rm 1PN}^{\rm Q}$ is given by Eq.~(\ref{limit_plus_Q}). 
Using $\left(\ve{\sigma} \times \ve{\nu}_{\rm 1PN}^{\rm Q}\right)^k = \epsilon^{ijk} \sigma^i \nu_{\rm 1PN}^{j\,{\rm Q}}$ one gets 
\begin{eqnarray}
	\left(\ve{\sigma} \times \ve{\nu}_{\rm 1PN}^{\rm Q} \right)^k &=& \frac{G \hat{M}_{ab}}{c^2} 
	\epsilon^{ijk} \sigma^i \left[\sum\limits_{n=5,7} {\cal D}^{j\,ab}_{\left(n\right)}\;\dot{\cal X}^{\infty}_{\left(n\right)} \right] ,  
\label{Appendix_Estimation_Q} 
\end{eqnarray}

\noindent
where the time-independent coefficients $D^{j\,ab}_{\left(n\right)}$ are given by
Eqs.~(\ref{coefficient_D5}) and (\ref{coefficient_D7}).
By inserting the limits (\ref{Limit_X_Y_n}) into (\ref{Appendix_Estimation_Q}) one obtains
\begin{eqnarray}
\left(\ve{\sigma} \times \ve{\nu}^{\rm Q}_{\rm 1PN}\right)^k &=&  
\frac{4}{3} \,\frac{G \hat{M}_{ab}}{c^2} \, \frac{\epsilon^{ijk} \sigma^i}{\left(d_{\sigma}\right)^4} 
\left[{\cal D}^{j\,ab}_{\left(5\right)} + \frac{4}{5}\,\frac{1}{\left(d_{\sigma}\right)^2}\,{\cal D}^{j\,ab}_{\left(7\right)}\right].
\label{Appendix_Estimation_Q_1} 
\end{eqnarray}

\noindent
The origin of the spatial coordinates $x^1,x^2,x^3$ is located at the center of mass of the body.
As stated in Section \ref{Total_Light_Deflection}, we assume the $x^3$-axis of coordinate system to be aligned with 
the symmetry axis $\ve{e}_3$ of the massive body. Hence, the quadrupole tensor (\ref{Quadrupole_Tensor}) takes the form 
\begin{eqnarray} 
        \hat{M}_{ab} = M\,J_2\,P^2 \left(\frac{1}{3}\,\delta_{ab} - \delta_{a3}\,\delta_{b3} \right) . 
        \label{Quadrupole_Tensor_in_z_axis} 
\end{eqnarray}

\noindent
By inserting (\ref{Quadrupole_Tensor_in_z_axis}) into (\ref{Appendix_Estimation_Q_1}) and taking into account that
\begin{eqnarray}
        \left| \ve{\sigma} \times \ve{\nu}\right| =  \left| \ve{\sigma} \times \left(\ve{\nu} \times \ve{\sigma} \right)\right|\,,
        \label{relation_absolute_value}
\end{eqnarray}

\noindent
which is valid since $\ve{\sigma}$ is a unit-vector, and using  
$\ve{\sigma} \times \left(\ve{d}_{\sigma} \times \ve{\sigma}\right) = \ve{d}_{\sigma}$ and
$\ve{\sigma} \times \left(\ve{x}_0\times \ve{\sigma}\right) = \ve{d}_{\sigma}$, one obtains 
\begin{eqnarray}
	\left|\ve{\sigma} \times \ve{\nu}^{\rm Q}_{\rm 1PN} \right| &=&  
        4\, \frac{G M}{c^2} \left| J_2 \right| \frac{P^2}{\left(d_{\sigma}\right)^3}
	\left| F_1 \,\frac{\ve{d_{\sigma}}}{d_{\sigma}} + F_2 \left(\ve{\sigma} \times \left(\ve{e}_3 \times \ve{\sigma}\right)\right) \right| ,
\label{Appendix_Estimation_Q_2} 
\end{eqnarray}

\noindent
where the functions read
\begin{eqnarray}
	F_1 &=& 1 - \left(\ve{\sigma} \cdot \ve{e}_3\right)^2 - 4 \,\frac{\left(\ve{d}_{\sigma} \cdot \ve{e}_3\right)^2}{\left(d_{\sigma}\right)^2} \;,
\label{Function_F1}
\\
	F_2 &=& 2 \,\frac{\left(\ve{d}_{\sigma} \cdot \ve{e}_3\right)}{d_{\sigma}} \;. 
\label{Function_F2}
\end{eqnarray}

\noindent
For determining the limit (\ref{Appendix_Estimation_Q_2}) it is necessary to take into account that $\ve{\sigma} \cdot \ve{d}_{\sigma} = 0$ 
which restricts the possible values for the angles $\delta\left(\ve{\sigma},\ve{e}_3\right)$ and $\delta\left(\ve{d}_{\sigma},\ve{e}_3\right)$ 
(see also endnote $\left[99\right]$). One obtains for the upper limit  
\begin{eqnarray}
	\left|\ve{\sigma} \times \ve{\nu}^{\rm Q}_{\rm 1PN}\right| \le 4 \frac{G M}{c^2} \left| J_2 \right| \frac{P^2}{\left(d_{\sigma}\right)^3} \;,  
\label{Appendix_Estimation_Q_3} 
\end{eqnarray}

\noindent
which has also been proven in our article \cite{Zschocke6} (cf. Eq.~(12) ibid.). Thus, the validity of relation (\ref{upper_limits_Q}) has been shown.

\section{Estimation of the expression in (\ref{upper_limits_M_M})}\label{Estimation_M_M}

In this appendix we want to proof the inequality in (\ref{upper_limits_M_M}).
That means we consider the upper limit of the absolute value $\big| \ve{\sigma} \times \ve{\nu}_{\rm 2PN}^{{\rm M} \times {\rm M}} \big|$
where $\ve{\nu}_{\rm 2PN}^{ {\rm M}\times {\rm M}}$ is given by Eq.~(\ref{limit_plus_M_M}).
Using $\left(\ve{\sigma} \times \ve{\nu}_{\rm 2PN}^{{\rm M} \times {\rm M}}\right)^k = \epsilon^{ijk} \sigma^i \nu_{\rm 2PN}^{j\,{\rm M} \times {\rm M}}$ one gets 
\begin{eqnarray}
	\left(\ve{\sigma} \times \ve{\nu}^{{\rm M} \times {\rm M}}_{\rm 2PN}\right)^k &=&
\frac{G^2 M^2}{c^4}\,
        \epsilon^{ijk} \sigma^i
        \left[\sum\limits_{n=2}^{6} {\cal F}^{j}_{\left(n\right)}\;\dot{\cal X}^{\infty}_{\left(n\right)}
        + {\cal G}^{j}_{\left(5\right)}\;\dot{\cal Y}^{\infty}_{\left(5\right)} \right] ,
\label{Appendix_Estimation_M_M}
\end{eqnarray}

\noindent
where the time-independent coefficients ${\cal F}^{j}_{\left(n\right)}$ and ${\cal G}^{j}_{\left(5\right)}$ are given by
Eqs.~(\ref{coefficient_F2}) - (\ref{coefficient_F6}) and (\ref{coefficient_G5}), respectively.
By inserting the limits (\ref{Limit_X_Y_n}) into (\ref{Appendix_Estimation_M_M}) one obtains
\begin{eqnarray}
\left[\sum\limits_{n=2}^{6} {\cal F}^{j}_{\left(n\right)}\;\dot{\cal X}^{\infty}_{\left(n\right)}
        + {\cal G}^{j}_{\left(5\right)}\;\dot{\cal Y}^{\infty}_{\left(5\right)} \right] 
	&=& 2\,\frac{{\cal F}^{j}_{\left(3\right)}}{\left(d_{\sigma}\right)^2} 
	+ \frac{4}{3}\,\frac{{\cal F}^{j}_{\left(5\right)}}{\left(d_{\sigma}\right)^4}
	+ \pi \left(\frac{{\cal F}^{j}_{\left(2\right)}}{d_{\sigma}} 
	+ \frac{1}{2}\,\frac{{\cal F}^{j}_{\left(4\right)}}{\left(d_{\sigma}\right)^3}   
	+ \frac{3}{8}\,\frac{{\cal F}^{j}_{\left(6\right)}}{\left(d_{\sigma}\right)^5}   
	- \frac{1}{6}\,\frac{{\cal G}^{j}_{\left(5\right)}}{\left(d_{\sigma}\right)^3} \right) .  
	\label{Appendix_Estimation_M_M_5} 
\end{eqnarray}

\noindent
By inserting the coefficients ${\cal F}^{j}_{\left(n\right)}$ and ${\cal G}^{j}_{\left(5\right)}$ into (\ref{Appendix_Estimation_M_M_5}) one obtains 
\begin{eqnarray}
\left[\sum\limits_{n=2}^{6} {\cal F}^{j}_{\left(n\right)} \dot{\cal X}^{\infty}_{\left(n\right)}
        + {\cal G}^{j}_{\left(5\right)} \dot{\cal Y}^{\infty}_{\left(5\right)} \right] 
        &=& 8 \,\frac{x_0 + \ve{\sigma} \cdot \ve{x}_0}{\left(d_{\sigma}\right)^4} d_{\sigma}^j 
        - \frac{15}{4} \,\pi\, \frac{d_{\sigma}^j}{\left(d_{\sigma}\right)^3}   
\label{Appendix_Estimation_M_M_10}
\end{eqnarray}

\noindent
where all terms proportional to $\sigma^j$ have been omitted since they do not contribute to the total 
light deflection because of $\epsilon^{ijk} \sigma^i \sigma^j = 0$. Inserting (\ref{Appendix_Estimation_M_M_10}) 
into (\ref{Appendix_Estimation_M_M}) yields for the absolute value 
\begin{eqnarray}
	\left|\ve{\sigma} \times \ve{\nu}^{{\rm M} \times {\rm M}}_{\rm 2PN}\right| &=& \frac{G^2 M^2}{c^4}\,
	\left| \frac{15}{4}\, \frac{\pi}{\left(d_{\sigma}\right)^2} 
	- 8\,\frac{x_0 + \ve{\sigma} \cdot \ve{x}_0}{\left(d_{\sigma}\right)^3} \right| \,,  
\label{Appendix_Estimation_M_M_15_A}
\end{eqnarray}

\noindent
which is the asserted relation (\ref{upper_limits_M_M}) and agrees with the 2PN terms in Eq.~(3.2.44) in \cite{Brumberg1991} 
as well as with the 2PN terms in Eq.~(65) in \cite{Article_Zschocke1}, if one uses the PPN parameters in the limit of general relativity. 
The upper limit (\ref{Appendix_Estimation_M_M_15_A}) is valid for any astrometric configuration of source, body, and observer, that means
for $\ve{\sigma} \cdot \ve{x}_0 \le 0$ as well as for $\ve{\sigma} \cdot \ve{x}_0 \ge 0$. We are mainly interested in 
astrometric configurations (cf. condition (\ref{configuration1}))
\begin{eqnarray}
        \ve{\sigma} \cdot \ve{x}_0 &\le& 0  \;,
        \label{appendix_configuration1}
\end{eqnarray}

\noindent
that means astrometric configurations where the massive body is located between the light source and the observer. 
In general, condition (\ref{appendix_configuration1}) holds for any light source far from the Solar System, like stellar light sources. 
If condition (\ref{appendix_configuration1}) is assumed, then the upper limit (\ref{Appendix_Estimation_M_M_15_A}) simplifies as follows:  
\begin{eqnarray}
        \left|\ve{\sigma} \times \ve{\nu}^{{\rm M} \times {\rm M}}_{\rm 2PN}\right| &=& \frac{15}{4}\,\pi\, 
	\frac{G^2 M^2}{c^4}\,\frac{1}{\left(d_{\sigma}\right)^2}\;, 
\label{Appendix_Estimation_M_M_15}
\end{eqnarray}

\noindent 
which is the asserted relation (\ref{upper_limits_M_M_configuration1}).

\section{Estimation of the expression in (\ref{upper_limits_M_Q})}\label{Estimation_M_Q}

In this appendix we want to proof the inequality in (\ref{upper_limits_M_Q}).  
That means we consider the upper limit of the absolute value $\big| \ve{\sigma} \times \ve{\nu}_{\rm 2PN}^{{\rm M} \times {\rm Q}} \big|$   
where $\ve{\nu}_{\rm 2PN}^{{\rm M} \times {\rm Q}}$ is given by Eq.~(\ref{limit_plus_M_Q}). 
Using $\left(\ve{\sigma} \times \ve{\nu}_{\rm 2PN}^{{\rm M} \times {\rm Q}}\right)^k = \epsilon^{ijk} \sigma^i \nu_{\rm 2PN}^{j\,{\rm M} \times {\rm Q}}$ one gets
\begin{eqnarray}
	\left(\ve{\sigma} \times \ve{\nu}^{{\rm M} \times {\rm Q}}_{\rm 2PN}\right)^k &=&  
\frac{G M}{c^2}\,\frac{G \hat{M}_{ab}}{c^2} 
        \epsilon^{ijk} \sigma^i 
	\left[\sum\limits_{n=2}^{10} {\cal L}^{j\,ab}_{\left(n\right)}\;\dot{\cal X}^{\infty}_{\left(n\right)}  
	+ \sum\limits_{n=7}^{9} {\cal M}^{j\,ab}_{\left(n\right)}\;\dot{\cal Y}^{\infty}_{\left(n\right)} \right] , 
\label{Appendix_Estimation_M_Q} 
\end{eqnarray}

\noindent
where the time-independent coefficients ${\cal L}^{j\,ab}_{\left(n\right)}$ and ${\cal M}^{j\,ab}_{\left(n\right)}$ are given by
Eqs.~(\ref{coefficient_L2}) - (\ref{coefficient_L10}) and (\ref{coefficient_M7}) - (\ref{coefficient_M9}), respectively.
By inserting the limits (\ref{Limit_X_Y_n}) into (\ref{Appendix_Estimation_M_Q}) one obtains
\begin{eqnarray}
	\left[\sum\limits_{n=2}^{10} {\cal L}^{j\,ab}_{\left(n\right)}\;\dot{\cal X}^{\infty}_{\left(n\right)}  
	+ \sum\limits_{n=7}^{9} {\cal M}^{j\,ab}_{\left(n\right)}\;\dot{\cal Y}^{\infty}_{\left(n\right)} \right] &=&  
	2\,\frac{{\cal L}^{j\,ab}_{\left(3\right)}}{\left(d_{\sigma}\right)^2} 
        + \frac{4}{3}\,\frac{{\cal L}^{j\,ab}_{\left(5\right)}}{\left(d_{\sigma}\right)^4} 
        + \frac{16}{15}\,\frac{{\cal L}^{j\,ab}_{\left(7\right)}}{\left(d_{\sigma}\right)^6} 
        + \frac{32}{35}\,\frac{{\cal L}^{j\,ab}_{\left(9\right)}}{\left(d_{\sigma}\right)^8}
        \nonumber\\
	&& \hspace{-3.0cm} + \pi \left(  
        \frac{{\cal L}^{j\,ab}_{\left(2\right)}}{d_{\sigma}}
        + \frac{1}{2}\,\frac{{\cal L}^{j\,ab}_{\left(4\right)}}{\left(d_{\sigma}\right)^3}
        + \frac{3}{8}\,\frac{{\cal L}^{j\,ab}_{\left(6\right)}}{\left(d_{\sigma}\right)^5}
        + \frac{5}{16}\,\frac{{\cal L}^{j\,ab}_{\left(8\right)}}{\left(d_{\sigma}\right)^7}
        + \frac{35}{128}\,\frac{{\cal L}^{j\,ab}_{\left(10\right)}}{\left(d_{\sigma}\right)^9}
        - \frac{3}{40}\,\frac{{\cal M}^{j\,ab}_{\left(7\right)}}{\left(d_{\sigma}\right)^5}
        - \frac{5}{112}\,\frac{{\cal M}^{j\,ab}_{\left(9\right)}}{\left(d_{\sigma}\right)^7} \right) .  
        \label{Appendix_Estimation_M_Q_1} 
        \end{eqnarray}

\noindent
Then, by inserting the time-independent coefficients ${\cal L}^{j\,ab}_{\left(n\right)}$ and ${\cal M}^{j\,ab}_{\left(n\right)}$ one obtains  
\begin{eqnarray}
	\left[ \sum\limits_{n=2}^{10} {\cal L}^{j\,ab}_{\left(n\right)}\;\dot{\cal X}^{\infty}_{\left(n\right)}  
	+ \sum\limits_{n=7}^{9} {\cal M}^{j\,ab}_{\left(n\right)}\;\dot{\cal Y}^{\infty}_{\left(n\right)} \right] 
	&=& \pi \left( \frac{465}{32}\,\frac{d_{\sigma}^a\,\delta^{bj}}{\left(d_{\sigma}\right)^5} 
        - \frac{855}{64}\,\frac{\sigma^a \sigma^b d_{\sigma}^j}{\left(d_{\sigma}\right)^5} 
	- \frac{2325}{64}\,\frac{d_{\sigma}^a d_{\sigma}^b d_{\sigma}^j}{\left(d_{\sigma}\right)^7} \right)  
        \nonumber\\ 
	&& \hspace{-5.0cm} + 16\,\frac{\sigma^a \delta^{bj}}{\left(d_{\sigma}\right)^4}
        - 8\,\frac{\ve{\sigma} \cdot \ve{x}_0}{x_0}\,\frac{\sigma^a \delta^{bj}}{\left(d_{\sigma}\right)^4}
        - 32\,\frac{\sigma^a d_{\sigma}^b d_{\sigma}^j}{\left(d_{\sigma}\right)^6}
        - 8\,\frac{\ve{\sigma} \cdot \ve{x}_0}{\left(x_0\right)^3}\,\frac{\sigma^a d_{\sigma}^b d_{\sigma}^j}{\left(d_{\sigma}\right)^4}
        - \frac{8}{x_0}\,\frac{d_{\sigma}^a \delta^{bj}}{\left(d_{\sigma}\right)^4}
	- 32 \left(x_0 + \ve{\sigma} \cdot \ve{x}_0\right) \frac{d_{\sigma}^a \delta^{bj}}{\left(d_{\sigma}\right)^6}
        \nonumber\\
        && \hspace{-5.0cm} - \frac{4}{x_0}\,\frac{\sigma^a \sigma^b d_{\sigma}^j}{\left(d_{\sigma}\right)^4}
	+ \frac{4}{\left(d_{\sigma}\right)^2}\,\frac{\sigma^a \sigma^b d_{\sigma}^j}{\left(x_0\right)^3}
        + 32 \left(x_0 + \ve{\sigma} \cdot \ve{x}_0\right) \frac{\sigma^a \sigma^b d_{\sigma}^j}{\left(d_{\sigma}\right)^6}
	- \frac{4}{\left(x_0\right)^3}\,\frac{d_{\sigma}^a d_{\sigma}^a d_{\sigma}^j}{\left(d_{\sigma}\right)^4}
        + 96\,\left(x_0 + \ve{\sigma} \cdot \ve{x}_0\right) \frac{d_{\sigma}^a d_{\sigma}^a d_{\sigma}^j}{\left(d_{\sigma}\right)^8}\;,
        \label{Appendix_Estimation_M_Q_F1_F2}
\end{eqnarray}

\noindent
where terms proportional to $\sigma^j$ have been omitted because the do not contribute to the total light deflection 
because of $\epsilon^{ijk} \sigma^i \sigma^j = 0$.  
Inserting the STF quadrupole moment (\ref{Quadrupole_Tensor_in_z_axis}) into Eq.~(\ref{Appendix_Estimation_M_Q_F1_F2}) 
yields for the absolute value of (\ref{Appendix_Estimation_M_Q}) 
\begin{eqnarray}
	\left|\ve{\sigma} \times \ve{\nu}^{{\rm M} \times {\rm Q}}_{\rm 2PN} \right| 
	&=& \frac{G^2 M^2}{c^4} \, \frac{\left|J_2\right| P^2}{\left(d_{\sigma}\right)^4} 
	\left| G_1 \frac{\ve{d_{\sigma}}}{d_{\sigma}} + G_2 \,\ve{\sigma} \times \left(\ve{e}_3 \times \ve{\sigma}\right)\right| ,
        \label{vector_total_light_deflection_M_Q} 
\end{eqnarray}

\noindent
where also relation (\ref{relation_absolute_value}) has been used. The functions in (\ref{vector_total_light_deflection_M_Q}) read
\begin{eqnarray}
	G_1 &=& - \frac{15}{32}\,\pi \left(25 - \frac{57}{2} \left(\ve{\sigma} \cdot \ve{e}_3 \right)^2
	- \frac{155}{2}\,\frac{\left(\ve{d}_{\sigma} \cdot \ve{e}_3\right)^2}{\left(d_{\sigma}\right)^2} \right) 
	+ 32 \left(1 - \left(\ve{\sigma} \cdot \ve{e}_3\right)^2 - 3\,\frac{\left(\ve{d}_{\sigma} \cdot \ve{e}_3\right)^2}{\left(d_{\sigma}\right)^2}\right) 
        \frac{x_0 + \ve{\sigma} \cdot \ve{x}_0}{d_{\sigma}} 
        \nonumber\\ 
	&& +\,32\,\frac{\ve{d}_{\sigma} \cdot \ve{e}_3}{d_{\sigma}}\,\left(\ve{\sigma} \cdot \ve{e}_3\right) 
	- 4\,\frac{d_{\sigma}}{x_0} \left(1 -\left(\ve{\sigma} \cdot \ve{e}_3\right)^2\right) 
	+\,8\,\left(d_{\sigma}\right)^2\,\frac{\ve{\sigma} \cdot \ve{x}_0}{\left(x_0\right)^3} 
	\left(\ve{\sigma} \cdot \ve{e}_3\right) \frac{\ve{d}_{\sigma} \cdot \ve{e}_3}{d_{\sigma}}   
        \nonumber\\
	&& + 4\,\frac{\left(d_{\sigma}\right)^3}{\left(x_0\right)^3} 
	\left(\frac{\left(\ve{d}_{\sigma} \cdot \ve{e}_3\right)^2}{\left(d_{\sigma}\right)^2} - \left(\ve{\sigma} \cdot \ve{e}_3\right)^2\right)\;,  
        \label{Function_G1}
        \\
	\nonumber\\ 
	G_2 &=& - \frac{465}{32} \pi \frac{\ve{d}_{\sigma} \cdot \ve{e}_3}{d_{\sigma}} 
        + 32\,\frac{x_0 + \ve{\sigma} \cdot \ve{x}_0}{d_{\sigma}} \,\frac{\ve{d}_{\sigma} \cdot \ve{e}_3}{d_{\sigma}} 
        - 16 \left(\ve{\sigma} \cdot \ve{e}_3\right) + 8 \frac{\ve{\sigma} \cdot \ve{x}_0}{x_0} \left(\ve{\sigma} \cdot \ve{e}_3\right)  
        + 8 \,\frac{\ve{d}_{\sigma} \cdot \ve{e}_3}{x_0} \;. 
\label{Function_G2}
\end{eqnarray}

\noindent
We note the relation   
\begin{eqnarray}
	\frac{\ve{\sigma} \cdot \ve{x}_0}{x_0} &=& \pm \sqrt{ 1 - {\left(\frac{d_{\sigma}}{x_0}\right)^2}} \quad 
	{\rm with} \quad 0 \le \frac{d_{\sigma}}{x_0} \le 1\;, 
        \label{Relation_5}
\end{eqnarray}

\noindent
which is a useful relation for the estimation of (\ref{vector_total_light_deflection_M_Q}). 
Then, by using the computer algebra system {\it Maple} \cite{Maple}, one obtains for astrometric configurations 
(\ref{appendix_configuration1}) the following upper limit: 
\begin{eqnarray}
	\left| \ve{\sigma} \times \ve{\nu}^{{\rm M} \times {\rm Q}}_{\rm 2PN}\right| &\le& 
	4\,\frac{G^2 M^2}{c^4}\,\left|J_2\right|\,\frac{P^2}{\left(d_{\sigma}\right)^2}\,
	\left| \frac{15}{4}\, \frac{\pi}{\left(d_{\sigma}\right)^2} + 8\,\frac{x_0 + \ve{\sigma} \cdot \ve{x}_0}{\left(d_{\sigma}\right)^3} \right| \,,   
	\label{Appendix_Estimation_M_Q_2_5_A}
\end{eqnarray}

\noindent 
which is the asserted relation in (\ref{upper_limits_M_Q}). 
For the estimation in (\ref{Appendix_Estimation_M_Q_2_5_A}) one has to take into account that $\ve{\sigma}$ and $\ve{d}_{\sigma}$ are perpendicular 
to each other, which reduces the possible areas of the angles $\delta\left(\ve{\sigma},\ve{e}_3\right)$ and $\delta\left(\ve{d}_{\sigma},\ve{e}_3\right)$.  
For instance, one obtains 
(see endnote \footnote{The terms on the left-hand side of relations (\ref{Relation_7}) and (\ref{Relation_10}) consist of scalar products of three-vectors, 
hence they are independent of the orientation of spatial coordinate axes. So, despite that we have previously assumed that $\ve{e}_3$ is parallel to the 
z-direction (cf. Eqs.~(\ref{Quadrupole_Tensor_in_z_axis}) and (\ref{Quadrupole_Tensor_in_z_axis_Product})), one may rotate these axes such that $\ve{\sigma}$ 
is aligned parallel to the x-direction and $\ve{d}_{\sigma}$ to the y-direction, while $\ve{e}_3 = \left(e_3^x,e_3^y,e_3^z\right)$ has three spatial components 
now. Then, $\ve{\sigma} \cdot \ve{e}_3 = e_3^x$ and $\displaystyle \frac{\ve{d}_{\sigma}}{d_{\sigma}} \cdot \ve{e}_3 = e_3^y$. Using these relations and the fact 
that the possible values for the components of $\ve{e}_3$ are restricted because $\ve{e}_3$ is a unit vector, one may show the validity of the 
inequalities (\ref{Relation_7}) and (\ref{Relation_10}).})
\begin{eqnarray}
\big|\ve{\sigma} \cdot \ve{e}_3 \big|\, \frac{\big| \ve{d}_{\sigma} \cdot \ve{e}_3 \big|}{d_{\sigma}} \le \frac{1}{2}\;, 
        \label{Relation_7}
\end{eqnarray}

\noindent 
and 
\begin{eqnarray}
        && \left|\!\left(\!1 - \left(\ve{\sigma} \cdot \ve{e}_3\right)^2 
        - 3 \frac{\left(\ve{d}_{\sigma} \cdot \ve{e}_3\right)^2}{\left(d_{\sigma}\right)^2}\!\right)\! 
        \frac{\ve{d}_{\sigma}}{d_{\sigma}} 
        + \frac{\ve{d}_{\sigma} \cdot \ve{e}_3}{d_{\sigma}}\, \ve{\sigma} \times \left( \ve{e}_3 \times \ve{\sigma}\!\right) \right| 
        \le 1\;.
        \label{Relation_10}
\end{eqnarray}

\noindent
The upper limit (\ref{Appendix_Estimation_M_Q_2_5_A}) is valid for any astrometric configuration of source, body, and observer, that means 
for $\ve{\sigma} \cdot \ve{x}_0 \le 0$ as well as for $\ve{\sigma} \cdot \ve{x}_0 > 0$. If one considers  
astrometric configurations with condition (\ref{appendix_configuration1}) then one obtains the following upper limit:
\begin{eqnarray}
       \left| \ve{\sigma} \times \ve{\nu}^{{\rm M} \times {\rm Q}}_{\rm 2PN}\right| &\le& 
        15\,\pi\,\frac{G^2 M^2}{c^4}\,\left|J_2\right|\,\frac{P^2}{\left(d_{\sigma}\right)^4}\,
        \label{Appendix_Estimation_M_Q_2_5}
\end{eqnarray}

\noindent 
which is the asserted relation (\ref{upper_limits_M_Q_configuration1}).

\section{Estimation of the expression in (\ref{upper_limits_Q_Q})}\label{Estimation_Q_Q}

In this appendix we want to proof the inequality in (\ref{upper_limits_Q_Q}).
That means we consider the upper limit of the absolute value $\big| \ve{\sigma} \times \ve{\nu}_{\rm 2PN}^{{\rm Q} \times {\rm Q}} \big|$ 
where $\ve{\nu}_{\rm 2PN}^{{\rm Q} \times {\rm Q}}$ is given by Eq.~(\ref{limit_plus_Q_Q}). 
Using $\left(\ve{\sigma} \times \ve{\nu}_{\rm 2PN}^{{\rm Q} \times {\rm Q}}\right)^k = \epsilon^{ijk} \sigma^i \nu_{\rm 2PN}^{j\,{\rm Q} \times {\rm Q}}$ one gets
\begin{eqnarray}
	\left(\ve{\sigma} \times \ve{\nu}^{{\rm Q} \times {\rm Q}}_{\rm 2PN}\right)^k &=&  
        \frac{G \hat{M}_{ab}}{c^2}\,\frac{G \hat{M}_{cd}}{c^2}\,  
        \epsilon^{ijk} \sigma^i 
	\left[\sum\limits_{n=4}^{14} {\cal Q}^{j\,abcd}_{\left(n\right)}\;\dot{\cal X}^{\infty}_{\left(n\right)}  
        \right] ,
\label{Appendix_Estimation_Q_Q} 
\end{eqnarray}

\noindent
where the time-independent coefficients ${\cal Q}^{j\,abcd}_{\left(n\right)}$ are given by
Eqs.~(\ref{coefficient_Q4}) - (\ref{coefficient_Q14}).
By inserting the limits (\ref{Limit_X_Y_n}) into (\ref{Appendix_Estimation_Q_Q}) one obtains
\begin{eqnarray}
	\left[ \sum\limits_{n=4}^{14} {\cal Q}^{j\,abcd}_{\left(n\right)}\;\dot{\cal X}^{\infty}_{\left(n\right)} \right] &=&   
        \frac{4}{3} \frac{{\cal Q}^{j\,abcd}_{\left(5\right)}}{\left(d_{\sigma}\right)^4}
        + \frac{16}{15}\,\frac{{\cal Q}^{j\,abcd}_{\left(7\right)}}{\left(d_{\sigma}\right)^6}
        + \frac{32}{35}\,\frac{{\cal Q}^{j\,abcd}_{\left(9\right)}}{\left(d_{\sigma}\right)^8}
\nonumber\\ 
	&& \hspace{-0.5cm} + \, \pi \left( \frac{1}{2}\,\frac{{\cal Q}^{j\,abcd}_{\left(4\right)}}{\left(d_{\sigma}\right)^3} 
        +  \frac{3}{8}\,\frac{{\cal Q}^{j\,abcd}_{\left(6\right)}}{\left(d_{\sigma}\right)^5}
        + \frac{5}{16}\,\frac{{\cal Q}^{j\,abcd}_{\left(8\right)}}{\left(d_{\sigma}\right)^7}
        + \frac{35}{128}\,\frac{{\cal Q}^{j\,abcd}_{\left(10\right)}}{\left(d_{\sigma}\right)^9}
        + \frac{63}{256}\,\frac{{\cal Q}^{j\,abcd}_{\left(12\right)}}{\left(d_{\sigma}\right)^{11}}
        + \frac{231}{1024}\,\frac{{\cal Q}^{j\,abcd}_{\left(14\right)}}{\left(d_{\sigma}\right)^{13}} \right) , 
        \label{Appendix_Estimation_Q_Q_1} 
\end{eqnarray}

\noindent
where ${\cal Q}^{j\,abcd}_{\left(11\right)} = {\cal Q}^{j\,abcd}_{\left(13\right)} = 0$ has been taken into account.
Then, by inserting the time-independent coefficients ${\cal Q}^{j\,abcd}_{\left(n\right)}$ as well as 
the product of two STF quadrupole moments in z-direction (\ref{Quadrupole_Tensor_in_z_axis}) 
\begin{eqnarray}
        \hat{M}_{ab} \, \hat{M}_{cd} &=& M^2\,\left|J_2\right|^2 P^4
        \left(\frac{1}{9}\,\delta_{ab} \delta_{cd} - \frac{1}{3} \delta_{ab} \delta_{c3} \delta_{d3}
        - \frac{1}{3} \delta_{a3} \delta_{b3} \delta_{cd}
        + \delta_{a3} \delta_{b3} \delta_{c3} \delta_{d3}\right)
        \label{Quadrupole_Tensor_in_z_axis_Product}
\end{eqnarray}

\noindent
into Eq.~(\ref{Appendix_Estimation_Q_Q_1}) yields for the absolute value of (\ref{Appendix_Estimation_Q_Q}) 
\begin{eqnarray}
	\left|\ve{\sigma} \times \ve{\nu}^{{\rm Q} \times {\rm Q}}_{\rm 2PN} \right| 
	&=& \frac{G^2 M^2}{c^4}\,\frac{\left|J_2\right|^2 P^4}{\left(d_{\sigma}\right)^6} 
	\left| H_1 \, \frac{\ve{d_{\sigma}}}{d_{\sigma}} + H_2\, \ve{\sigma} \times \left(\ve{e}_3 \times \ve{\sigma}\right) \right|,  
        \label{vector_total_light_deflection_Q_Q} 
\end{eqnarray}

\noindent
where relation (\ref{relation_absolute_value}) has been used. 
The functions $H_1$ and $H_2$ read  
\begin{eqnarray}
	H_1 &=& 
	- \frac{5}{128}\, \pi \left( 
	\frac{263}{3}  
	+ \frac{787}{4} \,\frac{\left(\ve{d}_{\sigma} \cdot \ve{e}_3\right)^2}{\left(d_{\sigma}\right)^2} 
	+ \frac{1323}{8} \left(\ve{\sigma} \cdot \ve{e}_3\right)^2 \frac{\left(\ve{d}_{\sigma} \cdot \ve{e}_3\right)^2}{\left(d_{\sigma}\right)^2} 
	- \frac{3969}{16} \frac{\left(\ve{d}_{\sigma} \cdot \ve{e}_3\right)^4}{\left(d_{\sigma}\right)^4} 
	- \frac{1265}{4} \left(\ve{\sigma} \cdot \ve{e}_3\right)^2  
	+ \frac{5175}{16} \left(\ve{\sigma} \cdot \ve{e}_3\right)^4  
	\right) 
	\nonumber\\
	&& + 24 \left(1 - 2 \left(\ve{\sigma} \cdot \ve{e}_3\right)^2 + \left(\ve{\sigma} \cdot \ve{e}_3\right)^4 \right) 
	\frac{x_0 + \ve{\sigma} \cdot \ve{x}_0}{d_{\sigma}} 
	- 48 \left(\ve{\sigma} \cdot \ve{e}_3\right) \frac{\ve{d}_{\sigma} \cdot \ve{e}_3}{d_{\sigma}}  
	\left(1 - \left(\ve{\sigma} \cdot \ve{e}_3\right)^2 \right) \frac{\ve{\sigma} \cdot \ve{x}_0}{x_0}
	\nonumber\\
	&& + 24 \,\left(d_{\sigma}\right)^2\,\frac{\ve{d}_{\sigma} \cdot \ve{e}_3}{d_{\sigma}} 
	\left(\ve{\sigma} \cdot \ve{e}_3\right) \, 
	\left(1 - \left(\ve{\sigma} \cdot \ve{e}_3\right)^2 \, 
	- 4 \,\frac{\left(\ve{d}_{\sigma} \cdot \ve{e}_2\right)^2}{\left(d_{\sigma}\right)^2}\right) 
	\frac{\ve{\sigma} \cdot \ve{x}_0}{\left(x_0\right)^3} 
	- 12\,\left(1 - \left(\ve{\sigma} \cdot \ve{e}_3\right)^2\right)^2\, \frac{d_{\sigma}}{x_0} 
	\nonumber\\ 
	&& + 12 \left( \frac{\left(\ve{d}_{\sigma} \cdot \ve{e}_3\right)^2}{\left(d_{\sigma}\right)^2} 
	+ 2\,\left(\ve{\sigma} \cdot \ve{e}_3\right)^2 \frac{\left(\ve{d}_{\sigma} \cdot \ve{e}_3\right)^2}{\left(d_{\sigma}\right)^2} 
	- 4\,\frac{\left(\ve{d}_{\sigma} \cdot \ve{e}_3\right)^4}{\left(d_{\sigma}\right)^4} 
	+ \left(\ve{\sigma} \cdot \ve{e}_3\right)^4 \right) 
	\frac{\left(d_{\sigma}\right)^3}{\left(x_0\right)^3}\;,  
	\label{H_1}
	\\
	\nonumber\\ 
	H_2 &=& 
	 \frac{105}{256}\,\pi\,\frac{\ve{d}_{\sigma} \cdot \ve{e}_3}{d_{\sigma}} 
	\left(1 - \frac{21}{2}\,\frac{\left(\ve{d}_{\sigma} \cdot \ve{e}_3\right)^2}{\left(d_{\sigma}\right)^2} 
	+ \frac{9}{2}\,\left(\ve{\sigma} \cdot \ve{e}_3\right)^2 \right) 
	- 16 \left(\ve{\sigma} \cdot \ve{e}_3\right) \left(1 - \left(\ve{\sigma} \cdot \ve{e}_3\right)^2\right)  
	+ 24 \left(\ve{\sigma} \cdot \ve{e}_3\right) \left(1 - \left(\ve{\sigma} \cdot \ve{e}_3\right)^2\right) 
	\frac{\ve{\sigma} \cdot \ve{x}_0}{x_0} 
	\nonumber\\ 
	&& + 48 \left(d_{\sigma}\right)^2 \left(\ve{\sigma} \cdot \ve{e}_3\right) \frac{\left(\ve{d}_{\sigma} \cdot \ve{e}_3\right)^2}{\left(d_{\sigma}\right)^2} 
	\frac{\ve{\sigma} \cdot \ve{x}_0}{\left(x_0\right)^3} 
        + 24\,\frac{\ve{d}_{\sigma} \cdot \ve{e}_3}{d_{\sigma}} 
        \left(\frac{\left(\ve{d}_{\sigma} \cdot \ve{e}_3\right)^2}{\left(d_{\sigma}\right)^2} - \left(\ve{\sigma} \cdot \ve{e}_3\right)^2\right) 
	\frac{\left(d_{\sigma}\right)^3}{\left(x_0\right)^3} \;. 
        \label{H_2}
\end{eqnarray}

\noindent 
Let us recall relation (\ref{Relation_5}) which is useful for the estimation of (\ref{vector_total_light_deflection_Q_Q}).  
By using the computer algebra system {\it Maple} \cite{Maple}, one obtains  
\begin{eqnarray}
        \left| \ve{\sigma} \times \ve{\nu}^{{\rm Q} \times {\rm Q}}_{\rm 2PN}\right| &\le&
        2\,\frac{G^2 M^2}{c^4}\,\left|J_2\right|^2\,\frac{P^4}{\left(d_{\sigma}\right)^4}\,
        \left| \frac{15}{4}\, \frac{\pi}{\left(d_{\sigma}\right)^2} + 12\,\frac{x_0 + \ve{\sigma} \cdot \ve{x}_0}{\left(d_{\sigma}\right)^3} \right| \,,  
        \label{Appendix_Estimation_Q_Q_2_5_A}
\end{eqnarray}

\noindent 
which is the asserted relation (\ref{upper_limits_Q_Q}). In order to show the upper limit in (\ref{Appendix_Estimation_Q_Q_2_5_A}) it is necessary to take 
into account that $\ve{\sigma} \cdot \ve{d}_{\sigma} = 0$ which restricts the possible values for the angles $\delta\left(\ve{\sigma},\ve{e}_3\right)$ and 
$\delta\left(\ve{d}_{\sigma},\ve{e}_3\right)$ (see also endnote $\left[99\right]$). 

The upper limit (\ref{Appendix_Estimation_Q_Q_2_5_A}) is valid for any astrometric configuration, that means
for $\ve{\sigma} \cdot \ve{x}_0 \le 0$ as well as for $\ve{\sigma} \cdot \ve{x}_0 > 0$. If one considers
astrometric configurations with condition (\ref{appendix_configuration1}) then one obtains the following upper limit:
\begin{eqnarray}
       \left| \ve{\sigma} \times \ve{\nu}^{{\rm Q} \times {\rm Q}}_{\rm 2PN}\right| &\le& 
	\frac{15}{2}\,\pi\,\frac{G^2 M^2}{c^4}\,\left|J_2\right|^2\,\frac{P^2}{\left(d_{\sigma}\right)^6}\,, 
        \label{Appendix_Estimation_Q_Q_2_5}
\end{eqnarray}

\noindent
which is the asserted relation (\ref{upper_limits_Q_Q_configuration1}).

\section{The expressions for the enhanced terms}\label{Appendix_Enhanced_Terms}

The expression of $\ve{n}_{\rm 1PN}\left(\ve{x}_{\rm N}\right)$ follows by inserting (\ref{First_Integration_1PN_M_Q}) - (\ref{First_Integration_1PN_Q}) 
into (\ref{Appendix_n_1PN_Enhanced}) and is given by  
\begin{eqnarray}
        n^i_{\rm 1PN}\left(\ve{x}_{\rm N}\right) &=&
        - 2\,\frac{G M}{c^2}\,\frac{d^i_{\sigma}}{\left(d_{\sigma}\right)^2} \left(1 + \frac{\ve{\sigma} \cdot \ve{x}_{\rm N}}{x_{\rm N}}\right)
        + 2\,\frac{G \hat{M}_{ab}}{c^2}\,\frac{\sigma^a \sigma^b \sigma^i - \sigma^a \delta^{bi}}{\left(x_{\rm N}\right)^3}
        + 6\,\frac{G \hat{M}_{ab}}{c^2}\,\frac{\sigma^a d_{\sigma}^b d_{\sigma}^i}{\left(x_{\rm N}\right)^5}
        \nonumber\\
        &&     + 8\,\frac{G \hat{M}_{ab}}{c^2}\,
        \frac{\left(d_{\sigma}\right)^2 \sigma^a \sigma^b d_{\sigma}^i
        - d_{\sigma}^a d_{\sigma}^b d_{\sigma}^i}{\left(d_{\sigma}\right)^2}
        \left[\frac{1}{\left(d_{\sigma}\right)^4}\left(1 + \frac{\ve{\sigma} \cdot \ve{x}_{\rm N}}{x_{\rm N}}\right)
        + \frac{1}{2}\,\frac{1}{\left(d_{\sigma}\right)^2}
   \frac{\ve{\sigma} \cdot \ve{x}_{\rm N}}{\left(x_{\rm N}\right)^3} + \frac{3}{8}\,\frac{\ve{\sigma} \cdot \ve{x}_{\rm N}}{\left(x_{\rm N}\right)^5}\right]
        \nonumber\\
        && + 2\,\frac{G \hat{M}_{ab}}{c^2}\,
    \frac{2\,d_{\sigma}^a \delta^{bi} - 5\,\sigma^a \sigma^b d_{\sigma}^i - 2\,\sigma^a d_{\sigma}^b \sigma^i}{\left(d_{\sigma}\right)^2}
        \left[\frac{1}{\left(d_{\sigma}\right)^2}\left(1 + \frac{\ve{\sigma} \cdot \ve{x}_{\rm N}}{x_{\rm N}}\right) + \frac{1}{2}\,
        \frac{\ve{\sigma} \cdot \ve{x}_{\rm N}}{\left(x_{\rm N}\right)^3} \right] , 
        \label{Appendix_n_N}
\end{eqnarray}

\noindent 
where the impact vector is given by (\ref{impact_vector_1}), that means 
\begin{eqnarray}
	\ve {d}_{\sigma} &=& \ve{\sigma} \times \left(\ve{x}_{\rm N} \times \ve{\sigma}\right) \quad {\rm and} \quad 
	d_{\sigma} = \left| \ve{d}_{\sigma} \right| .  
        \label{Appendix_impact_vector_x0}
\end{eqnarray}

\noindent
The expression $\ve{n}_{\rm 1PN}\left(\ve{x}_{\rm 1PN}\right)$ in (\ref{vector_n_2PN_B}) is obtained from (\ref{Appendix_n_N}) by replacing the argument 
$\ve{x}_{\rm N}$ by $\ve{x}_{\rm 1PN}$ everywhere. That means 
\begin{eqnarray}
        n^i_{\rm 1PN}\left(\ve{x}_{\rm 1PN}\right) &=&
        - 2\,\frac{G M}{c^2}\,\frac{\hat{d}^i_{\sigma}}{\left(\hat{d}_{\sigma}\right)^2} \left(1 + \frac{\ve{\sigma} \cdot \ve{x}_{\rm 1PN}}{x_{\rm 1PN}}\right)
        + 2\,\frac{G \hat{M}_{ab}}{c^2}\,\frac{\sigma^a \sigma^b \sigma^i - \sigma^a \delta^{bi}}{\left(x_{\rm 1PN}\right)^3}
        + 6\,\frac{G \hat{M}_{ab}}{c^2}\,\frac{\sigma^a \hat{d}_{\sigma}^b \hat{d}_{\sigma}^i}{\left(x_{\rm 1PN}\right)^5}
        \nonumber\\
        &&     + 8\,\frac{G \hat{M}_{ab}}{c^2}\,
        \frac{\left(\hat{d}_{\sigma}\right)^2 \sigma^a \sigma^b \hat{d}_{\sigma}^i
        - \hat{d}_{\sigma}^a \hat{d}_{\sigma}^b \hat{d}_{\sigma}^i}{\left(\hat{d}_{\sigma}\right)^2}
        \left[\frac{1}{\left(\hat{d}_{\sigma}\right)^4}\left(1 + \frac{\ve{\sigma} \cdot \ve{x}_{\rm 1PN}}{x_{\rm 1PN}}\right)
        + \frac{1}{2}\,\frac{1}{\left(\hat{d}_{\sigma}\right)^2}
   \frac{\ve{\sigma} \cdot \ve{x}_{\rm 1PN}}{\left(x_{\rm 1PN}\right)^3} + \frac{3}{8}\,\frac{\ve{\sigma} \cdot \ve{x}_{\rm 1PN}}{\left(x_{\rm 1PN}\right)^5}\right]
        \nonumber\\
        && + 2\,\frac{G \hat{M}_{ab}}{c^2}\,
    \frac{2\,\hat{d}_{\sigma}^a \delta^{bi} - 5\,\sigma^a \sigma^b \hat{d}_{\sigma}^i - 2\,\sigma^a \hat{d}_{\sigma}^b \sigma^i}{\left(\hat{d}_{\sigma}\right)^2}
	\left[\frac{1}{\left(\hat{d}_{\sigma}\right)^2}\left(1 + \frac{\ve{\sigma} \cdot \ve{x}_{\rm 1PN}}{x_{\rm 1PN}}\right) + \frac{1}{2}\,
	\frac{\ve{\sigma} \cdot \ve{x}_{\rm 1PN}}{\left(x_{\rm 1PN}\right)^3} \right] , 
        \label{Appendix_n_1PN}
\end{eqnarray}

\noindent
where the new impact vector 
\begin{eqnarray}
\hat{{\ve d}_{\sigma}} &=& \ve{\sigma} \times \left(\ve{x}_{\rm 1PN} \times \ve{\sigma}\right)  
        \label{Appendix_impact_vector_x1}
\end{eqnarray}

\noindent 
has been introduced; note that $\ve{\sigma} \cdot \hat{\ve{d}_{\sigma}} = 0$ and its absolute value $ \hat{d}_{\sigma} = \left| \hat{{\ve d}_{\sigma}} \right|$.  
In view of Eq.~(\ref{Second_Integration_1PN}) the new impact vector (\ref{Appendix_impact_vector_x1}) is determined by  
\begin{eqnarray}
	\hat{{\ve d}_{\sigma}} &=& \ve{d}_{\sigma} + \ve{\sigma} \times \left(\Delta \ve{x}_{\rm 1PN} \times \ve{\sigma}\right) 
        \label{Appendix_x_D}
\end{eqnarray}

\noindent
where $\Delta \ve{x}_{\rm 1PN}$ is given by Eq.(\ref{Second_Integration_1PN_M_Q}) with (\ref{Second_Integration_1PN_M}) and (\ref{Second_Integration_1PN_Q}).  
For the absolute value of (\ref{Appendix_x_D}) and the inverse powers one obtains in 1PN approximation  
\begin{eqnarray}
        \hat{d_{\sigma}} &=& d_{\sigma} + \frac{\ve{d}_{\sigma} \cdot \Delta \ve{x}_{\rm 1PN}}{d_{\sigma}}\;, 
        \label{Appendix_x_E}
        \\
        \frac{1}{\left(\hat{d}_{\sigma}\right)^n} &=& \frac{1}{\left(d_{\sigma}\right)^n} - \frac{n}{\left(d_{\sigma}\right)^n} \,
        \frac{\ve{d}_{\sigma} \cdot \Delta \ve{x}_{\rm 1PN}}{\left(d_{\sigma}\right)^2}\;.  
        \label{Appendix_x_F}
\end{eqnarray}

\noindent 
The relation (\ref{Appendix_x_F}) is actually the first term of a series expansion where the higher order terms are beyond 1PN approximation. This series 
expansion has a convergence radius determined by the condition $n\,\left| \ve{d}_{\sigma} \cdot \Delta \ve{x}_{\rm 1PN} \right|  \le \left(d_{\sigma}\right)^2$. 
Inserting the monopole term one finds $4\,n\,m\,x_1 < \left(d_{\sigma}\right)^2$; note that the inverse powers $2 \le n \le 6$ according to 
Eq.~(\ref{Appendix_n_1PN}). Using the extreme case of Jupiter or even the Sun one finds that this condition 
is well satisfied for any observer located in the Solar System, and is spoiled for observers if they would have distances of more than several hundred 
${\rm au}$ from the massive Solar System body. 

In order to determine the expression $\Delta \ve{n}_{\rm 2PN}$ in Eq.~(\ref{Delta_n_2PN}), 
\begin{eqnarray}
        \Delta \ve{n}_{\rm 2PN}\left(\ve{x}_{\rm N}\right) = \ve{n}_{\rm 1PN}\left(\ve{x}_{\rm N}\right) - \ve{n}_{\rm 1PN}\left(\ve{x}_{\rm 1PN}\right) 
        + {\cal O}\left(c^{-6}\right),  
        \label{Appendix_Delta_n_2PN}
\end{eqnarray}

\noindent
where $\ve{n}_{\rm 1PN}\left(\ve{x}_{\rm N}\right)$ and $\ve{n}_{\rm 1PN}\left(\ve{x}_{\rm 1PN}\right)$ are given by Eqs.~(\ref{Appendix_n_N}) and 
(\ref{Appendix_n_1PN}), respectively, one has to perform a series expansion of $\ve{n}_{\rm 1PN}\left(\ve{x}_{\rm 1PN}\right)$ up to terms of the 
order ${\cal O}\left(c^{-6}\right)$ to be consistent with the 2PN approximation. In this way, by means of relations (\ref{Appendix_x_A}) - (\ref{Appendix_x_C}) 
as well as (\ref{Appendix_x_D}) - (\ref{Appendix_x_F}), one obtains 
\begin{eqnarray}
        \Delta \ve{n}_{\rm 2PN} &=& \Delta \ve{n}_{\rm 2PN}^{{\rm M} \times {\rm M}}
        + \Delta \ve{n}_{\rm 2PN}^{{\rm M} \times {\rm Q}} + \Delta \ve{n}_{\rm 2PN}^{{\rm Q} \times {\rm Q}} \;,
	\label{Appendix_Enhanced_Terms_5}
\end{eqnarray}

\noindent
with
\begin{eqnarray} 
        \Delta \ve{n}_{\rm 2PN}^{{\rm M} \times {\rm M}}\left(\ve{x}_{\rm N}\right) &=&
         4\,\frac{G^2 M^2}{c^4}\,\frac{1}{\left(d_{\sigma}\right)^2}
	 \left(1 + \frac{\ve{\sigma} \cdot \ve{x}_{\rm N}}{x_{\rm N}}\right)
	 \frac{x_{\rm N} + \ve{\sigma} \cdot \ve{x}_{\rm N} - x_0 - \ve{\sigma} \cdot \ve{x}_0}{d_{\sigma}}\;, 
	\label{Appendix_Enhanced_Terms_5_MM}
\end{eqnarray} 

\noindent
and
\begin{eqnarray}
         \Delta \ve{n}_{\rm 2PN}^{{\rm M} \times {\rm Q}}\left(\ve{x}_{\rm N}\right)  &=&
         16\,\frac{G^2 M^2}{c^4}\,J_2\,\frac{P^2}{\left(d_{\sigma}\right)^4}
         \left(1 + \frac{\ve{\sigma} \cdot \ve{x}_{\rm N}}{x_{\rm N}}\right)
          \frac{x_{\rm N} + \ve{\sigma} \cdot \ve{x}_{\rm N} - x_0 - \ve{\sigma} \cdot \ve{x}_0}{d_{\sigma}}
         \nonumber\\
         && \times
        \Bigg[ \bigg(1 -  \left(\ve{\sigma} \cdot \ve{e}_3\right)^2 - 3\,\left(\frac{\ve{d}_{\sigma} \cdot \ve{e}_3}{d_{\sigma}}\right)^2 \bigg)
        \frac{\ve{d}_{\sigma}}{d_{\sigma}}
        + \frac{\ve{d}_{\sigma} \cdot \ve{e}_3}{d_{\sigma}} \; \ve{e}_3
        \Bigg] ,
	\label{Appendix_Enhanced_Terms_5_MQ}
\end{eqnarray}

\noindent
and
\begin{eqnarray}
         \Delta \ve{n}_{\rm 2PN}^{{\rm Q} \times {\rm Q}}\left(\ve{x}_{\rm N}\right) &=&
         12 \,\frac{G^2 M^2}{c^4} \,\left(J_2\right)^2\,\frac{P^4}{\left(d_{\sigma}\right)^6}
	\left(1 + \frac{\ve{\sigma} \cdot \ve{x}_{\rm N}}{x_{\rm N}}\right)
         \frac{x_{\rm N} + \ve{\sigma} \cdot \ve{x}_{\rm N} - x_0 - \ve{\sigma} \cdot \ve{x}_0}{d_{\sigma}}
        \left(1 - \left(\ve{\sigma} \cdot \ve{e}_3 \right)^2\right)^2 \,\frac{\ve{d}_{\sigma}}{d_{\sigma}}  \;,
	\label{Appendix_Enhanced_Terms_5_QQ}
\end{eqnarray}

\noindent 
where in (\ref{Appendix_Enhanced_Terms_5_MQ}) the quadrupole tensor (\ref{Quadrupole_Tensor_in_z_axis}) and in (\ref{Appendix_Enhanced_Terms_5_QQ}) the product 
of two quadrupole tensors (\ref{Quadrupole_Tensor_in_z_axis_Product}) have been inserted. 
In (\ref{Appendix_Enhanced_Terms_5_MM}) - (\ref{Appendix_Enhanced_Terms_5_QQ}) all terms which are proportional to $\frac{d_{\sigma}}{x_0}$ and 
$\frac{d_{\sigma}}{x_{\rm N}}$ have been omitted, because they contribute less than $1\,{\rm nas}$ even for grazing light rays at any massive Solar System body 
and can be neglected even for high-precision astrometry on the sub-\muas$\,$ and nas-level of accuracy. The calculation of (\ref{Appendix_Enhanced_Terms_5}) has 
revealed that the {\it enhanced 2PN terms} originate  from relations (\ref{Appendix_x_D}) - (\ref{Appendix_x_F}), that means they are solely caused by the 
new impact vector $\hat{\ve{d}_{\sigma}}$ in (\ref{Appendix_impact_vector_x1}).  

In (\ref{Appendix_Enhanced_Terms_5_MM}) - (\ref{Appendix_Enhanced_Terms_5_QQ}) one can replace the unperturbed position of the light signal 
at time of reception, $\ve{x}_{\rm N}\left(t_1\right)$, by the exact position of the observer, $\ve{x}_1$, in view of (\ref{Replacement_1PN}), 
that means such a replacement in (\ref{Appendix_Enhanced_Terms_5_MM}) - (\ref{Appendix_Enhanced_Terms_5_QQ}) would cause an error of the 
order ${\cal O}\left(c^{-6}\right)$ which is in line with the 2PN approximation. 
In this way one finally arrives at the expression as given by Eqs.~(\ref{Enhanced_Terms_5_MM}) - (\ref{Enhanced_Terms_5_QQ}). 
By evaluating the absolute value of the light deflection $\left|\ve{\sigma} \times \ve{n}\right|$ one finds the upper limits of 
these {\it enhanced terms} as given by Eqs.~(\ref{Enhanced_Terms_upper_limits_M_M}) - (\ref{Enhanced_Terms_upper_limits_Q_Q}).

\section{Proof of relations (\ref{upper_limit_n2_MM}) - (\ref{upper_limit_n2_QQ}) \label{Proof}} 

Finally we will proof the upper limits given in relations (\ref{upper_limit_n2_MM}) - (\ref{upper_limit_n2_QQ}). The expression for $\ve{n}_{\rm 2PN}$ is 
formally given by Eq.~(\ref{Appendix_n_2PN_Enhanced}), and the light deflection of this term reads 
\begin{eqnarray}
	\ve{\sigma} \times \ve{n}_{\rm 2PN} &=& \ve{\sigma} \times \frac{\Delta \dot{\ve x}_{\rm 2PN}}{c}  
	 - \left(\frac{\ve{\sigma} \cdot \Delta \dot{\ve x}_{\rm 1PN}}{c}\right)   
	\ve{\sigma} \times \frac{\Delta \dot{\ve x}_{\rm 1PN}}{c} \;. 
	\label{Appendix_Proof_n2_A}
\end{eqnarray}

\noindent
By inspection of Eqs.~(\ref{First_Integration_2PN_M_Q}) and (\ref{First_Integration_2PN_MM_MQ_QQ}) one obtains three terms 
\begin{eqnarray}
	\ve{\sigma} \times \ve{n}_{\rm 2PN} &=& \ve{\sigma} \times \ve{n}^{{\rm M} \times {\rm M}}_{\rm 2PN} 
	+ \ve{\sigma} \times \ve{n}^{{\rm M} \times {\rm Q}}_{\rm 2PN} + \ve{\sigma} \times \ve{n}^{{\rm Q} \times {\rm Q}}_{\rm 2PN}\;,  
	\label{Appendix_Proof_n2_B1} 
\end{eqnarray}

\noindent
with  
\begin{eqnarray}
	\ve{\sigma} \times \ve{n}^{{\rm M} \times {\rm M}}_{\rm 2PN} &=& 
	\ve{\sigma} \times \frac{\Delta \dot{\ve x}^{{\rm M} \times {\rm M}}_{\rm 2PN}}{c}  
	- \bigg(\!\frac{\ve{\sigma} \cdot \Delta \dot{\ve x}^{\rm M}_{\rm 1PN}}{c}\!\bigg)  
	\ve{\sigma} \times \frac{\Delta \dot{\ve x}^{\rm M}_{\rm 1PN}}{c}\,, 
        \label{Appendix_Proof_n2_MM}
	\\
	\ve{\sigma} \times \ve{n}^{{\rm M} \times {\rm Q}}_{\rm 2PN} &=& 
	\ve{\sigma} \times \frac{\Delta \dot{\ve x}^{{\rm M} \times {\rm Q}}_{\rm 2PN}}{c} 
	- 2 \bigg(\!\frac{\ve{\sigma} \cdot \Delta \dot{\ve x}^{\rm Q}_{\rm 1PN}}{c}\!\bigg) 
	\ve{\sigma} \times \frac{\Delta \dot{\ve x}^{\rm M}_{\rm 1PN}}{c} , 
	\label{Appendix_Proof_n2_MQ}
	\\
	\ve{\sigma} \times \ve{n}^{{\rm Q} \times {\rm Q}}_{\rm 2PN} &=& 
	\ve{\sigma} \times \frac{\Delta \dot{\ve x}^{{\rm Q} \times {\rm Q}}_{\rm 2PN}}{c}  
	- \bigg(\!\frac{\ve{\sigma} \cdot \Delta \dot{\ve x}^{\rm Q}_{\rm 1PN}}{c}\!\bigg)  
	\ve{\sigma} \times \frac{\Delta \dot{\ve x}^{\rm Q}_{\rm 1PN}}{c}\,, 
        \label{Appendix_Proof_n2_QQ}
\end{eqnarray}

\noindent
where the argument $\ve{x}_{\rm N}\left(t_1\right)$ can be replaced by the exact spatial position of the observer $\ve{x}_1$ in view of (\ref{Replacement_1PN}).  
The absolute value of (\ref{Appendix_Proof_n2_B1}) can be estimated by 
\begin{eqnarray}
	\left| \ve{\sigma} \times \ve{n}_{\rm 2PN} \right| &\le& \left| \ve{\sigma} \times \ve{n}^{{\rm M} \times {\rm M}}_{\rm 2PN}\right| 
        + \left| \ve{\sigma} \times \ve{n}^{{\rm M} \times {\rm Q}}_{\rm 2PN} \right| 
	+ \left| \ve{\sigma} \times \ve{n}^{{\rm Q} \times {\rm Q}}_{\rm 2PN}\right|\;.
        \label{Appendix_Proof_n2_B2}
\end{eqnarray}

\noindent 
By inserting Eqs.~(\ref{First_Integration_1PN_M}) - (\ref{First_Integration_1PN_Q}) and 
Eqs.~(\ref{First_Integration_2PN_Term_M_M}) - (\ref{First_Integration_2PN_Term_Q_Q}) into (\ref{Appendix_Proof_n2_MM}) - (\ref{Appendix_Proof_n2_QQ}) one might 
determine these upper limits in (\ref{Appendix_Proof_n2_B2}). Such an ambitious calculation is, however, not necessary, because of the following relations: 
\begin{eqnarray}
	\left| \ve{\sigma} \times \ve{n}^{{\rm M} \times {\rm M}}_{\rm 2PN} \right| &\le& 
	\left| \ve{\sigma} \times \ve{\nu}^{{\rm M} \times {\rm M}}_{\rm 2PN}\right|
	\le \frac{G^2 M^2}{c^4} \left| \frac{15}{4} \frac{\pi}{\left(d_{\sigma}\right)^2} 
        - 8 \frac{x_0 + \ve{\sigma} \cdot \ve{x}_0}{\left(d_{\sigma}\right)^3} \right|,   
        \label{Appendix_Proof_n2_D}
\end{eqnarray} 
\begin{eqnarray}
        \left| \ve{\sigma} \times \ve{n}^{{\rm M} \times {\rm Q}}_{\rm 2PN}\right| 
	&\le& \left| \ve{\sigma} \times \ve{\nu}^{{\rm M} \times {\rm Q}}_{\rm 2PN}\right|
	\le 4\,\frac{G^2 M^2}{c^4}\,\left|J_2\right|\,\frac{P^2}{\left(d_{\sigma}\right)^2}\,
        \left| \frac{15}{4}\, \frac{\pi}{\left(d_{\sigma}\right)^2} + 8\,\frac{x_0 + \ve{\sigma} \cdot \ve{x}_0}{\left(d_{\sigma}\right)^3} \right|, 
        \label{Appendix_Proof_n2_F}
\end{eqnarray} 
\begin{eqnarray}
        \left| \ve{\sigma} \times \ve{n}^{{\rm Q} \times {\rm Q}}_{\rm 2PN}\right| 
	&\le& \left| \ve{\sigma} \times \ve{\nu}^{{\rm Q} \times {\rm Q}}_{\rm 2PN}\right| 
	\le 2\,\frac{G^2 M^2}{c^4}\,\left|J_2\right|^2\,\frac{P^4}{\left(d_{\sigma}\right)^4}\,
        \left| \frac{15}{4}\, \frac{\pi}{\left(d_{\sigma}\right)^2} + 12\,\frac{x_0 + \ve{\sigma} \cdot \ve{x}_0}{\left(d_{\sigma}\right)^3} \right|,  
        \label{Appendix_Proof_n2_G}
\end{eqnarray}

\noindent
where we have used the upper limits of the 2PN terms in the total light deflection as given by relations (\ref{upper_limits_M_M}) - (\ref{upper_limits_Q_Q}). 
In view of relation (\ref{Relation_20}), these expressions coincide with the asserted relations in (\ref{upper_limit_n2_MM}) - (\ref{upper_limit_n2_QQ}).


\end{document}